\documentclass[traditabstract]{aa}
\usepackage{graphicx}
\usepackage{enumerate}
\usepackage{epsfig}
\usepackage{color}
\usepackage{txfonts}
\usepackage{multirow}
\usepackage{float}
\usepackage[export]{adjustbox}
\usepackage{amssymb}
\usepackage{scalefnt}

\newcommand\gaia{\textit{Gaia}}

\def\kms   {$\rm km\,s^{-1}$}
\def\para  {$\pi$}

\begin{document}

\title{Reviving old controversies: is the early Galaxy flat or round?}
\subtitle{Investigations into the early phases of the Milky Way's formation through stellar kinematics and chemical abundances\thanks{Based on observations collected  at the
European Organisation for Astronomical Research in the Southern
Hemisphere under ESO programmes 165.N-0276(A) (P.I.: R. Cayrel). }
}
\titlerunning{Reviving old controversies: is the early Galaxy flat or round?}

\author{
P.~Di Matteo\inst{1, 2},
M.~Spite\inst{1},
M.~Haywood\inst{1, 2},
P.~Bonifacio\inst{1},
A.~G\'omez\inst{1}, 
F.~Spite\inst{1},
E.~Caffau\inst{1}
}
 \authorrunning{P. Di Matteo et al.}
 
\institute {
GEPI, Observatoire de Paris, PSL Research University, CNRS, Place Jules Janssen, 92190 Meudon, France
\email{paola.dimatteo@obspm.fr}
\and Sorbonne Universit\'e, CNRS UMR 7095, Institut d'Astrophysique de Paris, 98bis bd Arago, 75014 Paris, France
}

\date{Received; accepted}

\abstract{We analysed a set of very metal-poor stars, for which accurate chemical abundances have been obtained
as part of the ESO Large Program 'First stars' in the light of the \gaia~DR2 data. The kinematics and orbital 
properties of the stars in the sample show they probably belong to the thick disc, partially heated to halo kinematics, and to the accreted 
Gaia Sausage-Enceladus satellite. 
The continuity of these properties with stars at both higher ($\rm [Fe/H]>-2$) and lower 
metallicities ($\rm [Fe/H]<-4.$) suggests that the Galaxy at $\rm [Fe/H] \lesssim -0.5$ and down to at least $\rm [Fe/H]\sim-6$ is dominated 
by these two populations. In particular, we show that the disc extends continuously from  $\rm [Fe/H] \le -4$ (where stars with disc-like kinematics have recently been discovered) up to  $\rm [Fe/H] \ge -2$, the metallicity regime of the Galactic  thick disc. An `ultra metal-poor thick disc' does indeed exist, constituting the extremely metal-poor tail of the canonical Galactic thick disc, and extending the latter from $\rm [Fe/H] \sim -0.5$ up to the most metal-poor stars discovered in the Galaxy to date. 
These results suggest that the disc may be the main, and possibly the only, stellar population that has 
formed in the Galaxy at these metallicities. This would mean that the dissipative collapse that led to the formation 
of the old Galactic disc must have been extremely fast. We also discuss these results in the light of 
recent simulation efforts made to reproduce the first stages of Milky Way-type galaxies.}


\maketitle
%
\section{Introduction}

Our definitions and understanding of the Galactic halo, and, more generally, of the old Galactic stellar populations have been 
strongly shaken by the results obtained from the \gaia~data releases \citep{brown16,GaiaDR2-Brown}. Because we like to cling to good ideas (as much as we like good stories), 
the vision that we had of the Galactic halo before \gaia~was built on two studies published more than half a century ago: the articles of 
\citet{eggen62} and \citet{searle78}.
After having elaborated for a few decades on these two studies, the halo was standardly described only a few years ago as the combination
of collapsed and accreted components, with relative weights not easily quantifiable  \citep[see, for example, ][]{helmi08,  carollo07}. The in-situ component was believed to have formed
from the collapsing gas, eventually brought into the Galaxy via gas-rich mergers, while the accreted component was thought to be formed from a multitude (several tens) of stellar sub-haloes \citep{gao10, griffen18}. The possibility that solar vicinity halo stars were dominated by the presence of possibly one only satellite has already been raised by \citet{brook03}, based on the stellar kinematics of halo stars, and by \citet{nissen10, nissen12, schuster12}, on the basis of  their chemical abundances, kinematics, and ages. The analysis of the  
 \gaia~DR1 and DR2 has allowed us to extend our view beyond the solar vicinity, and to show that the Galaxy at low metallicity  ($-2.5 \lesssim \rm [Fe/H] \lesssim -1$)  is, most probably, 
dominated by the remnant of a single accretion event that occurred 9-11~Gyr ago \citep{belokurov18, haywood18, helmi18, dimatteo18, mackereth18, myeong18, gallart19} and that the in-situ part of the Galactic halo may be attributed to heated disc stars \citep{bonaca17,haywood18, dimatteo18, belokurov19, gallart19} rather than a collapsing halo \citep{haywood18, dimatteo18}, thus supporting the predictions of N-body models \citep{zolotov10, purcell10, font11, qu11a, mccarthy12, jeanbaptiste17}.
Even more surprising, in this context, has been the result, from \citet{sestito19}, that about 20\% of all ultra-metal-poor ($\rm [Fe/H]<-4$)\footnote{We adopt the classical notations that, for each element X,~~ $\rm [X/H]=log(N_{X}/N_{H})_{star}-log(N_{X}/N_{H})_{Sun}$ ~~and~~~ [X/Fe]=[X/H]--[Fe/H].} stars 
known are on thin or thick disc orbits.
Stars at $\rm [Fe/H]<-4$ or even $\rm [Fe/H]<-6$ are expected to form  a few hundreds Myr after the Big Bang, raising two fundamental questions:
was the gaseous disc already the main, and possibly the only, structure of the Milky Way to form stars at these early epochs?; and, how could the fossil signatures of such kinematically cold stars have not been erased by the passage of time?
\\

The assembly sequence of our Galactic halo is encoded in spectra of its surviving low-metallicity stars, given that their elemental abundance ratios reflect the successive nucleosynthesis processes
and the nature of the stars creating them. 
In the frame of the ESO Large Program `First stars - First nucleosynthesis' (hereafter LP First stars) a sample of very metal-poor field stars,  giants and turnoff stars, with $\rm -4.2<[Fe/H]<-2$, was studied homogeneously from high-resolution and high S/N   spectra. 
Since these stars are very or extremely metal poor, it is supposed that they are very old and formed shortly after the Big Bang.\\
For many of these stars,
 the main astrometric parameters -- precise position, proper motion and parallax -- are now available in the second data release of the \gaia~ mission \citep[][]{GaiaDR2-Brown}.
The aim of the present paper is therefore to make a step toward a better understanding of the very metal-poor populations in the Galaxy and of their origin, by comparing the kinematics of the  LP First stars to the kinematics of different samples of metal-poor, as well as more metal-rich, stars.\\

The outline of the paper is as follows: in the following section, we describe the characteristics of our sample. 
In Section 3, we analyse and discuss the kinematics and orbital properties of our stars. In Section 4, we undertake an extensive comparison of the kinematic properties of our sample with  the higher metallicity sample studied by \citet{nissen10}, and also with the \gaia~DR2-APOGEE sample studied by \citet{dimatteo18} on the one side, and with the lower metallicity sample studied by \citet{sestito19} on the other side. In Section 5, we analyse 
the chemical properties of stars in the LP First stars, and compare it to the sample of r-rich stars at similar metallicities studied by  \citet{roederer18}. Finally, in Sects.~\ref{discussion} and \ref{conclusions}, we discuss our results and derive our conclusions.

\begin{figure}
\begin{center}
\includegraphics[clip=true, trim={0cm 0cm 0cm 0cm},width=0.45\textwidth]{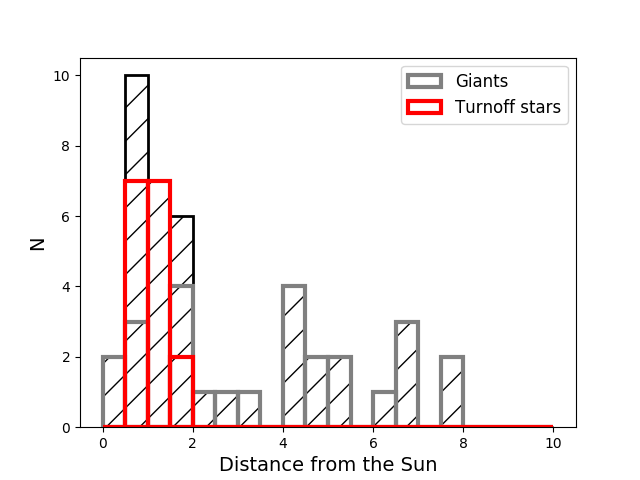}
\includegraphics[clip=true, trim={0cm 0cm 0cm 0cm},width=0.45\textwidth]{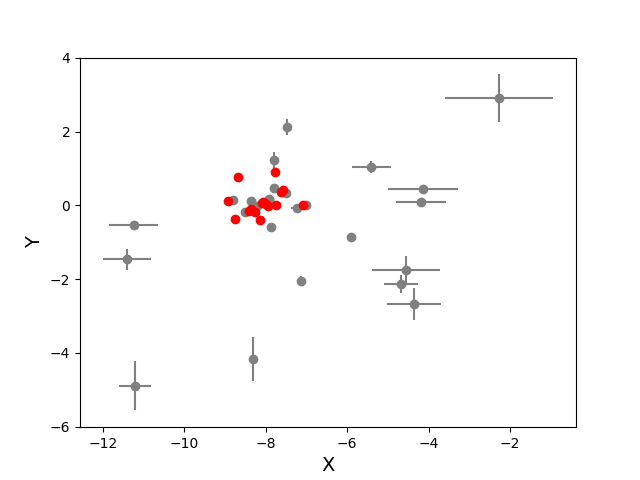}
\includegraphics[clip=true, trim={0cm 0cm 0cm 0cm},width=0.45\textwidth]{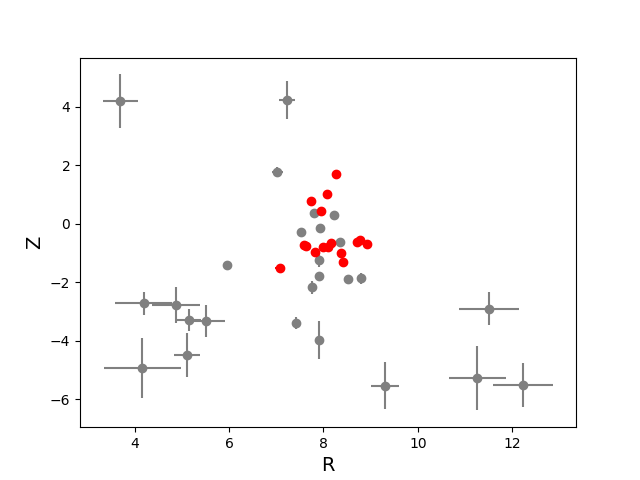}
\caption{\emph{Top  panel:} Histogram (hatched black) of the distances (in kpc) to the Sun, of the sample of  stars  studied in the frame of the ESO LP First Stars. The grey histogram shows the distance distribution of giant stars, the red histogram that of turnoff stars; \emph{Middle panel:} X-Y spatial distribution of stars in the sample (grey dots for giants, red dots for turnoff stars), and their uncertainties; \emph{Bottom panel:} R-Z spatial distribution of stars in the sample (grey dots for giants, red dots for turnoff stars) and their uncertainties. }
\label{histd}
\end{center}
\end{figure}

\begin{figure}
\begin{center}
\includegraphics[clip=true, trim={0cm 0cm 0cm 0cm},width=0.45\textwidth]{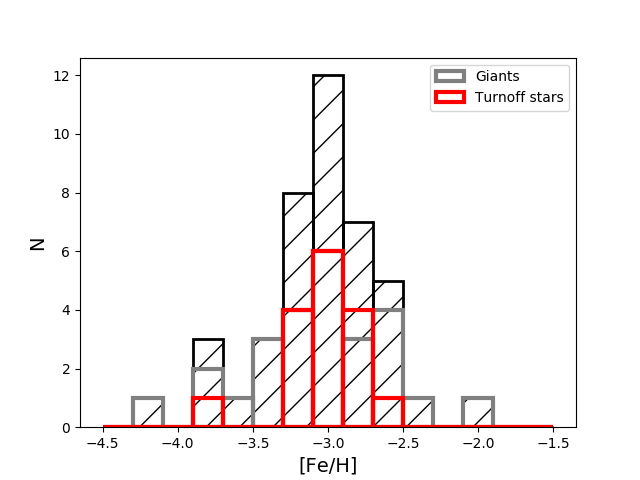}
\includegraphics[clip=true, trim={0cm 0cm 0cm 0cm},width=0.35\textwidth]{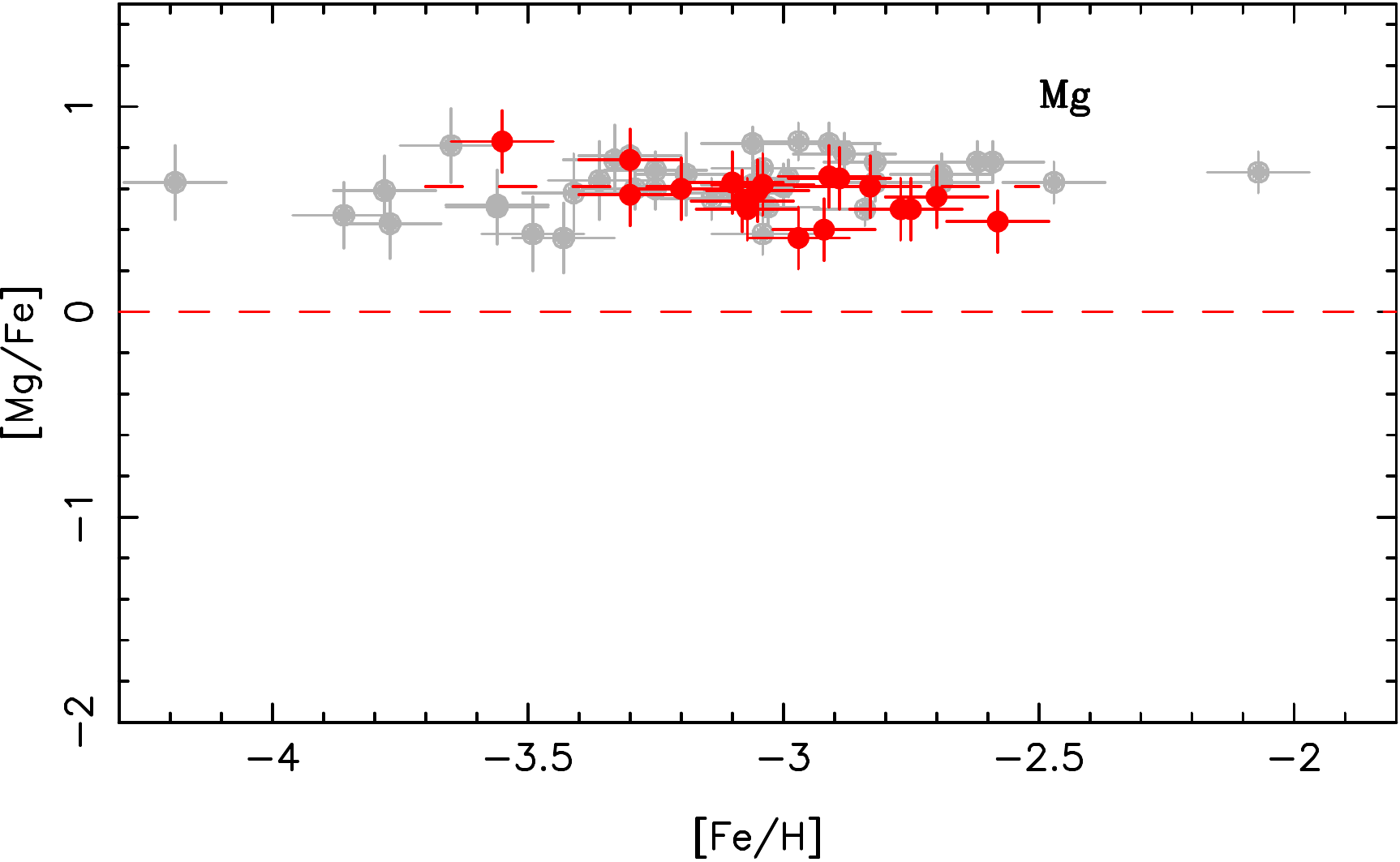}
\includegraphics[clip=true, trim={0cm 0cm 0cm 0cm},width=0.35\textwidth]{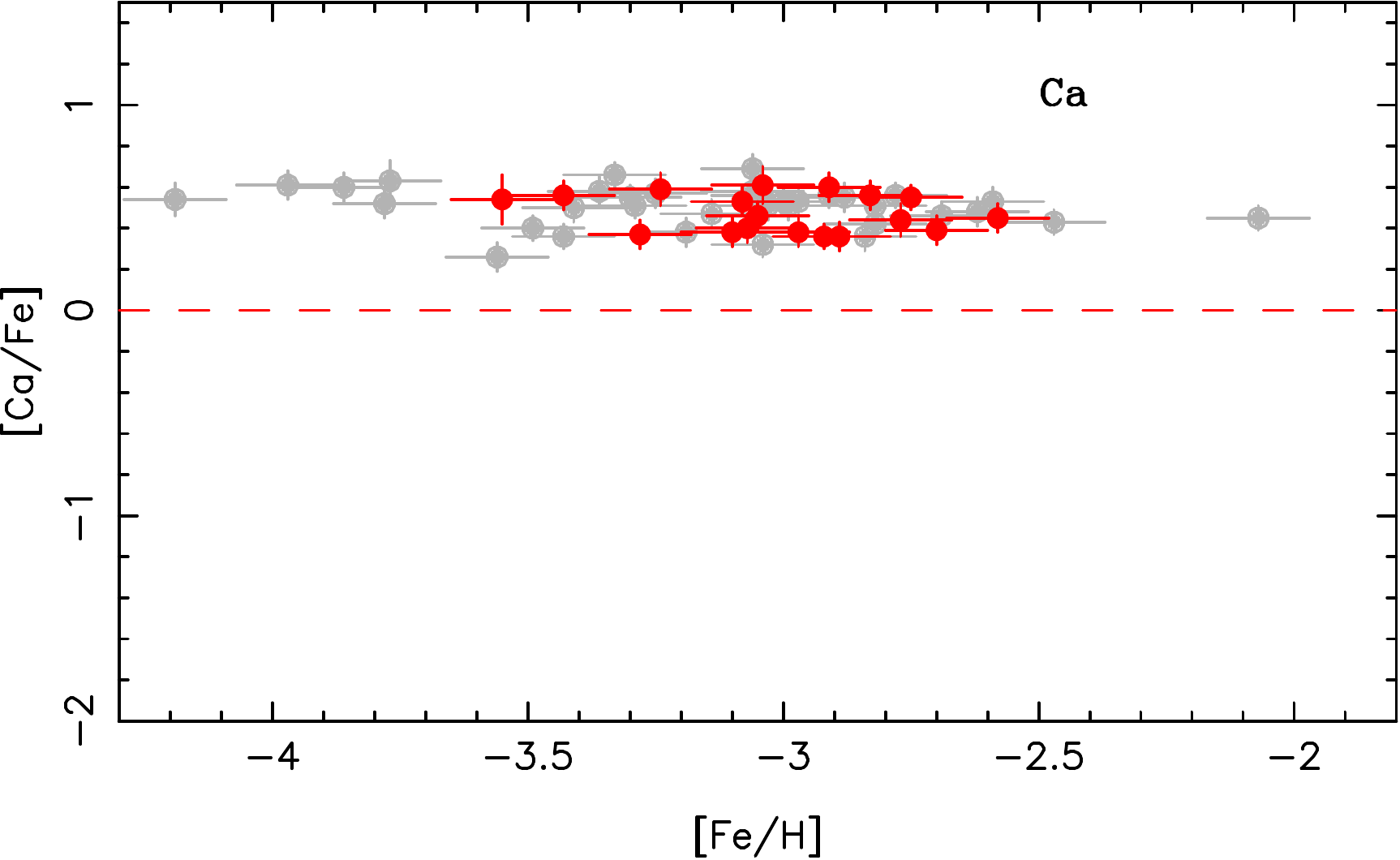}
\includegraphics[clip=true, trim={0cm 0cm 0cm 0cm},width=0.35\textwidth]{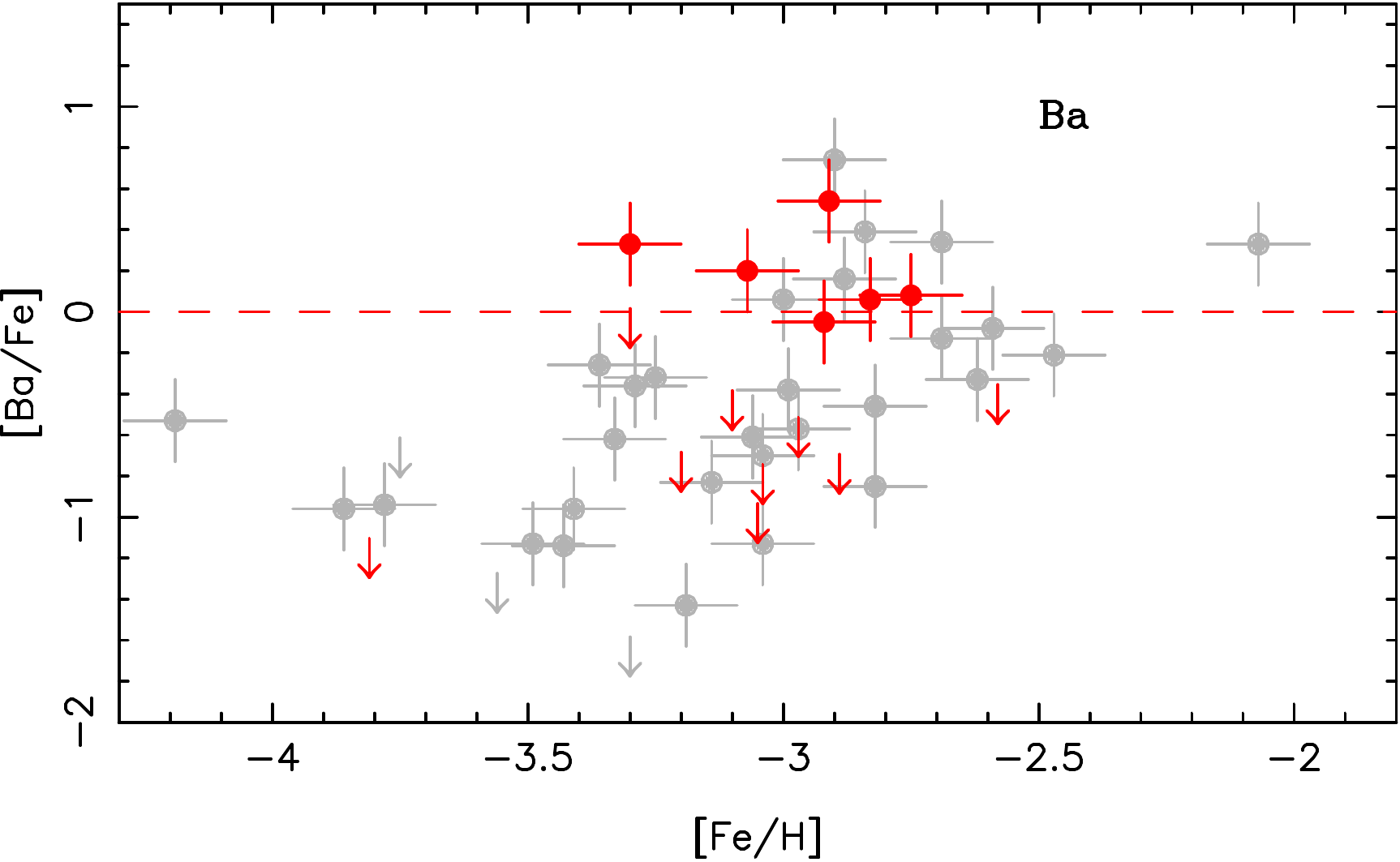}
\caption{\emph{From top to bottom:} [Fe/H] distribution (hatched black) of stars in sample. The grey histogram shows the [Fe/H] distribution of giant stars, the red histogram that of turnoff stars; $\rm [Mg/Fe]-[Fe/H]$ and $\rm [Ca/Fe]-[Fe/H]$ distributions of stars in the sample (grey dots for giants, red dots for turnoff stars); $\rm [Ba/Fe]-[Fe/H]$ distribution of stars in the sample (grey dots for giants, red dots for turnoff stars). Stars have only an upper limit estimate of the [Ba/Fe] ratio are indicated by an arrow.}
\label{histd2}
\end{center}
\end{figure}

\section{Data}\label{sec:data}

\subsection{LP First stars sample in the \gaia~ DR2 survey}

In the frame of the  LP First stars, a sample of 54 very metal-poor  field stars, giants, and turnoff stars was studied homogeneously from high-resolution ($R \approx 45\,000$), high-S/N  (S/N better than 200 per resolution element at 400\,nm) spectra. 
This sample of stars was selected from the  the HK survey  of \citet{beers85, BeersPS92} for their low metallicity after a medium resolution follow-up, without considering their kinematics. The metallicity of the stars was found to be in the range $\rm -4.2<[Fe/H]<-2$, with a peak at $\rm[Fe/H] \approx -3$. Following \citet{Beers-CosmicAb05}, these stars  are referred to as 'very metal-poor' (VMP)  if $\rm -3.0<[Fe/H]\leq-2.0$, or 'extremely metal-poor' (EMP) if $\rm-4.0<[Fe/H]\leq-3.0$. 
A complete LTE analysis using  OSMARCS models  \citep{GustafssonBE75,GustafssonEE03,GustafssonEE08} was carried out, based on the spectral analysis code {\tt turbospectrum} \citep{AlvarezP98}. The results of this analysis were published in a series of papers entitled First Stars\,I to First Stars\,XVI \citep[see, in particular: ][]{CayrelDS04,BonifacioSC09}. 

At low metallicities, many stars are carbon-enhanced compared to the normal EMP stars\footnote{The origin of this high C abundance is not always completely clear. It can be due to an enrichment of a normal EMP binary star by its more massive companion in its AGB phase, or the star could have been formed from a C-rich cloud. }. At $\rm [Fe/H]=-3$, for example, about 30\% of the stars are carbon-rich with $\rm [C/Fe] > +1.0$. They are referred to as `carbon-enhanced metal-poor' (CEMP) stars   \citep{BeersChristlieb05}, and their fraction increases when the metallicity decreases. A few CEMP stars (six in total) are also present among the LP First stars \citep[e.g.][]{DepagneHM02,SivaraniBM04,SivaraniBB06}. 
For the sake of homogeneity, we do not include these CEMP stars in our present sample. We did, however, verify that these CEMP stars  do not show any peculiarity in their kinematic properties when they are compared to the full LP First stars sample, as discussed in Appendix~\ref{CEMP}.

Most of the stars in the sample are too faint to have a line-of-sight velocity in the \gaia~DR2 \citep[][]{GaiaDR2-Brown, katz18}, but precise line-of-sight velocities were measured on each UVES spectrum of the stars and were published in \citet{BonifacioMS07,BonifacioSC09}. Taking into account the position of the telluric lines, the error is estimated to 0.5\,\kms for the turnoff stars and  0.3\,\kms~ for the giants. When they exist in \gaia~DR2, there is a good agreement, within the error bars, between the line-of-sight velocities measured by \gaia~and by UVES. 

Three turnoff stars have a variable line-of-sight velocity (CS 29499-060, CS 29527-015, CS 30339-069), and as a consequence, were not taken into account in this study.   
The characteristics of the remaining stars, angular position on the sky, line-of-sight velocity, proper motions, and parallaxes \para 
were extracted from the \gaia~DR2. We only kept  stars whose uncertainty on the parallax $\sigma_\pi/\pi$ was less than 20\%, after correcting by the zero-point offset of --0.03 \citep{ArenouLB18,GaiaDR2-Brown,LindegrenHB18}. This leaves us with a final sample of 42 stars (16 turnoff stars and 26 giants), whose characteristics are given in Appendix~\ref{tabdata} in Table \ref{gaia1}. For this final sample, distances were calculated by simply inverting the parallaxes.

In Fig.~\ref{histd}, we present the histogram of the distances from the Sun of the stars in the final sample, as well as their spatial coordinates. For a description of the method adopted to estimate distances, and to derive the spatial coordinates of the stars, as well as their uncertainties, we refer the reader to Sect.~\ref{orbitint}. The turnoff stars are, as expected, on average, closer to the Sun than the giants. The giants are almost regularly spaced between zero and 8~kpc. On the contrary, all the turnoff stars are concentrated between zero and only 2~kpc. The metallicity distribution of the final sample of stars (see Fig.~\ref{histd2}) peaks around $\rm [Fe/H]=-3.0$.\\

\subsection{Chemical properties of the LP First stars sample}\label{CP}

A surprise result of the spectroscopic study of the LP First stars sample (VMP and EMP stars) was the great homogeneity of their abundance ratios. From C to Zn, the abundance  ratios are very similar in all the stars, although the clouds from which they were formed have probably been enriched in metals by a very small number of massive SN\,II supernovae \citep[see][]{CayrelDS04,BonifacioSC09}.
In order to explain the abundance pattern of the elements, the mass of these supernovae should be  between 10 and 30 $M_{\odot}$  \citep{CayrelDS04}.
All these stars are, in particular, almost uniformly $\alpha$-rich \citep[see Fig.~\ref{histd2}, and][]{AndrievskySK10,SpiteAS12}, with a very small scatter after the non-LTE corrections.

This uniform $\alpha$~enhancement suggests that these stars were all formed from a matter only enriched by massive SN\,II supernovae, before the explosion of the first SN\,I ejected  lower $\rm N(\alpha)/N(Fe)$ ratio.  At higher metallicity ($\rm [Fe/H]> -1.6$), the ratio $\rm [\alpha/Fe]$ has often been used to distinguish different halo populations \citep{nissen10,hayes18}, but this does not currently seem possible in the VMP and EMP stars, since they all have about the same $\rm N(\alpha)/N(Fe)$ ratio (see Fig.~\ref{histd2}).

However, in these VMP and EMP stars, the abundance ratios of the elements heavier than Zn (hereafter 'heavy elements'), formed by neutron capture on iron seeds, are very variable (see in Fig.~\ref{histd2}, the large scatter of [Ba/Fe] vs. [Fe/H] and also \citet{GilroySP88,FrancoisDH07,SpiteSB18}). At the same metallicity, the ratio Ba/Fe varies by a factor of almost 100.
Since only the abundance ratios [X/Fe] of the heavy elements show significant differences from star to star, it can be interesting to study the kinematics of these VMP and EMP stars as a function of the abundance of these heavy elements.  Recently,  \citet{roederer18} studied the kinematic properties of a sample of only VMP and EMP stars rich in heavy elements, so it will be possible to compare our results with theirs.

\subsection{R-rich very metal-poor stars}
Elements heavier than Zn are formed by addition of neutrons on iron seeds, via various nucleosynthetic mechanisms, and predominantly via the r- and the s-processes. This addition may indeed be slow (compared to the $\beta$ decay) in the `s-process',  or very rapid, with an important flux of neutrons, in the `r-process'. 
The s-process mainly occurs in AGB stars, and it seems that the lifetime of their progenitors is too long to significantly contribute to the chemical evolution of the early Galaxy for  $\rm[Fe/H] \lesssim -1.5\,dex$ \citep{TravaglioGA04,KappelerGB11}. 
In the VMP and EMP stars, the heavy elements are thus mainly formed by the r-process. The different possible sites of the r-process (massive stars, neutron-stars, and neutron-star/black hole mergers) are reviewed by \citet{CowanSL19}.\\ 
A star is referred to as r-rich  when it is rich in europium relative to iron,  because the s-process builds very little Eu, and, as a consequence, the abundance of Eu is a good index of the r-process enrichment.  Following  \citet{roederer18}, a star is referred to as r-rich when  [Eu/Fe] > 0.7 .

Unfortunately, in most of the turnoff metal-poor stars, the Eu lines are too weak to be measured, and the abundance of Eu cannot be directly estimated. However, we showed \citep{SpiteSB18} that, in all the `normal' VMP and EMP stars (i.e. not C rich), there is an excellent correlation between the abundance of Eu and the abundance of Ba, with $\rm[Eu/Ba]\approx +0.5\,dex$.\\
At higher metallicities ($\rm[Fe/H] \gtrsim -1.5\,dex$), the pattern of the heavy elements in the stars can be affected by the s-process, and the ratio [Ba/Fe] increases \citep[see for example ][]{RoedererKP16}.\\
In order to include the turnoff stars in this study, since all the stars have a metallicity lower than --2.0\,dex, we used the  [Ba/Fe] ratio as a proxy of [Eu/Fe].
 The values of [Fe/H] and [Ba/Fe] are given for each star of the LP First stars sample in the Table \ref{gaia1}.

\section{Kinematic and orbital properties of the LP First stars}

\begin{figure*}[h!]
\begin{center}
\includegraphics[clip=true, trim={0cm 0cm 0cm 0cm},width=0.45\textwidth]{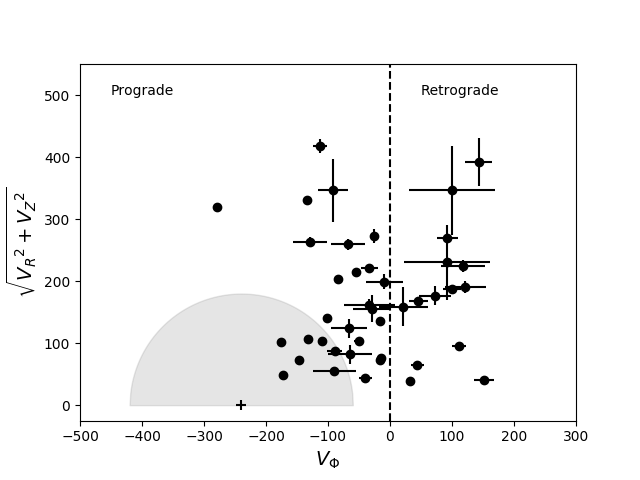}
\includegraphics[clip=true, trim={0cm 0cm 0cm 0cm},width=0.45\textwidth]{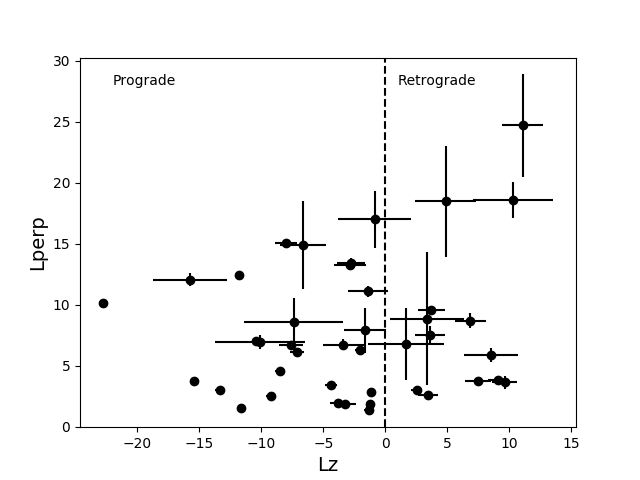}
\includegraphics[clip=true, trim={0cm 0cm 0cm 0cm},width=0.45\textwidth]{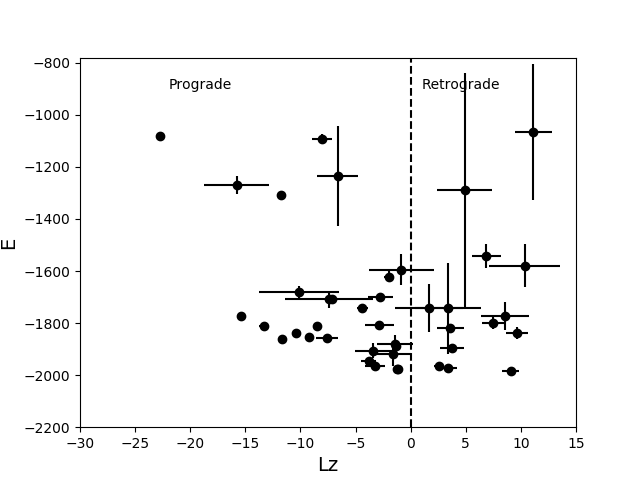}
\includegraphics[clip=true, trim={0cm 0cm 0cm 0cm},width=0.45\textwidth]{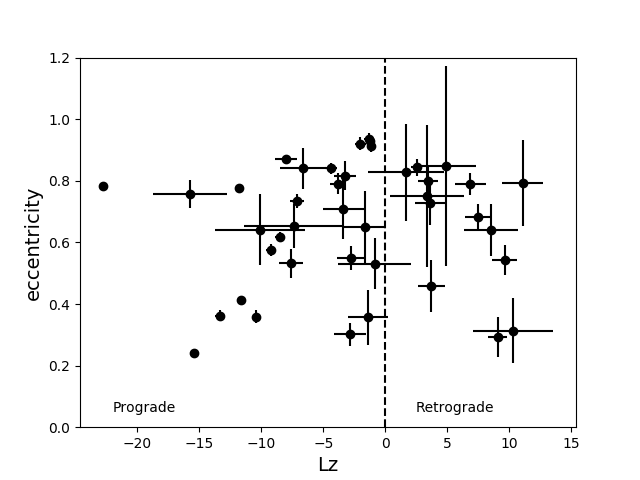}
\includegraphics[clip=true, trim={0cm 0cm 0cm 0cm},width=0.45\textwidth]{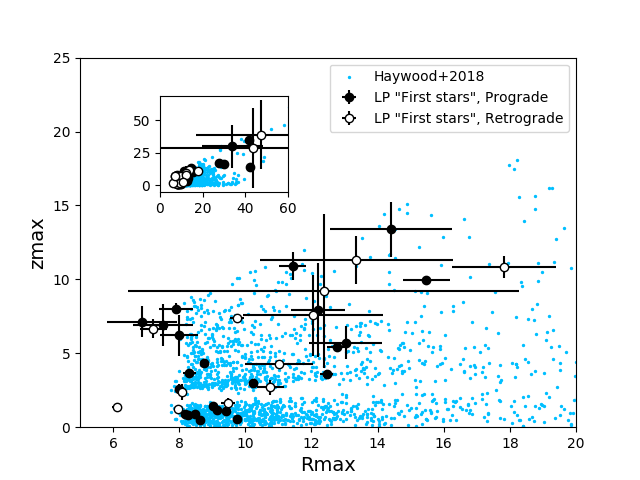}
\includegraphics[clip=true, trim={0cm 0cm 0cm 0cm},width=0.45\textwidth]{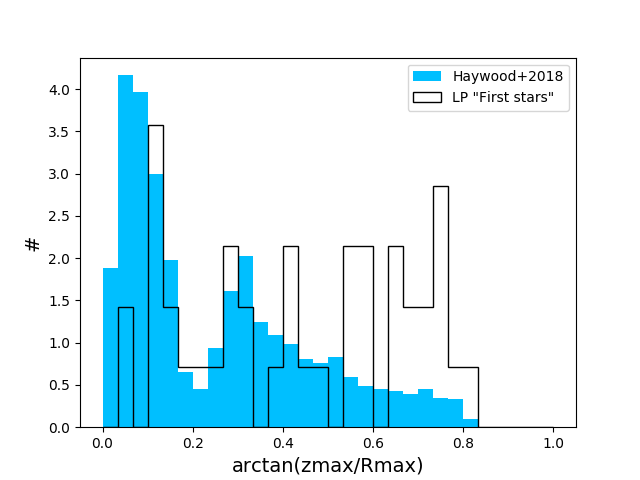}
\caption{\emph{Top-left panel:} Toomre diagram for stars in LP sample. The vertical dashed line separates prograde from retrograde motions. Velocities are in units of km/s. The grey  area separates stars with       $\sqrt{{V_R}^2+{(V_\Phi-V_{LSR})}^2+{V_Z}^2} \le 180$~km/s from stars with higher relative velocities with respect to the LSR. \emph{Top-right panel:} Distribution of the LP First stars in the $L_z-L_{perp}$ plane. Angular momenta are in units of 100~kpc~km/s. \emph{Middle-left panel:} Distribution of LP First stars in the $E-L_z$ plane. The  vertical dashed line separates prograde from retrograde motions. Angular momenta are in units of 100~kpc~km/s, energies in units of $\rm 100~km^2/s^2$. \emph{Middle-right panel}: Distribution of LP First stars in the eccentricity-$L_z$ plane.  \emph{Bottom-left panel}: Distribution of stars in the $z_{max}-R_{max}$ plane. The inset shows the whole distribution, over 60~kpc from the Galactic centre, while the main panel shows a zoom in the innermost 20~kpc. Filled symbols indicate stars with prograde orbits,  empty symbols stars with retrograde orbits. Blue points show the comparison with the sample of stars analysed by  \citet{haywood18}. Units are in kpc. \emph{Bottom-right panel}: Normalised distribution of the arctan($z_{max}/R_{max}$) for stars in the LP sample (black histogram), compared to stars studied by \citet{haywood18} (blue histogram). In all panels, uncertainties  have been estimated as described in Sect.~\ref{orbitint}. }
\label{kinorb_MP}
\end{center}
\end{figure*}

We start our analysis by deriving the orbits of stars in our sample, the associated parameters, and their uncertainties, to establish the kinematic properties of very metal-poor and extremely metal-poor stars.

\subsection{Orbit integration}\label{orbitint}

For calculating positions and velocities in the galactocentric rest frame, we assumed an in-plane distance of the Sun from the Galactic centre, $R_\odot$ = 8.34 kpc \citep{reid14}, a height of the Sun above the Galactic plane, $z_\odot$ = 27 pc \citep{chen01}, a velocity for the local standard of rest (LSR), $V_{LSR}$= 240 km/s \citep{reid14}, and a peculiar velocity of the Sun with respect to the LSR, $U_\odot$ = 11.1 km/s, $V_\odot$=12.24 km/s, $W_\odot$=7.25 km/s \citep{SchonrichBD10}.  
We note that in our choice of the galactocentric coordinate system, the Sun lies on the x-axis with a negative value of $x=-8.34$~kpc, and the $V_{\odot}$ is positive, that is parallel to the y axis. This implies that the disc rotates clockwise,
and, as a consequence, the z-component of the disc angular momentum, $L_z$, and the disc azimuthal velocity, $V_{\Phi}$ are negative\footnote{We remind the reader that the azimuthal velocity  $\rm V_{\Phi}$  is defined as the z-component of the angular momentum $L_z$ divided by the (in-plane) distance $R$ of the star from the Galactic centre, that is $\rm V_{\Phi}=(XV_Y-YV_Z)/R$. The radial velocity, $\rm V_R,$ is in turn defined as: $\rm V_R=(XV_X+YV_Y)/R$}.
Thus, negative $V_{\Phi}$ corresponds to prograde motion, and positive $V_{\Phi}$ to retrograde motion.

For each star, we took into account its errors on parallax, proper motions, and line-of-sight velocity, by assuming gaussian distributions of the errors, and by generating 100 random realisations of these parameters. The corresponding errors on positions and velocities are given in Table~\ref{DataOrb1}.

Finally, in this section and in the following, to integrate the orbits of stars, we made use of the axisymmetric Galactic potential 'PII' described in \citep{pouliasis17}, which consists of a thin and of a thick stellar disc and a spherical dark matter halo, and which reproduces a number of characteristics of the Milky Way \citep[see][for details]{pouliasis17}. Starting from the current positions and velocities of stars in the galactocentric rest frame, derived as described above, we have integrated their orbits backward in time for 6~Gyr, by making use of a leap-frog algorithm with fixed time step $\Delta t=10^5$~yr. For each star, we can thus quantify its total energy $E$ (that is the sum of its kinetic and potential energy), reconstruct its orbit in the Galactic potential adopted, and hence estimate its eccentricity, $ecc$, the maximum height from the plane, $z_{max}$, it reaches, as well as its (in-plane) apocentre $R_{max}$.  To estimate the uncertainties on the orbital parameters, for each star, we compute 100 realisations of its orbit, by making use of the 100 random realisations of its parallax, proper motions, and line-of-sight velocity, as described above. All these realisations are also integrated in  the same Galactic potential and for the same total time interval.
The orbits of all stars in the LP First stars  are given in Appendix~\ref{orbits}. The corresponding orbital parameters, and their uncertainties are given in Appendix~\ref{tabdata},  Table~\ref{DataOrb2}  .

\subsection{Kinematic and orbital properties of the LP First stars}

We start the analysis of the kinematics of stars in our sample by showing the Toomre diagram -- that is the $V_\Phi-\sqrt{{V_R}^2+{V_Z}^2}$ plane, with $V_R$ and $V_Z$ being, respectively, the radial and vertical components of the velocity of stars -- in Fig~\ref{kinorb_MP} (top-left panel). 
In this plane, most of the stars in the sample seem  kinematically associated to the halo, the absolute value of their velocity, $\sqrt{(V_\phi-V_{LSR})^2+{V_R}^2+{V_Z}^2}$, relative to the LSR, being larger than 180~km/s, a threshold often used to distinguish stars with disc-like kinematics from stars with kinematics more akin to the halo \citep[see, for example][]{nissen10}. However, eight out of the 42 LP First stars, that is nearly 20\% of the sample, redistribute in the grey area of the Toomre diagram in Fig~\ref{kinorb_MP}, which represents the locus of stars with  disc-like kinematics, where the absolute value of the star velocity, relative to the LSR, is lower than 180~km/s. \\
Because stars in the LP sample have a wide spatial distribution, extending to distances of several kpc from the Sun (see Fig.~\ref{histd2}), differences in velocities can simply reflect differences in the positions of the stars, rather than intrinsic differences in their kinematic properties. A more robust comparison can be done by analysing integral-of-motion spaces, like the $L_z-L_{perp}$ or $E-L_z$ space. Those spaces, and in particular the clumpiness of the stellar distribution in those spaces, were suggested by \citet{helmi99, helmi00} as efficient diagnostics to infer the (accreted) origin of stars in the Galaxy. It was later shown that this approach has severe limitations \citep{jeanbaptiste17} -- and indeed in the folllowing, we avoid using these diagnostics, alone, to infer the nature of stars in our sample. It is nevertheless more reliable to compare the kinematic properties of a spatially extended sample of stars in those planes, rather than in velocity-only planes, since velocities can change within the spatial volume covered by our data because of velocity gradients. \\ The $L_z-L_{perp}$ plane (see Fig.~\ref{kinorb_MP}, top-right panel) shows $L_z$, which is the $z-$component of the orbital angular momentum of a star, versus $L_{perp}=\sqrt{{L_x}^2+{L_y}^2}$, which is the perpendicular angular momentum component. We note that, while in an axisymmetric potential, $L_z$ is conserved, $L_{perp}$ is not. In all plots shown in the rest of the paper, the adopted value of $L_{perp}$  is thus the time-averaged value, calculated over 6~Gyr of orbital evolution. Figure~\ref{kinorb_MP} shows that the LP sample has a broad distribution in the $L_z-L_{perp}$ plane: stars with the most retrograde motions also have the highest values of $L_{perp}$ , while, among stars with prograde motions, some have values of $L_z$ very similar to those of stars of the LSR ($L_{z,LSR}=-20$, in units of 100~kpc~km/s).\\
The distribution of our sample stars in the energy, $E$, versus the vertical component of the angular momentum, $L_z$ (middle-left panel of Fig~\ref{kinorb_MP}) shows that some of the stars, both on prograde and retrograde orbits, can have very high energies, of which the absolute value is about twice as high as those of stars with disc-like kinematics (the latter can be identified in this plane as the stars with $L_z \lesssim -5$ and $-1900 \lesssim E \lesssim -1800$).\\ Interestingly, LP stars on prograde and on retrograde orbits show similar eccentricities (middle-right panel of Fig~\ref{kinorb_MP}): few stars on retrograde orbits have eccentricities lower than 0.4. In this plot, as in the following, the eccentricity of a star in defined as:
\begin{equation}
ecc=\frac{R_{\rm 3D,max}-R_{\rm 3D,min}}{R_{\rm 3D,max}+R_{\rm 3D,min}},
\end{equation}
$R_{\rm 3D,max}$ and $R_{\rm 3D,min}$ being, respectively, the (3D) apocentre and pericentre of the star's orbit.

\begin{figure}[h!]
\begin{center}
\includegraphics[clip=true, trim={0cm 0cm 0cm 0cm},width=0.45\textwidth]{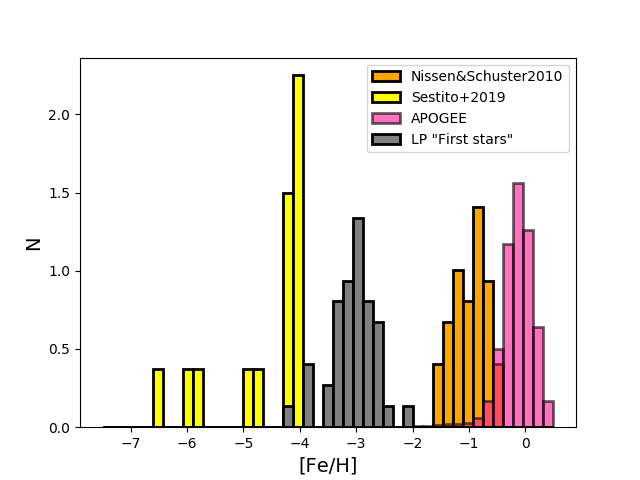}
\includegraphics[clip=true, trim={0cm 0cm 0cm 0cm},width=0.45\textwidth]{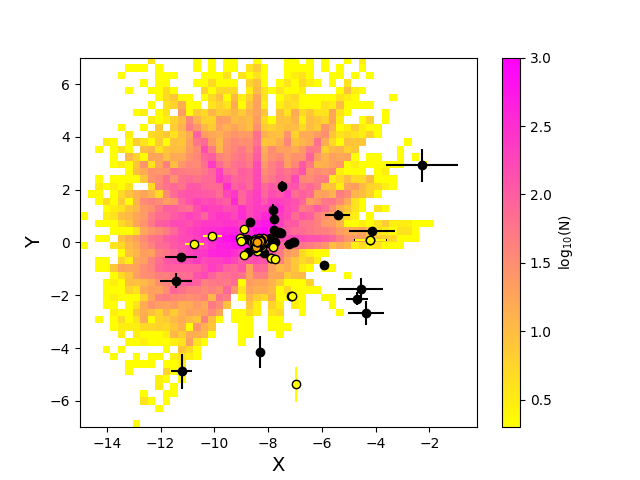}
\includegraphics[clip=true, trim={0cm 0cm 0cm 0cm},width=0.45\textwidth]{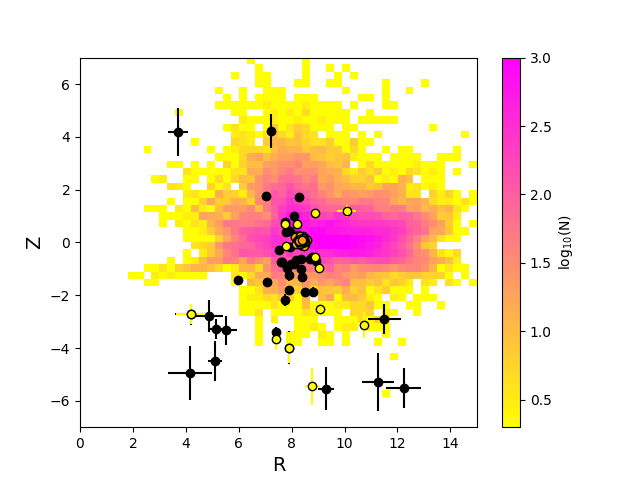}
\caption{\emph{Top panel:} Normalised metallicity distribution of the LP First stars sample (grey histogram), compared to the ultra-metal-poor sample studied by Sestito et al 2019 (yellow histogram),  the Nissen \& Schuster 2010 sample (orange histogram) and the \gaia~~DR2-APOGEE sample (pink histogram). The four samples span about 7~dex in [Fe/H]. \emph{Middle panel:} Distribution in the XY plane of stars from the \gaia~~DR2-APOGEE (stellar number density map in logarithmic scale), the LP First stars (black dots), the Nissen \& Schuster 2010 sample (orange dots), and the Sestito et al 2019 sample (yellow dots). Units are in kpc. \emph{Bottom panel:} Distribution in the RZ plane of stars in the different samples. Colours and units are the same as those adopted in the middle panel. }
\label{MP+S+NS+APO}
\end{center}
\end{figure}

Finally, the maximal radial excursion of the orbits of stars as projected onto the Galactic plane, $R_{max}$, versus their maximal height from the plane,  $z_{max}$, is reported in Fig~\ref{kinorb_MP} (bottom-left panel). While the majority of stars in the sample are confined inside 20~kpc from the Galactic centre, some can reach much larger distances, as high as 50~kpc. The  striking feature of this plot, however, is that VMP and EMP stars are not distributed homogeneously in this plane, but they tend to group along (at least) two different structures: one confined close to the Galactic plane (i.e. low $z_{max}$), and one characterised by larger $z_{max}$, for all values of $R_{max}$.  These structures, or wedges, were already noted in the sample of about 2000 \gaia~DR2 stars with high tangential velocities relative to the LSR studied by \citet{haywood18}. Also, stars in \citet{haywood18} are indeed redistributed along at least two wedges (see discussion in that paper), the one with low $z_{max}$ populated by stars with disc-like orbits, which mean they are always confined close to the Galactic plane. The stars studied by \citet{haywood18}, however, have a mean [Fe/H] $\sim -1$~dex, that is between 1 and 3~dex higher than the [Fe/H] of  LP First stars. The same orbital structure found by \citet{haywood18} at high metallicities thus seems to persist over several orders in metallicities. We investigate this point further in the next section. We conclude by noting that the distribution of stars in the $R_{max}-z_{max}$ plane is independent of them being on prograde or retrograde orbits: stars with retrograde or prograde rotation are redistributed rather homogeneously in this plane, and some retrograde stars are clearly confined to disc-like kinematics (i.e. low $z_{max}$). As in \citet{haywood18}, it is possible to quantify the amount of stars with disc-like orbits, and distinguish them from those with halo-like orbits, by estimating the arctangent of the ratio $z_{max}/R_{max}$ (see Fig~\ref{kinorb_MP}, bottom-right panel), which represents the inclination of the wedges in this plane. The lack of stars with arctan($z_{max}/R_{max}$)$\sim$~0.2, noticed by \citet{haywood18} for stars at  [Fe/H] $\sim -1$~dex, and found also in this VMP and EMP sample, does separate the two samples well. By making use of $\rm arctan(z_{max}/R_{max})=0.2$ as separating value, we find disc-like orbits for 10 out of the  42 stars in the LP First stars, that is for about 20\% of the stars in the sample, a fraction comparable to that derived using the Toomre diagram and of a discriminating value for disc-like orbits of $\sqrt{(V_\phi-V_{LSR})^2+{V_R}^2+{V_Z}^2}=180$~km/s (see discussion at the beginning of this section). We note that, compared to the \citet{haywood18} stars, the distribution of the arctan($z_{max}/R_{max}$) for stars of the LP sample shows a lower fraction of stars with disc-like kinematics. This is a natural consequence of the fact that the \citet{haywood18} sample is dominated by thick disc stars at $\rm [Fe/H] \ge -1$ with halo-like kinematics, which are the dominant contributor to the kinematically defined halo population at few kpc from the Sun \citep[see also][]{dimatteo18}.

\section{Comparison with samples of stars at lower and higher metallicities}

In this section, we aim  to compare the kinematic and orbital properties of stars in the LP sample with those of samples that cover different metallicity ranges, as detailed below. The reason for this comparison is twofold. Firstly, we want to understand whether the properties described in the previous section are found also among stars of lower and higher metallicities.   Secondly, by comparing with other samples (in particular with those at higher metallicities, where a distinction between in-situ and accreted stars is possible on the basis of their chemical abundances), we can try to interpret the kinematic and orbital characteristics of our stars in terms of their in-situ/accreted origin.

In the following part of this section, we compare the kinematic and orbital properties of stars in the LP First stars to three different samples, which are listed below in order of increasing [Fe/H].
\begin{itemize}
\item The sample of ultra iron-poor stars (UIP) studied by \citet{sestito19}. This sample consists of 
42 stars with $\rm [Fe/H] \le -4$. \citet{sestito19} already derived orbital parameters for these stars using \gaia~DR2 parameters, and a bayesian estimate of the distances. For coherence with the approach used for the analysis of the LP sample, and throughout the rest of this paper, we applied the same selection applied to the LP sample to the \citet{sestito19} sample. After correcting the parallaxes for the zero-point offset, we only retain stars with positive parallaxes, and a relative error on the latter smaller than 20\%. We then estimate the distances of stars by simply inverting the parallaxes, and integrate their orbits in the same potential and with the same numerical method adopted for the LP sample (see Sect.~\ref{orbitint}). Uncertainties on positions, velocities, and resulting orbital parameters are also estimated as was done for our sample (see Tables~\ref{DataOrb1} and \ref{DataOrb2}). The advantage of this approach is the choice of the same methods and corrections for all stars, the disadvantage, of course, is the reduced statistics. The selection of the quality of parallaxes indeed severely reduces the sample from 42 to 15 members. Two stars of the 15, with \gaia~id  5000753194373767424 and 6692925538259931136, are common to the LP First stars sample.
\item The sample of metal-poor stars studied by \citet{nissen10} \citep[see also][]{nissen11, schuster12, nissen12}. This sample consists of  94 dwarf stars in the metallicity range $\rm -1.6 < [Fe/H] < -0.4$. The kinematics and orbital properties of stars in this sample were discussed in \citet{nissen10} and \citet{schuster12}. We recalculated these properties using \gaia~DR2 astrometry and by employing the same selection of parallaxes and their relative errors, as adopted for the LP First stars and \citet{sestito19} samples, as well as by making use of the same Galactic potential for orbit integration. The final sample contains 84 stars. 
\item The \gaia~DR2-APOGEE sample studied by \citet{dimatteo18}. This sample is the result of cross-matching the \gaia~DR2 \citep{GaiaDR2-Brown} with APOGEE data from DR14 \citep{majewski17}, using the CDS X-match service. To construct this sample, we selected stars in the two catalogues with a position mismatch tolerance of of 0.5 arcsec, and retained only those with positive parallaxes, relative error on parallaxes less than 20\%, and a signal-to-noise ratio in the APOGEE spectra, SNR> 100. Also in this case, parallaxes have been corrected for the zero-point parallax offset. All line-of-sight velocities used for this sample are from APOGEE. Following the study of \citet{fernandez18}, we applied additional selection criteria only retaining stars with effective temperatures, Teff >4000, and gravities, 1<log(g)<3.5. Finally, we also removed all APOGEE stars with ASCAPFLAG and STARFLAG warning of any problems with the determinations of the atmospheric parameters (specifically those with a warning about the reliability of the effective temperature, log(g), rotation and having a very bright neighbour). After applying all these selection criteria, our final sample consists of 61789 stars. Also for this 
sample, we have derived positions and velocities of stars as described in Sect.~\ref{orbitint} and integrated the orbits for 6~Gyr in the PII Galactic potential described in \citet{pouliasis17}.
\end{itemize}

Figure~\ref{MP+S+NS+APO} shows the normalised [Fe/H] distributions of stars in these samples, as compared to the LP distribution, and their spatial distribution in the galactic plane (XY), as well as in the meridional plane (RZ). 
The \citet{nissen10} sample is very local (distances from the Sun less than about 300~pc), the other samples span a larger range of distances from the Sun. In particular, we note that the APOGEE sample lacks stars in the fourth quadrant, and that this gap is partially filled by the LP First stars and samples from \citet{sestito19}. While the differences in the spatial extension and coverage of all these samples must be taken into account, the advantage of this approach is to compare  the kinematics of stars in the Galaxy over a range of nearly 7~dex in [Fe/H], something that, to our knowledge, was done here for the first time. 

\subsection{Spanning 7~dex in [Fe/H]: the ubiquity of the Galactic disc}

\begin{figure*}
\begin{center}
\includegraphics[clip=true, trim={0cm 0cm 0cm 0cm},width=0.33\textwidth]{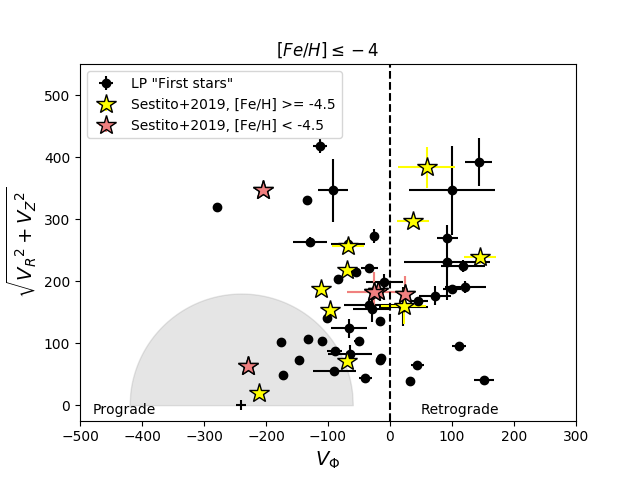}
\includegraphics[clip=true, trim={0cm 0cm 0cm 0cm},width=0.33\textwidth]{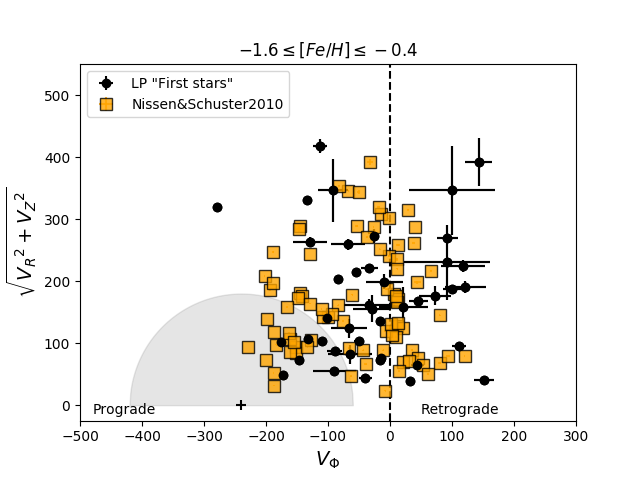}
\includegraphics[clip=true, trim={0cm 0cm 0cm 0cm},width=0.33\textwidth]{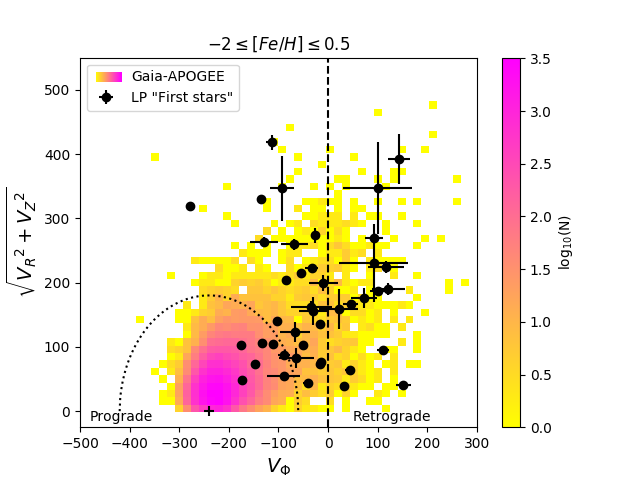}\\
\includegraphics[clip=true, trim={0cm 0cm 0cm 0cm},width=0.33\textwidth]{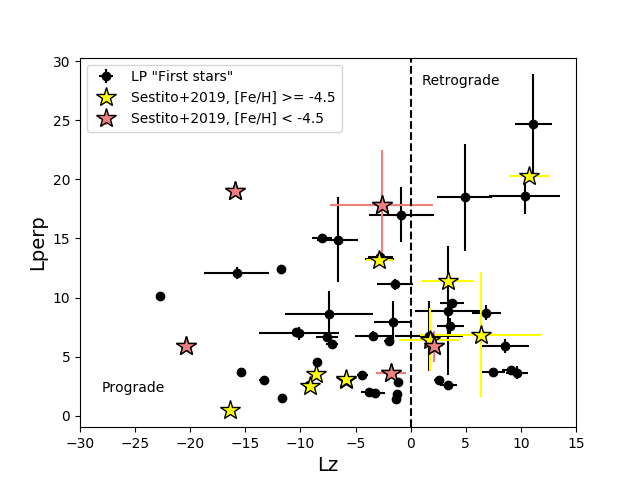}
\includegraphics[clip=true, trim={0cm 0cm 0cm 0cm},width=0.33\textwidth]{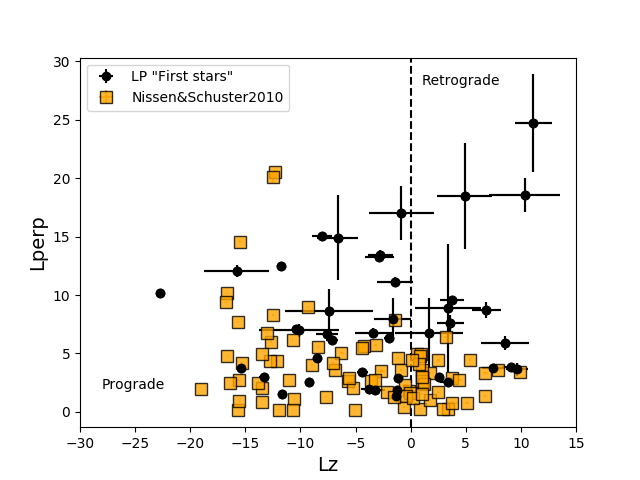}
\includegraphics[clip=true, trim={0cm 0cm 0cm 0cm},width=0.33\textwidth]{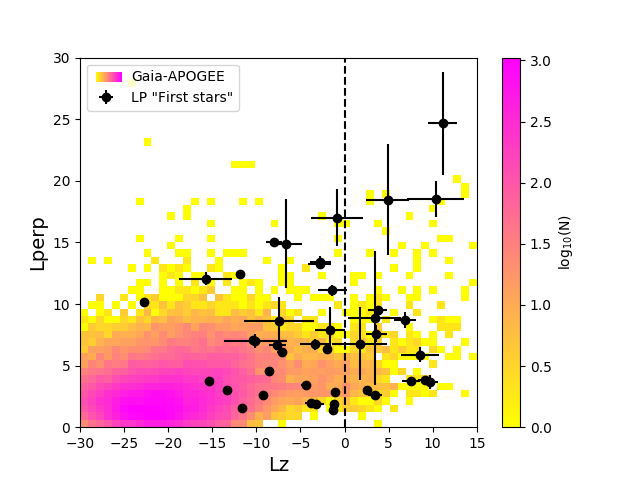}\\
\includegraphics[clip=true, trim={0cm 0cm 0cm 0cm},width=0.33\textwidth]{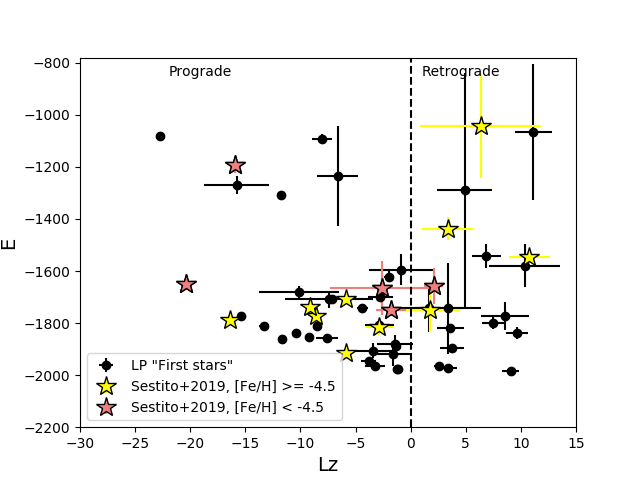}
\includegraphics[clip=true, trim={0cm 0cm 0cm 0cm},width=0.33\textwidth]{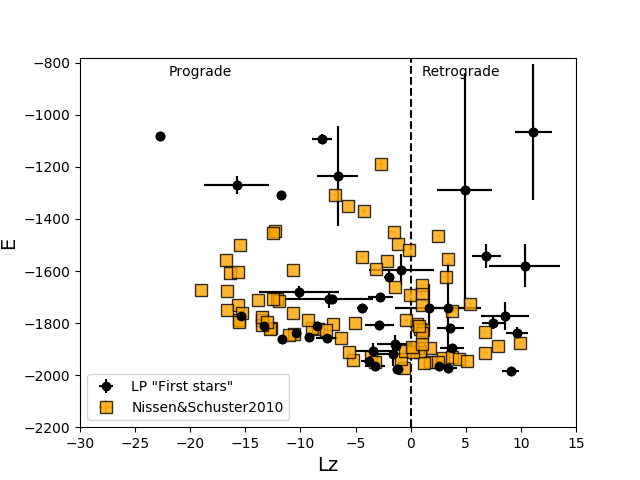}
\includegraphics[clip=true, trim={0cm 0cm 0cm 0cm},width=0.33\textwidth]{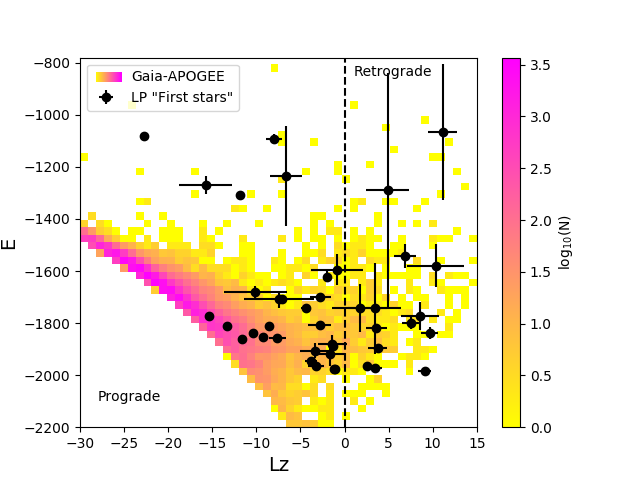}\\
\includegraphics[clip=true, trim={0cm 0cm 0cm 0cm},width=0.33\textwidth]{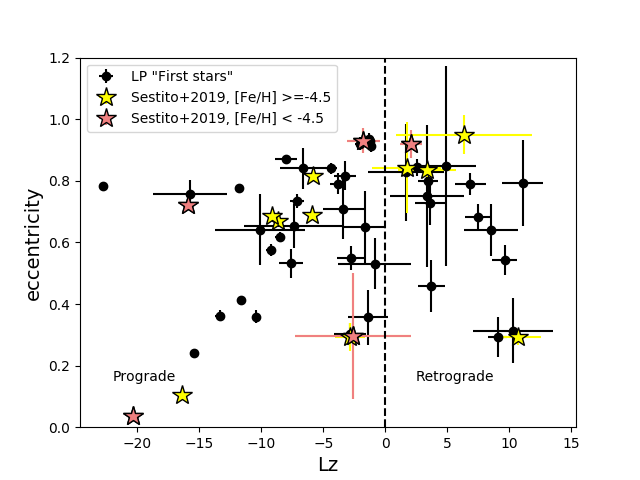}
\includegraphics[clip=true, trim={0cm 0cm 0cm 0cm},width=0.33\textwidth]{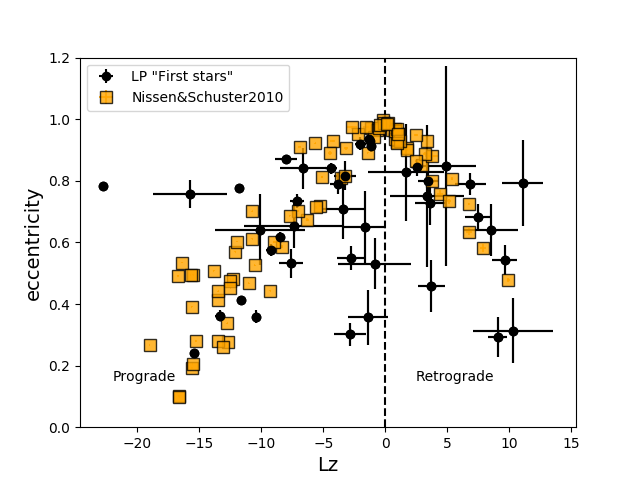}
\includegraphics[clip=true, trim={0cm 0cm 0cm 0cm},width=0.33\textwidth]{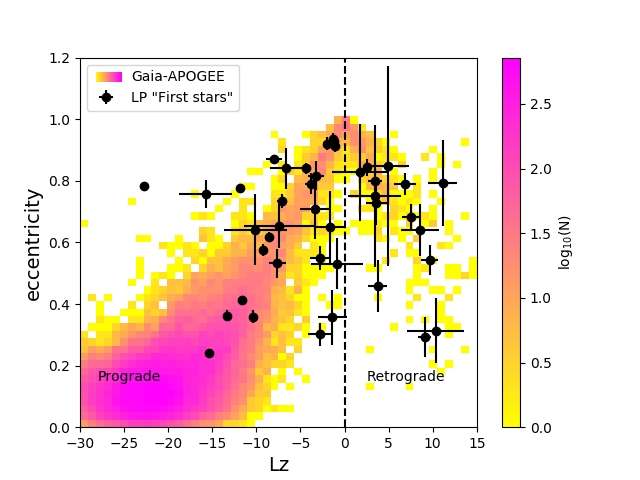}\\
\includegraphics[clip=true, trim={0cm 0cm 0cm 0cm},width=0.33\textwidth]{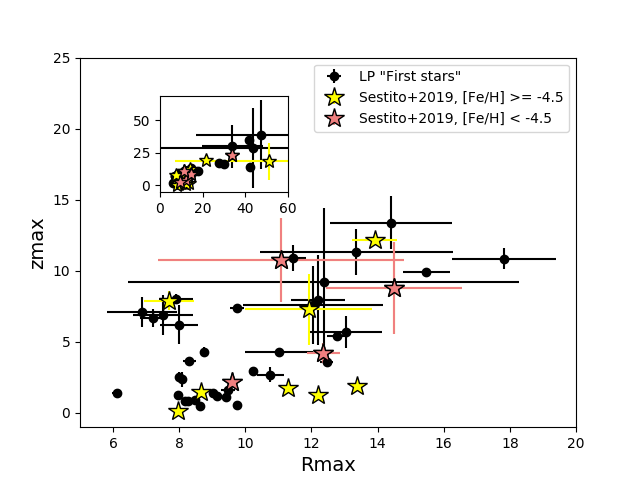}
\includegraphics[clip=true, trim={0cm 0cm 0cm 0cm},width=0.33\textwidth]{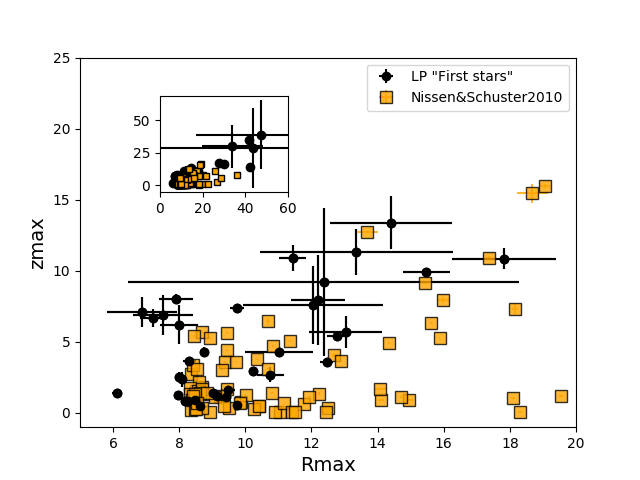}
\includegraphics[clip=true, trim={0cm 0cm 0cm 0cm},width=0.33\textwidth]{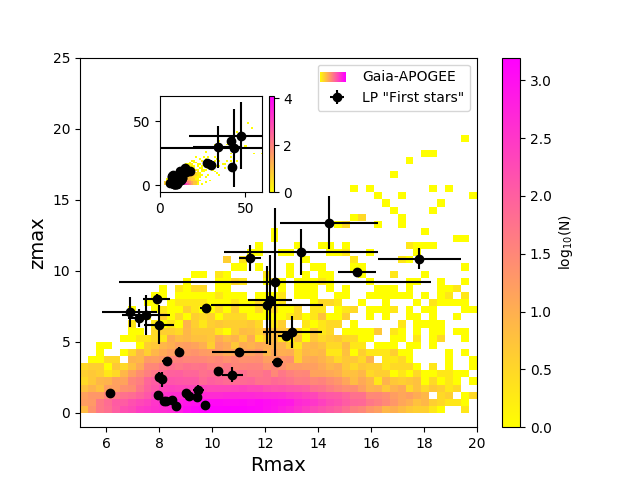}
\caption{\emph{From top to bottom: } Toomre diagram, $E-L_z$, $ecc-L_z$ and $R_{max}-z_{max}$ planes for stars of LP First stars (black dots), compared to the Sestito et al 2019 sample (yellow stars, first column), the Nissen \& Schuster 2010 sample (orange squares, middle column), and the \gaia~~DR2-APOGEE sample (density map, right column). The vertical dashed lines  separate prograde from retrograde motions. Velocities are in units of $\rm km/s$, distances are in units of kpc, angular momenta are in units of 100~kpc~km/s, and energies in units of $\rm 100~km^2/s^2$.}
\label{kinorbALL}
\end{center}
\end{figure*}

Figure~\ref{kinorbALL} shows the Toomre diagram, the $E-L_z$ plane, the $ecc-L_z$ plane and the $R_{max}-z_{max}$ plane for all stars in the LP sample, compared, respectively,  to the Sestito et al 2019 sample, the Nissen \& Schuster 2010 sample, and the \gaia~~DR2-APOGEE sample. 
The VMP and EMP stars of our sample show striking similarities in all these planes with  the UIP stars from Sestito et al 2019. The two samples essentially show the same distribution in all these spaces. In both samples, about 20\% of the stars have disc-like kinematics. This fraction is comparable to that derived by \citet{sestito19}, using all 42 stars, and not our restricted sample of 15, thus suggesting that our cut on parallax errors did not introduce any bias in the relative fraction of stars with disc/halo kinematics. As already noticed by \citet{sestito19}, some stars in the sample have very low eccentricities, like 2MASS J18082002-5104378 \citep{melendez16}, which is on a circular orbit ($ecc=0.1$). Stars on disc-like orbits are thus found all along the [Fe/H] sequence, from the most metal-poor stars up to the most metal-rich samples \citep[][\gaia~DR2-APOGEE]{nissen10}, suggesting that, despite their different abundances and iron contents, a fraction of the UIP, EMP, VMP, and metal-poor stars can all share the same common origin, tracing the early phases of the Milky Way disc formation \citep[see also the recent work by][for the finding of a EMP star with disc-like kinematics]{venn19}.\\

While a non-negligible fraction of stars have disc-like kinematics, the majority of the UIP stars, as well as the VMP and EMP stars, have halo-like kinematics. This does not necessarily mean that they are all accreted, since a fraction of the halo can be  made of stars formerly in the disc, but later kinematically heated to halo kinematics by one or several satellite accretions. This has indeed proven to be the dominant in-situ mode of formation of the Galactic halo for stars at higher metallicities, and at few kpc from the Sun \citep[see][]{dimatteo18}, as we discuss more extensively in the next section. It is, however, interesting to note that, compared to stars at higher metallicities, such as stars in the Nissen \& Schuster sample and stars in the \gaia~~DR2-APOGEE sample (middle and right columns in Fig.~\ref{kinorbALL}), in the Toomre diagram, the UIP and LP samples seem to lack stars with null angular momentum, meaning along the $L_z=0$ line. At values of $\sqrt{({V_R}^2+{V_Z}^2)} \gtrsim 200$~km/s, stars with $\rm [Fe/H] < -4$ have rather prograde or retrograde motions, but none seem to lie along the sequence of accreted halo stars discovered by Nissen \& Schuster 2010, and later confirmed in \gaia~ DR1 and DR2 data by \citet{belokurov18} (Gaia Sausage), \citet{haywood18}, \citet{helmi18} (Gaia Enceladus). The Gaia Sausage, which is this group of halo stars with very radial orbits, and null $\rm V_{\Phi}$, seems to indeed disappear at $\rm [Fe/H] < -2$ in this plane. We emphasise that this apparent difference between the kinematics of VMP and EMP stars, on the one side, and stars with $\rm [Fe/H] > -2$, on the other side, is simply the consequence of these stars  probing different regions and distances from the Galactic centre. Indeed, when  one compares the kinematics of these different samples of stars in the quasi-integral-of-motion space $L_z-L_{perp}$ plane (see Fig~\ref{kinorbALL}, second row), rather than in the Toomre diagram,  the kinematic properties of these samples are the same, over the whole [Fe/H] interval. Before moving further, we need, however, to emphasise two points of the comparison with samples at higher metallicity. Firstly, when compared to the \citet{nissen10} sample, which, we remind the reader, is a kinematically selected sample of thick disc and halo stars, the VMP and EMP stars in the LP First stars sample show an excess of stars at retrograde motions (positive $L_z$, i.e. $L_z > 5$ ) and high values of $L_{perp}$ ($L_{perp} > 15$). None of the  \citet{nissen10} stars occupy this region of the $L_z-L_{perp}$ diagram, and we suggest this is a consequence of the `local' character of stars in the  \citet{nissen10}  study, which are all limited to a few hundred parsecs from the Sun. Indeed, when VMP and EMP stars are compared to stars in the  \gaia~~DR2-APOGEE sample, one can see that stars with $L_z$ and $L_{perp}$ as extreme as $L_z > 5$ and $L_{perp} > 15$ are found also in the latter. Secondly, because in the comparison shown in this figure we used all stars in the \gaia~~DR2-APOGEE sample, not restricting ourself to stars from the kinematically defined thick disc and halo, the reader will not be surprised to find that the majority of stars in \gaia~~DR2-APOGEE sample are stars with cold (i.e. thin) disc-like kinematics, their distribution peaking at $L_z \sim 20$ and $L_{perp} \le 5$. At this stage, what is important to retain is that the region occupied by all these samples, independently of their [Fe/H] ratio, is the same: the relative fraction of stars in one or in another region of the space under analysis can change from one sample to another, but not their overall distribution.\\

Finally, the last two rows of Fig.~\ref{kinorbALL} show the comparison between the LP First stars sample with the other datasets in the $ecc-L_z$ and $R_{max}-z_{max}$ plane. The reader may notice that the similarity in the kinematic properties of stars, along 7~dex in [Fe/H] is also remarkable in these planes. When compared to the \citet{sestito19} sample, the distribution is similar both in the $ecc-L_z$ and $R_{max}-z_{max}$ planes. It is remarkable that the two stars with the lowest eccentricity among all the stars with $\rm [Fe/H] \le -2$ are stars of the \citet{sestito19} sample, and have eccentricities below 0.2. The comparison with samples at higher metallicities shows, overall, a good agreement, even if we note the absence (in the LP sample) of stars with low $z_{max}$ and $R_{max} > 10$~kpc (see comparison with the \citet{nissen10} sample and with the  \gaia~DR2-APOGEE sample. For the time being, it is difficult to say whether this difference is real or not. We note, however, that this difference seems peculiar to the LP sample, and it is not evident when the high-metallicity samples are compared to stars from \citet{sestito19}. In this latter case, some stars have limited $z_{max}$ but in-plane apocentres $R_{max} > 10$.

\subsection{Comparing with in-situ and accreted stars at $\rm [Fe/H] > -2$}

\begin{figure*}
\begin{center}
\includegraphics[clip=true, trim={0cm 0cm 0cm 0cm},width=0.33\textwidth]{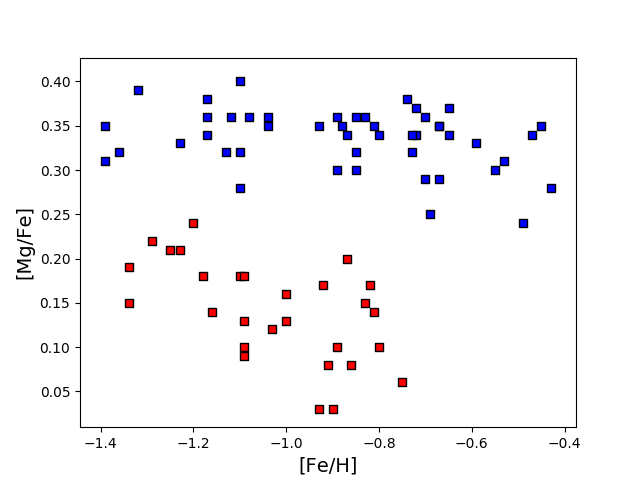}
\includegraphics[clip=true, trim={0cm 0cm 0cm 0cm},width=0.33\textwidth]{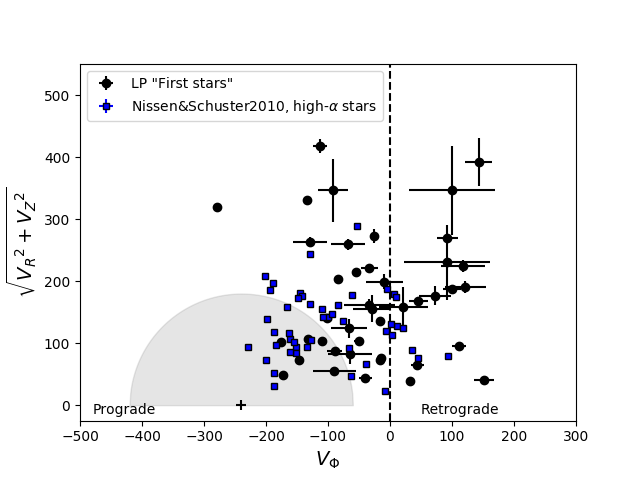}
\includegraphics[clip=true, trim={0cm 0cm 0cm 0cm},width=0.33\textwidth]{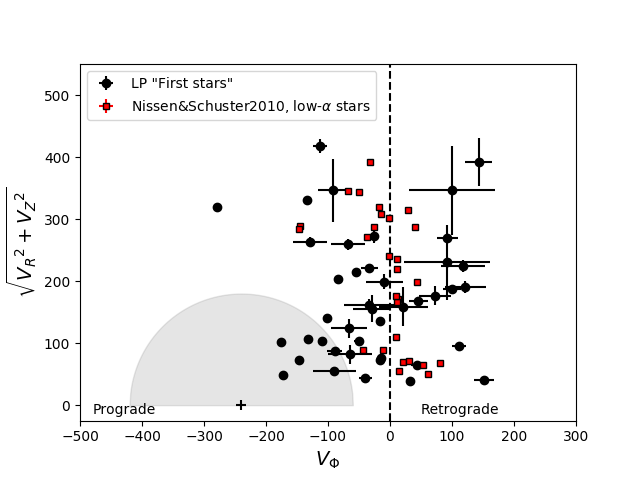}\\
\includegraphics[clip=true, trim={0cm 0cm 0cm 0cm},width=0.33\textwidth]{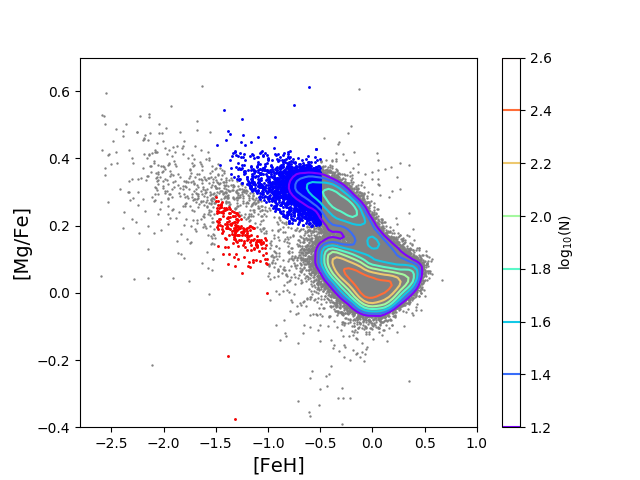}
\includegraphics[clip=true, trim={0cm 0cm 0cm 0cm},width=0.33\textwidth]{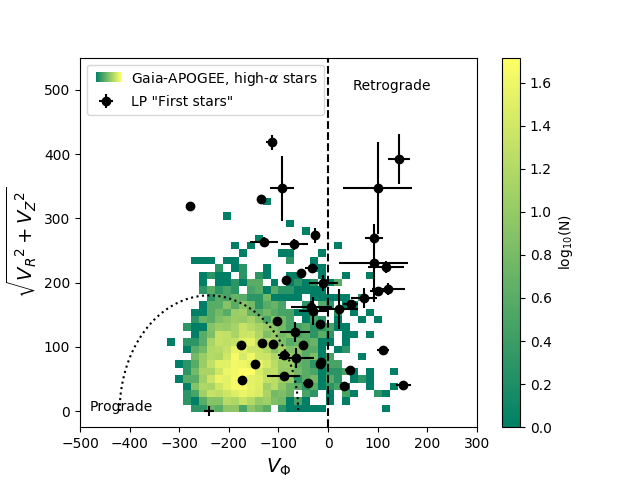}
\includegraphics[clip=true, trim={0cm 0cm 0cm 0cm},width=0.33\textwidth]{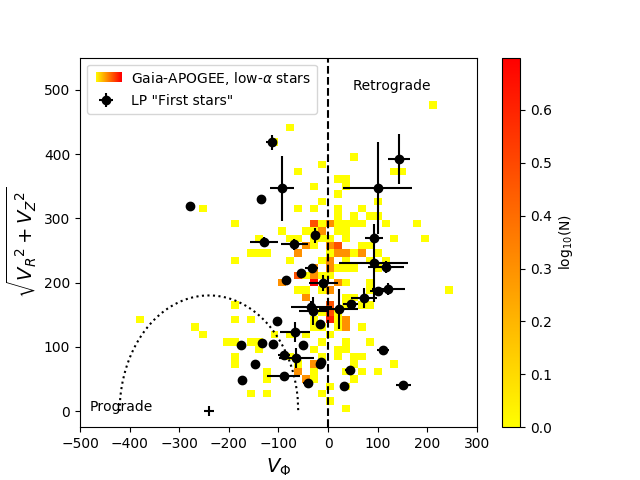}
\caption{\emph{First row, left panel: } [Mg/Fe] versus [Fe/H] distribution for stars in Nissen \& Schuster 2010 sample. Red colours indicate low-$\alpha$ stars, blue colours indicate high-$\alpha$ (thick disc and halo) stars, following the classification given in Nissen \& Schuster 2010. \emph{First row, middle panel: } Toomre diagram of LP First stars (black dots), and of the high-$\alpha$ (thick disc and halo) stars in the Nissen \& Schuster 2010 sample (blue squares).  \emph{First row, right panel:} Toomre diagram of LP First stars (black dots), and of the low-$\alpha$ stars in the Nissen \& Schuster 2010 sample (red squares). The grey  area in this panel and in the previous one separates stars with $\sqrt{{V_R}^2+{(V_\Phi-V_{LSR})}^2+{V_Z}^2} \le 180$~km/s from stars with higher relative velocities with respect to the LSR. The vertical dashed lines  separate prograde from retrograde motions. \emph{Second row, left panel: } [Mg/Fe] versus [Fe/H] distribution for stars in the \gaia~~DR2-APOGEE sample. The distribution of the whole sample is shown by grey dots and contours, while the distribution of stars selected as in-situ, high-$\alpha$ stars and accreted, low-$\alpha$ stars are shown, respectively, with blue and red dots.  \emph{Second row, middle panel: } Toomre diagram of LP First stars (black dots), and of the high-$\alpha$ stars in the \gaia~~DR2-APOGEE sample (density map).   \emph{Second row, right panel: } Toomre diagram of LP First stars (black dots), and of the low-$\alpha$ stars in the \gaia~DR2-APOGEE sample (density map).  The grey dashed curve in this panel and in the previous one separates stars with $\sqrt{{V_R}^2+{(V_\Phi-V_{LSR})}^2+{V_Z}^2} \le 180$~km/s from stars with higher relative velocities with respect to the LSR. The vertical dashed lines  separate prograde from retrograde motions.}
\label{Toomre_insituacc}
\end{center}
\end{figure*}

While in the previous section we compared the kinematic and orbital properties of the LP sample to stars of lower and higher metallicities, in this section we push the comparison with the higher metallicity samples (Nissen \& Schuster 2010, \gaia~~DR2-APOGEE) further. The metallicity range $\rm -2 \lesssim [Fe/H] \lesssim-0.5$ is particularly interesting for stellar population studies because stars in this [Fe/H] interval are grouped into two separate chemical sequences: a high-$\alpha$ sequence, made of thick disc and in-situ halo stars, and a low-$\alpha$ sequence, interpreted as made of accreted stars (see Fig.~\ref{Toomre_insituacc}, first column). These two distinct chemical sequences, discovered by \citet{nissen10}, on the basis of their [$\alpha$/Fe] content, have since been confirmed as two distinct sequences based on a number of other elemental abundance ratios \citep[seee][]{nissen11, schuster12, nissen12}. As shown by \citet{hayes18}, in a study based on the analysis of APOGEE data, and by \citet{haywood18} on the basis of \gaia~DR2 data, these stars represent the sampling at the solar vicinity  of a much extended structure, visible up to several kpc from the Sun, now referred to as the Gaia Sausage \citep{belokurov18} or Gaia Enceladus \citep{helmi18}. While the Nissen \& Schuster sample is limited in terms of statistics, it does have an exquisite spectroscopic quality that makes it ideal for a first comparison with our samples of VMP and EMP stars. The \gaia~~DR2-APOGEE sample, in turn, provides a much larger statistics, and extends the comparison to regions beyond the solar vicinity, up to several kpc from the Sun. 

\begin{figure*}
\begin{center}
\includegraphics[clip=true, trim={0cm 0cm 0cm 0cm},width=0.45\textwidth]{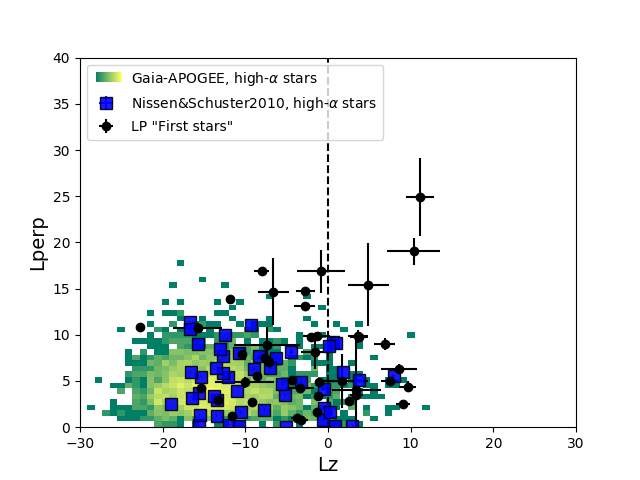}
\includegraphics[clip=true, trim={0cm 0cm 0cm 0cm},width=0.45\textwidth]{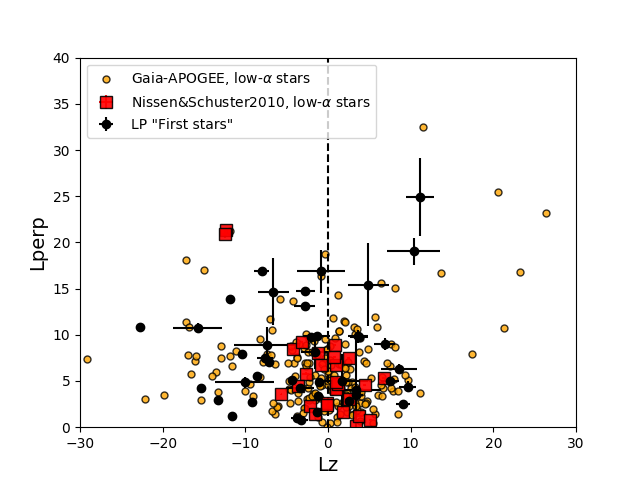}\\
\includegraphics[clip=true, trim={0cm 0cm 0cm 0cm},width=0.45\textwidth]{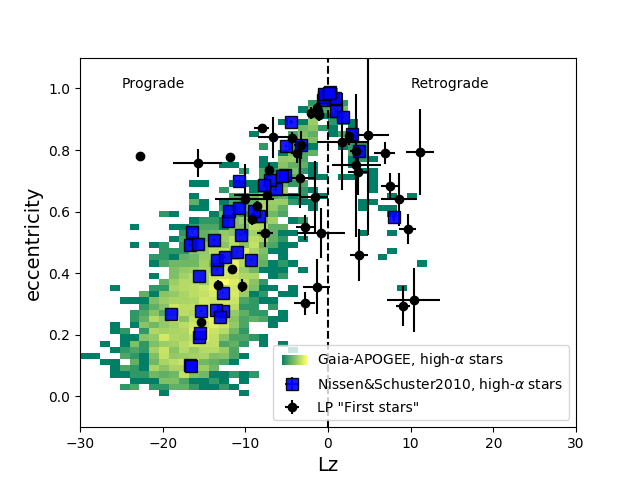}
\includegraphics[clip=true, trim={0cm 0cm 0cm 0cm},width=0.45\textwidth]{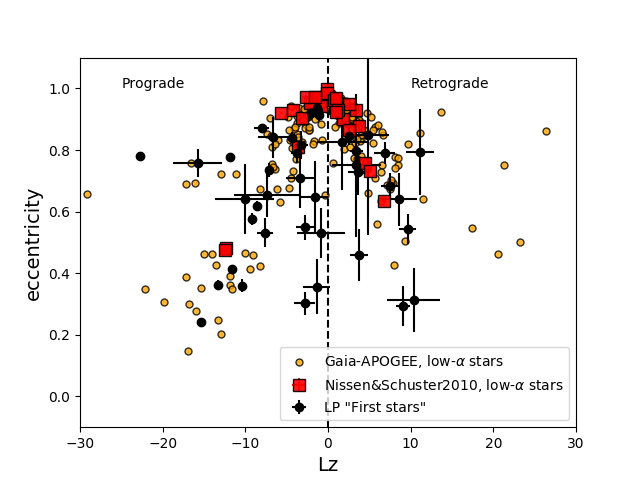}\\
\includegraphics[clip=true, trim={0cm 0cm 0cm 0cm},width=0.45\textwidth]{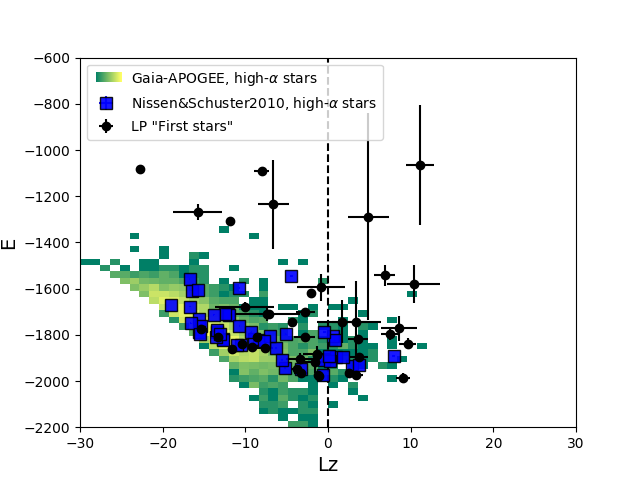}
\includegraphics[clip=true, trim={0cm 0cm 0cm 0cm},width=0.45\textwidth]{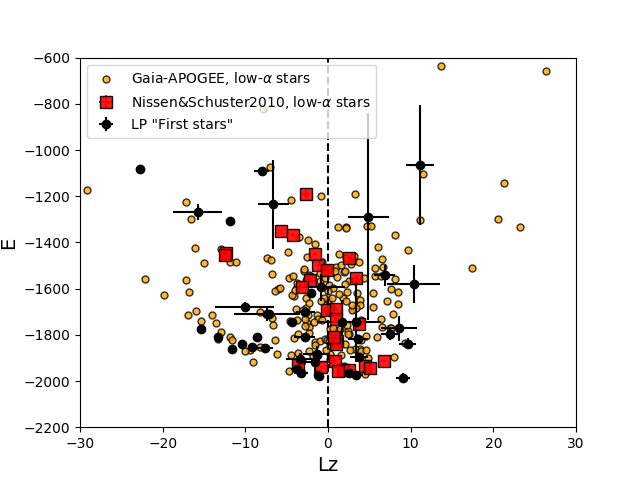}

\caption{\emph{Left column, from top to bottom:} $L_{z}-L_{perp}$ plane, $ecc-L_z$ plane and $E-L_z$ plane for stars in LP sample (black dots), compared to high-$\alpha$ (thick disc and halo)  stars of Nissen \& Schuster sample (blue squares) and high-$\alpha$ from \gaia~DR2-APOGEE sample (green density maps). \emph{Right column, from top to bottom:}  $L_{z}-L_{perp}$ plane,  $ecc-L_z$ plane and $E-L_z$ plane  for stars in the LP sample (black dots), compared to low-$\alpha$ stars of the Nissen \& Schuster sample (red squares) and low-$\alpha$ stars from the \gaia~DR2-APOGEE sample (orange points). }
\label{LzLperp_insituacc}
\end{center}
\end{figure*}

\begin{figure}
\begin{center}
\includegraphics[clip=true, trim={0cm 0cm 0cm 0cm},width=0.45\textwidth]{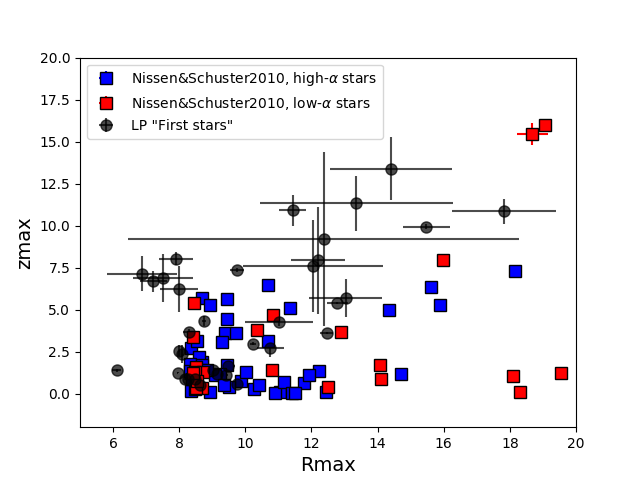}
\includegraphics[clip=true, trim={0cm 0cm 0cm 0cm},width=0.45\textwidth]{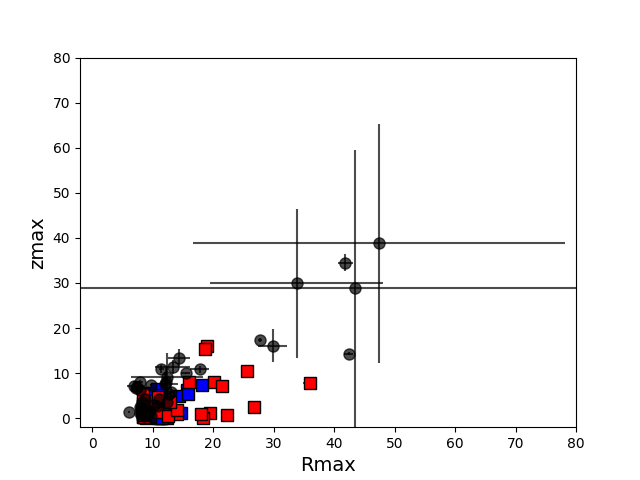}
\caption{$R_{max}-z_{max}$ plane for stars in LP sample (black dots), compared to high- and low-$\alpha$ stars of Nissen \& Schuster sample (respectively, blue and red squares). The top panel shows the $R_{max}-z_{max}$ distribution inside 20~kpc, the bottom panel show the distribution for the whole samples.  }
\label{Rmaxzmax_insituacc_MP+NS}
\end{center}
\end{figure}

\begin{figure}
\begin{center}
\includegraphics[clip=true, trim={0cm 0cm 0cm 0cm},width=0.5\textwidth]{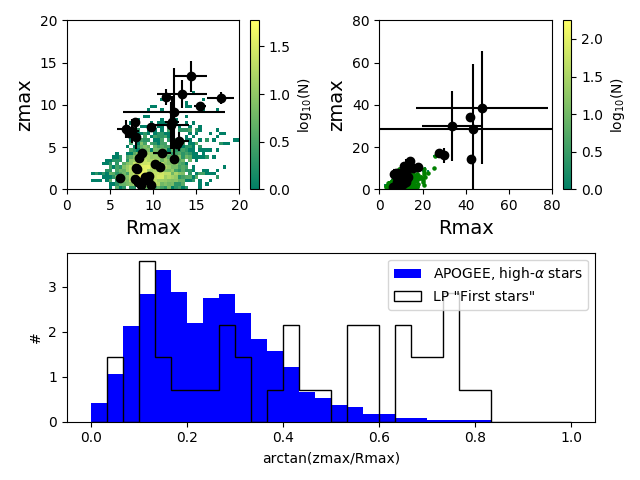}
\includegraphics[clip=true, trim={0cm 0cm 0cm 0cm},width=0.5\textwidth]{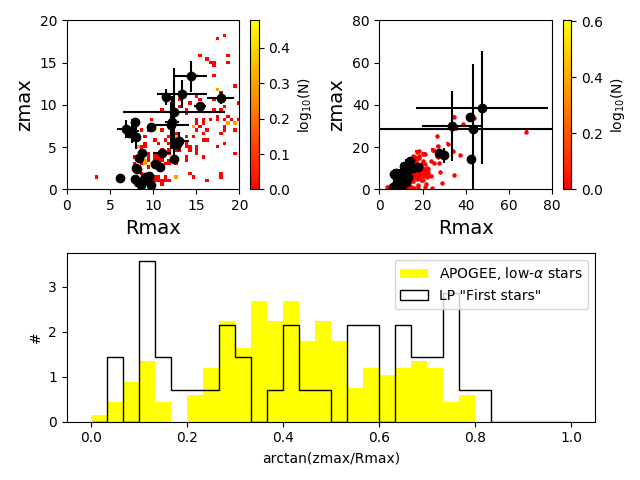}
\caption{$R_{max}-z_{max}$ plane for stars in LP sample (black dots), compared to high-$\alpha$ (first row) and low-$\alpha$ stars (third row) of \gaia~~DR2-APOGEE sample. In the first and third rows,  the left panel shows a zoom for stars with Rmax $\le 20$~kpc, the right panel the whole distribution.\emph{Second and fourth row:}  Normalised histogram of the acrtangent of the ratio zmax/Rmax for high-$\alpha$ stars (blue histogram, second row), and low-$\alpha$ stars (yellow histogram, second row) in \gaia~~DR2-APOGEE. In both panels, the distribution is compared to that of the LP sample (black histogram).}
\label{Rmaxzmax_insituacc}
\end{center}
\end{figure}

\begin{figure}
\begin{center}
\includegraphics[clip=true, trim={0cm 0cm 0cm 0cm},width=0.45\textwidth]{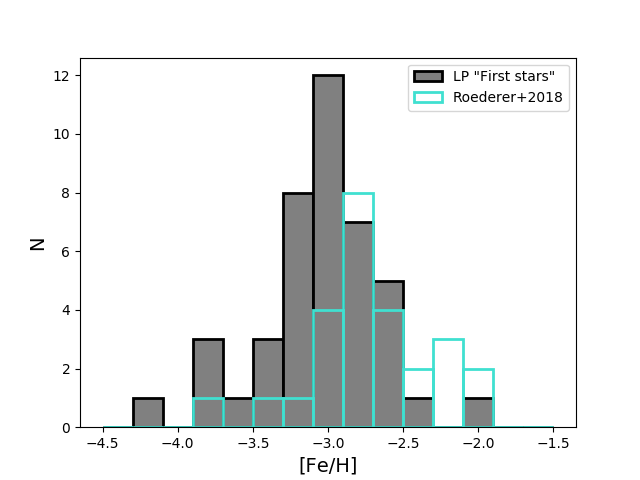}
\includegraphics[clip=true, trim={0cm 0cm 0cm 0cm},width=0.45\textwidth]{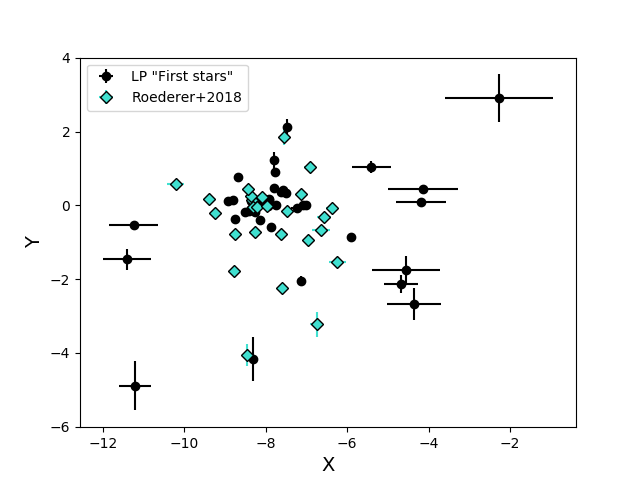}
\includegraphics[clip=true, trim={0cm 0cm 0cm 0cm},width=0.45\textwidth]{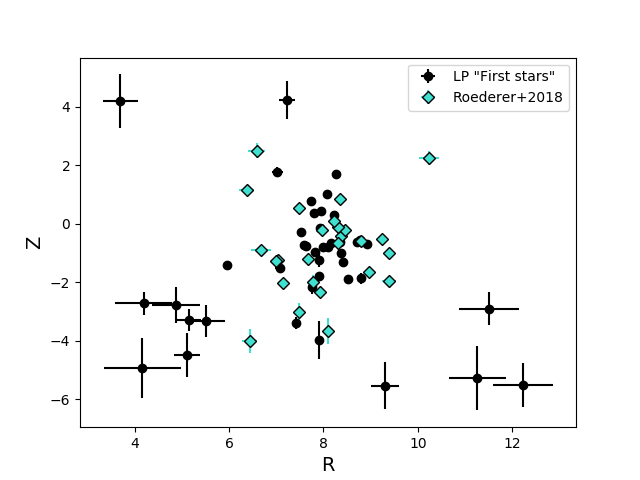}
\caption{\emph{From top to bottom:} [Fe/H] distribution of stars in LP First stars sample (grey histogram), compared to [Fe/H] distribution of stars in the Roederer et al 2018 sample (turquoise); distribution in the $X-Y$ plane of stars in LP First stars sample (black points) and of stars in Roederer et al 2018 sample (turquoise diamonds);  distribution in meridional, $R-Z$, plane of stars in LP First stars sample (black points), and of stars in Roederer et al 2018 sample (turquoise diamonds). }
\label{histoFeH_MP+RO.png}
\end{center}
\end{figure}

\begin{figure*}
\begin{center}
\includegraphics[clip=true, trim={0cm 0cm 1cm 1cm},width=0.3\textwidth]{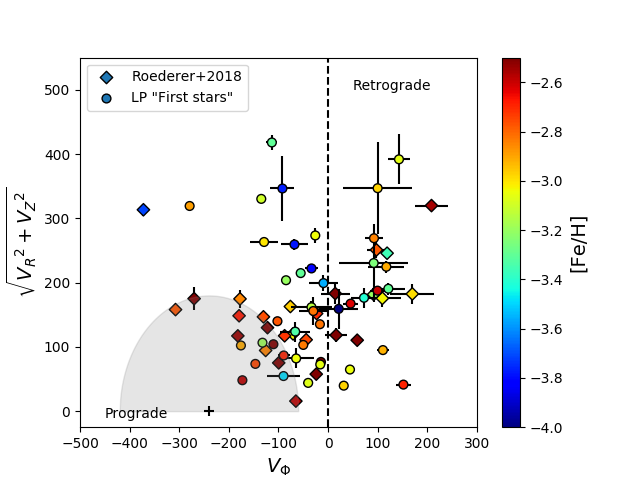}
\includegraphics[clip=true, trim={0cm 0cm 1cm 1cm},width=0.3\textwidth]{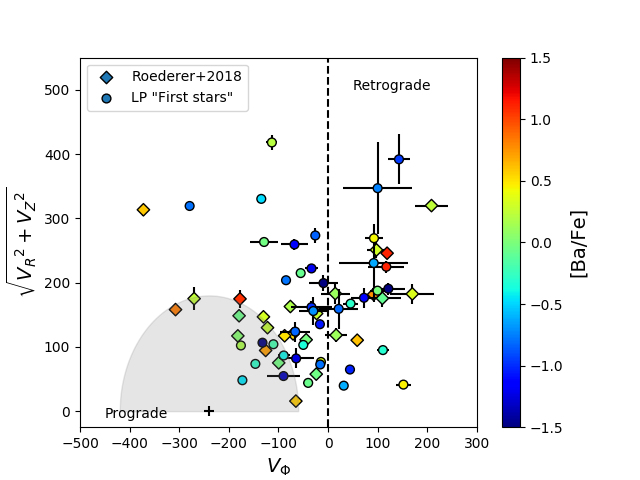}\\
\includegraphics[clip=true, trim={0cm 0cm 1cm 1cm},width=0.3\textwidth]{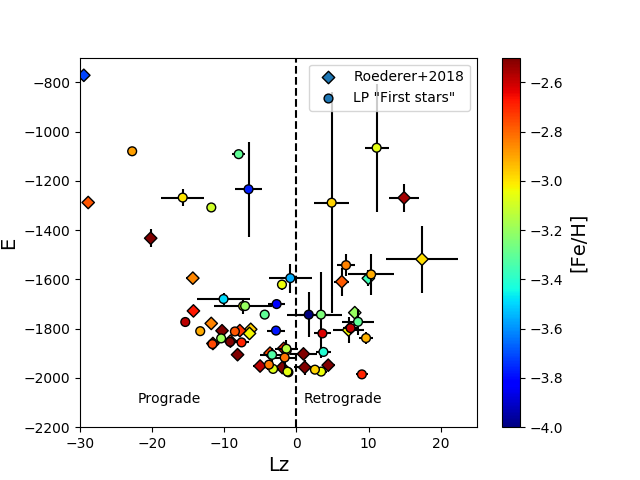}
\includegraphics[clip=true, trim={0cm 0cm 1cm 1cm},width=0.3\textwidth]{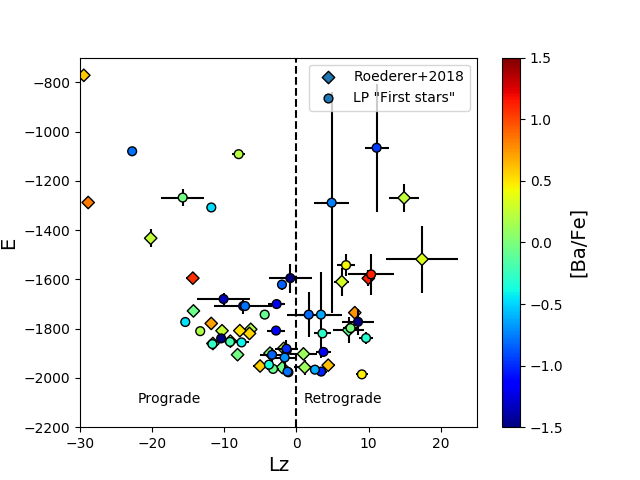}\\
\includegraphics[clip=true, trim={0cm 0cm 1cm 1cm},width=0.3\textwidth]{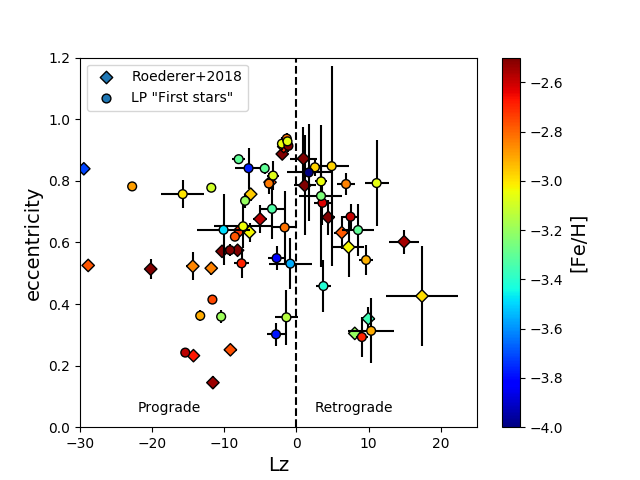}
\includegraphics[clip=true, trim={0cm 0cm 1cm 1cm},width=0.3\textwidth]{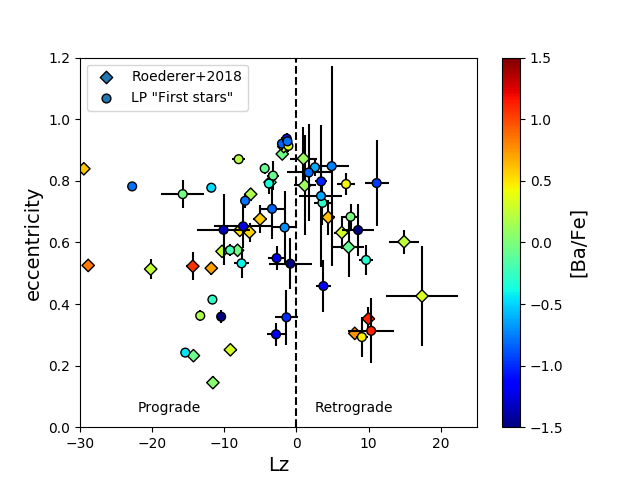}\\
\includegraphics[clip=true, trim={0cm 0cm 1cm 1cm},width=0.3\textwidth]{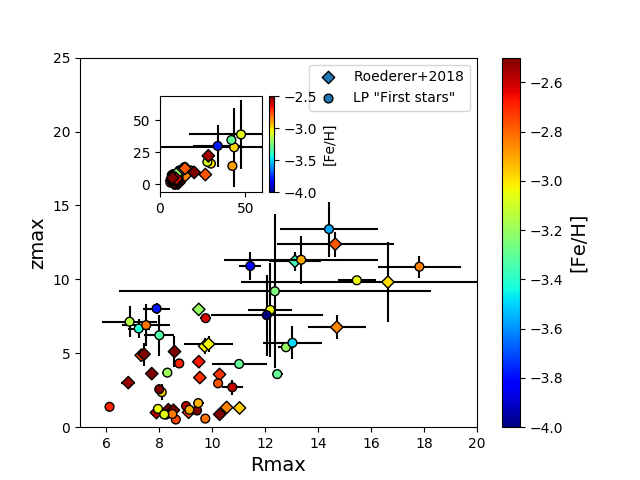}
\includegraphics[clip=true, trim={0cm 0cm 1cm 1cm},width=0.3\textwidth]{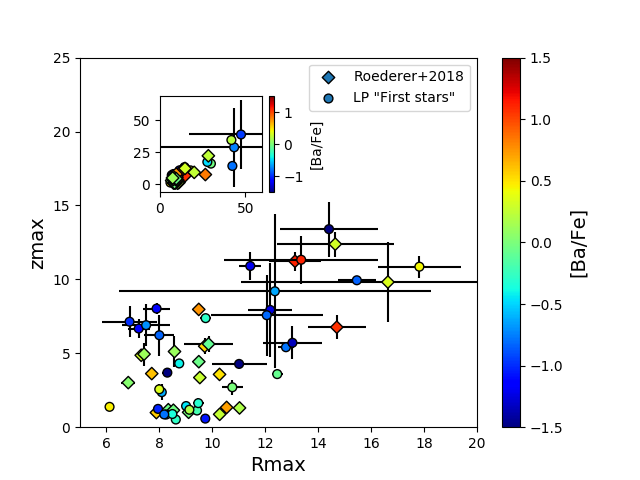}

\caption{\emph{From top to bottom: }Toomre diagram, $E-L_z$, $ecc-L_z$ and $R_{max}-z_{max}$ planes colour-coded by [Fe/H] (\emph{first column}), and  [Ba/Fe] (\emph{second column}). }
\label{kinorbchem}
\end{center}
\end{figure*}

\begin{figure*}
\begin{center}
\includegraphics[clip=true, trim={0cm 0cm 1cm 1cm},width=0.3\textwidth]{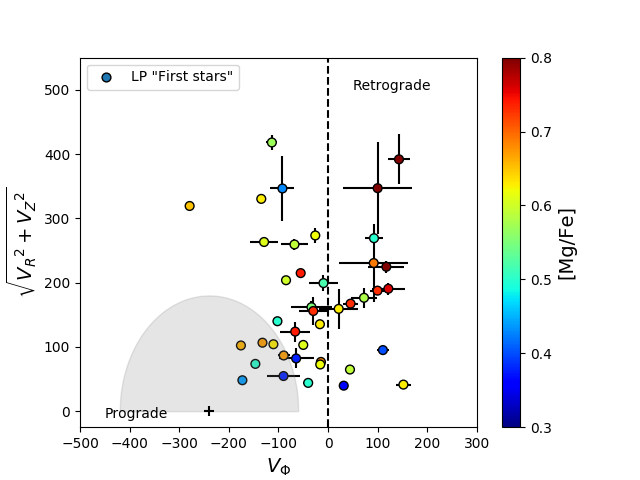}
\includegraphics[clip=true, trim={0cm 0cm 1cm 1cm},width=0.3\textwidth]{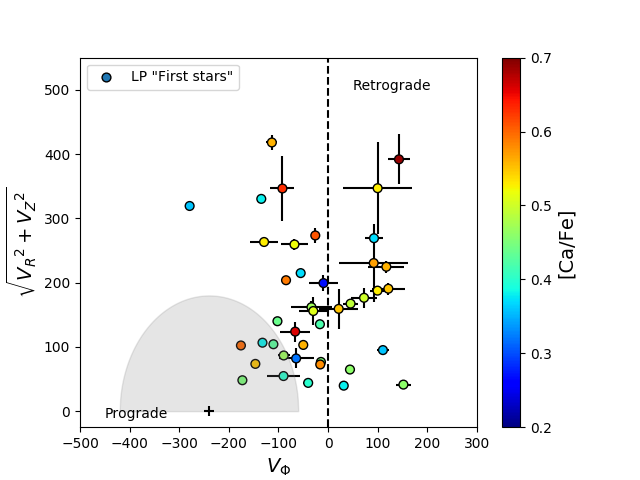}\\
\includegraphics[clip=true, trim={0cm 0cm 1cm 1cm},width=0.3\textwidth]{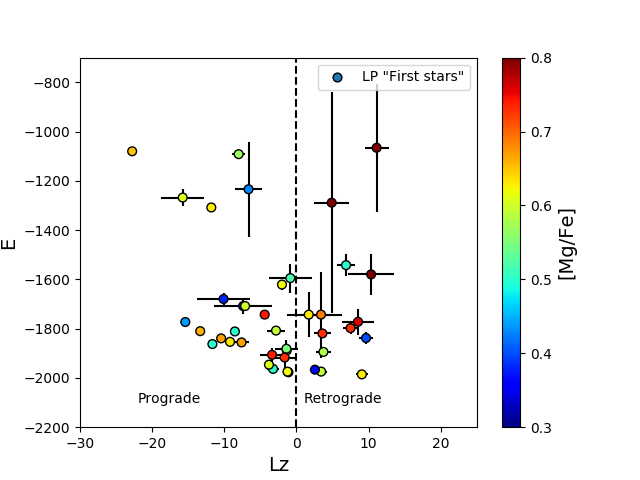}
\includegraphics[clip=true, trim={0cm 0cm 1cm 1cm},width=0.3\textwidth]{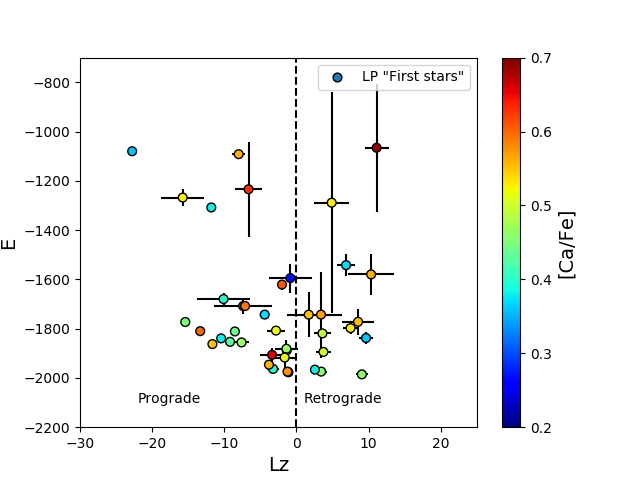}\\
\includegraphics[clip=true, trim={0cm 0cm 1cm 1cm},width=0.3\textwidth]{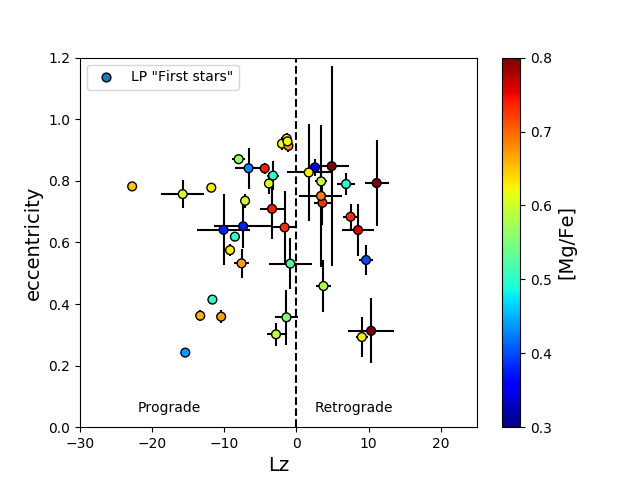}
\includegraphics[clip=true, trim={0cm 0cm 1cm 1cm},width=0.3\textwidth]{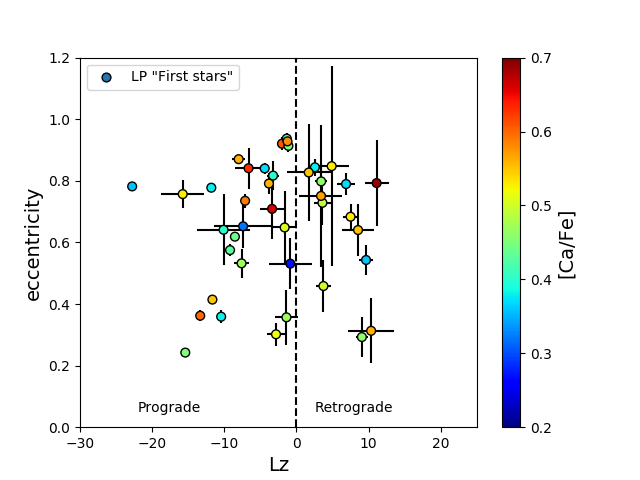}\\
\includegraphics[clip=true, trim={0cm 0cm 1cm 1cm},width=0.3\textwidth]{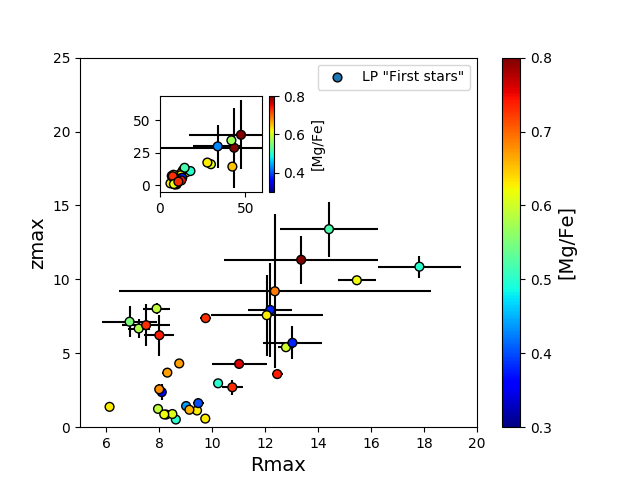}
\includegraphics[clip=true, trim={0cm 0cm 1cm 1cm},width=0.3\textwidth]{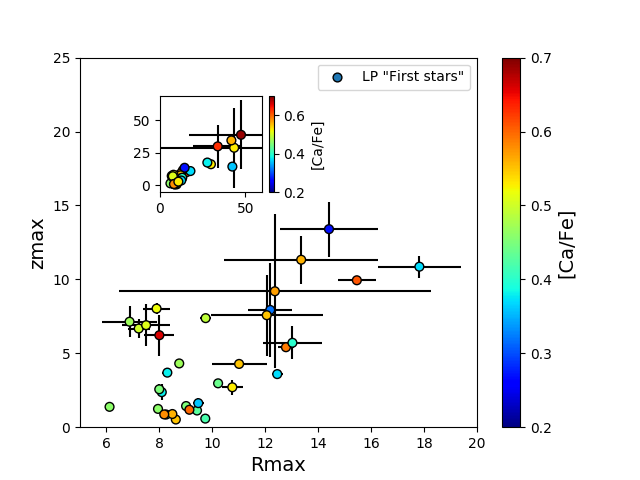}
\caption{\emph{From top to bottom: }Toomre diagram, $E-L_z$, $ecc-L_z$ and $R_{max}-z_{max}$ planes colour-coded by [Mg/Fe] (\emph{first column}), [Ca/Fe]  (\emph{second column}). }
\label{kinorbchem2}
\end{center}
\end{figure*}

\begin{figure*}
\begin{center}
\includegraphics[clip=true, trim={0cm 0cm 1cm 1cm},width=0.33\textwidth]{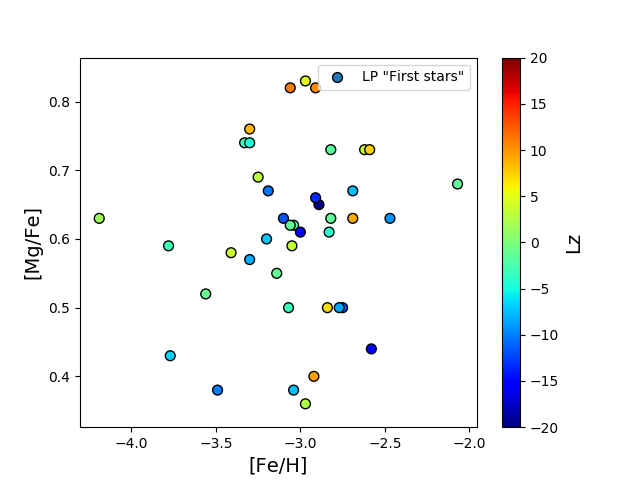}
\includegraphics[clip=true, trim={0cm 0cm 1cm 1cm},width=0.33\textwidth]{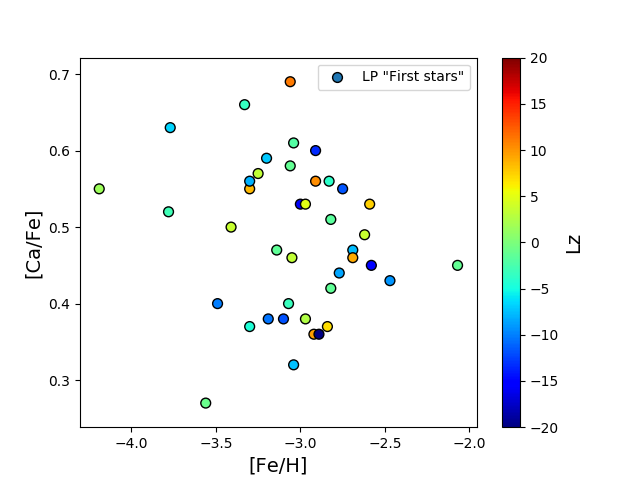}
\includegraphics[clip=true, trim={0cm 0cm 1cm 1cm},width=0.33\textwidth]{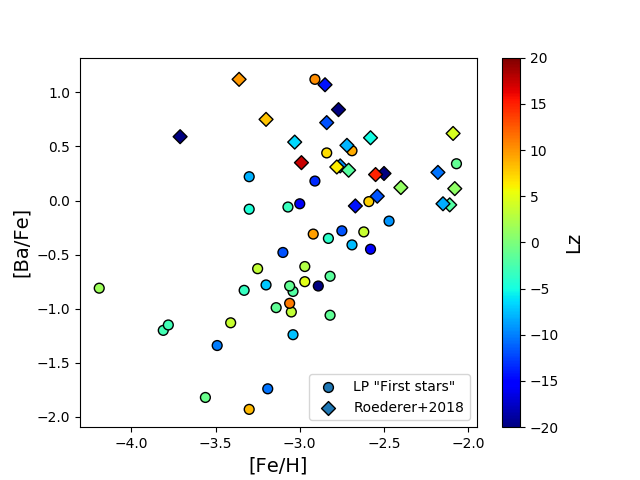}\\
\includegraphics[clip=true, trim={0cm 0cm 1cm 1cm},width=0.33\textwidth]{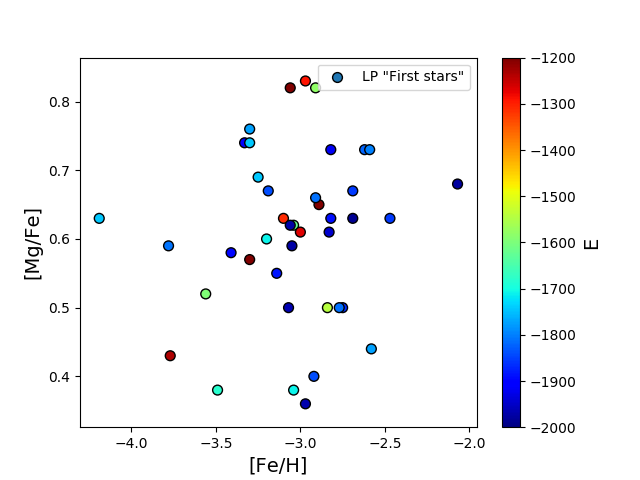}
\includegraphics[clip=true, trim={0cm 0cm 1cm 1cm},width=0.33\textwidth]{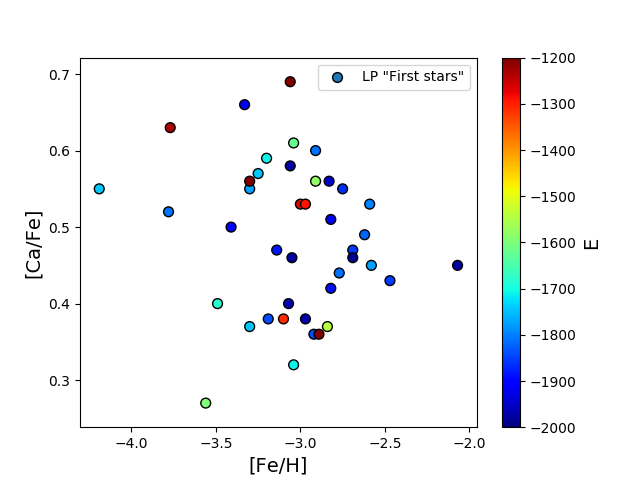}
\includegraphics[clip=true, trim={0cm 0cm 1cm 1cm},width=0.33\textwidth]{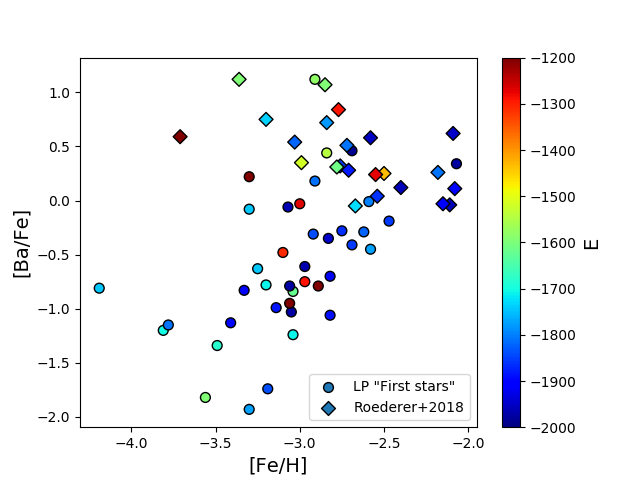}\\
\includegraphics[clip=true, trim={0cm 0cm 1cm 1cm},width=0.33\textwidth]{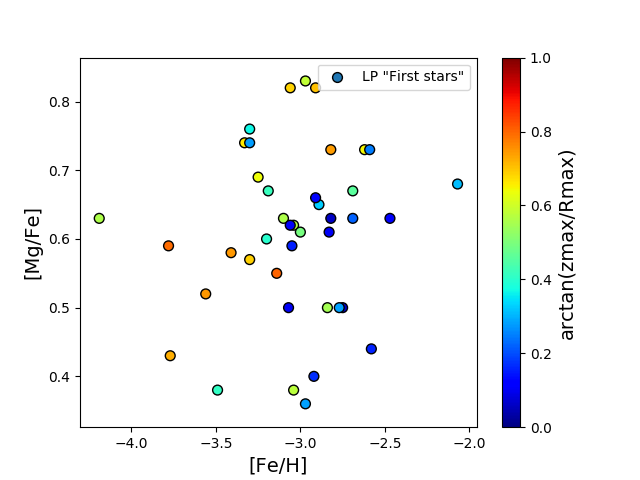}
\includegraphics[clip=true, trim={0cm 0cm 1cm 1cm},width=0.33\textwidth]{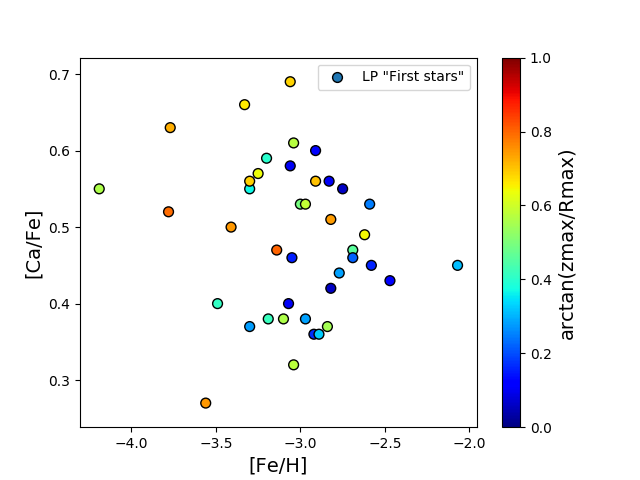}
\includegraphics[clip=true, trim={0cm 0cm 1cm 1cm},width=0.33\textwidth]{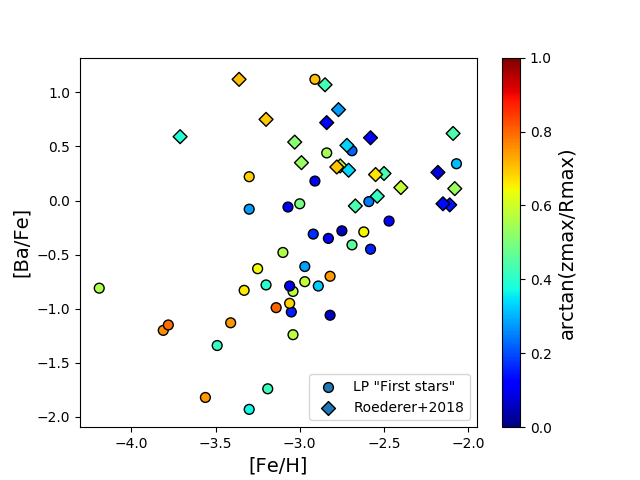}\\
\caption{\emph{From left to right:} Stars in [Mg/Fe]-[Fe/H],  [Ca/Fe]-[Fe/H] and  [Ba/Fe]-[Fe/H] planes colour-coded by their angular momentum $L_z$ (first row), their orbital energy $E$ (second row), and arctangent of the ratio $z_{max}/R_{max}$ (third row). Stars from Roederer et al 2018 are also plotted in the Ba/Fe]-[Fe/H] plane.}
\label{BaMgCaFe}
\end{center}
\end{figure*}

Figure~\ref{Toomre_insituacc} (middle and right columns) shows the Toomre diagram of stars in the in-situ and accreted sequences of the \citet{nissen10} and \gaia~~DR2-APOGEE samples. For the former sample, we used the classification given in the work by  \citet{nissen10} to distinguish in-situ (high-$\alpha$, thick disc and halo) from accreted (low-$\alpha$) stars. For the latter sample, to have a clean separation between these two populations in the [Mg/Fe]-[Fe/H] plane, we adopt a separation similar to that already used by \citet{dimatteo18}, defining in-situ and accreted stars, respectively, as stars above and below a separating line $\rm [Mg/Fe] = -0.26\times [Fe/H])$. In particular, accreted stars are defined as those with a [Mg/Fe] content at least 0.1 dex below this line, for any given value of [Fe/H], in order to minimise any contamination between the two samples.
For the in-situ population, we restrict the search to metallicities $\rm -1.5 \le [Fe/H] \le -0.5$, thus using a metallicity interval similar to that of the Nissen \& Schuster stars, while to also remove the contamination from the metal-poor thin disc, accreted stars from the low-$\alpha$ sequence are selected only if their [Fe/H] is lower than -0.7 dex. These choices of course drastically reduce the statistics of the accreted sequence, but have the advantage of minimising any contamination from the thick disc and thin disc populations.
The comparison between these two sets of in-situ and accreted stars and our LP sample shows that the accreted versus in-situ nature of each star in the sample is difficult to establish on the basis of the Toomre diagram alone. Stars with disc kinematics (i.e. stars with $\sqrt{{V_R}^2+{(V_\Phi-V_{LSR})}^2+{V_Z}^2} \le 180$~km/s) in the LP sample are mostly possibly in-situ, since all stars in the \citet{nissen10} sample with disc-like kinematics have an in-situ origin (see top-middle panel of Fig.~\ref{Toomre_insituacc}).\\ 
A second avenue for understanding the accreted or in-situ nature of stars in our sample is to compare them to the \citet{nissen10} and  \gaia~DR2-APOGEE samples in the $L_z-L_{perp}$ plane (see Fig.~\ref{LzLperp_insituacc}, top panels). Stars in the LP sample with $L_z \lesssim -10$ (8 out of 42) lie on a region mostly (but not excusively) occupied by in-situ stars, both of the \citet{nissen10} and of the   \gaia~DR2-APOGEE sample. The region with with $L_{perp} \gtrsim 13$ and $L_z \gtrsim -10$ (8 out of 42) is  mostly, if not only, populated by accreted stars both in the  \citet{nissen10} and in the   \gaia~DR2-APOGEE sample, but not by their in-situ counterpart. On the basis of this argument, we suggest that stars in the LP sample in this region of the $L_z-L_{perp}$ plane are mostly accreted. However, the nature of the majority of the sample, that is of LP First stars  with $L_z \gtrsim -10$ and $L_{perp} \lesssim 13$ is more difficult to establish: this region is the locus where stars of the Gaia-Sausage-Enceladus are redistributed, but also the locus of in-situ stars heated to halo-like kinematics by the accretion(s), the so-called `plume' identified in the $\rm v_{\Phi}-[Fe/H]$ plane \citep[see][]{dimatteo18}. \\ 
Some additional information comes from the comparison in the $ecc-L_z$ and $E-L_z$ planes (see Fig.~\ref{LzLperp_insituacc}, second and third rows): while we note, once again, the significant overlap of in-situ and accreted stars in kinematic spaces, and thus the difficulty of discriminating, overall, the accreted or in-situ nature of stars in our sample, the comparison with stars of the \citet{nissen10} sequences and the high and low-$\alpha$ sequences in the \gaia~DR2-APOGEE samples  allows us to understand the nature of some of the stars in the LP First stars sample. The two stars with $ecc < 0.4$ and low, negative $L_z$ that we already discussed in Fig.~\ref{kinorbALL}, are most probably in-situ (they lie within $2\sigma$ from the $ecc-L_z$ distribution of high-$\alpha$ stars in the  \gaia~DR2-APOGEE sample). Stars with energies $E > -1500$ are most probably accreted, because they lie in a region populated only by accreted stars of the \gaia~DR2-APOGEE  and \citet{nissen10} samples.\\
Finally, we conclude with a discussion of the $R_{max}-z_{max}$ plane (see Fig.~\ref{Rmaxzmax_insituacc_MP+NS} and Fig.~\ref{Rmaxzmax_insituacc}).
The existence of structures in the $z_{max}-R_{max}$ plane among stars with $\rm -1.5 \lesssim [Fe/H] \lesssim -0.5$ was already noted by \citet{schuster12} and has since been confirmed by the analysis of \gaia~~DR2 data thanks to the amount of available statistics and excellent quality of its astrometry \citep[see][]{haywood18}. The comparison with the \citet{nissen10} sample allows us to understand that the wedges found in the $z_{max}-R_{max}$ plane for LP First stars (see Fig.~\ref{kinorb_MP}) cannot be univocally linked to a different nature of stars than those that make it. One may be tempted to interpret stars in the lowest wedge, meaning those with disc-like kinematics, made exclusively by in-situ populations, and stars with halo-like kinematics made only of accreted material. The comparison with Nissen \& Schuster stars, however,  discourage this conclusion, since low-$\alpha$ sequence (accreted) stars are found in both samples, among stars with disc-like kinematics and among stars with halo-like kinematics as well (see Fig.~\ref{Rmaxzmax_insituacc_MP+NS}). In particular, some of the stars of the \citet{nissen10} sample, with low $z_{max}$, and classified as accreted by \citet{nissen10} on the basis of their low-$\alpha$ content, have $R_{max}$ inside 10~kpc, which is in the same region where all the LP First stars with low-$z_{max}$ are found.\\
To further discuss the in-situ/accreted nature of stars in our sample, we compare their distribution in the $z_{max}-R_{max}$ plane to that of in-situ and accreted stars from \gaia~~DR2-APOGEE (see Fig.~\ref{Rmaxzmax_insituacc}). As already suggested by the analysis of the Nissen \& Schuster sample, in-situ stars at few kpc from the Sun tend to populate a very narrow region of the $z_{max}-R_{max}$ plane, with 5~kpc $\lesssim R_{max} \lesssim$~15 kpc and $z_{max} \lesssim 10$~kpc. 
Only a handful of in-situ stars in this sample have indeed $R_{max} > 20$~kpc. Accreted stars, in turn, have a much broader distribution in this plane, significantly extending  to  $R_{max}$ and  $z_{max} $ above 20~kpc. Another way to appreciate the difference in orbital parameters of these two samples is by comparing the distribution of the arctan ($z_{max} / R_{max} $) (see second and fourth panels of Fig.~\ref{Rmaxzmax_insituacc}): the distribution of the arctangent of the $z_{max} / R_{max} $ ratio of in-situ stars shows a first local maximum at about 0.15, a dip at   0.2, a second local maximum at 0.3, and declines very rapidly at larger arctangent values. In-situ stars with disc-like orbits constitute 40\% of the total in-situ sample at these metallicities. The distribution of the arctangent of $z_{max} / R_{max} $ is, in turn, different for the accreted population: it shows no rapid decline above 0.2, but rather a  flat distribution,  and only very few stars (less than 10\%) are on disc-like orbits ($ arctan(z_{max} / R_{max} < 0.2$). By comparing the distribution of extremely metal-poor stars to those derived for in-situ and accreted stars in the APOGEE sample, we can conclude  that:
\begin{itemize}
\item LP stars with high values of the $\rm arctan(z_{max} / R_{max})$ (above $\sim$~0.5) are most probably accreted;
\item the significant fraction of stars with disc-like orbits ($\sim 24\%$ of the LP First stars sample have $\rm arctan(z_{max} / R_{max}) < 0.2$ ) cannot be explained if they all have an accreted origin. If all stars in our samples were indeed accreted, we would expect the fraction of stars with disc-like orbits to be less than 10\% -- by analogy with the fraction of accreted stars on disc-like orbits in the \gaia~DR2-APOGEE sample -- while this fraction is 2-2.5 times larger for stars in the LP. Based on the comparison with stars at higher metallicities, we thus conclude that $50-60\%$ of the VMP and EMP stars on disc-like orbits are indeed in-situ stars,  formed in the very early phases of the Milky Way disc assembly; 
\item stars with intermediate values of $\rm arctan(z_{max} / R_{max})$ (0.2--0.5) are probably a mixture of in-situ and accreted populations. By analogy with the \citet{nissen10} and  \gaia~DR2-APOGEE samples, we suggest that the in-situ stars in this intermediate range of $\rm arctan(z_{max} / R_{max})$ values are also stars of the early Milky Way disc (thus, they have the same origin of in-situ stars with $\rm arctan(z_{max} / R_{max}) < 0.2$), but they were heated to halo kinematics by the major accretion(s) experienced by the Galaxy over its lifetime. The relative weights of in-situ versus accreted stars in this range are currently  difficult to establish: they indeed have comparable kinematics, and they are indistinguishable both in [Fe/H], [Mg/Fe], [Ca/Fe], and [Ba/Fe] abundances, as discussed in the next section.
\end{itemize}

\section{Adding chemical properties to the analysis: linking $\alpha$ and $r-$abundances to kinematics}

Chemical information always provides important clues, either because it shows homogeneity, 
significant grouping, or dispersion in chemical abundance spaces. The sample of the LP First stars has been shown to be
 chemically homogeneous (see Sect.~\ref{CP} and Fig.~\ref{histd2}). The new information brought by \gaia~ is that these stars 
probably have different origins, and it is worth reconsidering these properties in the light of 
the analysis presented in the previous section. In this section, we thus aim to understand whether any of the kinematic properties discussed in the previous sections show any dependence on the chemical abundances of VMP and EMP stars. For stars in the LP First stars,  abundance ratios  of the $\alpha$ elements [Mg/Fe] and [Ca/Fe] corrected for non-LTE effects are available \citep[][and Fig.~\ref{histd2}]{AndrievskySK10,SpiteAS12}, together with [Ba/Fe] and the metallicity [Fe/H] (Table 1 and Fig.~\ref{histd2}).
Together with $\alpha-$abundances, which have been shown to be good discriminants between in-situ and accreted stars at higher metallicities \citep[see][]{nissen10, hayes18, haywood18, helmi18, dimatteo18, mackereth18}, we are also interested in analysing the dependency of kinematics on the [Ba/Fe] abundances. This because it has recently been suggested  that VMP and EMP r-rich objects may have an accreted origin \citep[see][]{roederer18}, and we would like to re-investigate this claim in the framework of our analysis.
R-rich objects, however, are relatively rare. In a sample of about 260 metal-poor stars studied within the framework of the Hamburg/ESO R-process enhanced star survey,  \citet{ChristliebBB04} and \citet{BarklemCB05} found only  24 stars meeting the criterion chosen by \citet{roederer18}: [Eu/Fe]>0.7\,dex, and there are only five such stars in our sample of extremely metal-poor stars. To enrich our sample of r-rich stars, we thus decided to add the stars from the r-rich sample of \citet{roederer18} as long as they meet our specifications: not C-rich, not binaries, $\rm [Fe/H]<-2 $ and $\sigma_\pi/\pi <0.2$. Stars in the \citet{roederer18}  sample that meet this criteria are 26.  As mentioned is Section~2, barium is a good proxy for europium. Because our sample lacks europium measurement for turn-off star, we rely on barium instead, and likewise for stars analysed by  \citet{roederer18}.
Their main characteristics are presented in Table~\ref{gaia2};  in Fig.\ref{histoFeH_MP+RO.png}, the histogram of their metallicity is compared to the histogram of the stars of the LP First stars; a comparison of the positions in the $XY$ and $RZ$ plane among these two samples is also given. For these `Roederer' stars, we adopted the radial velocity  deduced from high resolution spectra given by \citet{roederer18}. The computed distances are slightly different from the values given by \citet{roederer18}, since they  adopted  a Bayesian estimation of the distances of the stars, while we calculate the distances by inverting the parallaxes (corrected for the zero-point offset), as done for the previous samples. \\

In Figs.~\ref{kinorbchem} and \ref{kinorbchem2}, we analyse the same kinematic and orbital parameter spaces discussed in the previous sections (Toomre diagram, $E-L_z$, $ecc-L_z$ space and $R_{max}-z_{max}$ spaces), this time adding the abundance information (for positions, velocities, orbital parameters, and corresponding errors of the  \citet{roederer18} stars, see Tables~\ref{DataOrb1} and \ref{DataOrb2}). When analysing the dependence of the results on the [Fe/H] and [Ba/Fe] content, we also add the data from Roederer et al 2018 to increase the statistics. Whatever the abundance ratio analysed in  Fig.~\ref{kinorbchem}, no clear dependency of the kinematics on the chemistry of the stars appears evident. \\

It is however interesting to see that in the Toomre diagram, as well as in the $R_{max}-z_{max}$ plane, some of the stars in the  \citet{roederer18} sample have disc-like kinematics, thus confirming the results found for the LP First stars: eight out of the 42 stars in our sample, and 10 of the 26 stars from  \citet{roederer18} have low kinetic energies, compared to the LSR, with $\sqrt{{V_R}^2+{(V_\Phi-V_{LSR})}^2+{V_Z}^2} \le 180$~km/s, within 1-$\sigma$ error; 10 of the 42 stars in our sample, and seven of the 26 stars from  \citet{roederer18} have $\rm arctan(z_{max}/R_{max}) < 0.2$. This result does not confirm the conclusion of  \citet{roederer18}, who claimed that their sample did not include stars with disc-like kinematics, and we go on to discuss the reasons behind these different conclusions.\\

Evidence that there is no clear correlation between the [Fe/H], [Mg/Fe], [Ca/Fe] and  [Ba/Fe] content and the stellar kinematics is also provided in Fig.~\ref{BaMgCaFe}, where stars in our samples are plotted in the [Mg/Fe]-[Fe/H], [Ca/Fe]-[Fe/H], and [Ba/Fe]-[Fe/H] planes and colour-coded by their angular momenta $L_z$, their orbital energy $E$ and the arctangent of the $z_{max}/R_{max}$ ratio. Stars from  \citet{roederer18}  are also plotted in the [Ba/Fe]-[Fe/H] plane. 
In Roederer et al. (2018), the authors conclude that r-enhanced stars in particular are probably accreted because they do not find  r-enhanced objects with disc kinematics. 
The key difference with our analysis, however, is how \citet{roederer18} define disc stars, as those with $\sqrt{{V_R}^2+{(V_\Phi-V_{LSR})}^2+{V_Z}^2} \le 100$~km/s. That is, their definition of the disc kinematics is much more restrictive than ours, and more restrictive than definitions usually adopted in the literature. Clearly, among stars in the \citet{roederer18} sample, some have (thick) disc-like kinematics.
Moreover, because of such a restrictive definition of the disc kinematics, and because none of their stars fit this definition, \citet{roederer18}   conclude that their sample must exclusively contain  halo-like objects (either being genuine in-situ halo stars, or having been kinematically heated from the disc, or accreted).
However, they favour the accreted origin of stars in their sample because they are able to assign most of the stars to groups, or overdensities, in kinematic spaces. This approach is fraught with errors, because it has been shown that accretions with the mass ratio of the Gaia Sausage ($\gtrsim 1:10$) leave substantial grouping in kinematic spaces even for stars already present in the Galaxy at the time of accretion \citep{jeanbaptiste17}, and that several distinct groups can all have the same in-situ or accreted origin. In other words, belonging to a kinematic group is not in itself an indication that a star has been accreted \citep[again, see the results by][]{jeanbaptiste17}.
Among the 14 stars (out of 26) in the sample of  \citet{roederer18} that have $z_{max}<5$~kpc, all have $R_{max}<12$~kpc (and in fact most have $R_{max}<10$~kpc), and while this is not a guarantee that they are all (thick) disc objects, at least half must belong to this population (see previous section). \\We therefore cannot conclude with Roederer that disc stars are not r-enhanced, and that r-enhancement is a signature of an extragalactic origin of a star. This is discussed further in the next section.

\section{Discussion}\label{discussion}

We now summarise the results obtained in the previous sections. 
\begin{itemize}
\item Firstly, we showed that the kinematic and orbital properties of stars at $\rm [Fe/H]<\sim -0.4$
are surprisingly similar at all metallicities (see Fig.~6), as already emphasised in Sect.~4.1. By this we mean that the different samples occupy similar regions in kinematic spaces, independently of their [Fe/H]. This does not imply, of course, that their density distributions in any of those kinematic spaces are the same (because the latter property depends, at least, on the selection function of each sample). 
At high metallicities ($\rm -1.5 \lesssim [Fe/H] \lesssim -0.4$), these properties are dominated by two populations
that can clearly be identified as accreted stars from the so-called Gaia Sausage-Enceladus event \citep{belokurov18, haywood18, helmi18},
 and the $\alpha$-enhanced thick disc, partially heated to halo kinematics \citep{bonaca17, haywood18, dimatteo18, gallart19, belokurov19}. There is no evidence, from our analysis, that the data studied here contain other populations in significant proportions.
\item Secondly, the chemical abundances investigated here (barium, calcium, and magnesium) demonstrate no obvious dependence on kinematics, and in particular no obvious difference among possibly in-situ and accreted stars.
As long as it was assumed that an in-situ halo  was formed in the Galactic halo, the match found 
with accreted stars would have not been surprising: they could have formed from the same kind 
of environment, or Galactic sub-haloes. However, with the increasing evidence that in-situ stars at $\rm [Fe/H]< -0.5$
have more likely  formed in a (massive) disc, then the similarity in the abundance ratios between 
in-situ and accreted stars raises new questions.
\end{itemize}

\subsection{Evidence of the ubiquity of the Galactic thick disc, from $\rm [Fe/H]\sim -6$ to nearly solar $\rm [Fe/H]$}

\citet{sestito19} discuss the implication of their findings on the existence of ultra metal-poor stars on disc-like orbits for the first stages of the formation of the Milky Way. 
Among the different scenarios they propose, they envisage that stars in their sample with disc-like orbits could belong to the thick disc population, and
the link we make in this paper with populations at higher metallicities put this hypothesis on a more robust basis. \citet{sestito19} discuss three different scenarios for the formation of their ultra metal-poor disc stars: (1) they originate in the early Galactic disc; (2) they could have been accreted, even if the authors themselves cast doubts on this hypothesis, since the only (up to now) evident massive merger experienced by the Galaxy is represented by Gaia Sausage-Enceladus, which brought mainly stars on retrograde or low $L_z$ orbits; 
(3) they could be the remnants of massive building block(s), or clumps, of the proto-Milky Way that formed the backbone of the Milky Way disc.\\
While we note that scenarios (1) and (3) are not necessarily distinct and may have some overlap, we think that the continuity found here in the orbital and kinematic properties of  stars on the whole metallicity range support the conclusions of \citet{sestito19} that stars with prograde motion exist with similar characteristics from the highest metallicities ($\rm -1. \le [Fe/H] \le -0.3$), 
where they can formally be identified as $\alpha$-enhanced thick disc stars, to metallicities in the range $\rm -2 \le [Fe/H] \le -1$ \citep[the so-called metal-weak thick disc, see][]{norris85, morrison90, chiba00, beers02, reddy08, brown08, kordopatis13, hawkins15, li17, hayes18, dimatteo18} all the way down to metallicities as low as $\rm [Fe/H]\sim -6$.
So the simplest deduction that can be made is that very metal-poor, extremely metal-poor and ultra iron-poor stars with disc-like kinematics (or at least the majority, see Sect.~4.2) are stars born in the Galaxy itself in the very early phases of its formation. These stars all experienced the same heating events culminating with the end of the accretion of Gaia Sausage-Enceladus, visible at $\rm [Fe/H]\sim -0.3, and~[Mg/Fe]\sim 0.2$, that is about 10~Gyr ago \citep{dimatteo18, gallart19, belokurov19}. This ultra iron-poor thick disc is of fundamental importance to trace the disc formation back in time, up to the most metal-poor and first stars discovered up to date. We note that \citet{beers95} suggested that the metal-weak thick disc extended
below metallicity --2.0, and in fact, one of the metal-weak thick disc candidates,
observed at spectral high resolution by \citet{bonifacio99}, CS\,29529-12 with [Fe/H]=--2.27,
is indeed on a thick disc orbit, according to its \gaia~DR2 parallax and proper motions.\\

How do these results compare with state-of-the-art simulations? Simulations have indicated consistently over the last 15 years that the most ancient stars are concentrated in the inner part of Milky-Way-type galaxies, \citep[see, for example][]{diemand05, gao10,  tumlinson10, ishiyama16, griffen18}, and are usually made of stars formed in sub-haloes that merge to form a central concentration. Neither of the two populations of VMP and EMP discussed seems to be consistent with this picture. The accreted population, like its counterpart at high metallicity, could be explained by a single event, the Gaia Sausage, whose stars redistribute over a large range of kinematic and orbital properties. The results presented here and in \citet{sestito19} indicate that at least a fraction of the most metal-poor and possibly oldest in-situ stars formed in the Galaxy have similar kinematic properties to what is known nowadays, at higher metallicities, like the thick disc population. Several studies \citep[e.g.][]{purcell10, qu11a, mccarthy12} have found that an in-situ halo may originate from a disc of stars heated by interactions and accretions. As a matter fact, \citet{mccarthy12} find that their in-situ halo stars have formed at relatively late times $z<2$. The most recent simulations may point in the right direction. For instance, \citet{pillepich19} present simulations where galaxies  are rotationally supported very early on, with $\rm V_{rot}/\sigma > 2-3$ already at redshift $z=5$ and below. However, at this redshift, stars are predicted to have metallicities around $\rm [Fe/H] \sim-2$ \citep{tumlinson10, starkenburg17}, and so are at the upper limit of the sample studied here. The data therefore indicate that the Milky Way disc probably settled at redshift $z>5$, with stars at metallicities of about $\rm [Fe/H]=-4$ or $-5$  formed at redshift $z>6$. This is yet to be found in simulations and in observations of disc morphologies at these redshifts.

\subsection{Accreted stars at $\rm [Fe/H] \le -2$ and the difficulty of using chemical abundances to discriminate the nature of very metal-poor and extremely metal-poor  populations}

Regarding stars with halo kinematics, \citet{sestito19} divide those in their sample in inner (apocentres inside 30~kpc) and 
outer (apocentres greater than 30~kpc) halo, suggesting a possible different origin.
The comparison we made in Sect.~4.1 with higher metallicity samples, and in particular with the \gaia~DR2-APOGEE stars, 
suggests that stars with high apocentres in the  \citet{sestito19} and in our samples could all be related to the same accretion event known as Gaia Sausage-Enceladus. The homogeneity of the $\alpha$-element abundance ratios 
in the metallicity range covered by the LP First stars sample also supports this view\footnote{Here we mean that the LP First stars  are homogeneous given the uncertainties on the abundance estimates (of the order of 0.15 dex for [Ca/Fe] and 0.2 dex for [Mg/Fe], see Fig.~2). In these very metal-poor stars with a metallicity far from the Sun, the individual uncertainties do not come from the observed spectra, but mainly from the uncertainties on the model parameters.}. Inner halo stars appear, in turn, as a $melange$ of stars from the Gaia  Sausage-Enceladus and of stars of the early Milky Way disc  heated to halo kinematics.\\
The situation for enhanced r-process elements is more complex.
We do not confirm the conclusions of  \citet{roederer18} about the nature of stars with enhanced r-process elements: as traced here by barium,  there is no evidence that these stars have all an accreted origin. More precisely, the barium-rich stars studied 
by \citet{roederer18}  show no sign, from their kinematics and orbital properties, that they originate from any 
other population than those associated with the Gaia Sausage-Enceladus or the early Galactic disc, partially heated to halo kinematics. We emphasise once more that the distribution of stars in several $N$ -independent groups in kinematic spaces is not an indication either of their accreted origin or of their  belonging to $N$ -distinct satellites \citep{jeanbaptiste17}. Some of the stars studied by  \citet{roederer18} clearly have (thick) disc-like orbits,  or lie, in kinematic spaces, in regions occupied at higher metallicities, by the Gaia Sausage.  Before making the hypothesis that these stars are associated with low-mass dwarfs,or ultra-faint galaxies, as suggested by  \citet{roederer18},  we first need to robustly demonstrate that they are neither stars of the disc or of the in-situ halo (that is the heated disc), nor are they associated with Gaia Enceladus.\\

\section{Conclusions}\label{conclusions}

In this work, we analyse a set of very metal-poor and extremely metal-poor stars for which accurate chemical abundances and radial velocities have been obtained
as part of the ESO LP First stars. Combining spectroscopic information with the astrometric data available from the \gaia~DR2, and comparing this sample to stars at lower \citep{sestito19} and higher \citep{nissen10, dimatteo18} metallicities, we make the following conclusions:
\begin{itemize}
\item At all metallicities,  from $\rm [Fe/H]\sim -6$ to $\rm [Fe/H] \sim -0.4$, stars show very similar kinematic properties. By analogy with stars at higher metallicities ($\rm [Fe/H]\sim -2$ to $\rm [Fe/H] \sim -0.4$), these kinematic properties can be interpreted as the presence of two dominant populations at  $\rm -6 \lesssim [Fe/H] \lesssim -2$: a disc, partially heated to halo kinematics, and the low-metallicity stars of the Gaia Sausage-Enceladus satellite.
\item The Galactic disc extends not only to the metal-poor regime (the so-called ``metal-weak thick disc"), but in fact down to  metallicities as low as  $\rm [Fe/H]\sim -6$.  In other words, an ultra metal-poor thick disc exists,  which constitutes the extremely metal-poor tail of the 'canonical' Galactic thick disc and of the metal-weak thick disc.
\item At all metallicities from $\rm [Fe/H]\sim -6$, up to  $\rm [Fe/H] \sim -0.4$, we suggest that these early disc stars have similar kinematic properties, because they experienced the same violent heating processes, that ended with the accretion of the Gaia Sausage-Enceladus satellite, about 10~Gyr ago \citep{dimatteo18, gallart19, belokurov19}. We note that it is still possible that the disc experienced some cooling in the first 2-3~Gyr of its formation \citep[see, for example][]{samland03}, but the signatures of this process are currently difficult to identify, on the one side, because the samples of stars at  $\rm [Fe/H] < -2$ still suffer from low statistics, and, on the other side, because of the concurrent kinematic heating the disc experienced, which partially hides the cooling signature, if present.
\item Besides the disc, the halo population that is present all over the $\rm -6 \lesssim [Fe/H] \lesssim -0.5$ range is a mix of disc stars heated to halo kinematics \citep[the same phenomenon experienced by canonical thick disc stars, see][]{dimatteo18, belokurov19}, and accreted stars, possibly all associated with Gaia Sausage-Enceladus. There is no evidence of other significant  populations in this sample of stars, which, we remind the reader, extends up to 8~kpc from the Sun.
\item Surprisingly, we find no clear relation between the kinematics and the chemical abundances of stars at $\rm [Fe/H] < -2$: no dependence on $\alpha$-elements as Mg and Ca, and no dependence on r-process elements, as barium. In this respect, we cannot confirm the results by  \citet{roederer18}, regarding the possible exclusively accreted nature of r-rich stars at $\rm [Fe/H] < -2$. R-rich stars indeed appear to be a mixture of disc stars, partially heated to halo kinematics, and accreted stars with kinematic properties compatible to those of Gaia Sausage-Enceladus.
\end{itemize}

These results raise a number of questions. In-situ and accreted stars at $\rm [Fe/H] < -2$ seem to share the same chemical abundances, both in Mg, Ca and Ba. This is surprising given the stochasticity of the star formation process at those early epochs of Galaxy formation, and given the different sites where these stars originated (Galaxy versus Gaia Sausage-Enceladus).\\ 
\citet{limongi05}  showed that the average abundances
  of the LP giant stars \citep{CayrelDS04} can be reproduced rather
well with a single zero metallicity supernova of moderate mass (20-50 solar
masses), or by a population of zero-metallicity stars with
a standard Salpeter Initial Mass Function (IMF) with index --2.35.
More recently, \citet{ishigaki18}, using the chemical abundances of 
about 200 EMP stars, concluded that the masses of the first generation
of stars were predominantly below 40 solar masses.
The general conclusion is that there is no need for a top-heavy
IMF of zero-metallicity stars to reproduce the abundance ratios
observed in VMP and EMP stars. 
Our finding that the LP First stars  were probably formed in two distinct galaxies,
the Milky Way and Gaia Sausage-Enceladus, implies that this conclusion
holds for both galaxies. This strongly supports the notion
that the IMF is universal, even at zero metallicity. 
It is, of course, difficult to reconcile these observations
with the theoretical predictions of zero-metallicity
star formation that require a top-heavy IMF \citep[e.g.][and references therein]{bromm04}. We note, however, that recent high-resolution simulations of star formation predict an IMF that 
is not necessarily top heavy \citep[e.g.][]{greif07, clark11}.\\
Finally, to get insights into the nature of stars at $\rm [Fe/H] < -2$, it has been necessary to compare their kinematic properties to those of more metal-rich stars ($\rm -2 \lesssim [Fe/H] \lesssim -0.5$). This latter interval is fundamental, because here we clearly see two distinct main populations, the ($\alpha$-enhanced) thick disc, partially heated to halo kinematics, and stars from the Gaia Sausage-Enceladus. 
The possibility to distinguish these two chemical sequences, at $\rm -1.5 \lesssim [Fe/H] \lesssim -0.5$, and to study their corresponding kinematic properties is thus also vital  to interpreting the nature and origin of stars at lower metallicities, where $\alpha-$abundance patterns appear very homogeneous, and [Ba/Fe] seems not to be discriminant. 
In this context, we need to be aware that we still need to robustly establish that the low-$\alpha$ sequence discovered by \citet{nissen10} and then discussed in a number of subsequent works  is made of stars from one unique satellite (i.e. Gaia Sausage-Enceladus) and it is not hiding multiple accreted populations, possibly of similar masses \citep[see, for example][]{snaith16}. The link between this sequence and the metal-poor thin disc, that is the outer disc of the Milky Way, is also yet to be completely understood \citep[see the recent work by][] {buck19}. Digging into this low-$\alpha$ sequence, its connection with the outer disc, and its constituents, is necessary for a deeper understanding of stars over the whole range of metallicities,  from $\rm [Fe/H]\sim -6$ to $\rm [Fe/H] \sim -0.5$, thus well beyond the limited [Fe/H] range where this low-$\alpha$ sequence is currently found.

\begin {acknowledgements} 
The authors wish to thank C. Gallart, the referee of this paper, for a prompt and very constructive report. PDM acknowledges enriching discussions with C.~Brook, B.~Gibson, and D.~Katz on a first  version of this manuscript.
This work has been supported by the ANR (Agence Nationale de la Recherche) through the MOD4Gaia project (ANR-15- CE31-0007, P.I.: P. Di Matteo).\\
This work has made use of data
from the European Space Agency (ESA) mission Gaia
(https://www.cosmos.esa.int/gaia), processed by the
Gaia Data Processing and Analysis Consortium (DPAC,
https://www.cosmos.esa.int/web/gaia/dpac/consortium).
Funding for the DPAC has been provided by national
institutions, in particular the institutions participating
in the Gaia Multilateral Agreement. \\This research has
made use of the SIMBAD database, operated at CDS,
Strasbourg, France.\\ Funding for the Sloan Digital Sky
Survey IV has been provided by the Alfred P. Sloan
Foundation, the U.S. Department of Energy Office of
Science, and the Participating Institutions. SDSS-IV
acknowledges support and resources from the Center
for High-Performance Computing at the University of
Utah. The SDSS web site is www.sdss.org. SDSS-IV
is managed by the Astrophysical Research Consortium
for the Participating Institutions of the SDSS Collaboration
including the Brazilian Participation Group,
the Carnegie Institution for Science, Carnegie Mellon
University, the Chilean Participation Group, the French
Participation Group, Harvard-Smithsonian Center for
Astrophysics, Instituto de Astrof$\rm \acute{i}$sica de Canarias,
The Johns Hopkins University, Kavli Institute for the
Physics and Mathematics of the Universe (IPMU)
/ University of Tokyo, Lawrence Berkeley National
Laboratory, Leibniz Institut f$\rm\ddot u$r Astrophysik Potsdam
(AIP), Max-Planck-Institut f$\rm\ddot u$r Astronomie (MPIA
Heidelberg), Max-Planck-Institut f$\rm\ddot u$r Astrophysik (MPA
Garching), Max-Planck-Institut f$\rm\ddot u$r Extraterrestrische
Physik (MPE), National Astronomical Observatories of
China, New Mexico State University, New York University,
University of Notre Dame, Observat$\rm\acute{a}$rio Nacional /
MCTI, The Ohio State University, Pennsylvania State
University, Shanghai Astronomical Observatory, United
Kingdom Participation Group, Universidad Nacional
Aut$\rm\acute{o}$noma de M$\rm\acute{e}$xico, University of Arizona, University
of Colorado Boulder, University of Oxford, University of
Portsmouth, University of Utah, University of Virginia,
University of Washington, University of Wisconsin,
Vanderbilt University, and Yale University.
\end{acknowledgements}

\bibliographystyle{aa}
\bibliography{biblio}

\appendix

\section{On the kinematic properties of the carbon enhanced metal-poor stars in the LP First stars sample}\label{CEMP}

\begin{figure*}
\begin{flushleft}
\includegraphics[clip=true, trim = {0cm 0cm 0cm 0cm},width=0.4\textwidth]{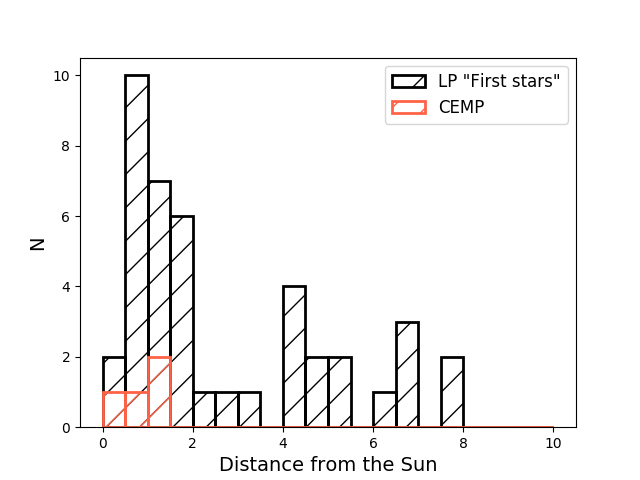}\\
\includegraphics[clip=true, trim = {0cm 0cm 0cm 0cm},width=0.4\textwidth]{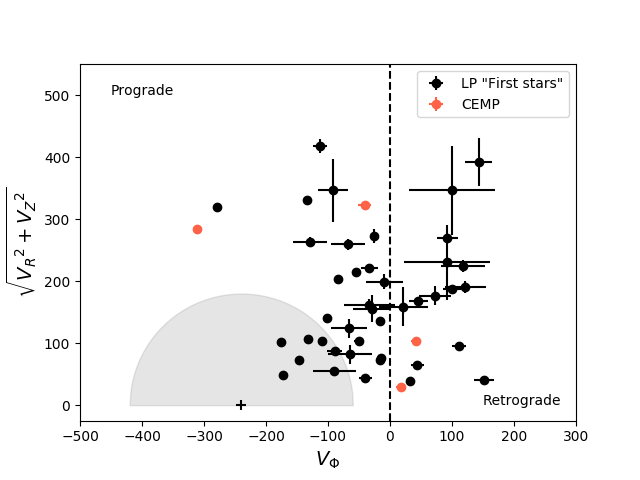}
\includegraphics[clip=true, trim = {0cm 0cm 0cm 0cm},width=0.4\textwidth]{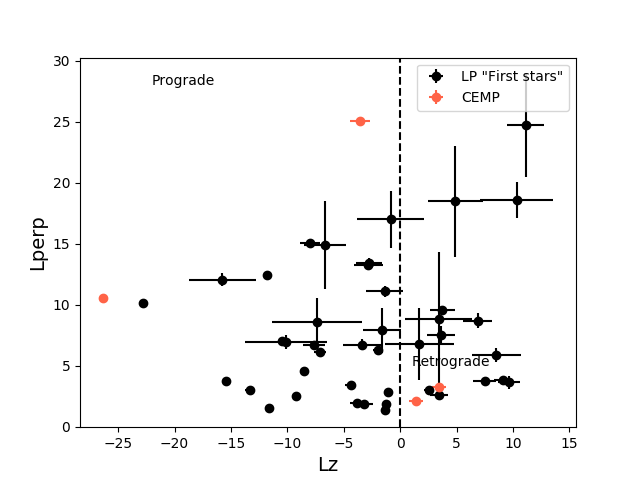}\\
\includegraphics[clip=true, trim = {0cm 0cm 0cm 0cm},width=0.4\textwidth]{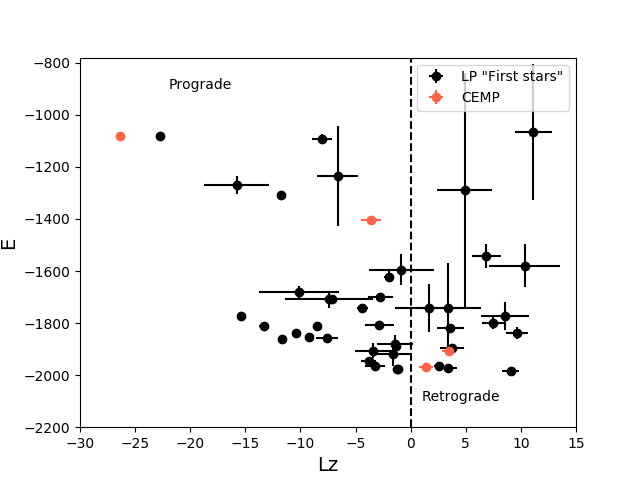}
\includegraphics[clip=true, trim = {0cm 0cm 0cm 0cm},width=0.4\textwidth]{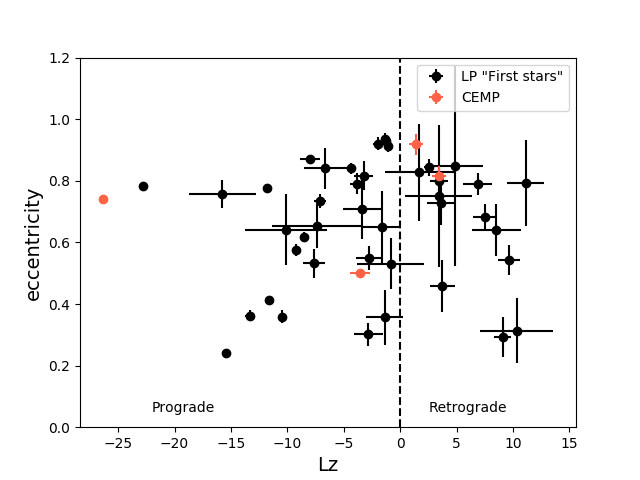}
\end{flushleft}
\caption{\emph{Top row:} Histogram (hatched orange) of distances (in kpc) to the Sun, of the sample of CEMP stars in the ESO LP First stars, with \gaia~DR2 relative errors on parallaxes smaller  than 20\%. For comparison, the histogram of the  distances to the Sun of stars in the LP First stars studied in this paper, is also shown (hatched black). \emph{Middle row, left panel:} Toomre diagram for CEMP stars (orange dots).  Stars from the LP First stars sample studied in this paper are also shown (black dots). The vertical dashed line separates prograde from retrograde motions. Velocities are in units of km/s. The grey  area separates stars with $\sqrt{{V_R}^2+{(V_\Phi-V_{LSR})}^2+{V_Z}^2} \le 180$~km/s from stars with higher relative velocities with respect to the LSR. \emph{Middle row, right panel:} Distribution of the CEMP stars (orange dots) in the $\rm L_z-L_{perp}$ plane.  Stars from the LP First stars sample studied in this paper are also shown (black dots). Angular momenta are in units of 100~kpc~km/s. \emph{Bottom row, left panel:} Distribution of CEMP stars (orange dots) in the $\rm E-L_z$ plane.  Stars from the LP First stars sample studied in this paper are also shown (black dots). The vertical dashed line separates prograde from retrograde motions. Angular momenta are in units of 100~kpc~km/s, energies in units of 100~$\rm km^2/s^2$.  \emph{Bottom row, right panel:} Distribution of CEMP stars (orange dots) in the $\rm eccentricity-L_z$ plane.  Stars from the LP First stars sample studied in this paper are also shown (black dots).\label{CEMP_kins}}
\end{figure*}

Among the stars in  the LP First stars sample, six are carbon-enhanced metal-poor stars \citep[CEMP, see][]{BeersChristlieb05}, and, for the sake of homogeneity, were excluded from our study. These stars are: \mbox{CS22949-37},  \mbox{CS22892-52} (\gaia~DR2 ID=\mbox{6826025986350385280}),  \mbox{CS29497-30} (\gaia~DR2 ID=\mbox{2347402354016302208}),  \mbox{CS31080-95} (\gaia~DR2 ID=\mbox{4790354875531489024}),  \mbox{CS22958-42} (\gaia~DR2 ID=\mbox{4718427642340545408}),  \mbox{CS29528-41} (\gaia~DR2 ID=\mbox{5131487257219876864}). Of those six stars, the first does not have a measured parallax in \gaia~DR2, while the second has a parallax relative error larger than 20\%. We are thus left with four stars only, whose distances have been estimated by inverting their parallaxes, and whose galactocentric positions and velocities have been derived as described in Sect.~3.1. For all four stars, proper motions are from \gaia~DR2, while their line-of-sight velocities are taken from the work of  \citet{SivaraniBB06}, except for CS 22892-052, whose adopted line-of-sight velocity is taken from the work of \citet{roederer14}.
The derived distances from the Sun, and distributions in the Toomre diagram, $L_z-L_{perp}$, $E-L_z$ and $ecc-L_z$ planes are reported in Fig.~\ref{CEMP_kins}, and there they are compared to the corresponding properties of the LP First stars sample analysed in this paper.  As shown in this figure, the CEMP stars in the sample do not show different kinematic properties to those of the main LP sample. We emphasise, however, that the statistics of CEMP stars in the LP sample is too low to derive more robust conclusions. 

\section{Tabular data}\label{tabdata}
 In this Appendix, we report the main parameters (parallaxes, proper motions, radial velocities, [Fe/H], and [Ba/Fe]) for stars of the LP First stars programme (Table~\ref{gaia1}), for stars of the \citet{roederer18} sample (Table~\ref{gaia2}) and for stars of the \citet{sestito19} sample  (Table~\ref{gaia3}). Table~\ref{desig} gives the correspondence between the   name of the star and its \gaia~DR2 ID. Tables~\ref{DataOrb1} and ~\ref{DataOrb2} give the galactocentric positions and velocities, orbital parameters and relative errors for stars of the LP First stars
programme, of the \citet{roederer18}  and \citet{sestito19} samples. We note that two stars among the 15 of the \citet{sestito19} sample analysed in this paper are common to the LP First stars: CS 22885-0096  (\gaia~DR2 ID=6692925538259931136) and HE 0044-3755 (\gaia~DR2 ID=5000753194373767424). As a consequence, they are not reported in the list of \citet{sestito19} stars given in Tables~\ref{DataOrb1} and~\ref{DataOrb2}, but only in the list of LP First stars.

\begin{table*}
\caption{Main \gaia~data of the stars studied in the frame of the LP First stars. In this table, we state the [Fe/H], the metallicity of the star, [Ba/Fe] (a proxy for [Eu/Fe]), the \gaia~equatorial coordinates ra and dec (epoch 2015.5), the proper motions pmra and pmdec (in mas), the line-of-sight velocity (in \kms) measured by \gaia~and measured on the UVES spectra (Gaia RVel and UVES RVel) with the corresponding errors, the \gaia~DR2 parallax corrected by the zero-point offset, the error  $\sigma_\pi$ and $\sigma_\pi/\pi$, and finally the distances of the stars $d$ (in kpc) deduced from the parallax. The stars were kept only when $\sigma_\pi/\pi < 0.20$.
}      
\label{gaia1}    
\scalefont{0.85}
\centering                          

\end{table*}

\begin{table*}
\caption{Main \gaia~data of the stars selected in the sample  of \citet{roederer18}.
The columns are the same as in Table \ref{gaia1} 
}      
\label{gaia2}    
\scalefont{0.85}
\centering                          
\end{table*}

\begin{table*}
\caption{Main \gaia~data of the stars selected in the sample  of \citet{sestito19}. Five of these stars are ultra iron-poor stars with [Fe/H]<-4.5.
The other stars have metallicities close to the metallicity of the LP First stars sample, but many of them are C rich.
The columns are the same as in Table \ref{gaia1} 
}      
\label{gaia3}    
\scalefont{0.85}
\centering                          
\end{table*}

\begin{table*}
\caption{Correspondance name of the star, Gaia Designation; and Gaia Designation, name of the star, for the stars of the different samples. All these stars used in the calculations have a parallax error less than 20\% in the \gaia~DR2.}
\label{desig}
\scalefont{0.68}
\centering                       
\begin{tabular}{lllllrr}
\hline
\\
\multicolumn{5}{c}{{\bf ~~Dwarfs ESO LP "First stars"}}\\
BS 16023-046               & DR2 1256787527655339520    &~~|~~~&DR2 1154852693702722432  & BS 16968-061         \\
BS 16076-006               & DR2 3954415903126795136    &~~|~~~&DR2 1256787527655339520  & BS 16023-046         \\
BS 16968-061               & DR2 1154852693702722432    &~~|~~~&DR2 2329078133745332608  & CS 22966-011         \\
BS 17570-063               & DR2 2801174135693760256    &~~|~~~&DR2 2503556060544507008  & CS 31061-032         \\
CS 22177-009               & DR2 4890881265153979904    &~~|~~~&DR2 2676443097097288704  & CS 22965-054         \\
CS 22888-031               & DR2 6553564535381928320    &~~|~~~&DR2 2801174135693760256  & BS 17570-063         \\
\hline  
\end{tabular}
\end{table*}

\begin{table*}
\caption{Galactic coordinates X, Y, Z are in pc and the velocities in \kms. }      
\label{DataOrb1}    
\scalefont{0.68}
\centering                          
\end{table*}

\begin{table*}
\caption{Calculated Orbital Parameters}      
\label{DataOrb2}    
\scalefont{0.67}
\centering                          
\end{table*}

\section{Orbits}   \label{orbits}

In this Appendix, we show the orbits of the 42 stars in the LP First stars sample. Both the meridional plane, $R-Z$, and the projection on the $X-Y$ plane, are shown.

\begin{figure*}
\begin{centering}
\includegraphics[clip=true, trim = {0cm 0cm 1cm 0cm},width=4.5cm]{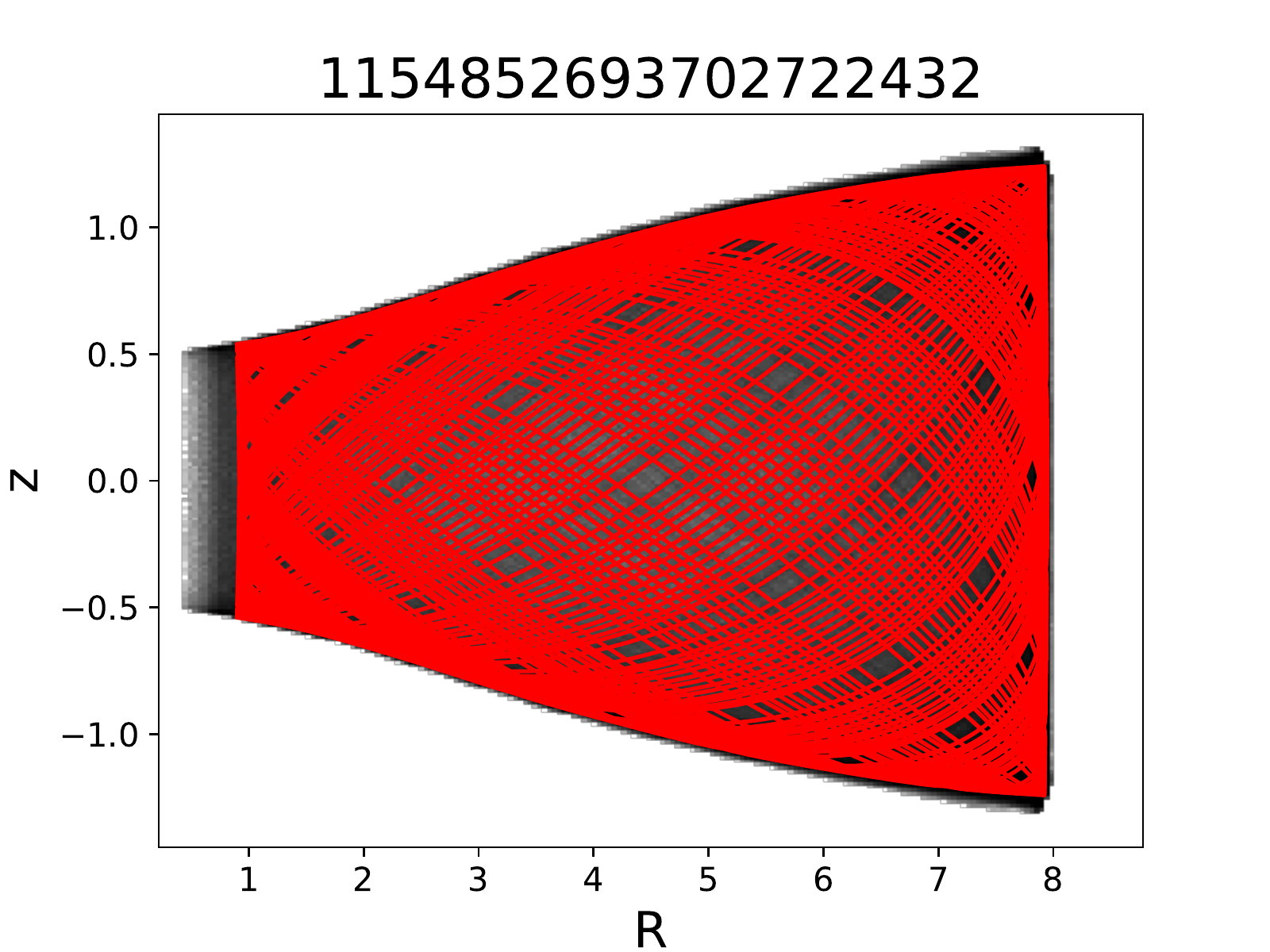}
\includegraphics[clip=true, trim = {0cm 0cm 1cm 0cm},width=4.5cm]{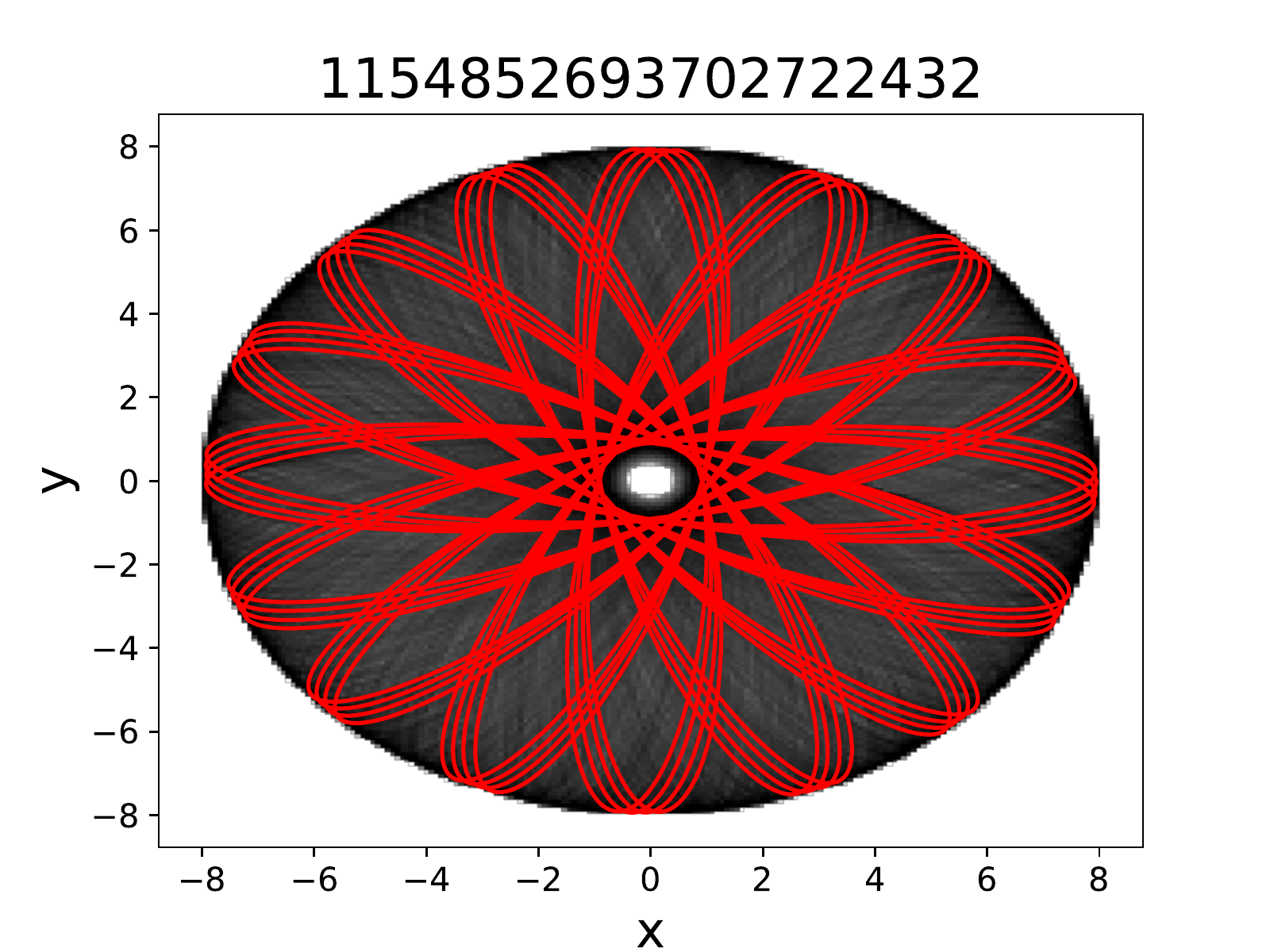}
\includegraphics[clip=true, trim = {0cm 0cm 1cm 0cm},width=4.5cm]{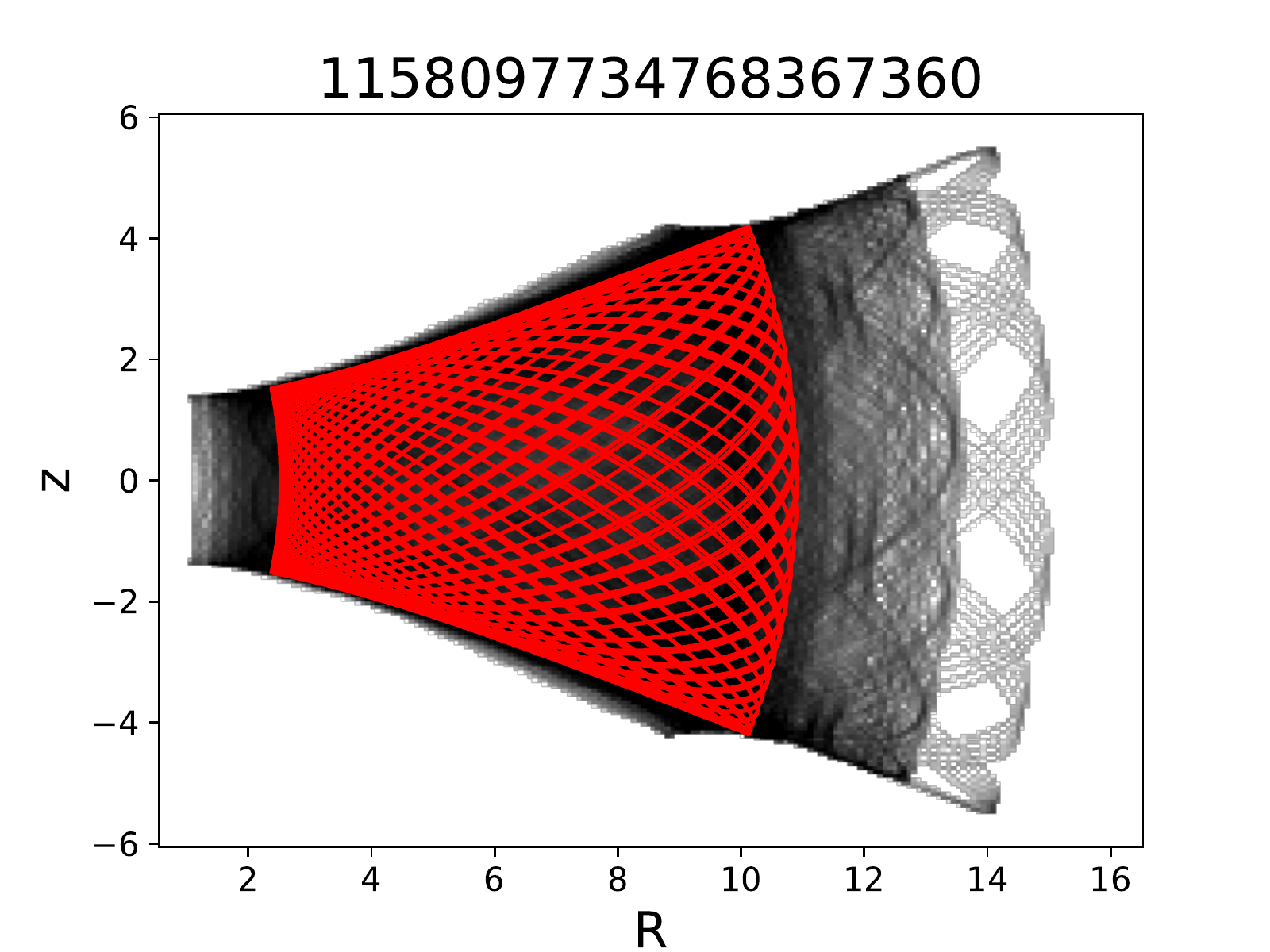}
\includegraphics[clip=true, trim = {0cm 0cm 1cm 0cm},width=4.5cm]{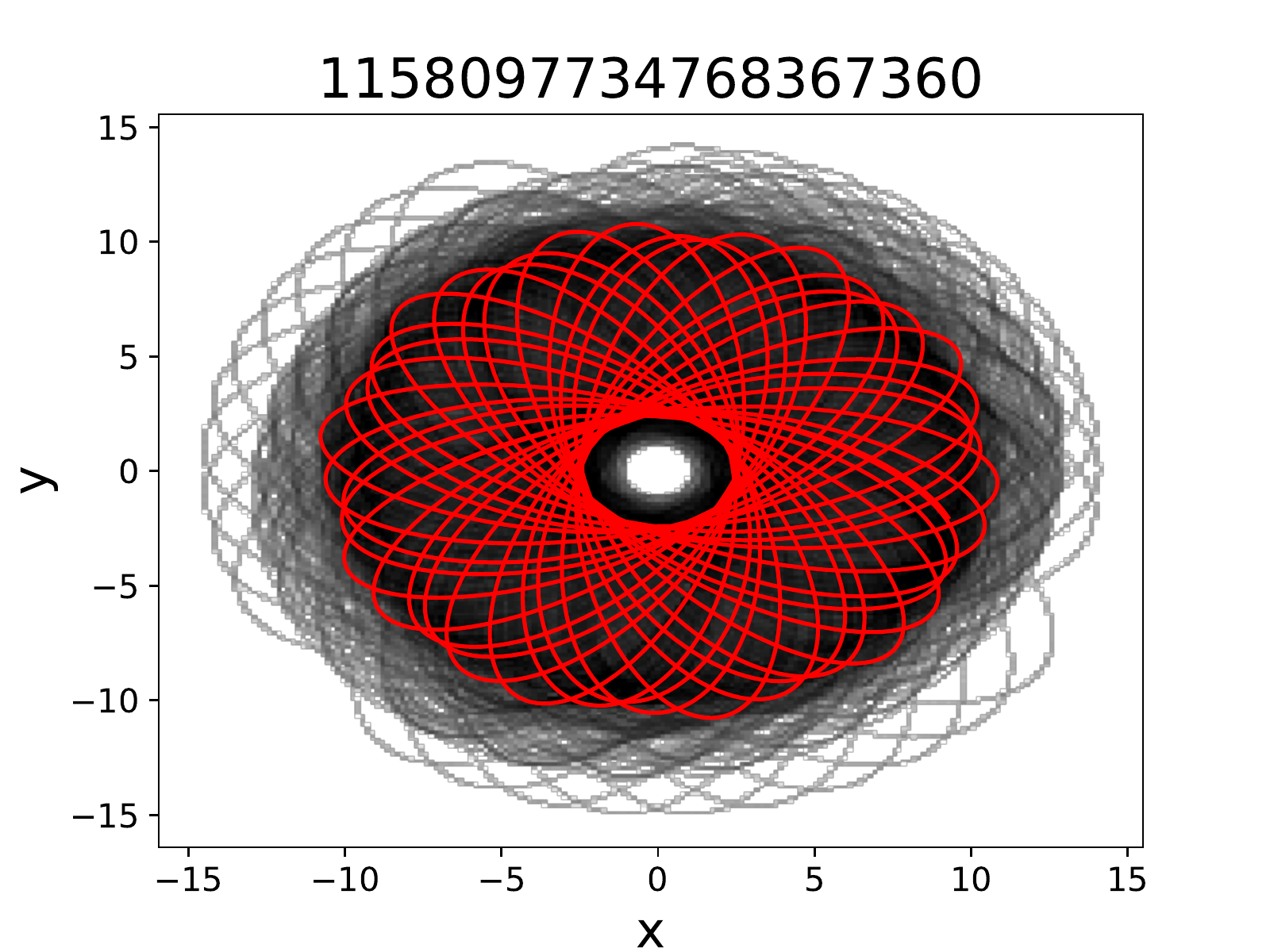}
\includegraphics[clip=true, trim = {0cm 0cm 1cm 0cm},width=4.5cm]{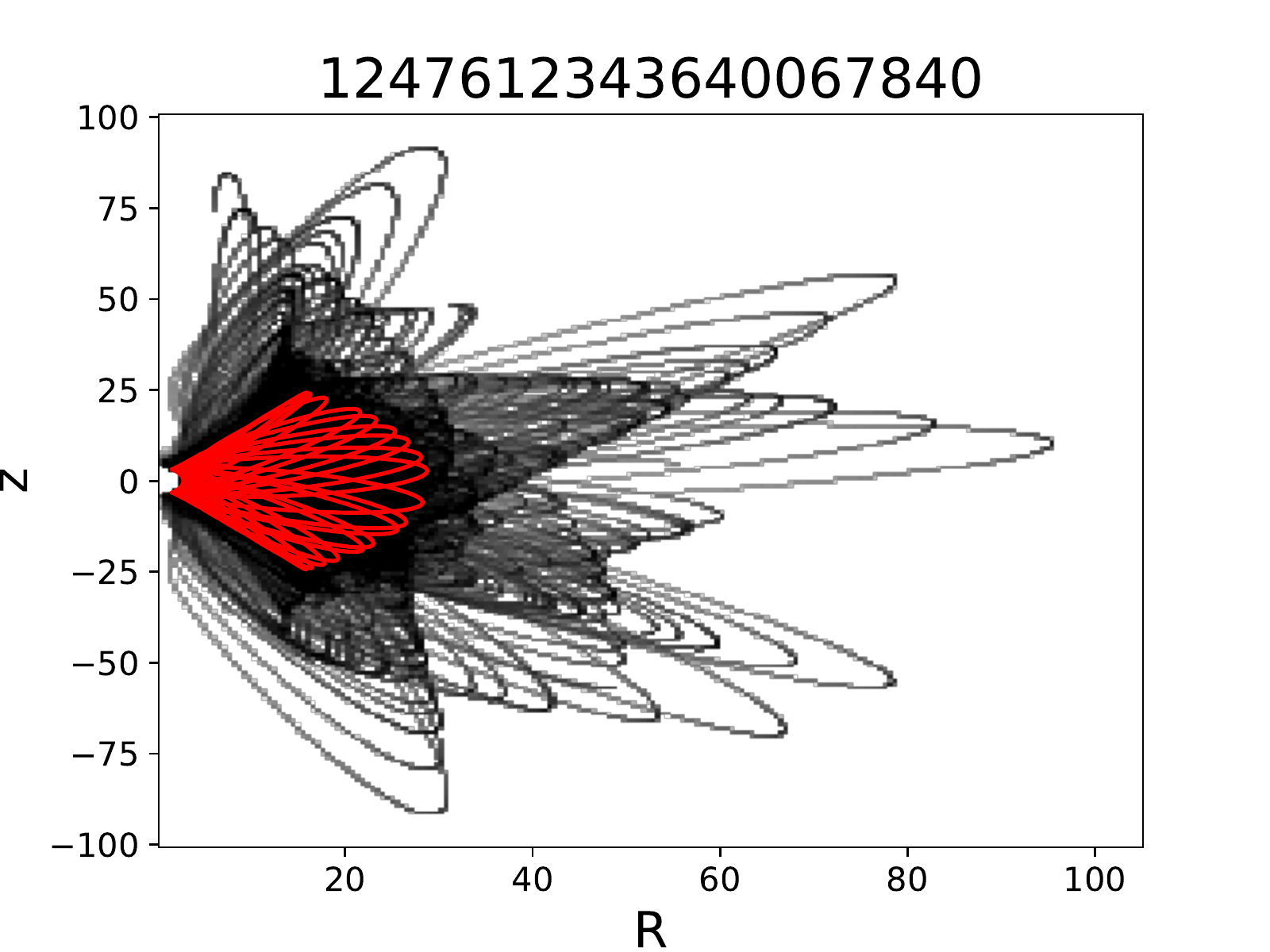}
\includegraphics[clip=true, trim = {0cm 0cm 1cm 0cm},width=4.5cm]{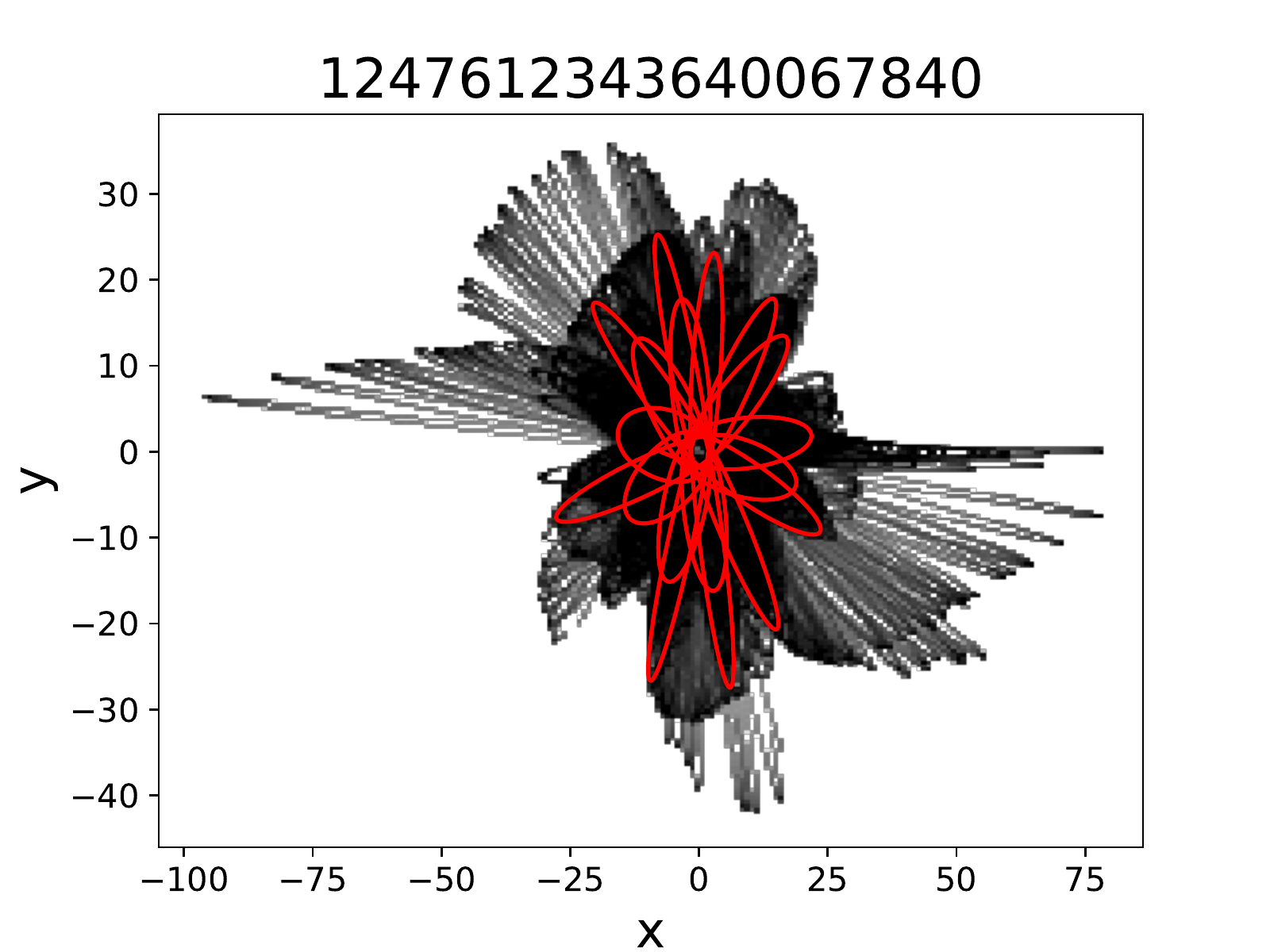}
\includegraphics[clip=true, trim = {0cm 0cm 1cm 0cm},width=4.5cm]{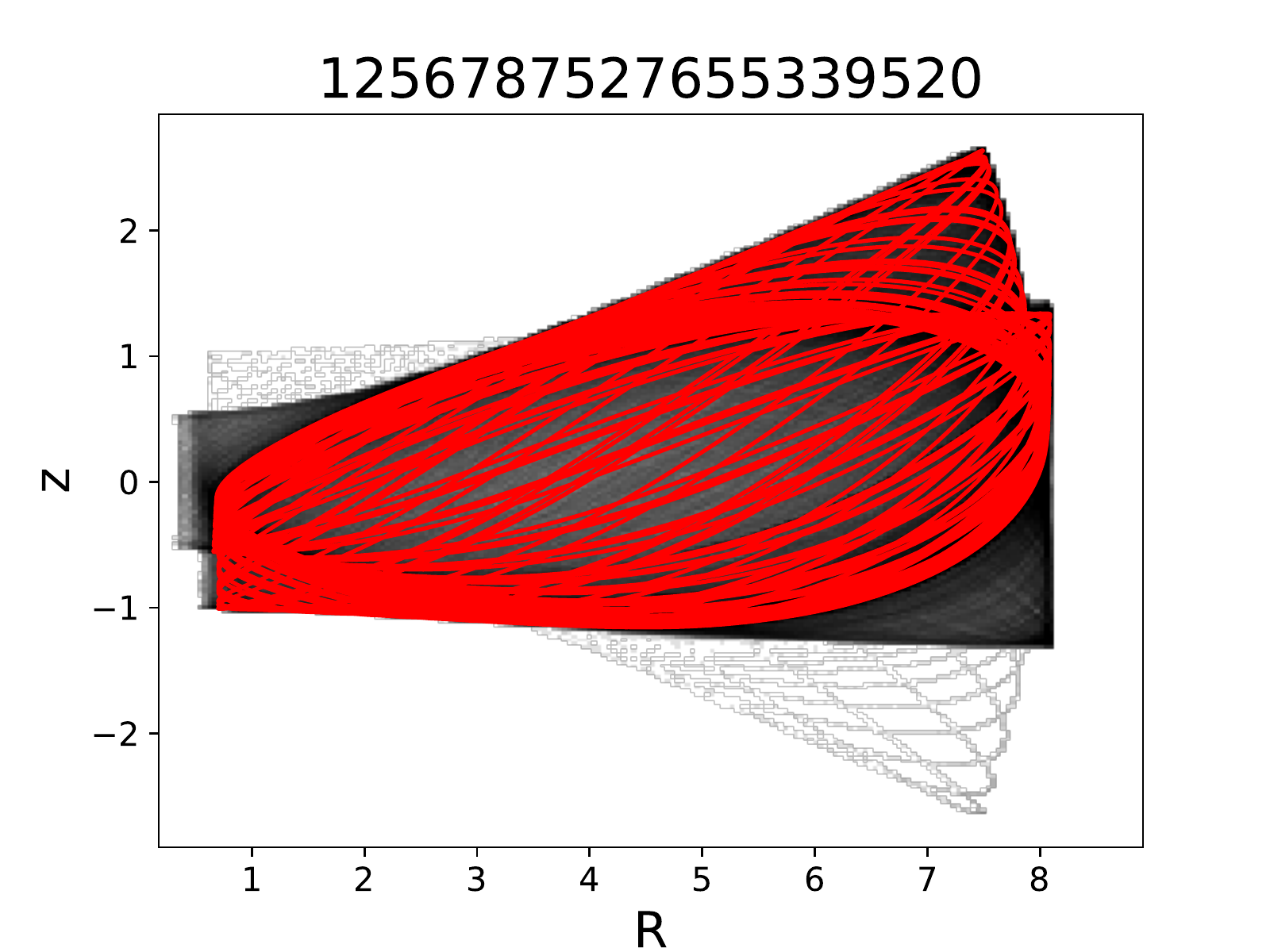}
\includegraphics[clip=true, trim = {0cm 0cm 1cm 0cm},width=4.5cm]{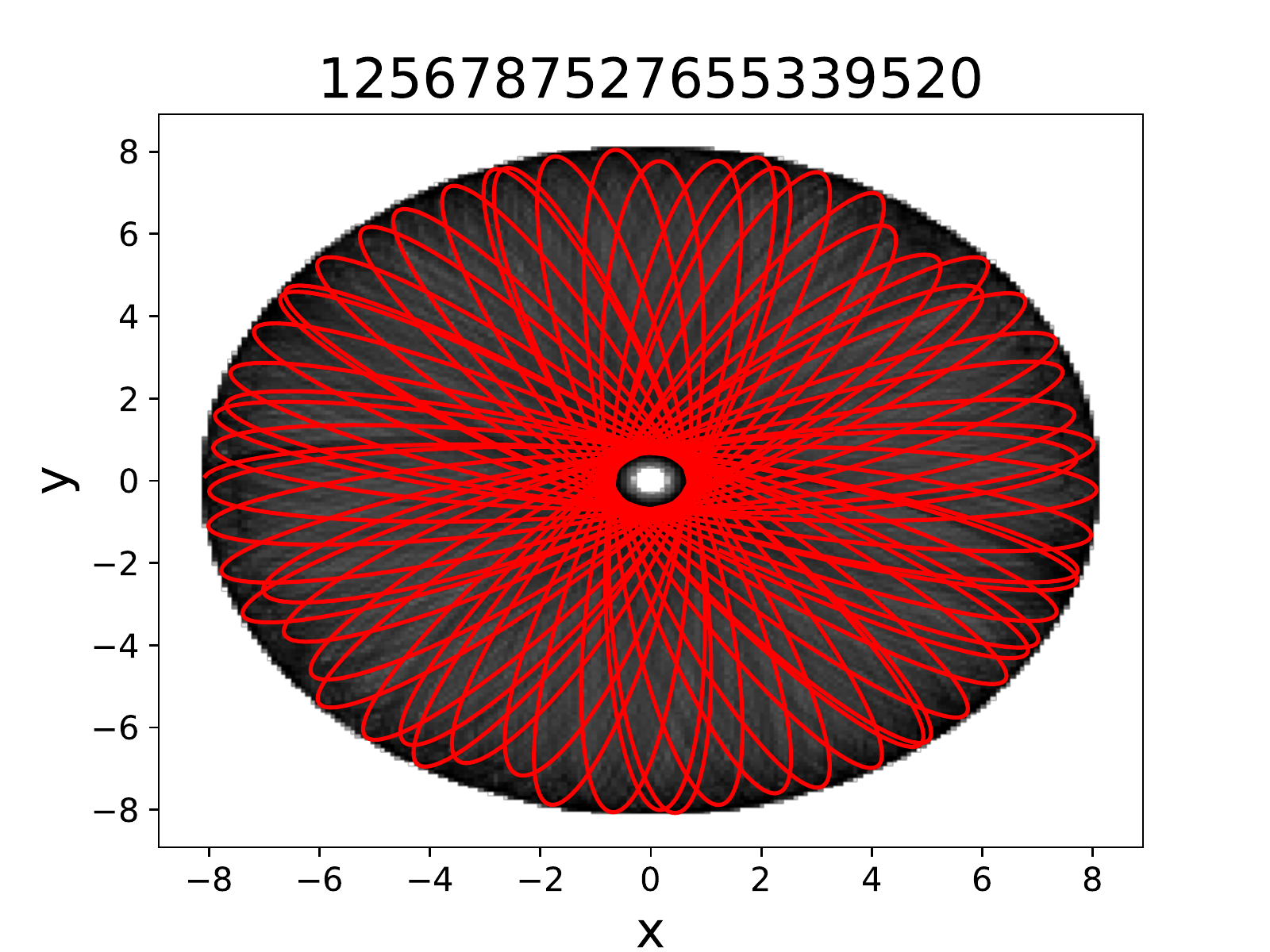}
\includegraphics[clip=true, trim = {0cm 0cm 1cm 0cm},width=4.5cm]{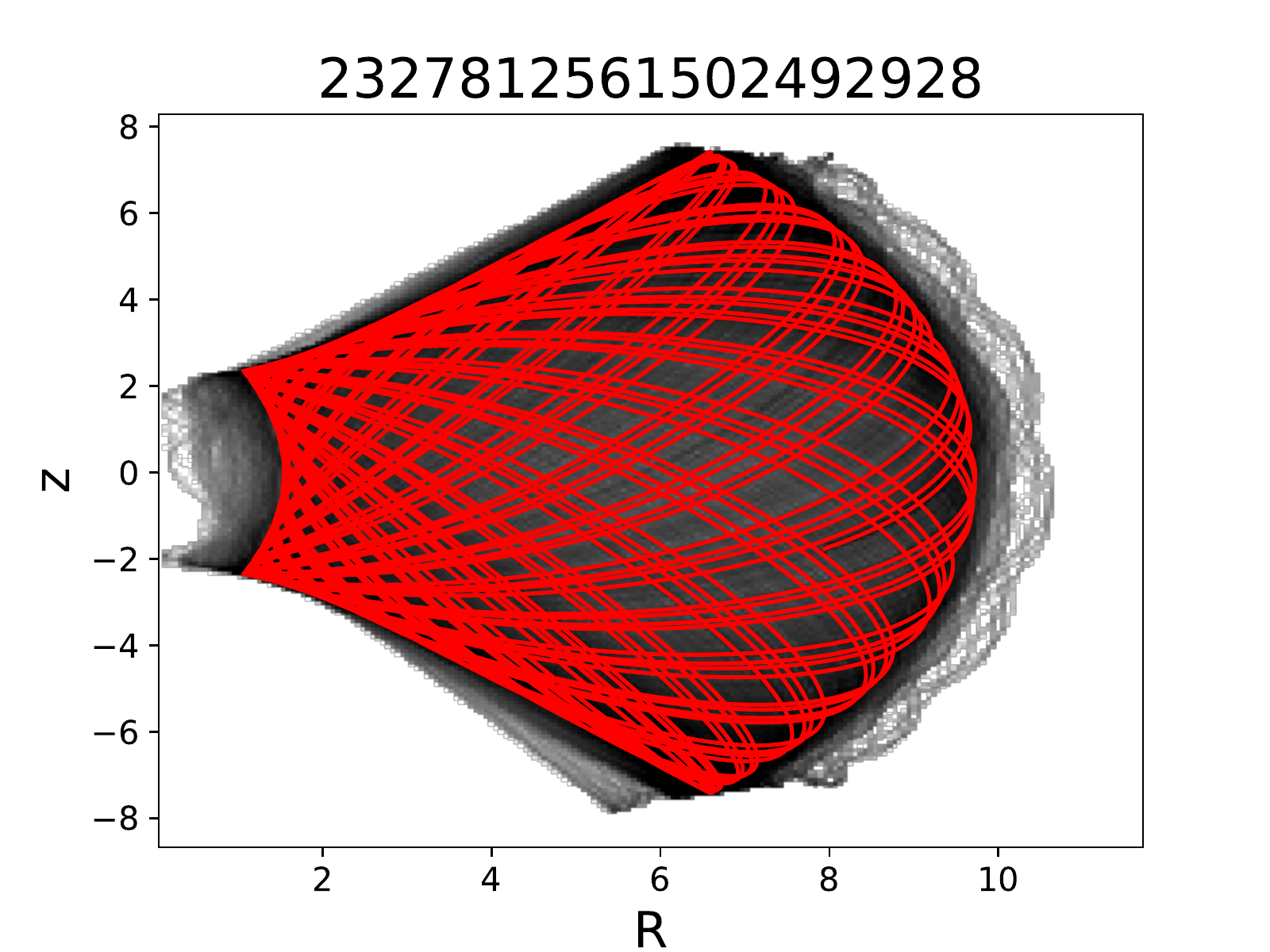}
\includegraphics[clip=true, trim = {0cm 0cm 1cm 0cm},width=4.5cm]{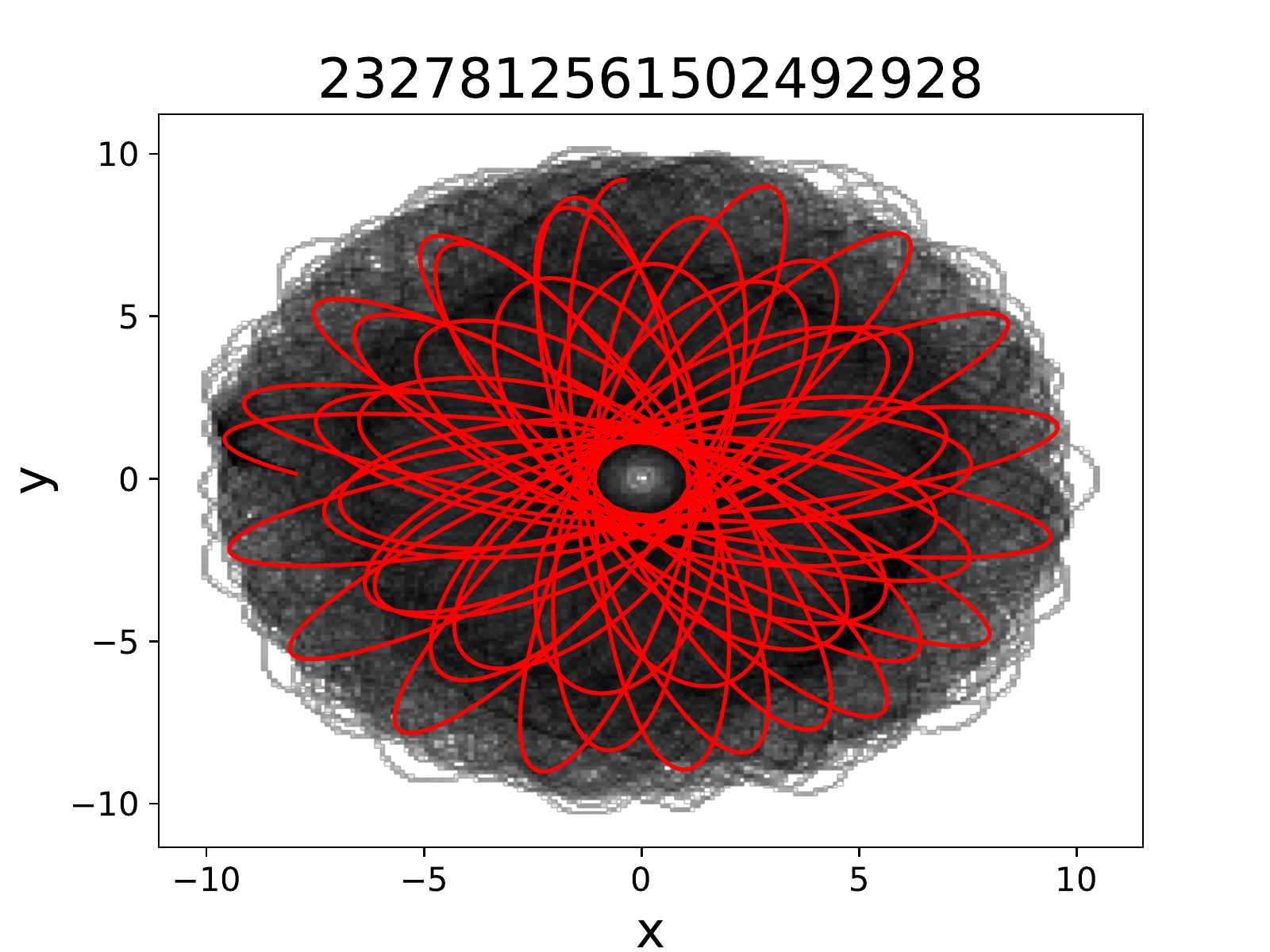}
\includegraphics[clip=true, trim = {0cm 0cm 1cm 0cm},width=4.5cm]{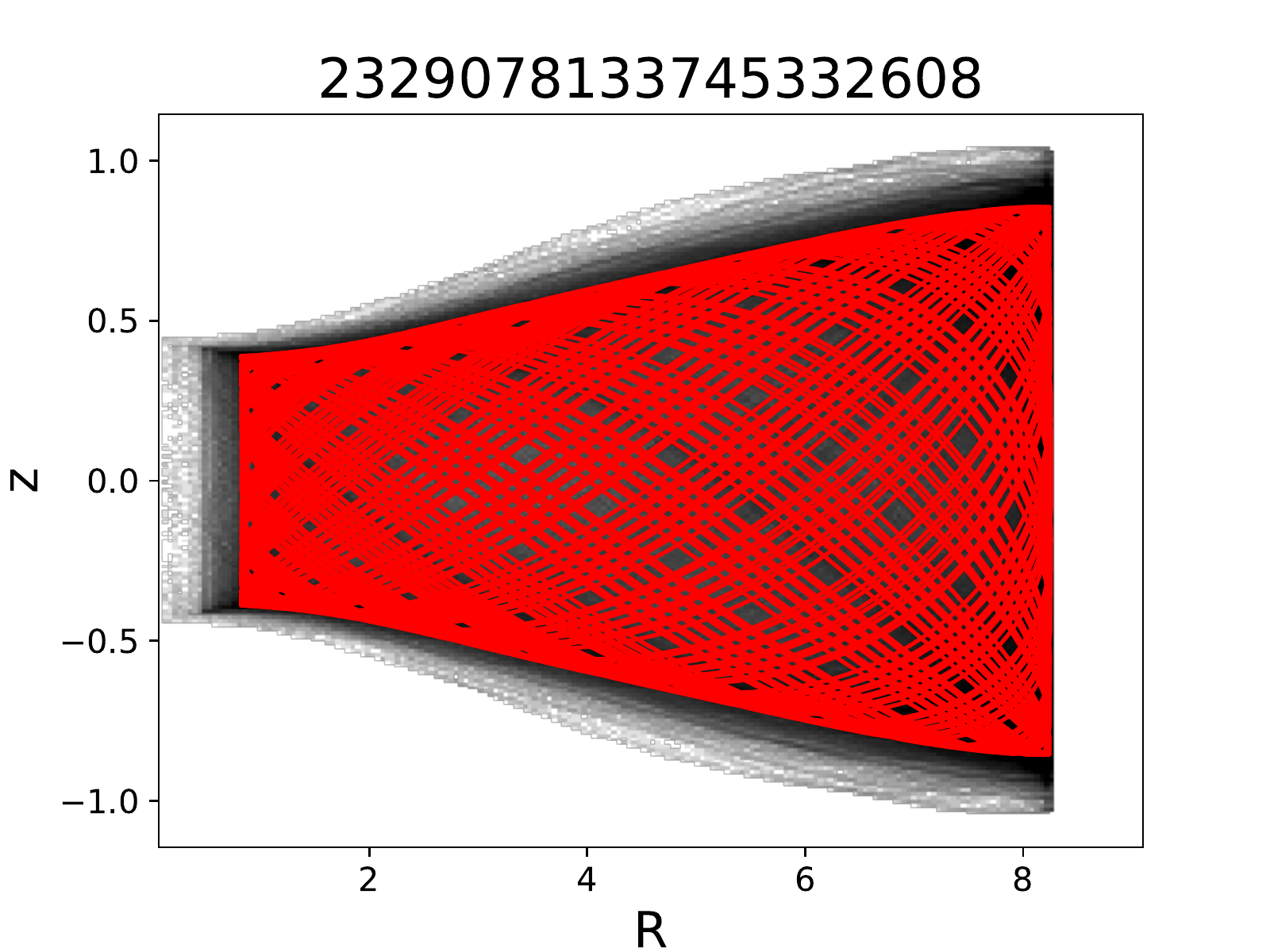}
\includegraphics[clip=true, trim = {0cm 0cm 1cm 0cm},width=4.5cm]{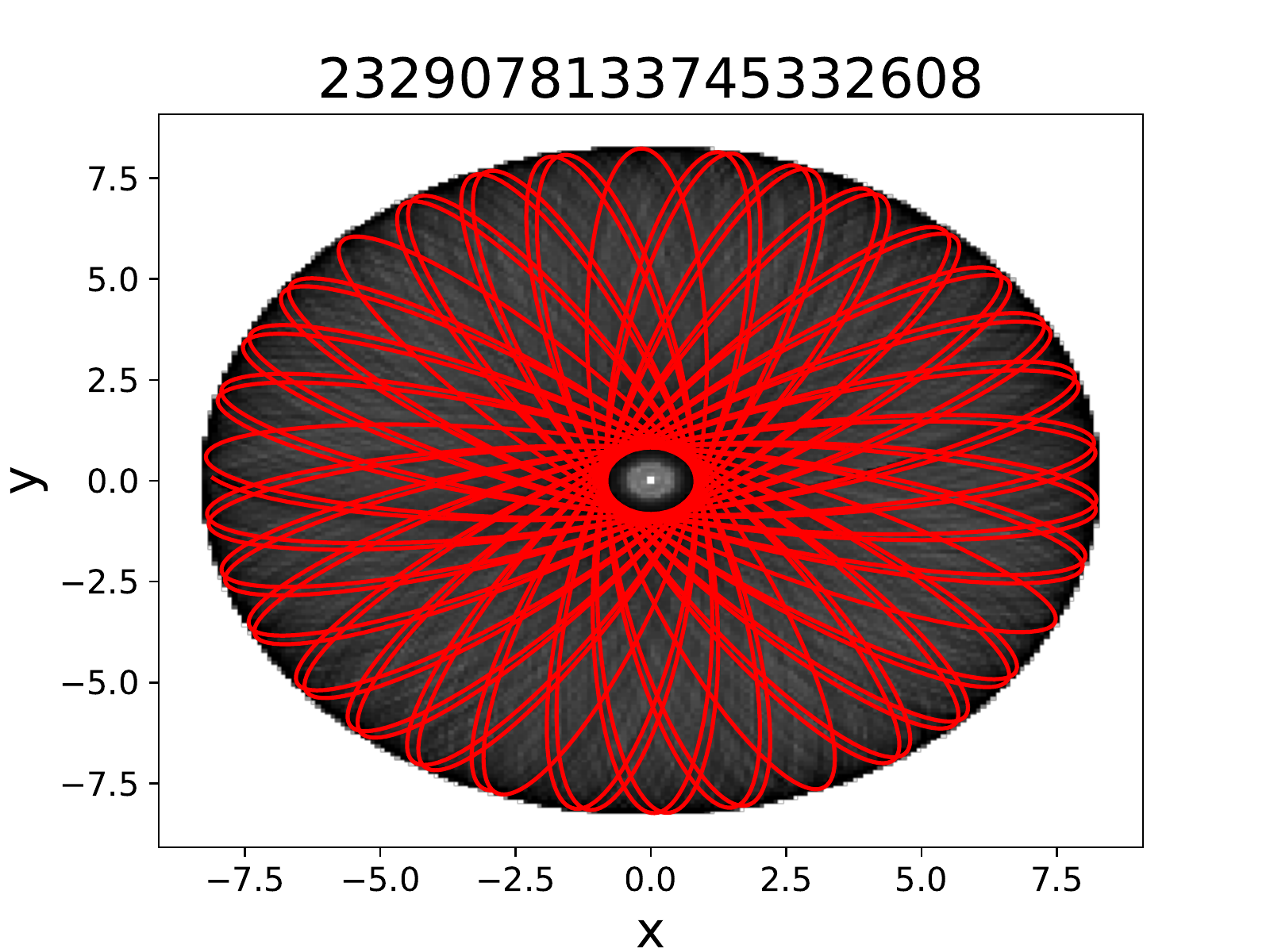}
\includegraphics[clip=true, trim = {0cm 0cm 1cm 0cm},width=4.5cm]{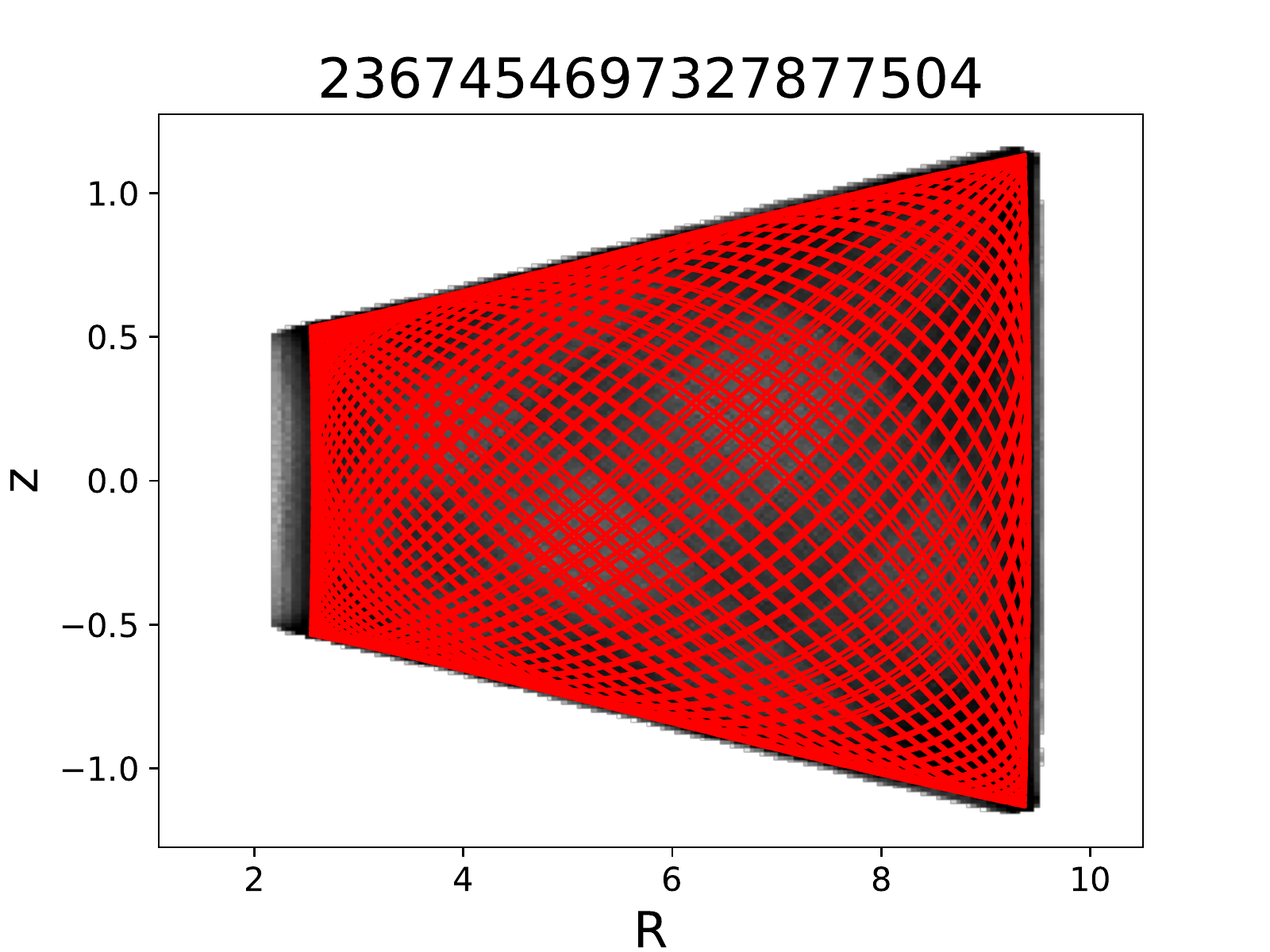}
\includegraphics[clip=true, trim = {0cm 0cm 1cm 0cm},width=4.5cm]{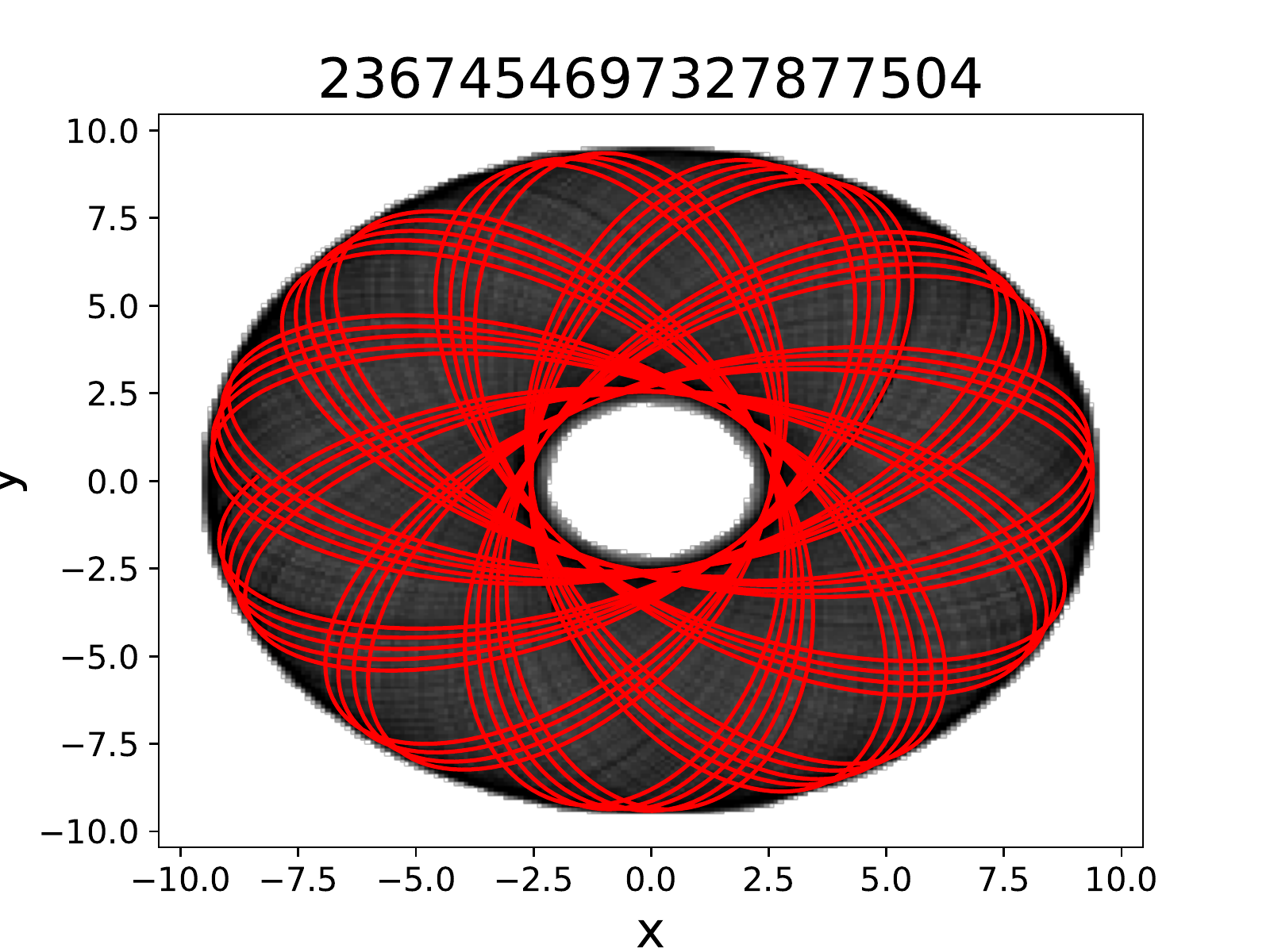}
\includegraphics[clip=true, trim = {0cm 0cm 1cm 0cm},width=4.5cm]{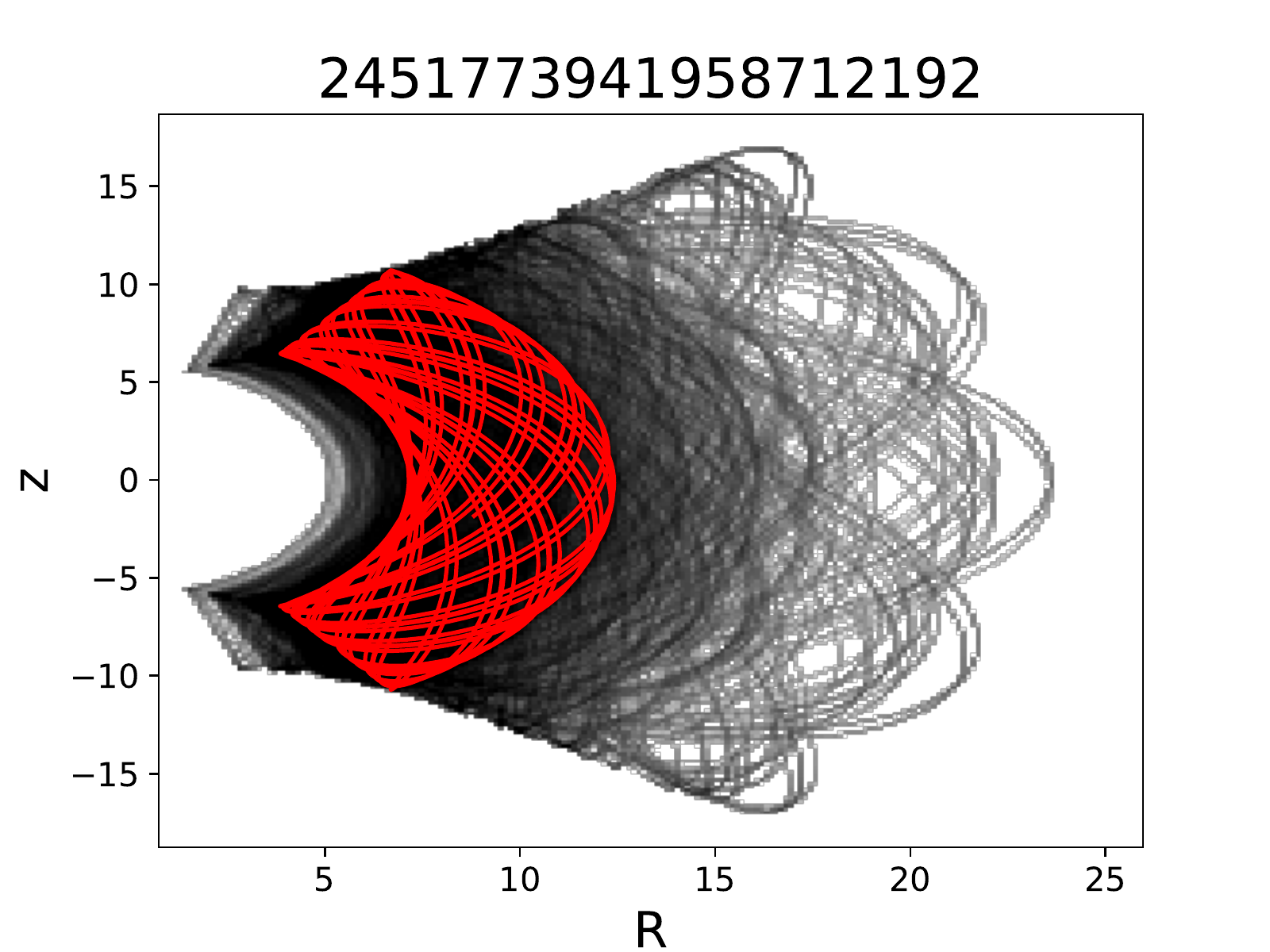}
\includegraphics[clip=true, trim = {0cm 0cm 1cm 0cm},width=4.5cm]{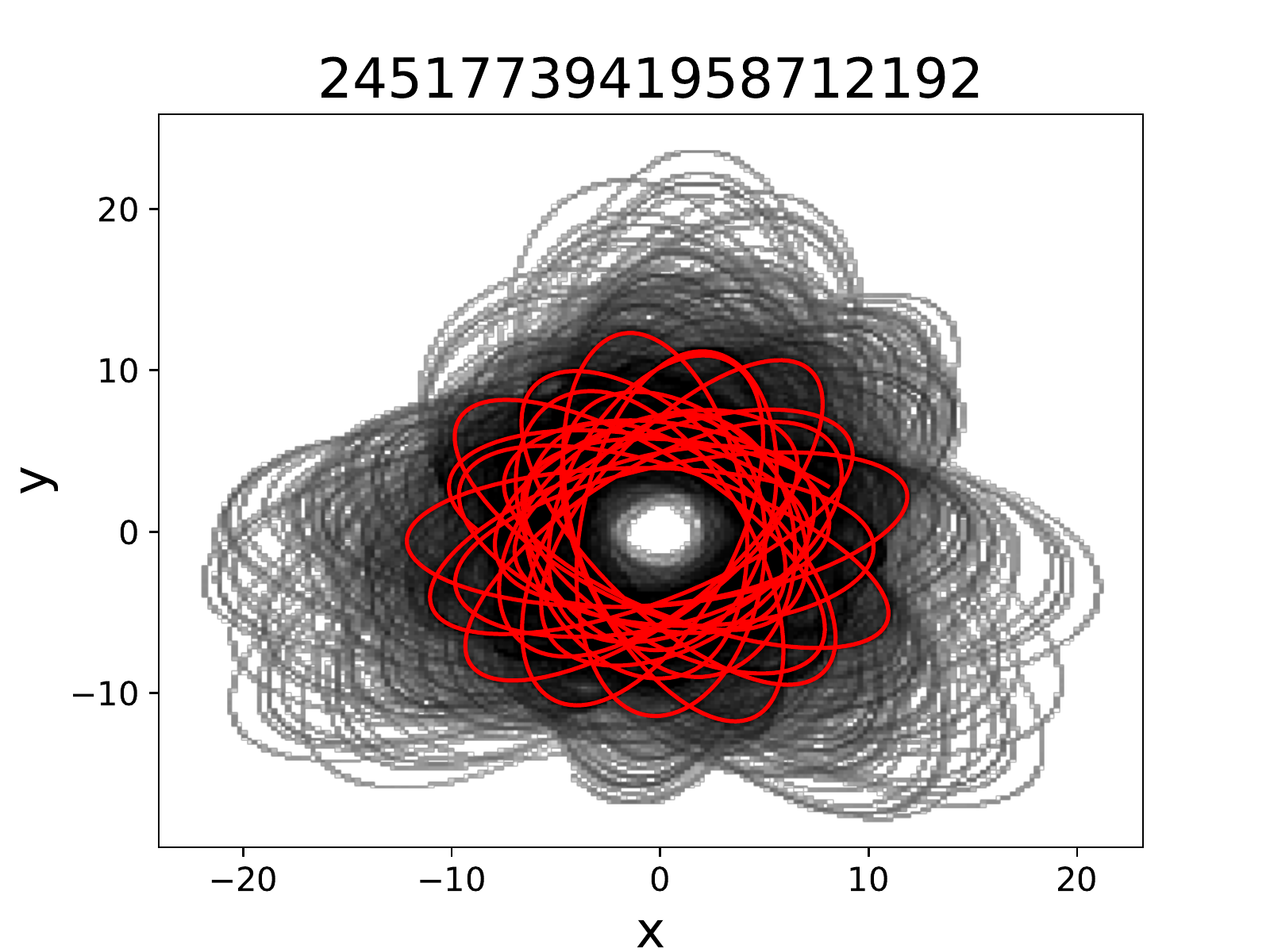}
\includegraphics[clip=true, trim = {0cm 0cm 1cm 0cm},width=4.5cm]{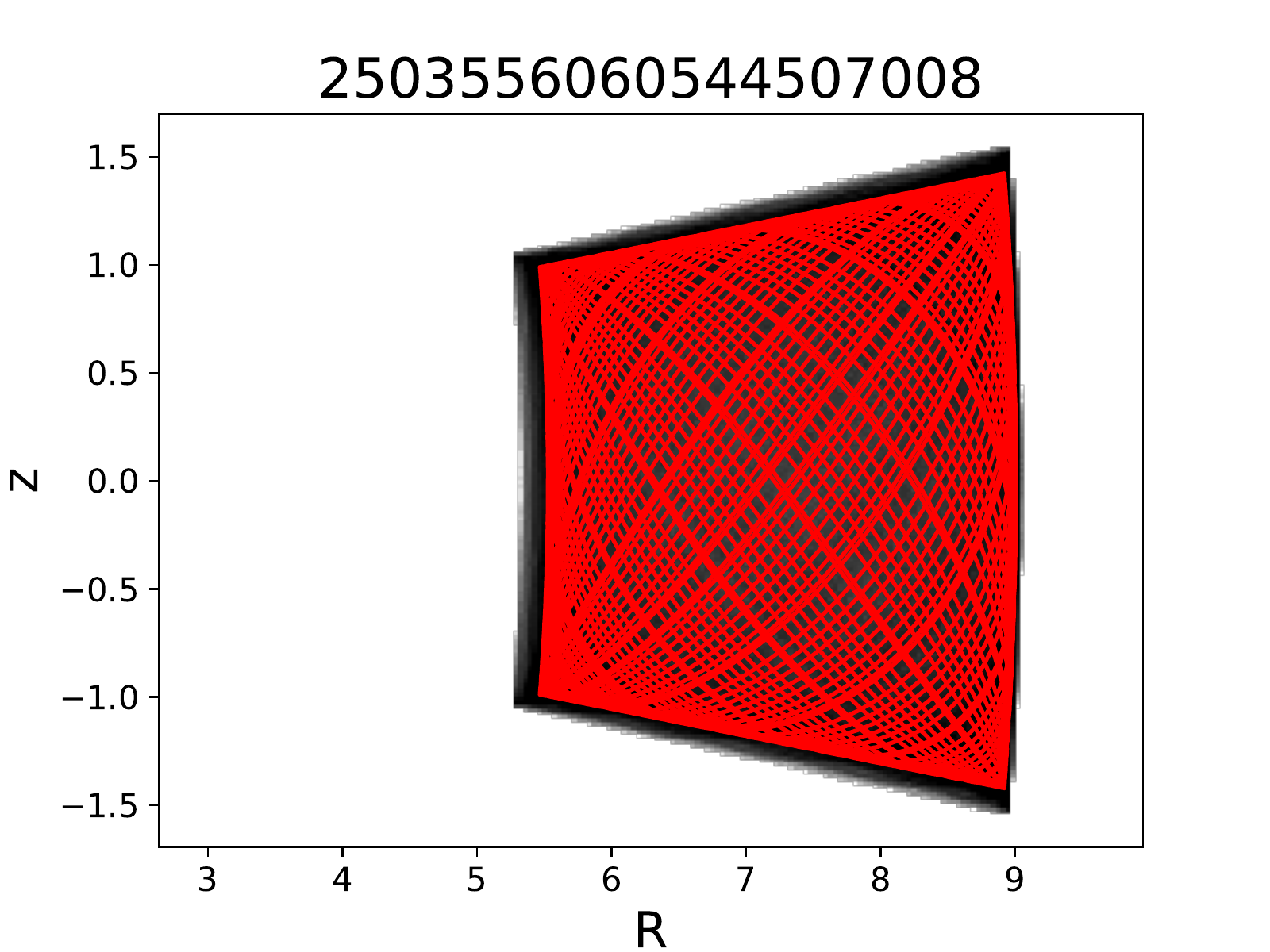}
\includegraphics[clip=true, trim = {0cm 0cm 1cm 0cm},width=4.5cm]{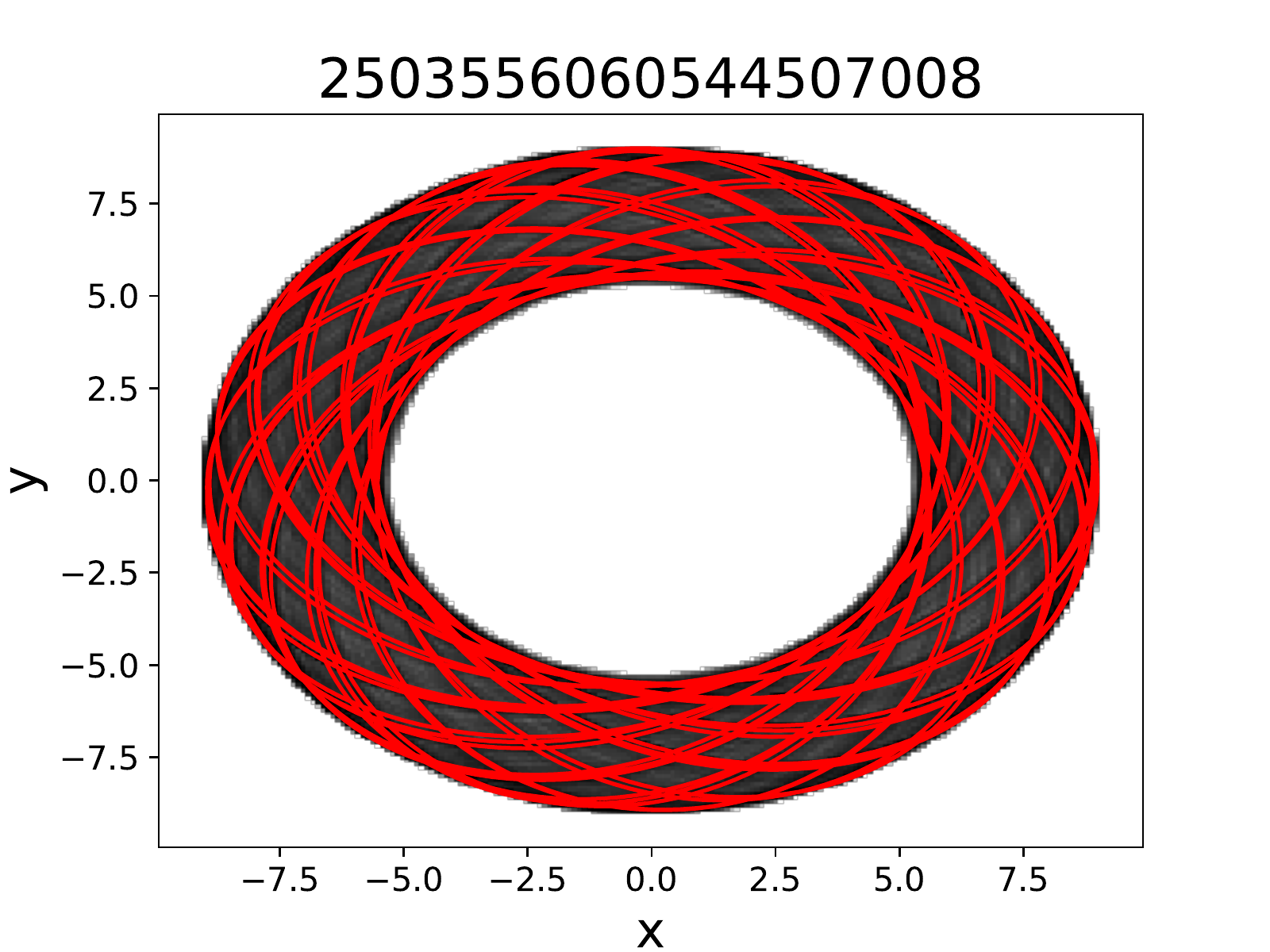}
\includegraphics[clip=true, trim = {0cm 0cm 1cm 0cm},width=4.5cm]{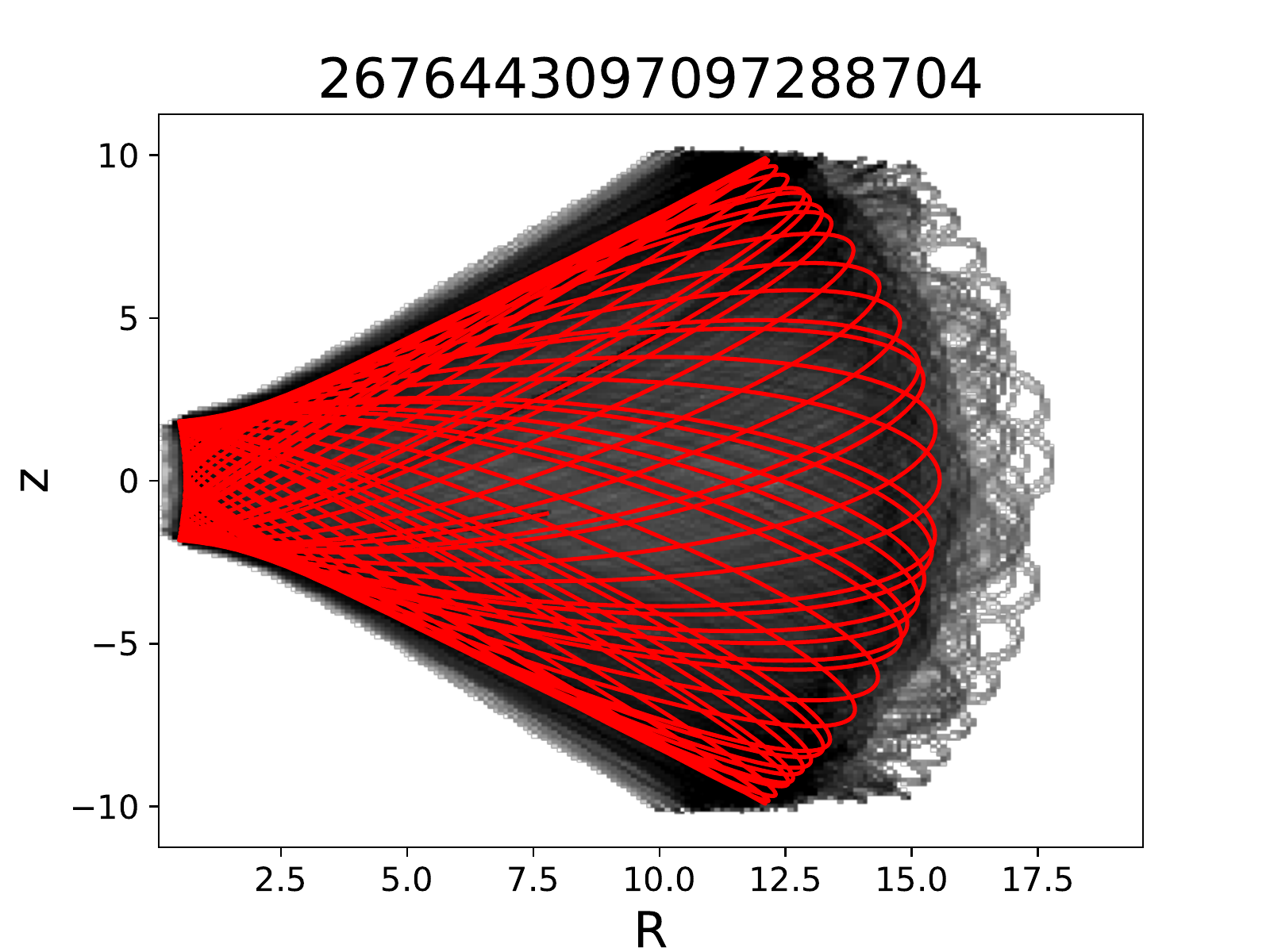}
\includegraphics[clip=true, trim = {0cm 0cm 1cm 0cm},width=4.5cm]{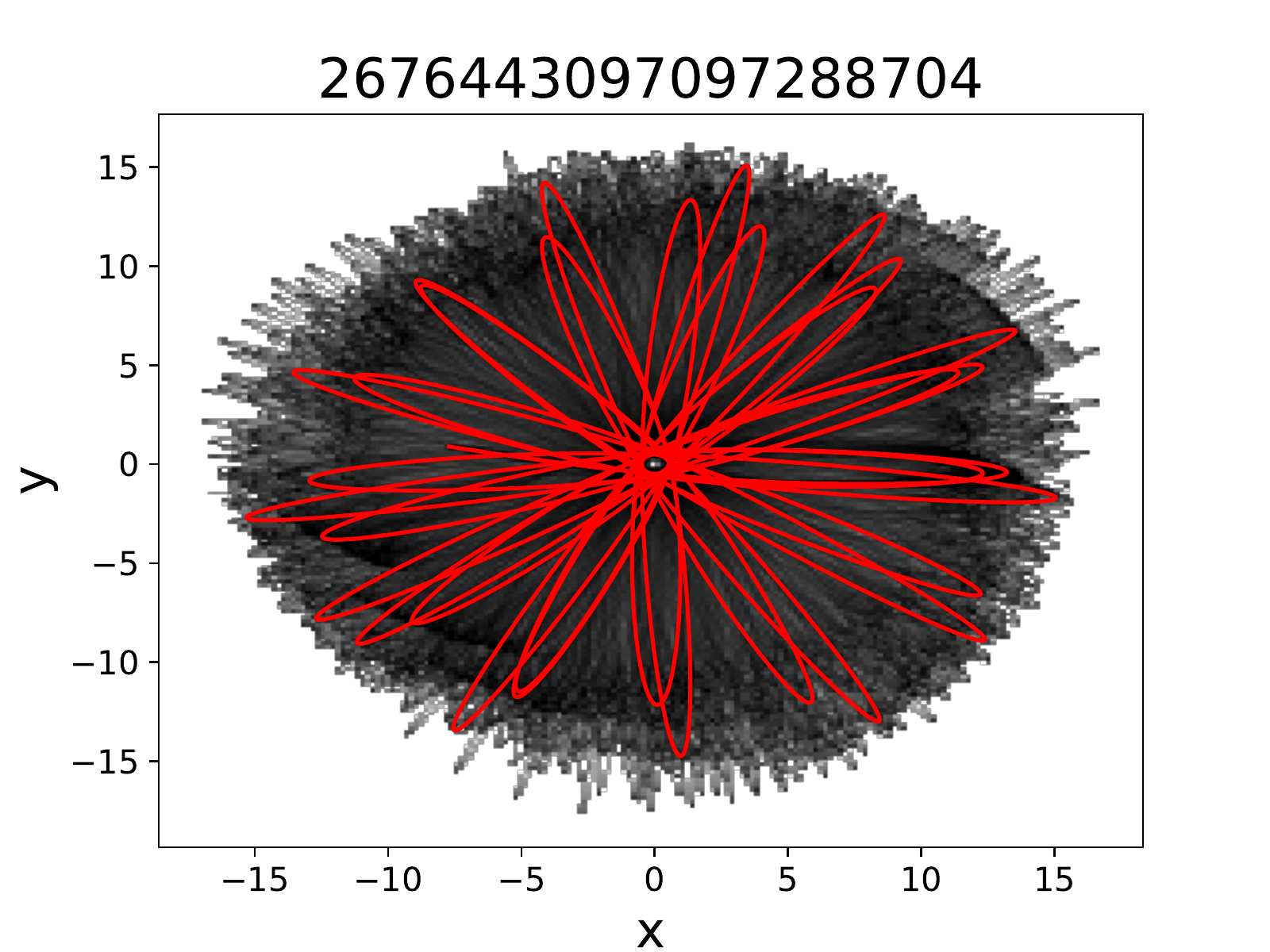}
\includegraphics[clip=true, trim = {0cm 0cm 1cm 0cm},width=4.5cm]{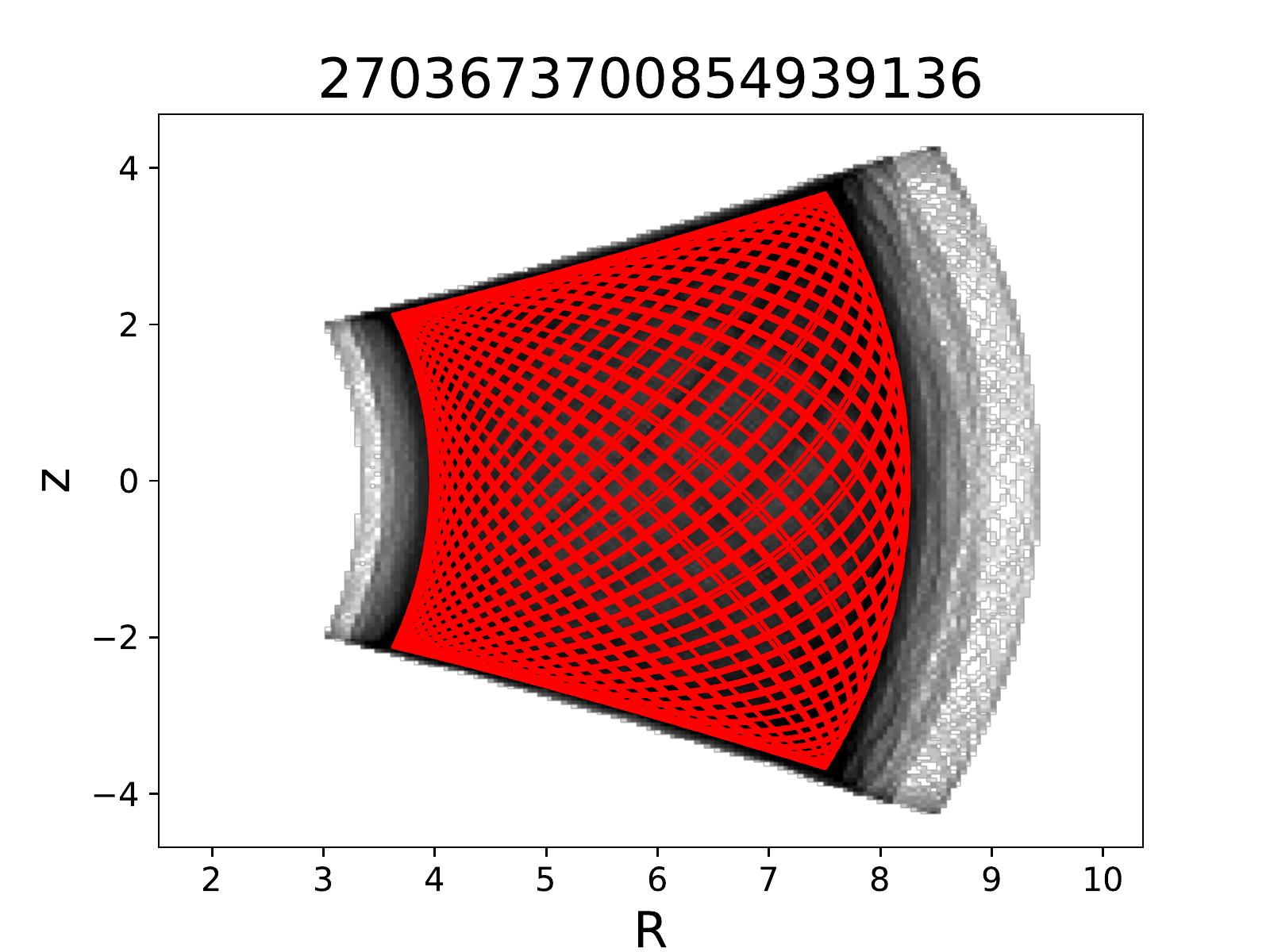}
\includegraphics[clip=true, trim = {0cm 0cm 1cm 0cm},width=4.5cm]{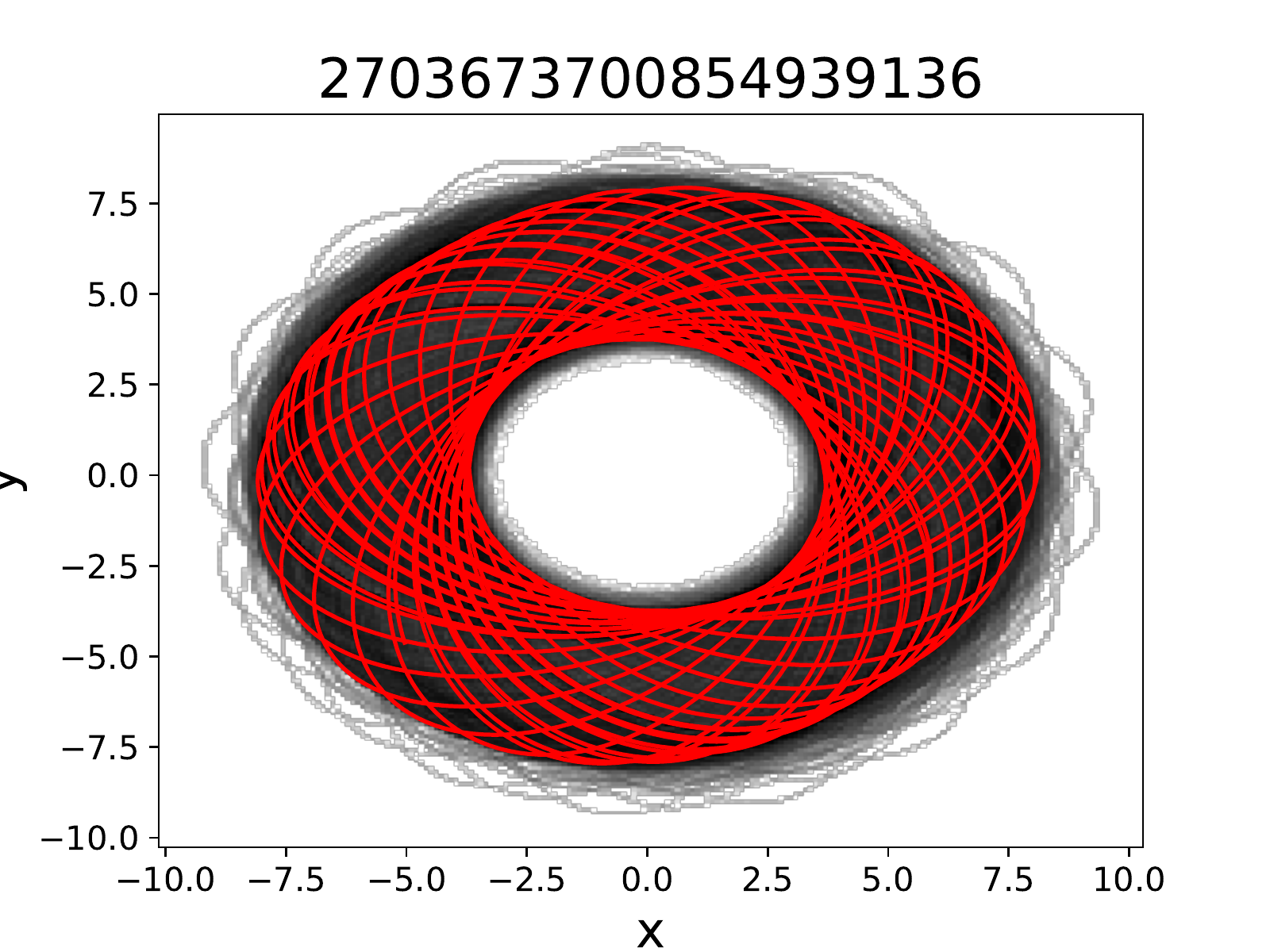}
\includegraphics[clip=true, trim = {0cm 0cm 1cm 0cm},width=4.5cm]{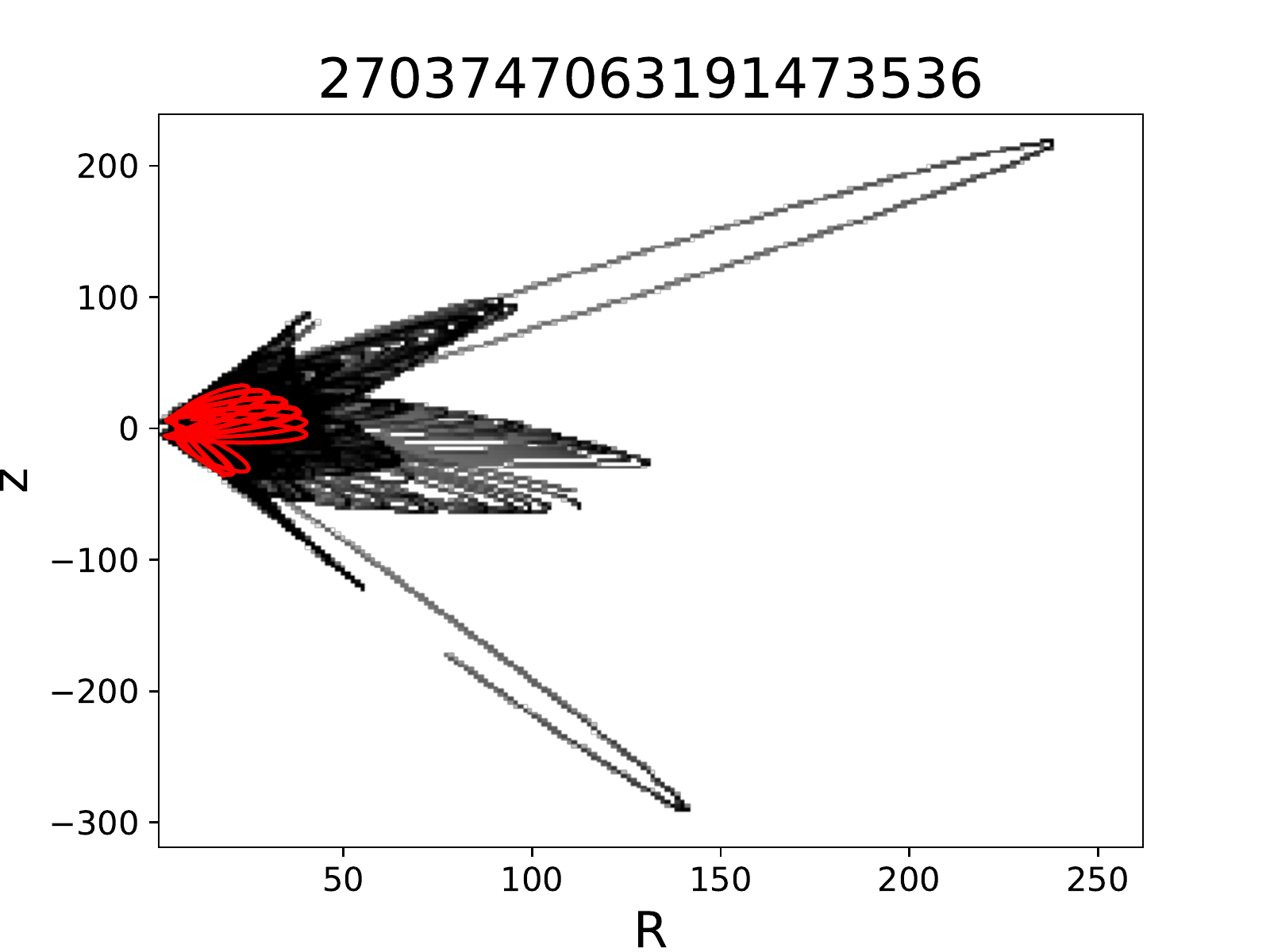}
\includegraphics[clip=true, trim = {0cm 0cm 1cm 0cm},width=4.5cm]{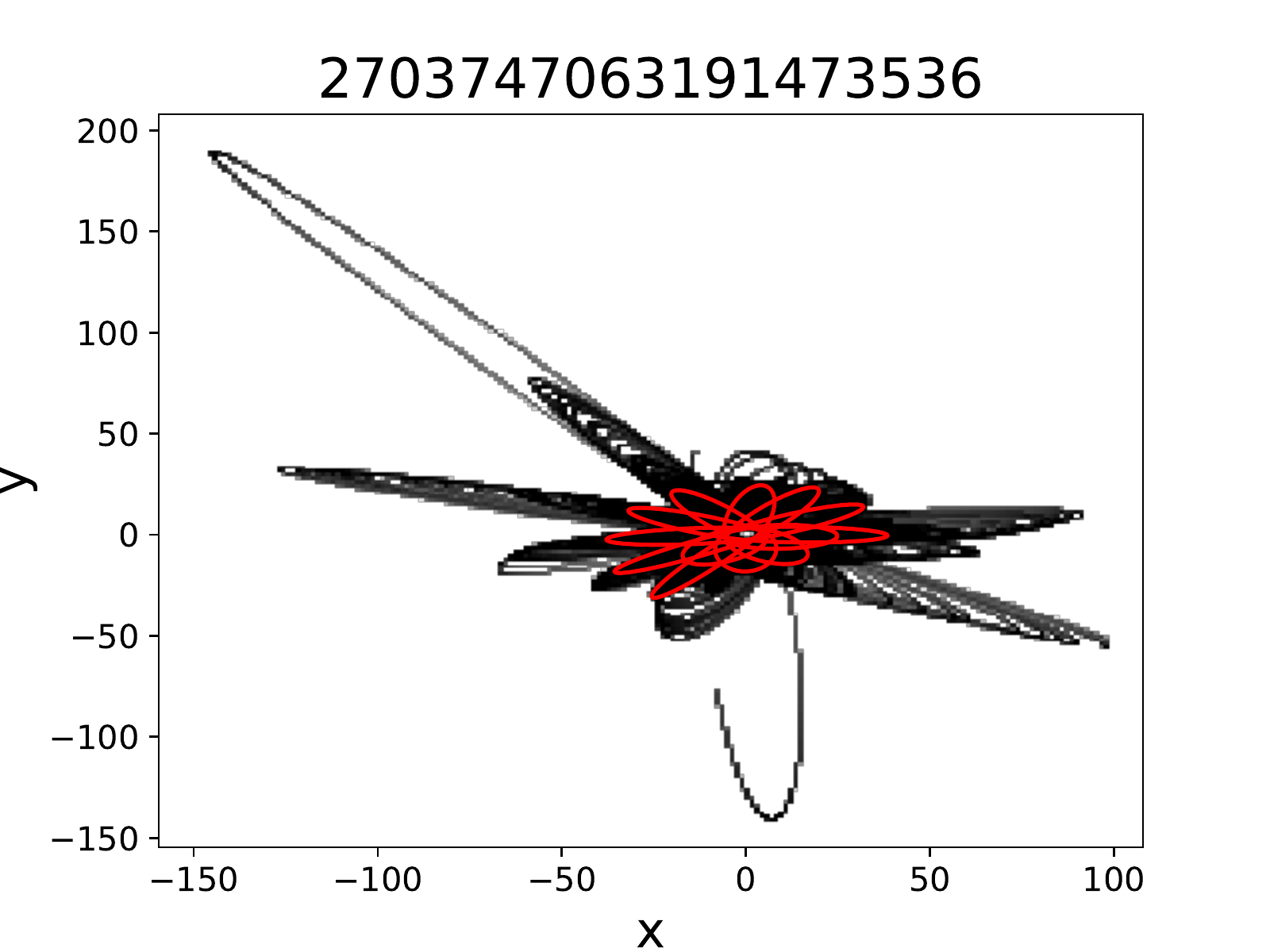}
\caption{Projection on $R-Z$ plane and on $X-Y$ plane of orbits of stars in the  LP First stars sample. The red line corresponds to the mean orbit of each star, the grey lines to the 100 realisations of this orbit, once errors on the observables (parallaxes, proper motions, line-of-sight velocities) are taken into account. The Gaia~DR2~ID of each star is reported on the top of each panel.} 
\label{T1}
\end{centering}
\end{figure*}

\begin{figure*}
\begin{centering}
\includegraphics[clip=true, trim = {0cm 0cm 1cm 0cm},width=4.5cm]{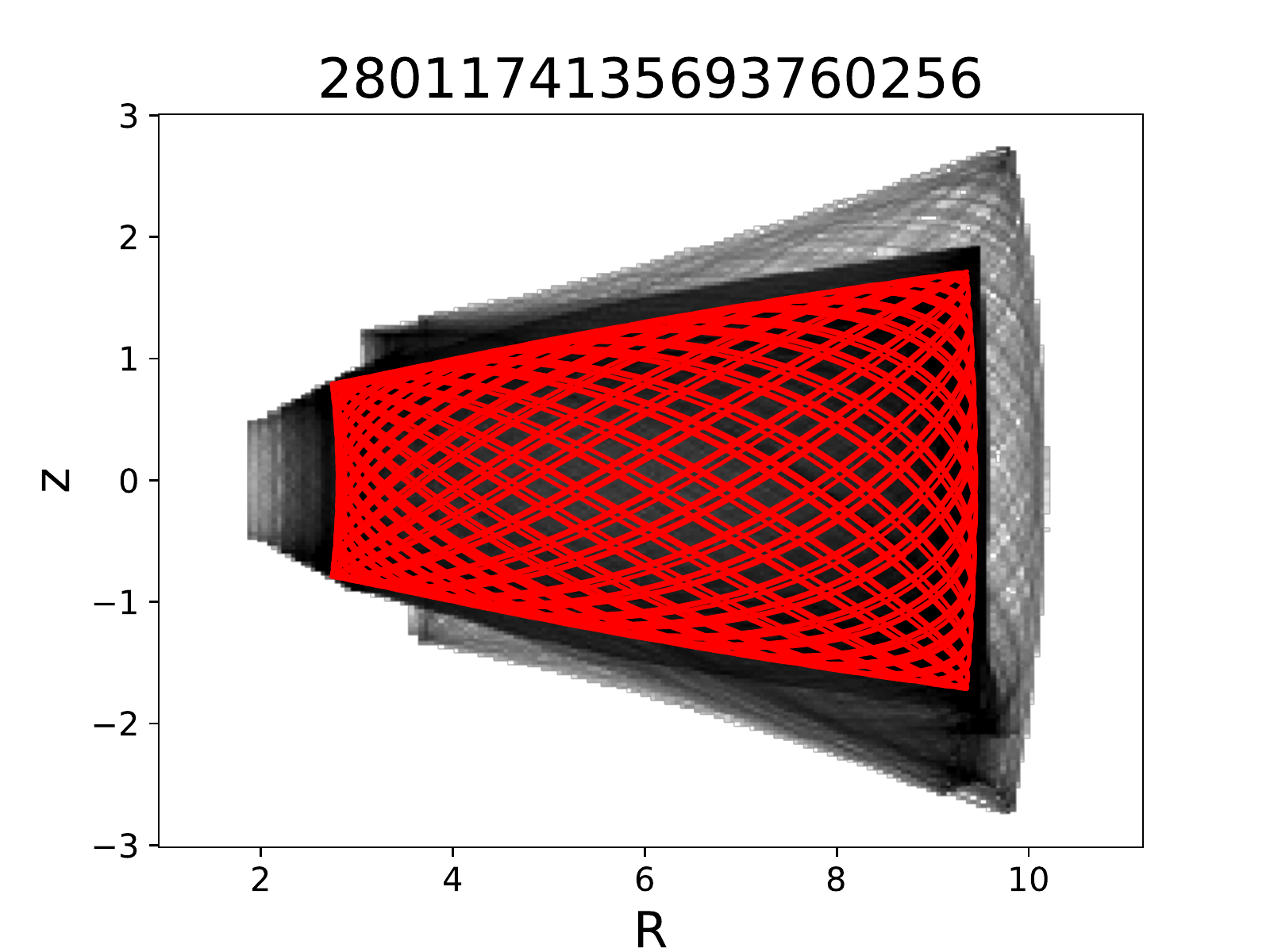}
\includegraphics[clip=true, trim = {0cm 0cm 1cm 0cm},width=4.5cm]{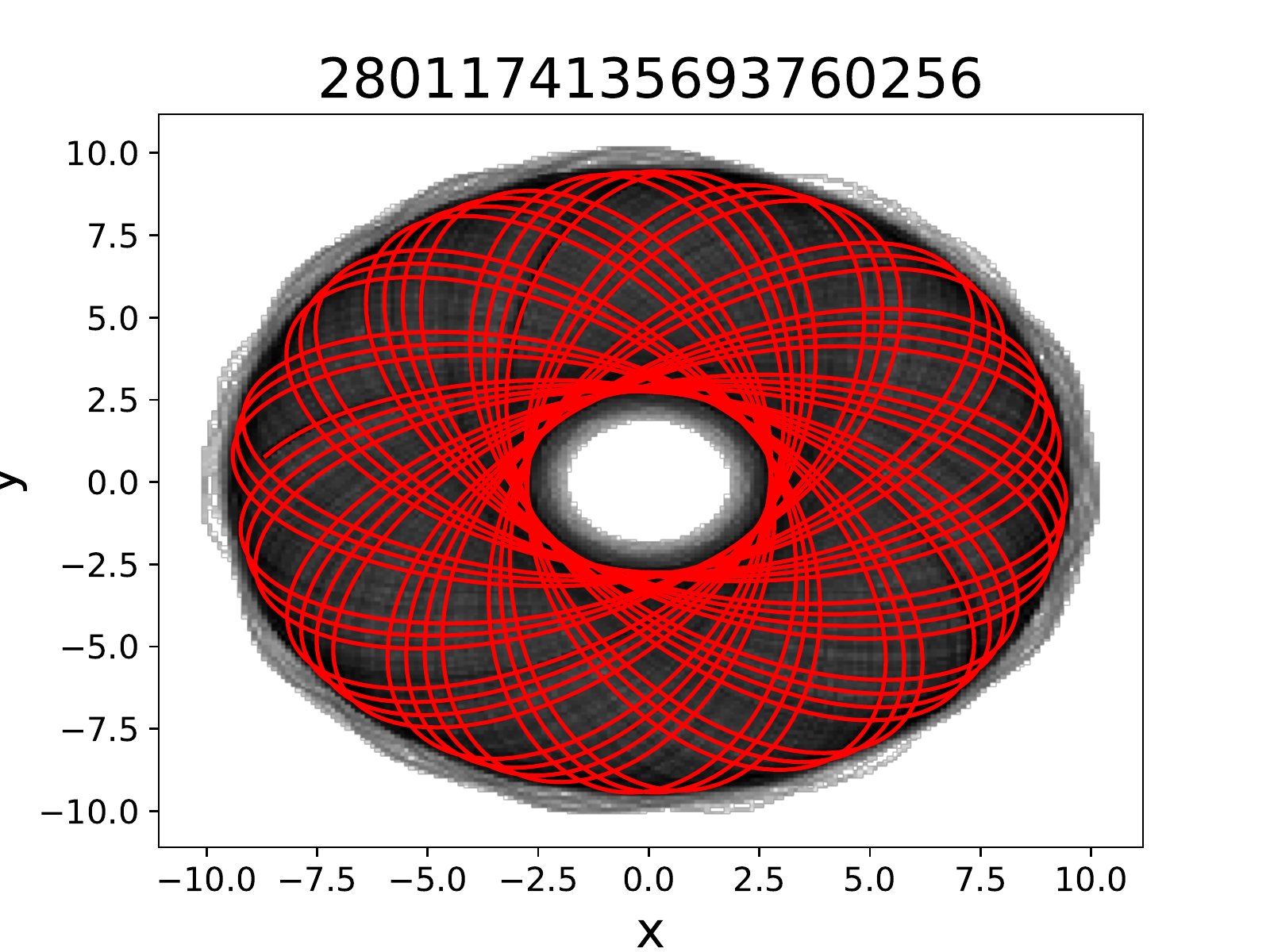}
\includegraphics[clip=true, trim = {0cm 0cm 1cm 0cm},width=4.5cm]{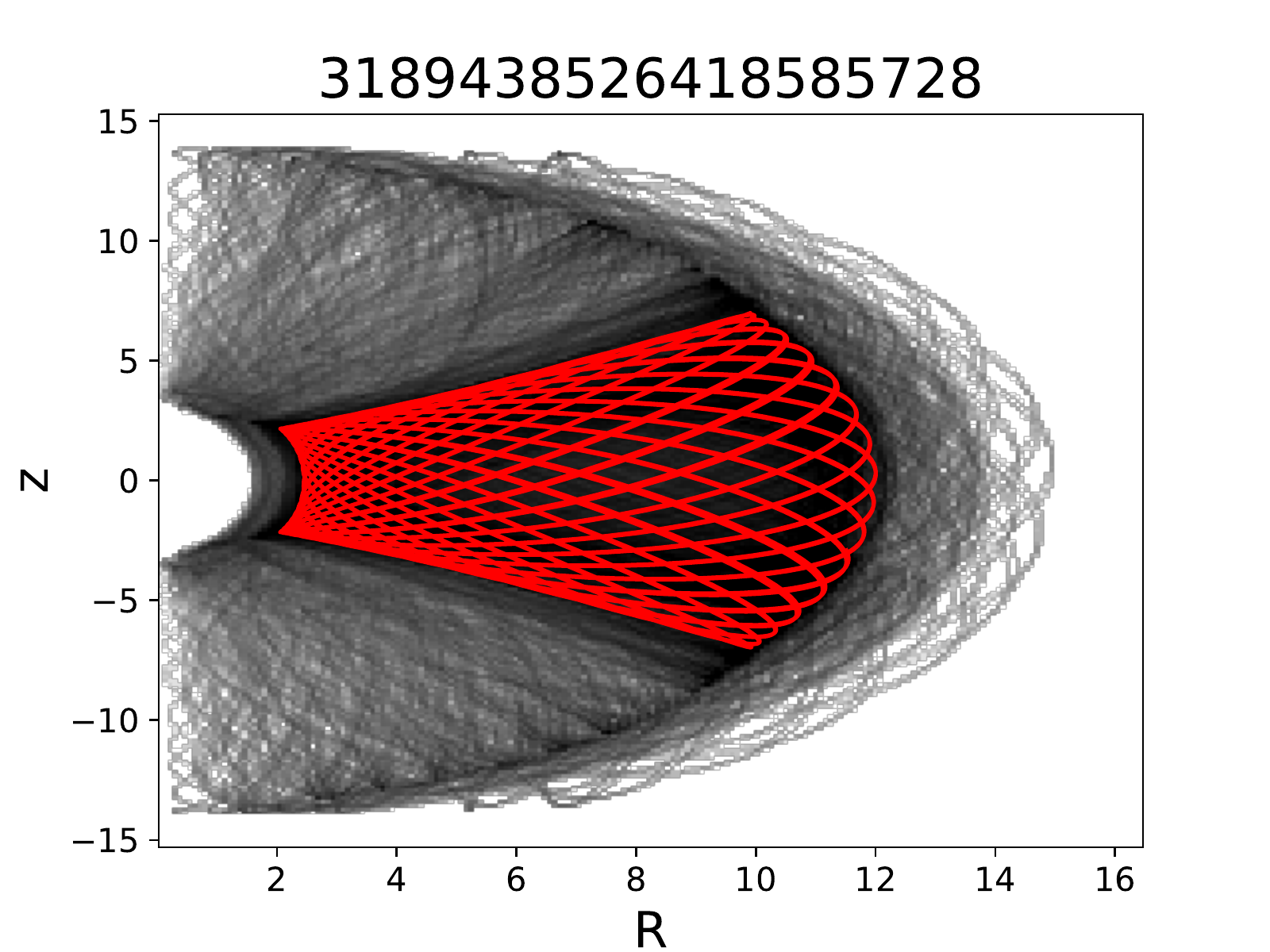}
\includegraphics[clip=true, trim = {0cm 0cm 1cm 0cm},width=4.5cm]{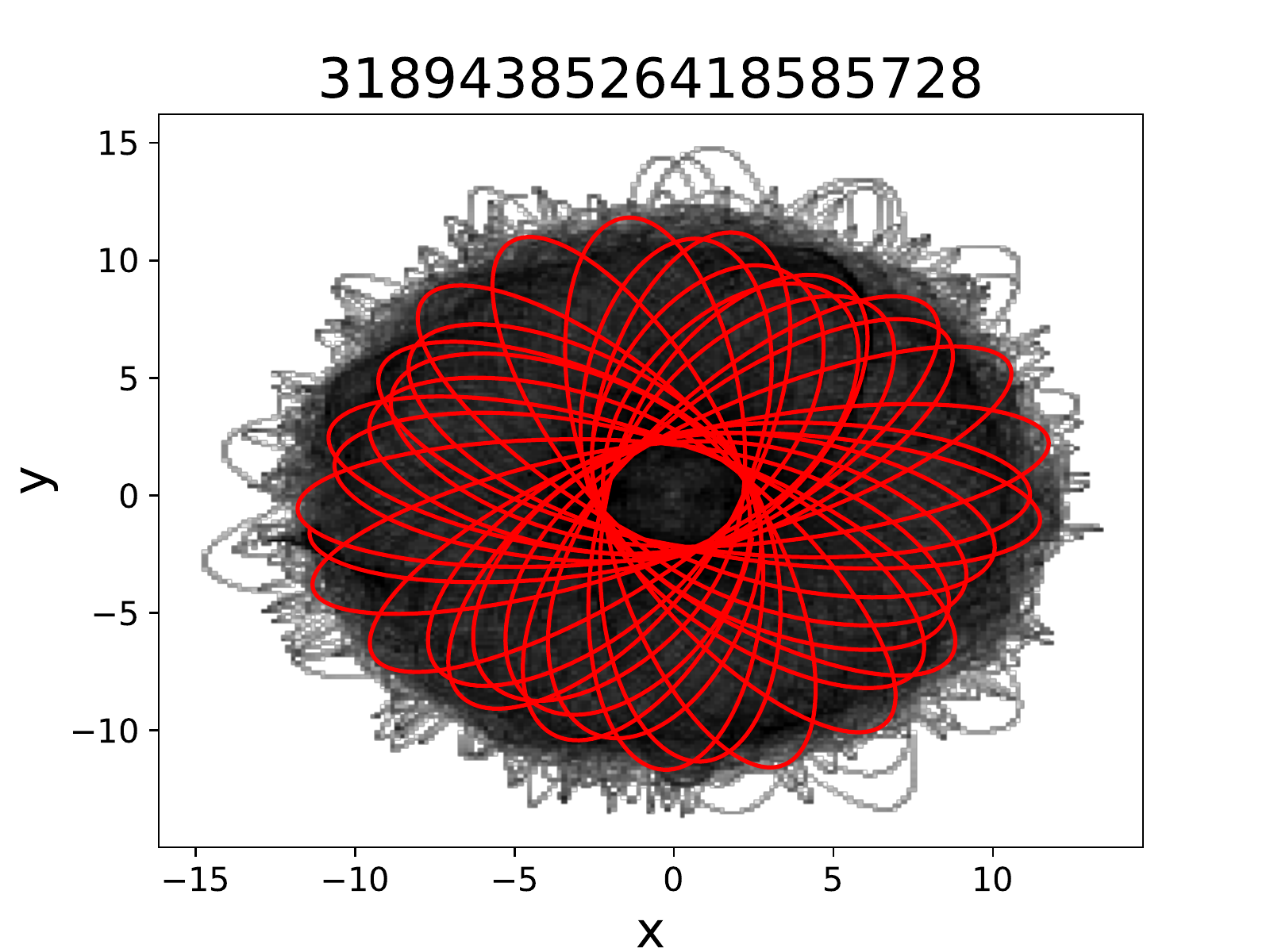}
\includegraphics[clip=true, trim = {0cm 0cm 1cm 0cm},width=4.5cm]{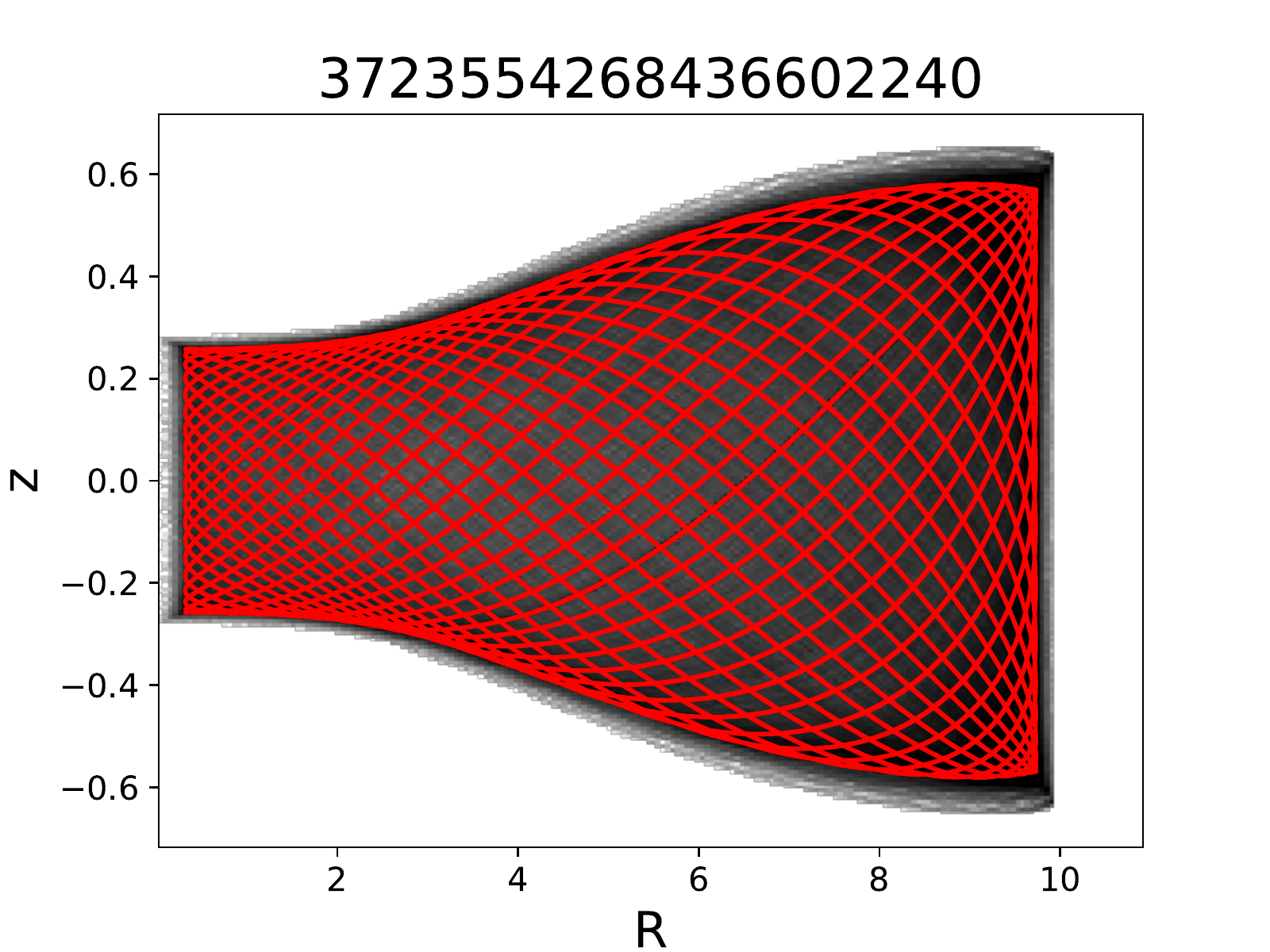}
\includegraphics[clip=true, trim = {0cm 0cm 1cm 0cm},width=4.5cm]{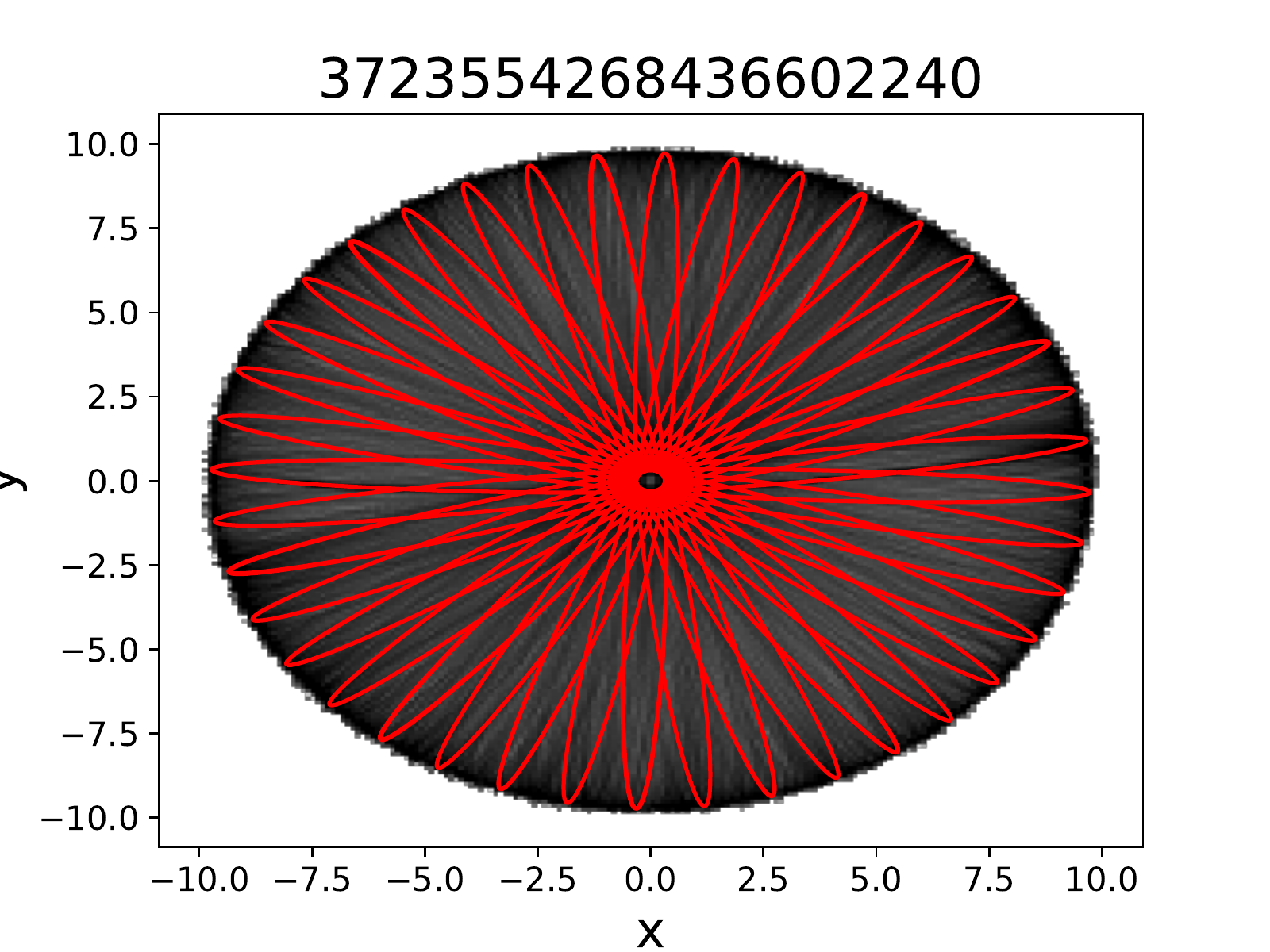}
\includegraphics[clip=true, trim = {0cm 0cm 1cm 0cm},width=4.5cm]{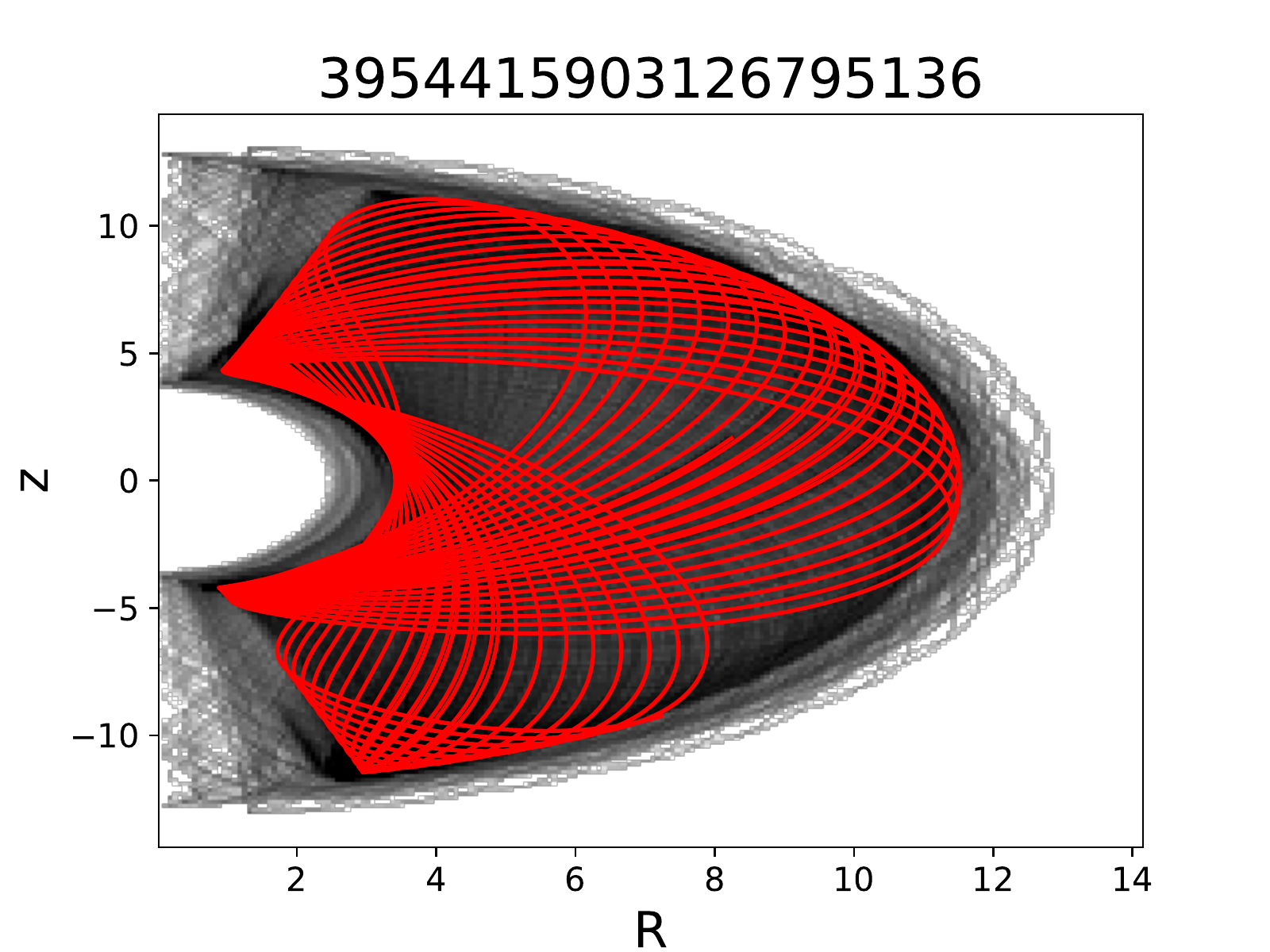}
\includegraphics[clip=true, trim = {0cm 0cm 1cm 0cm},width=4.5cm]{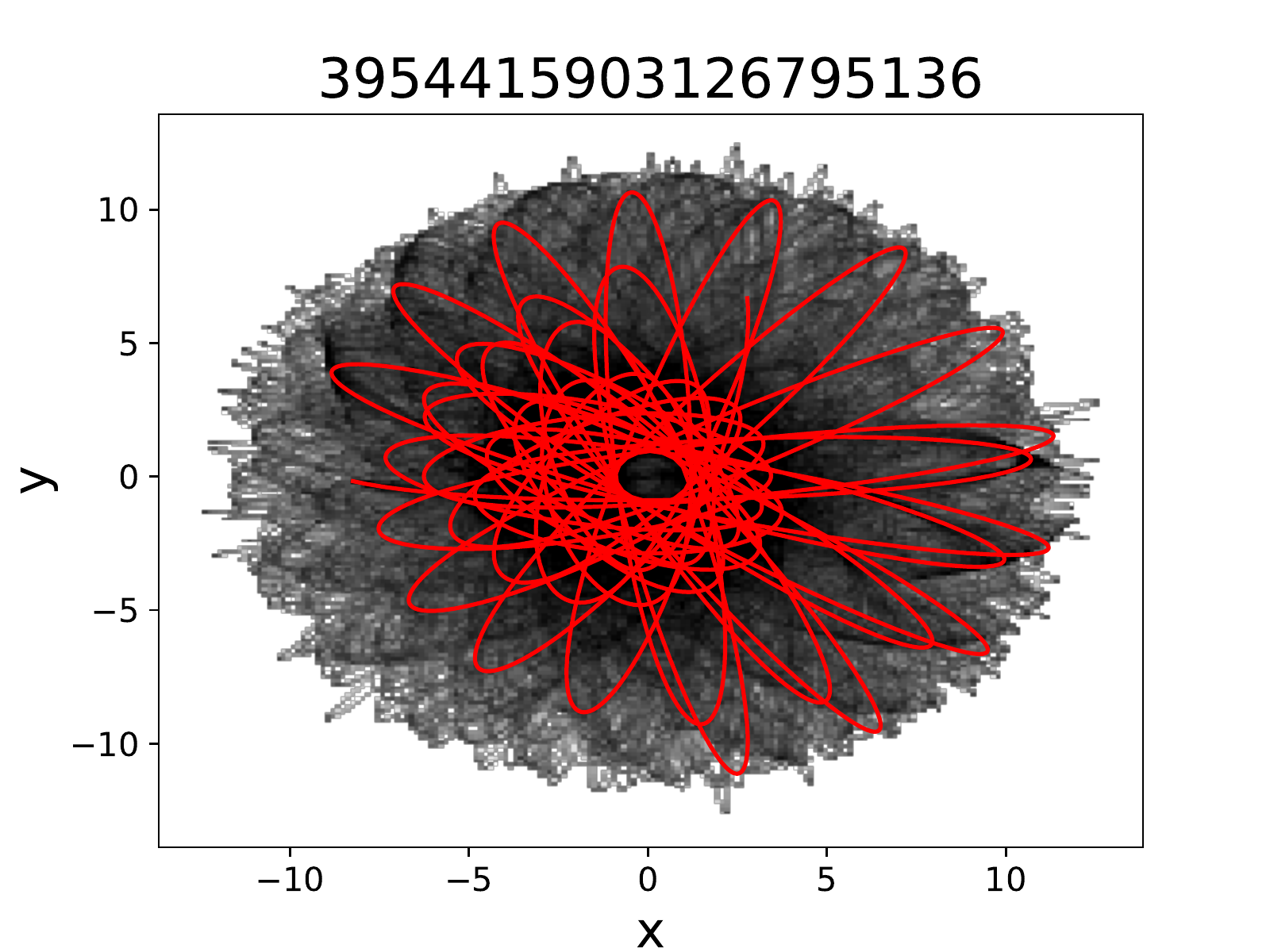}
\includegraphics[clip=true, trim = {0cm 0cm 1cm 0cm},width=4.5cm]{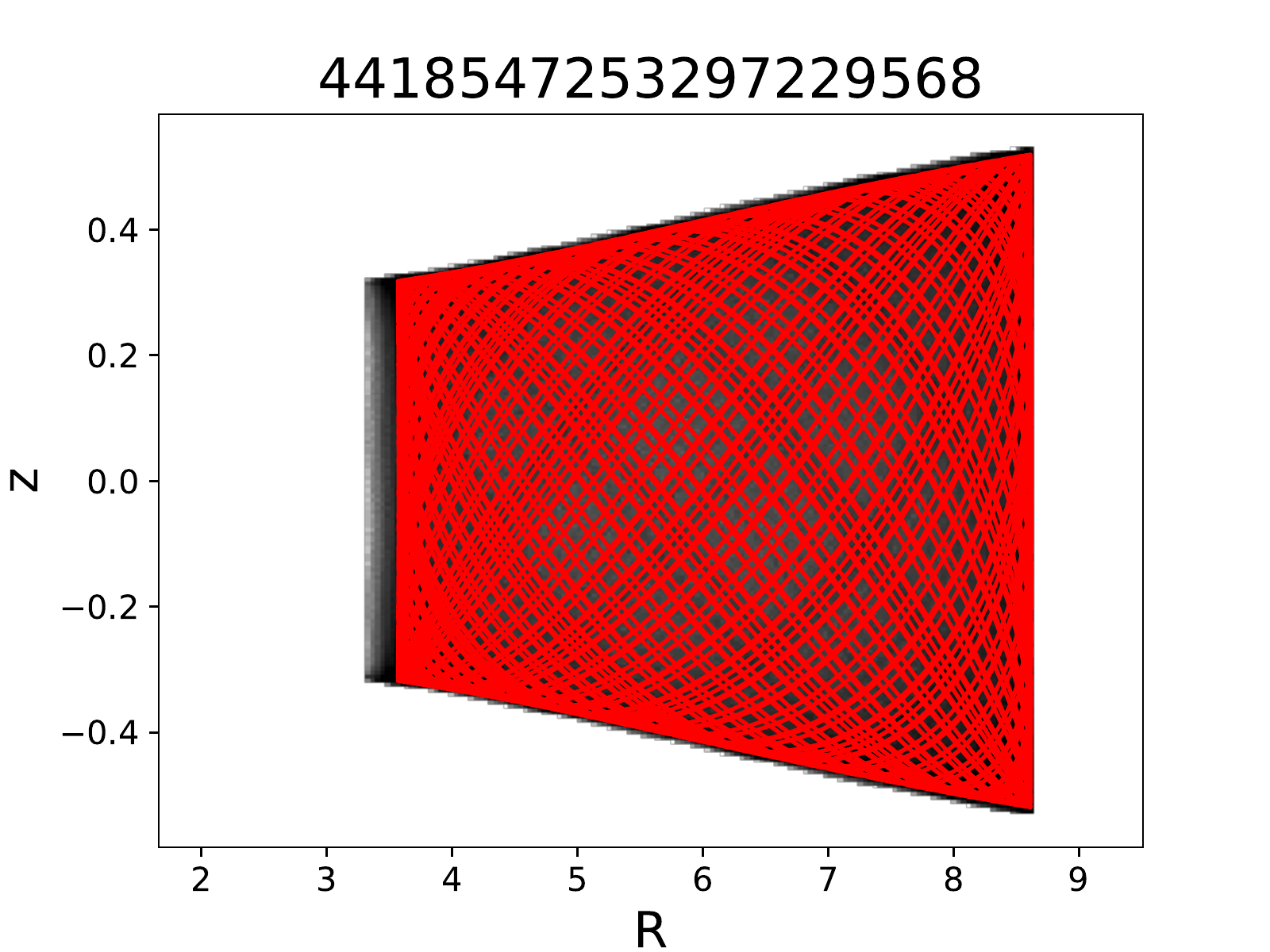}
\includegraphics[clip=true, trim = {0cm 0cm 1cm 0cm},width=4.5cm]{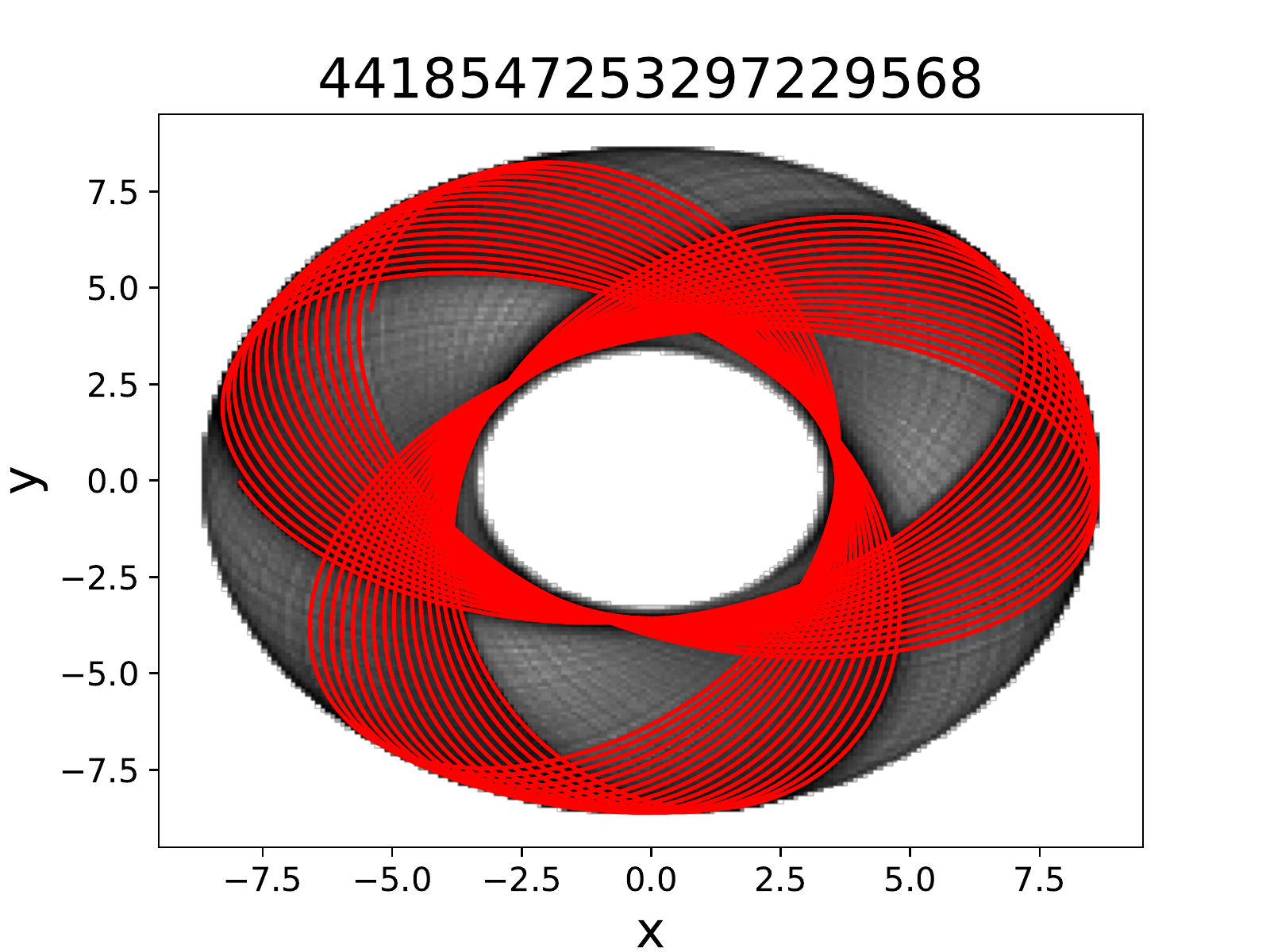}
\includegraphics[clip=true, trim = {0cm 0cm 1cm 0cm},width=4.5cm]{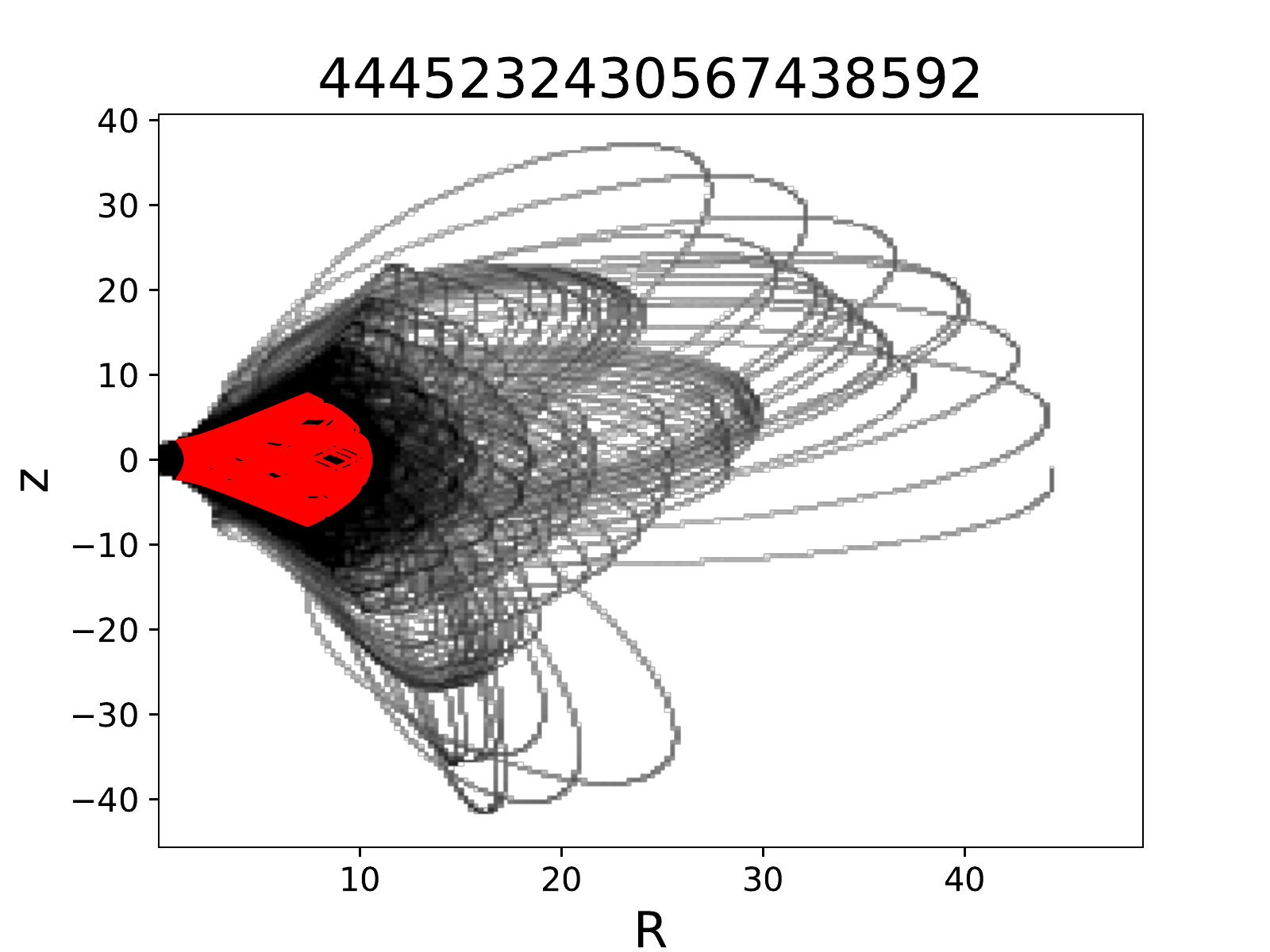}
\includegraphics[clip=true, trim = {0cm 0cm 1cm 0cm},width=4.5cm]{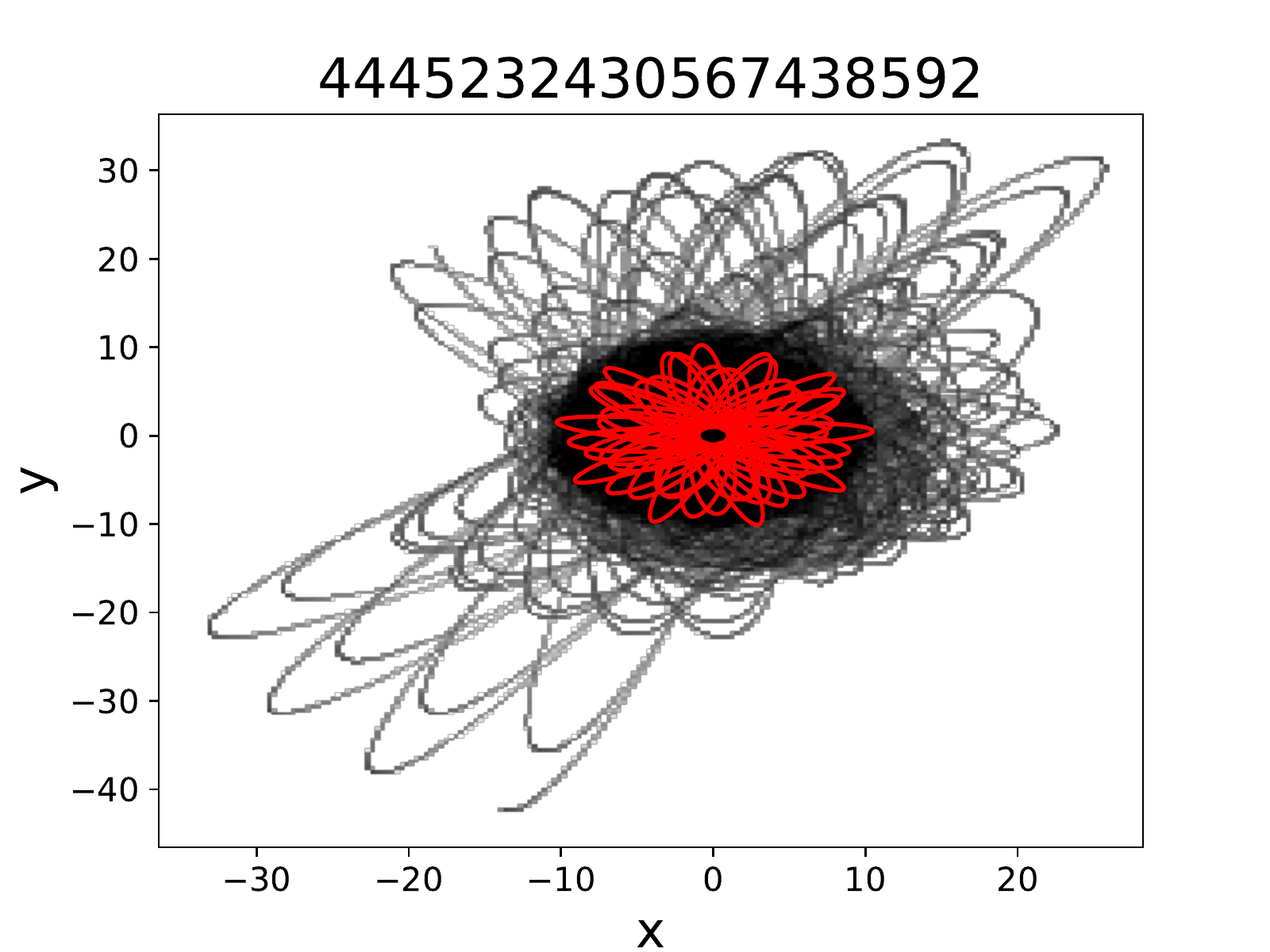}
\includegraphics[clip=true, trim = {0cm 0cm 1cm 0cm},width=4.5cm]{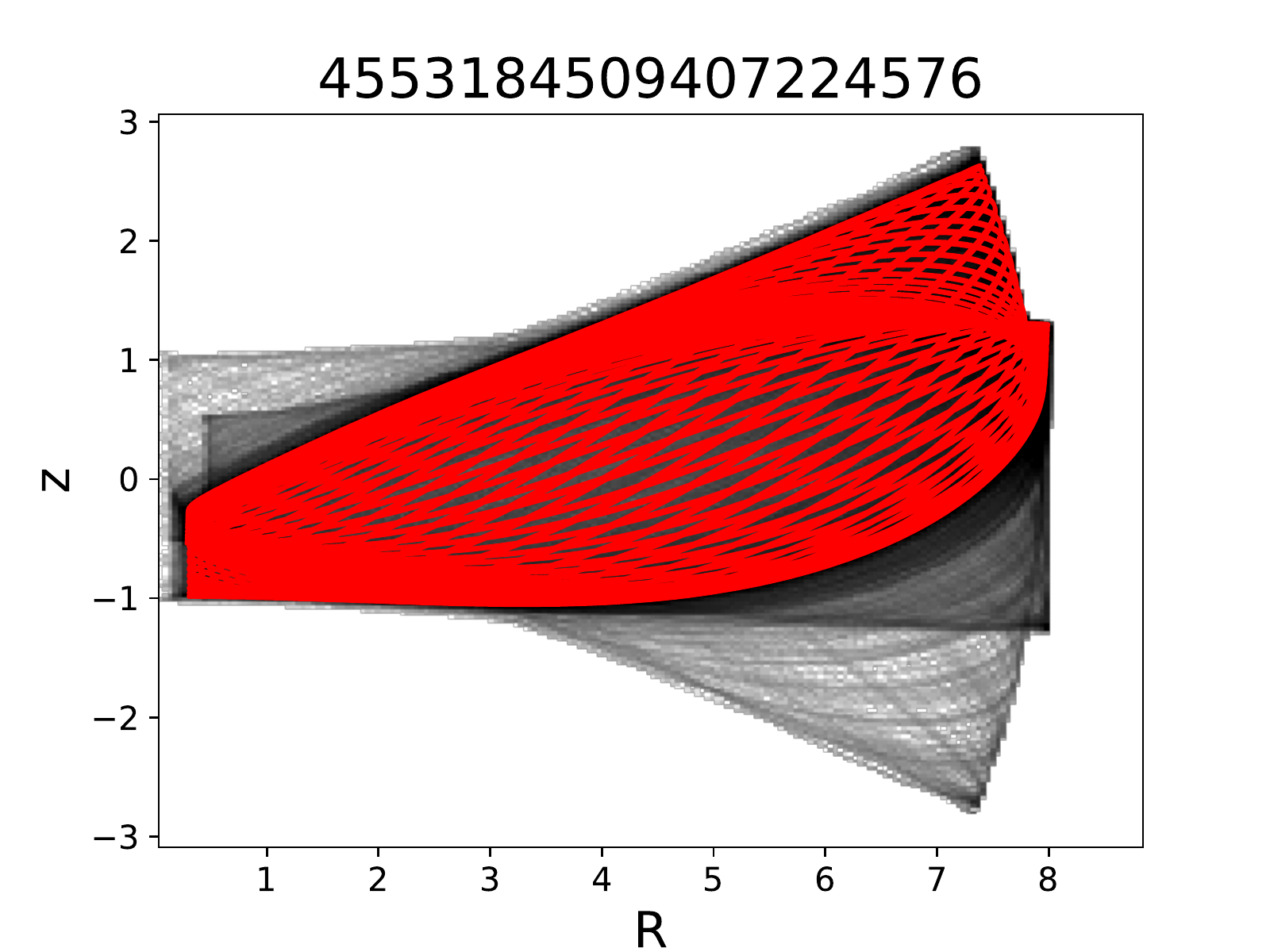}
\includegraphics[clip=true, trim = {0cm 0cm 1cm 0cm},width=4.5cm]{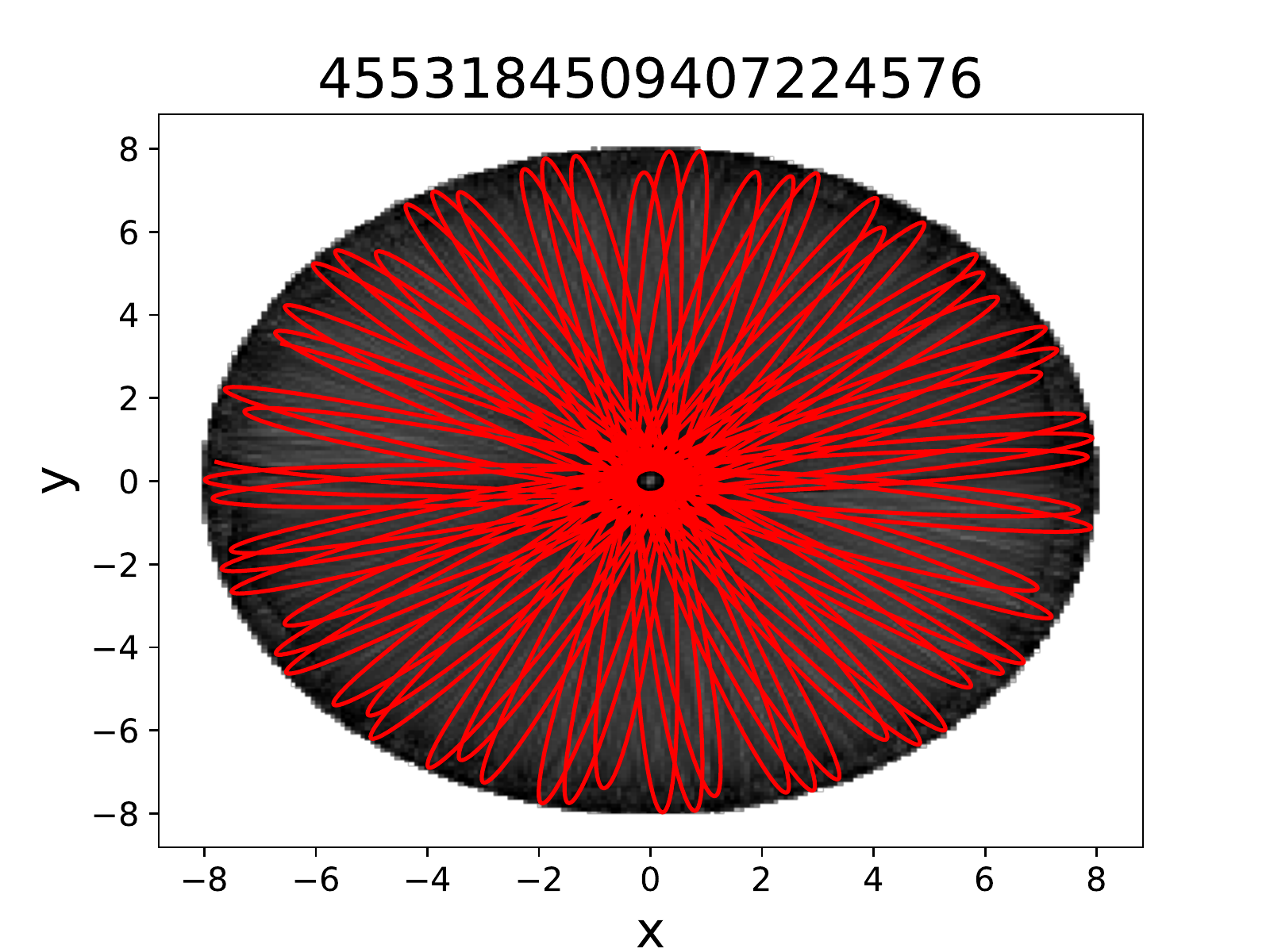}
\includegraphics[clip=true, trim = {0cm 0cm 1cm 0cm},width=4.5cm]{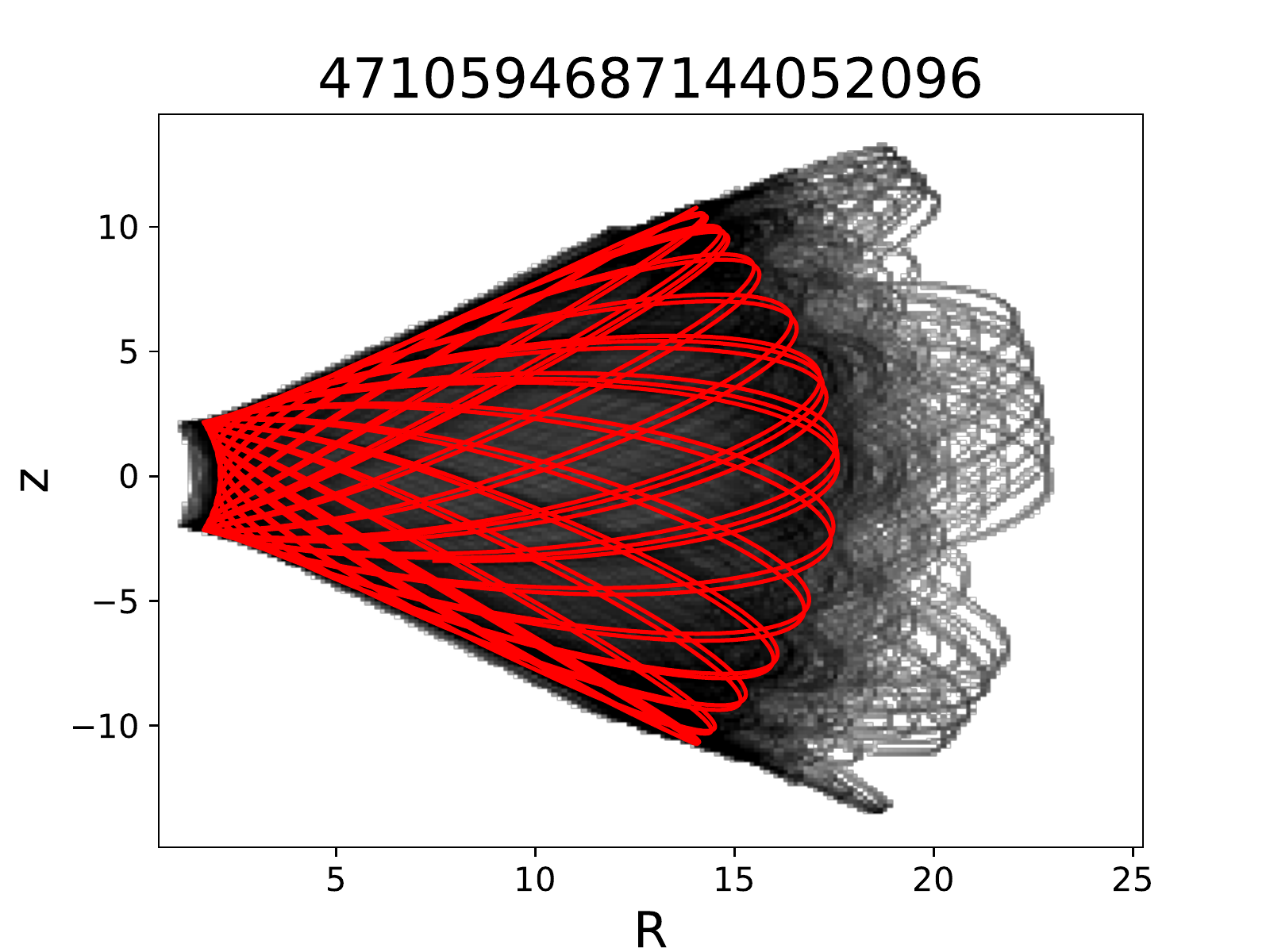}
\includegraphics[clip=true, trim = {0cm 0cm 1cm 0cm},width=4.5cm]{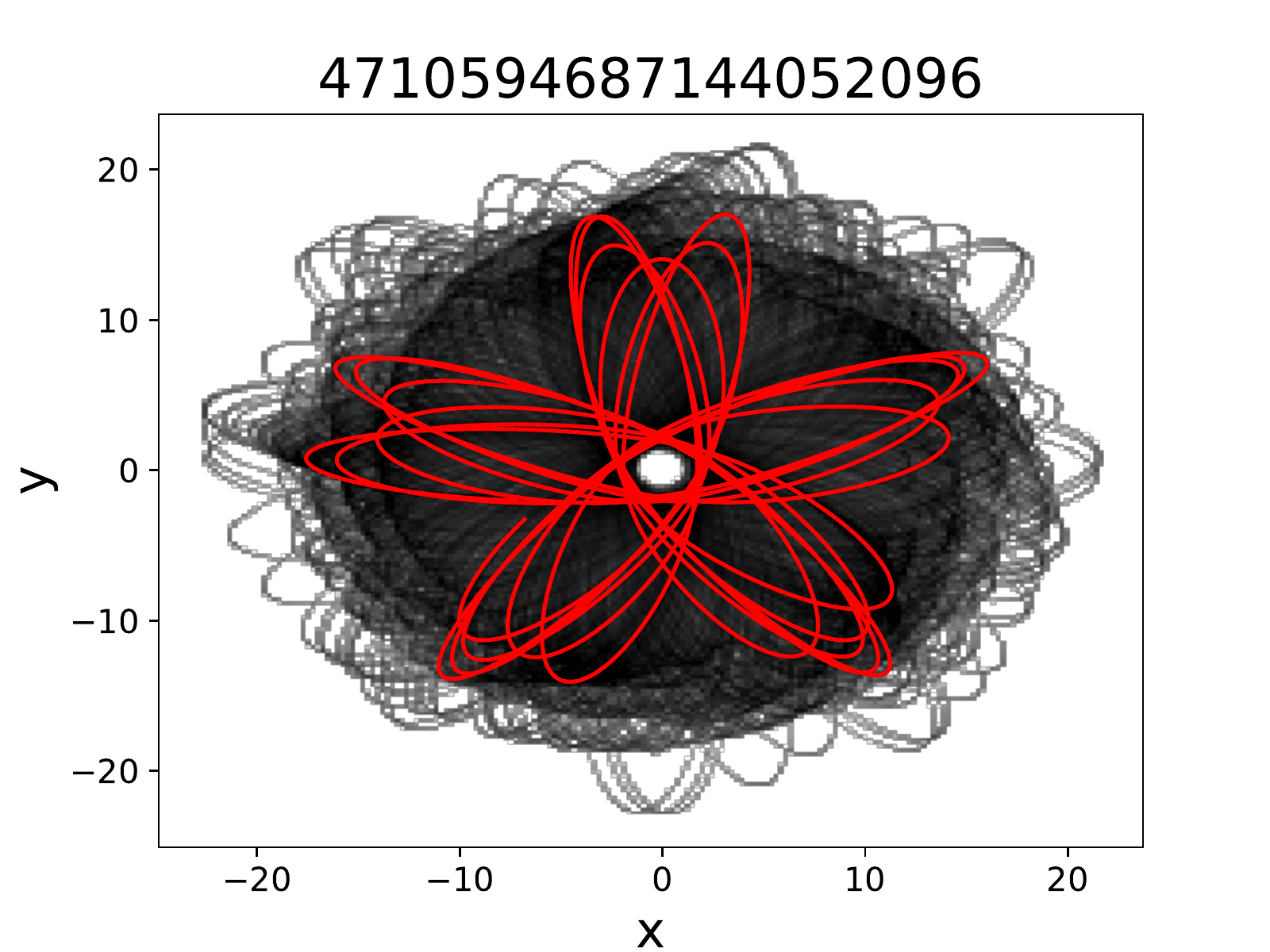}
\includegraphics[clip=true, trim = {0cm 0cm 1cm 0cm},width=4.5cm]{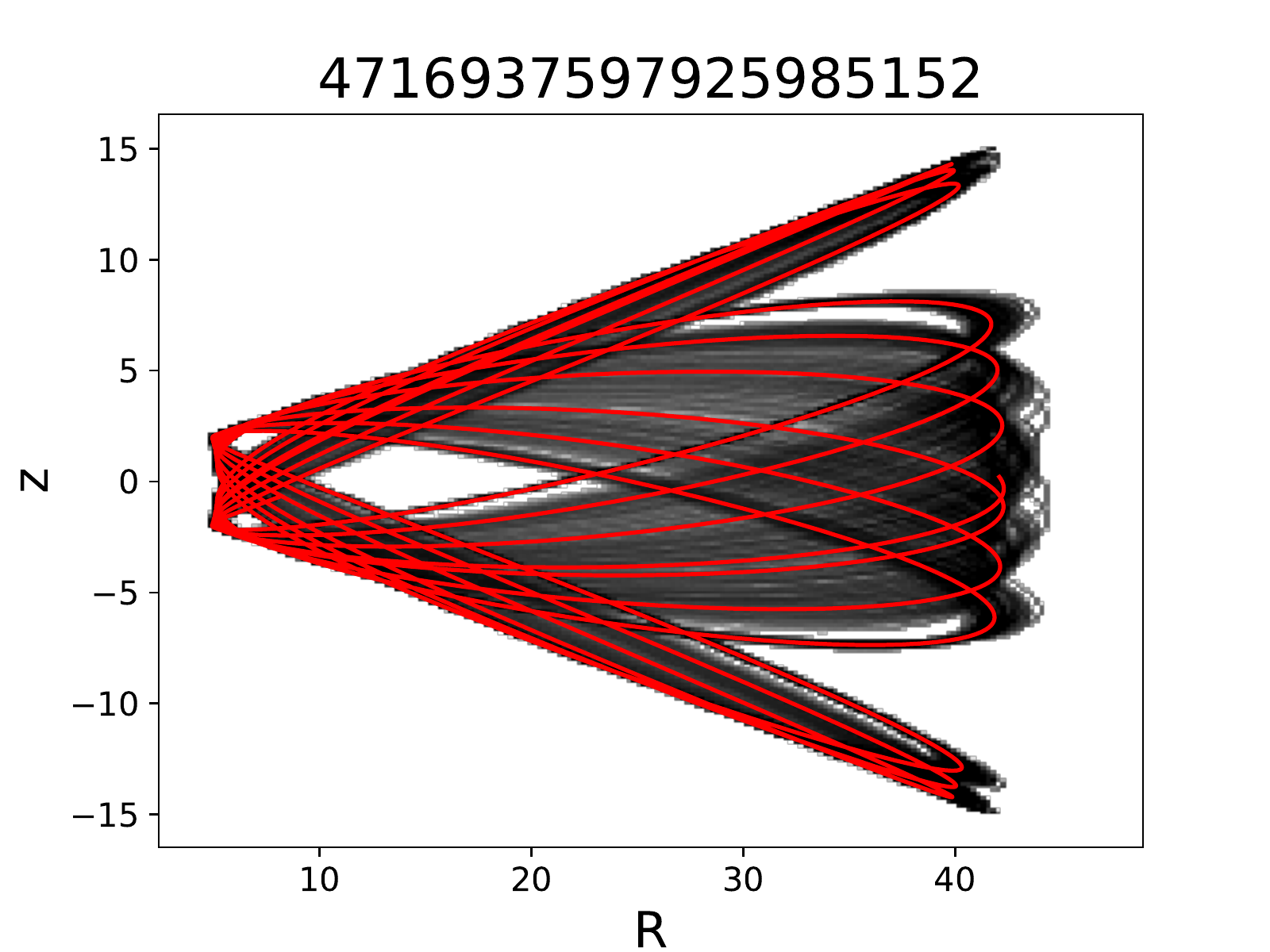}
\includegraphics[clip=true, trim = {0cm 0cm 1cm 0cm},width=4.5cm]{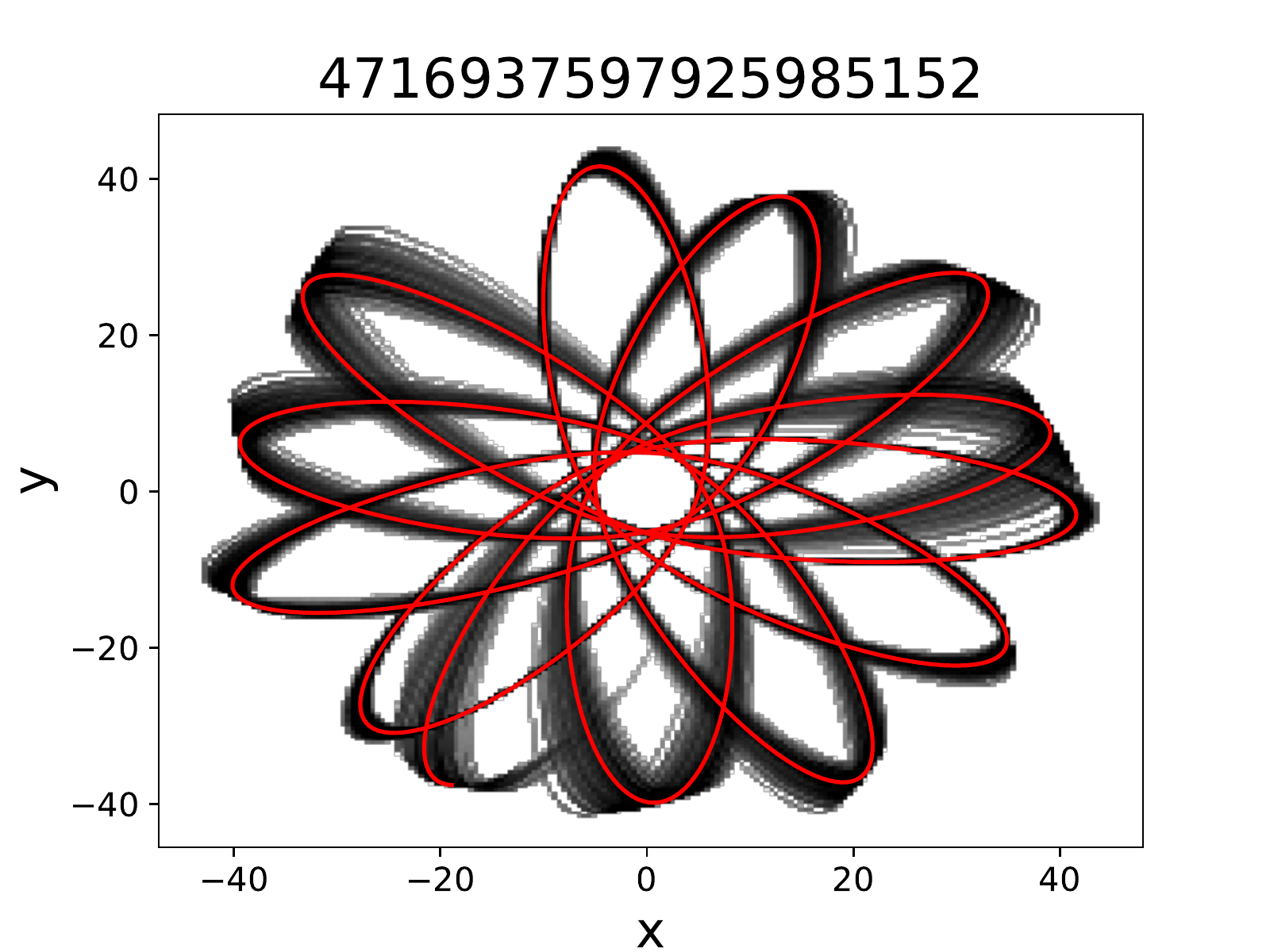}
\includegraphics[clip=true, trim = {0cm 0cm 1cm 0cm},width=4.5cm]{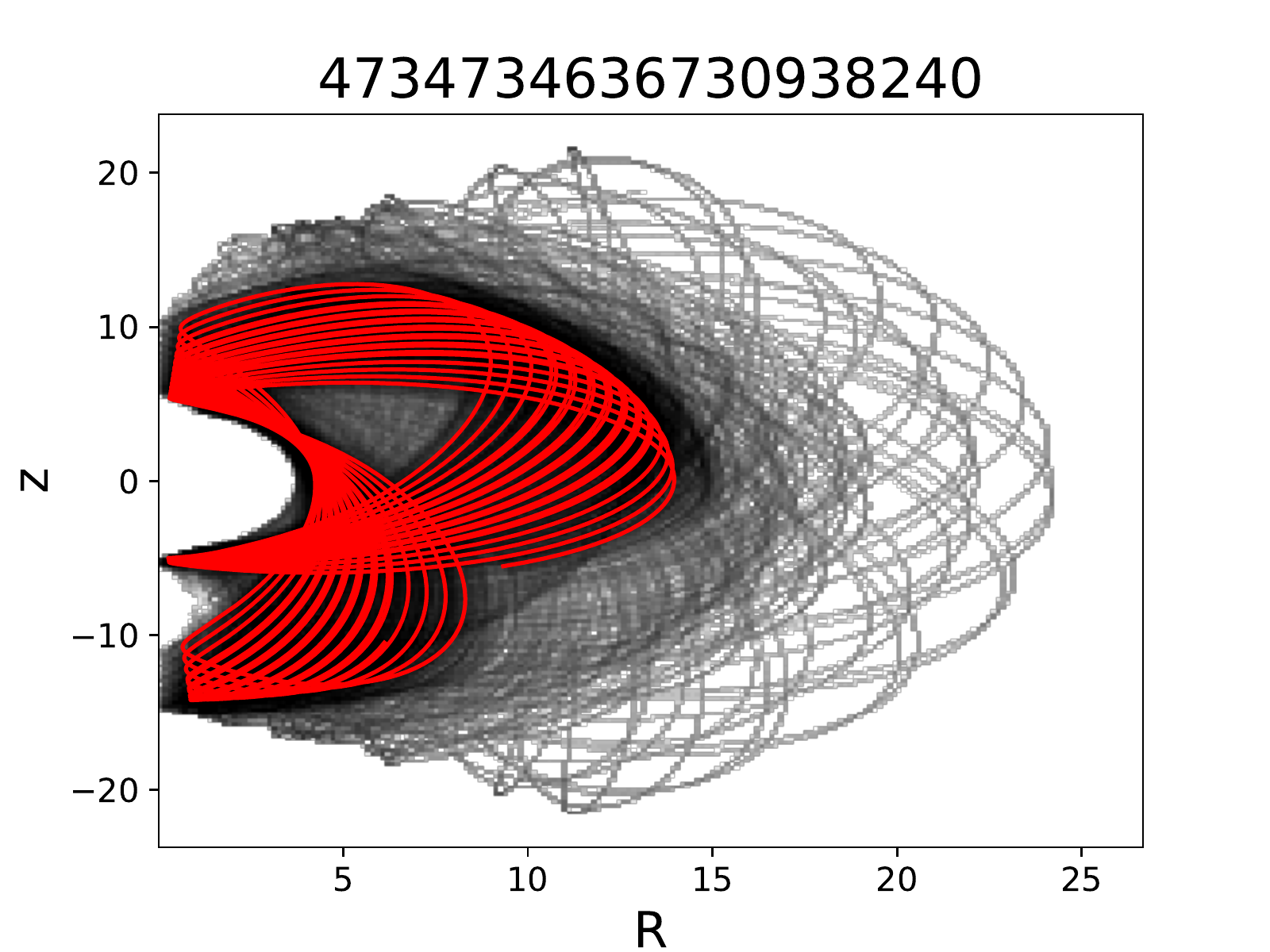}
\includegraphics[clip=true, trim = {0cm 0cm 1cm 0cm},width=4.5cm]{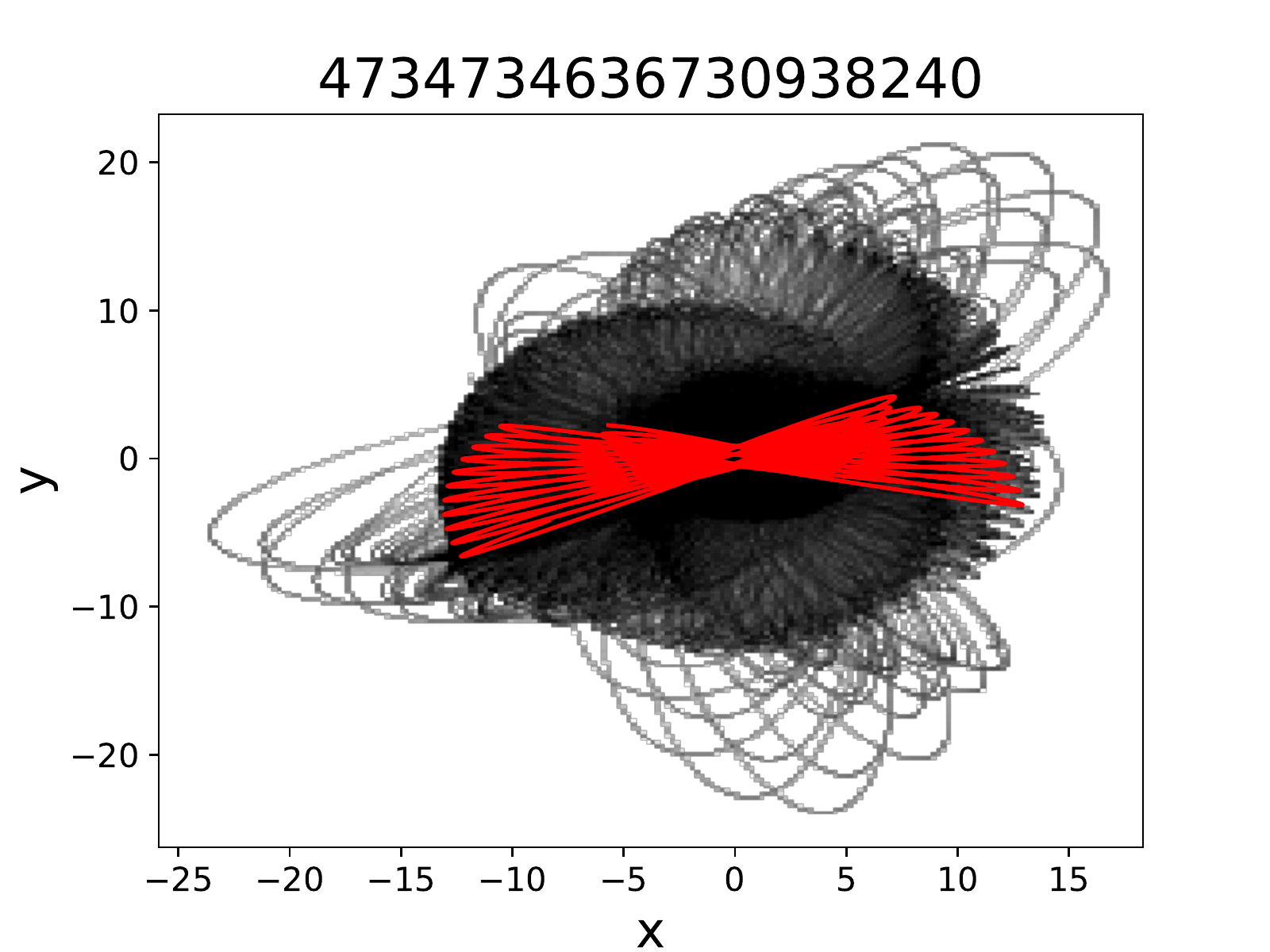}
\includegraphics[clip=true, trim = {0cm 0cm 1cm 0cm},width=4.5cm]{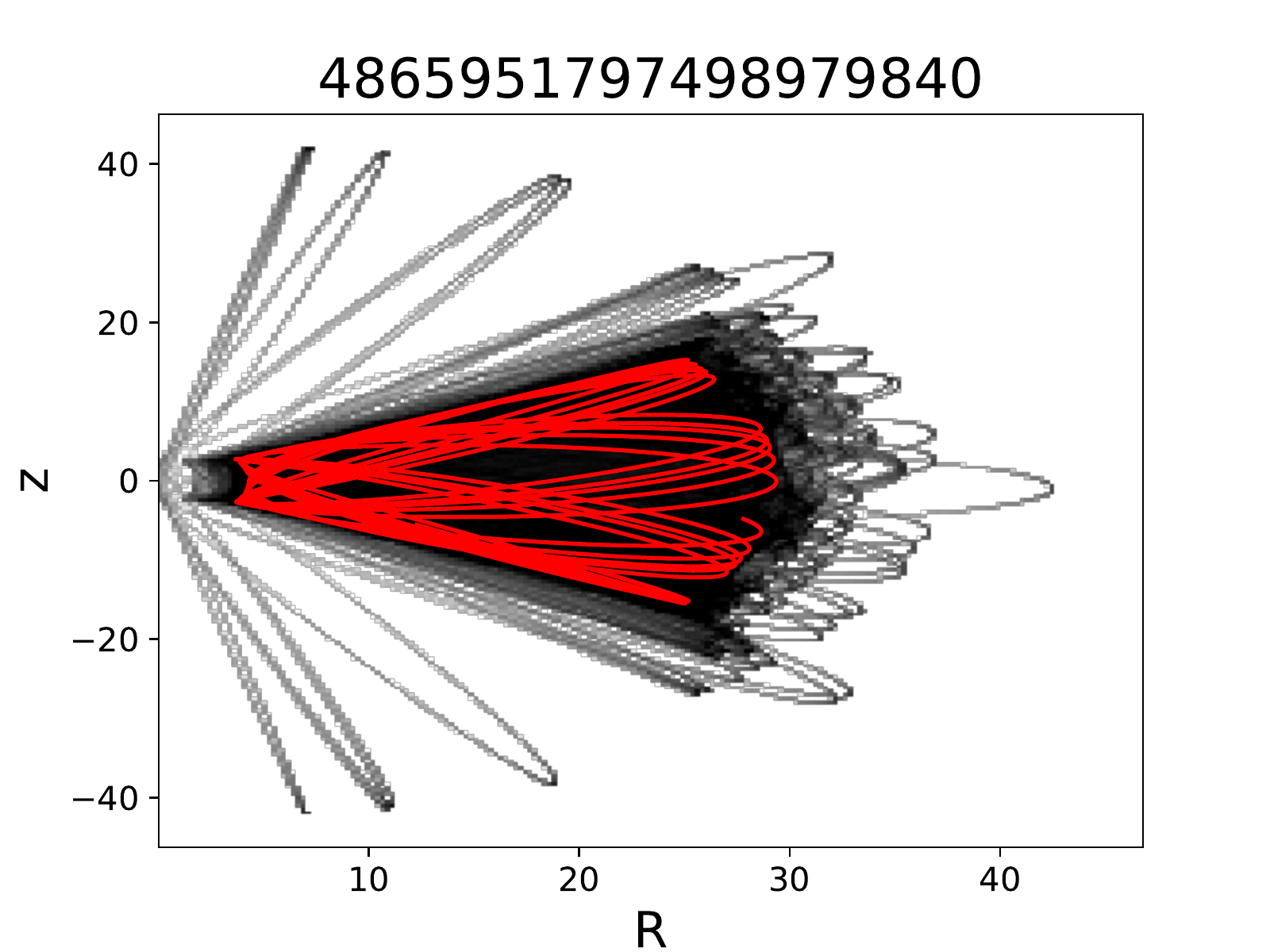}
\includegraphics[clip=true, trim = {0cm 0cm 1cm 0cm},width=4.5cm]{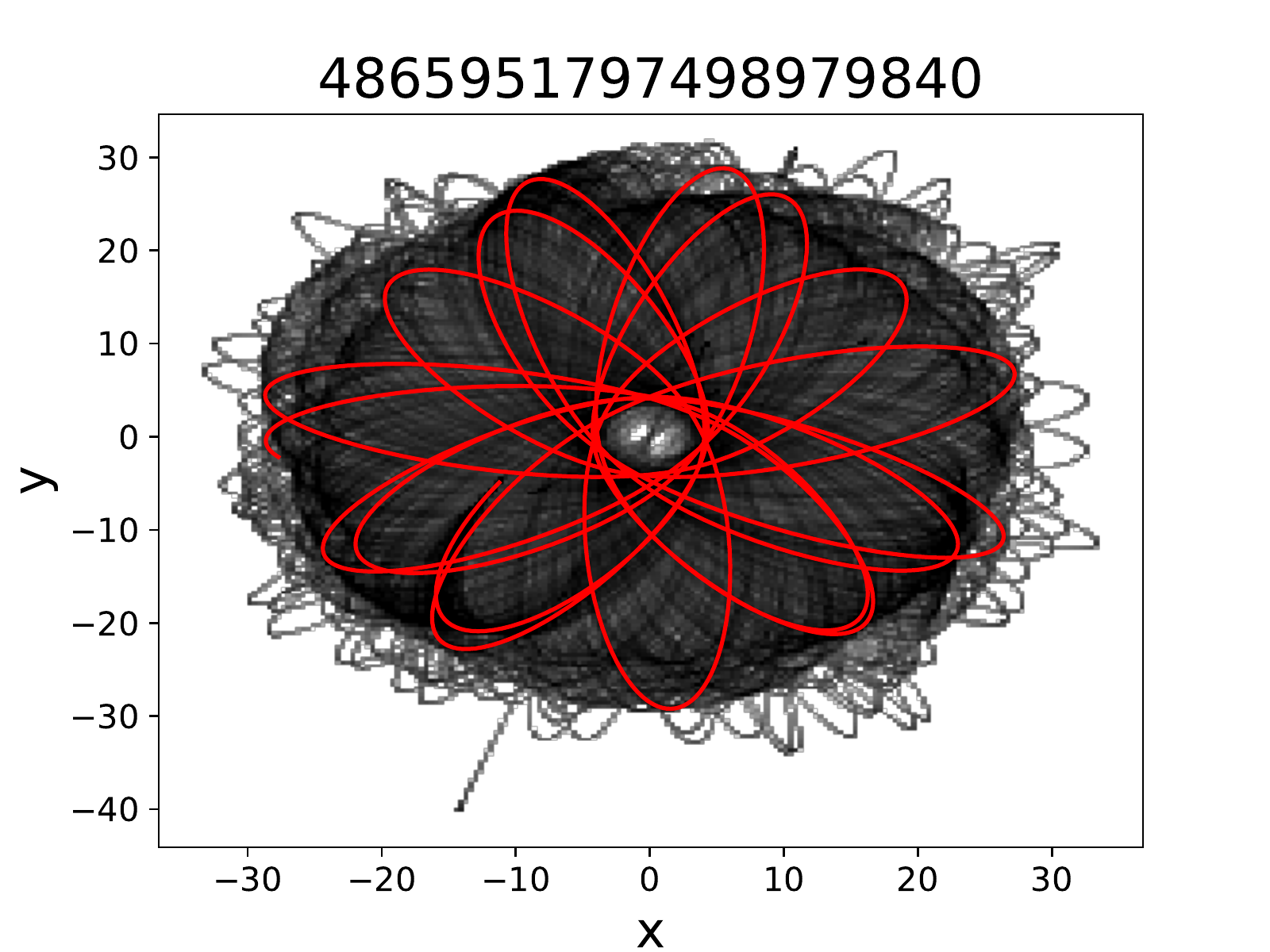}
\includegraphics[clip=true, trim = {0cm 0cm 1cm 0cm},width=4.5cm]{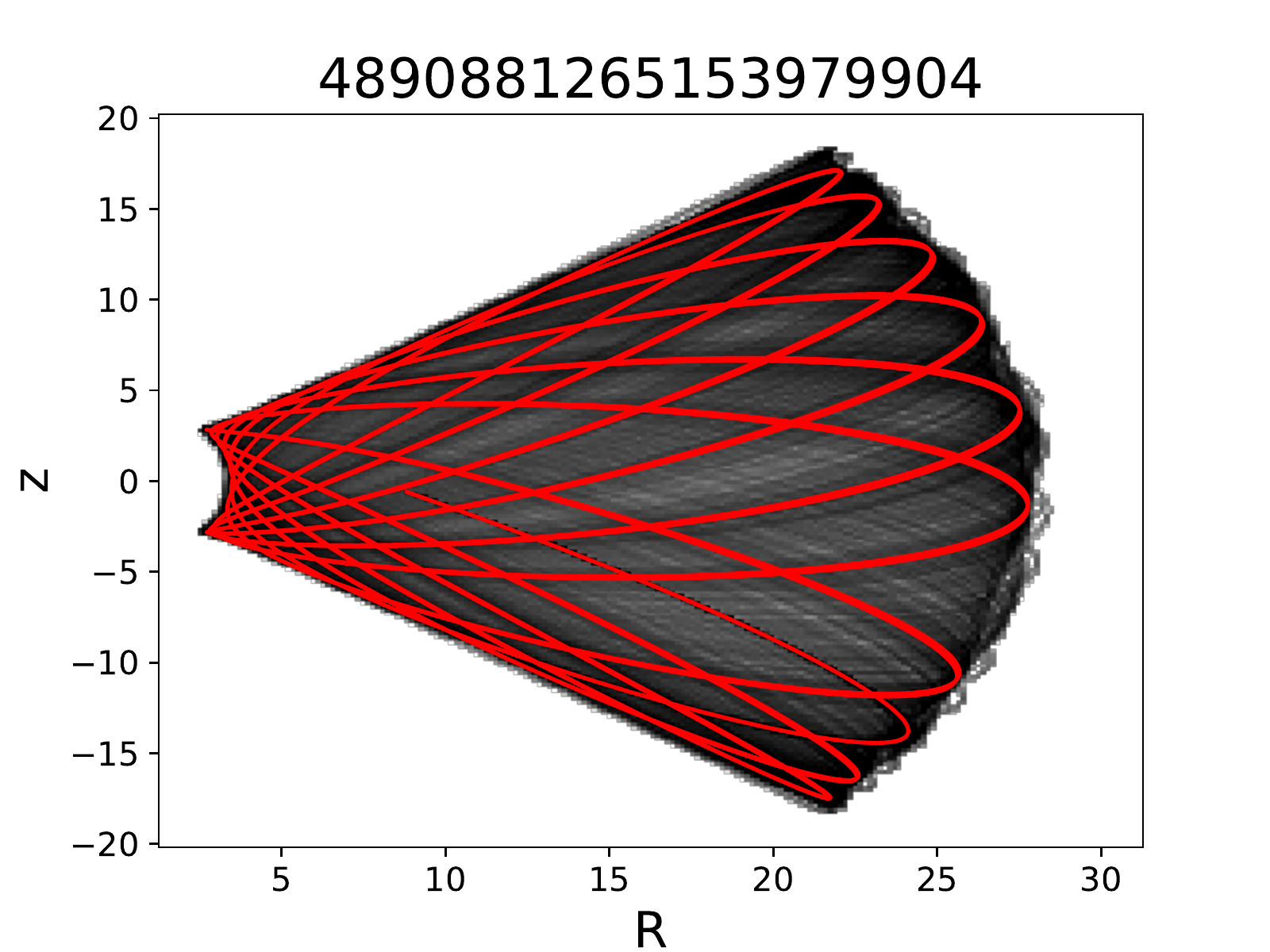}
\includegraphics[clip=true, trim = {0cm 0cm 1cm 0cm},width=4.5cm]{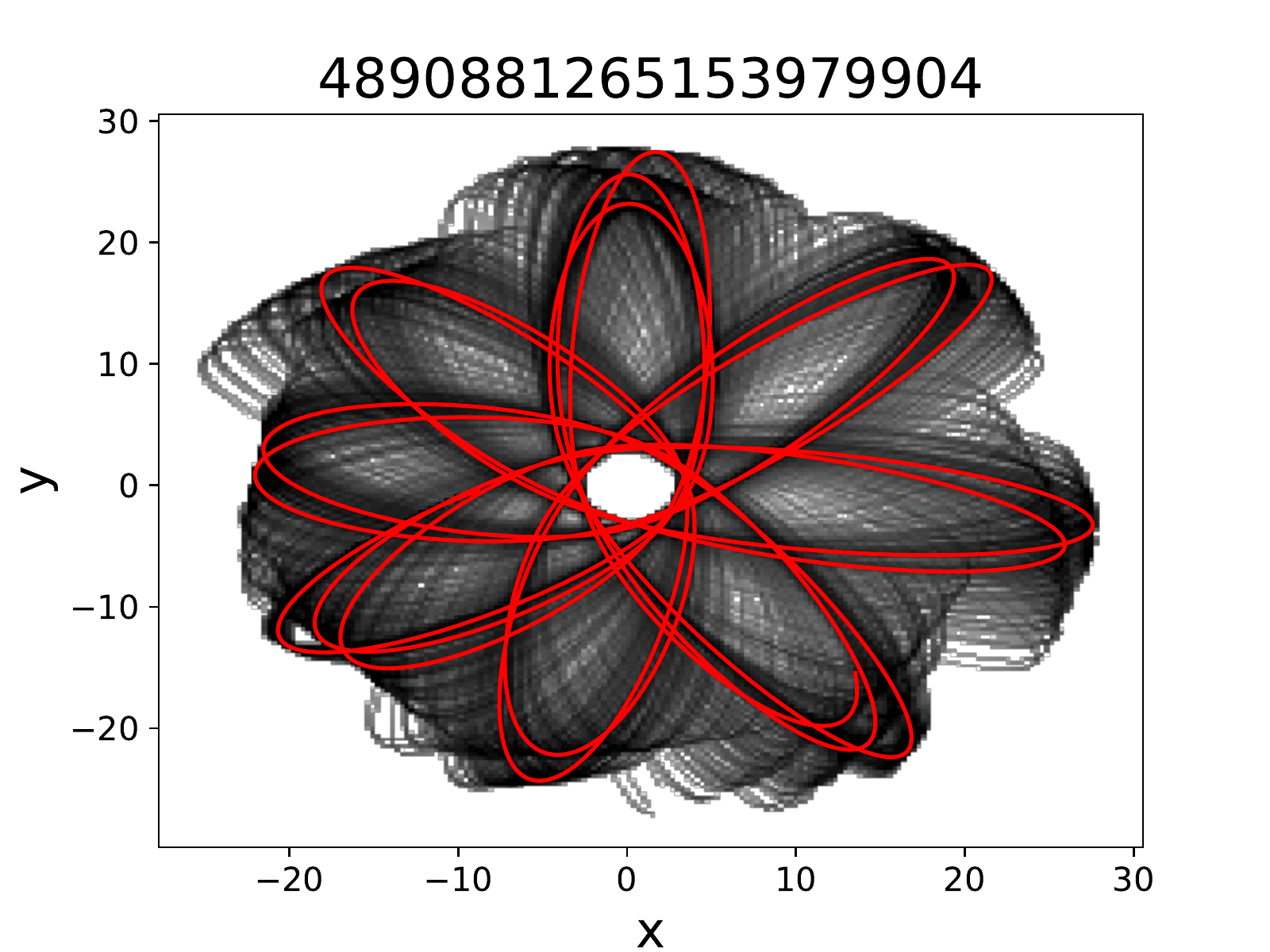}
\caption{Continued from Fig.~\ref{T1}.}
\end{centering}
\end{figure*}

\begin{figure*}
\begin{centering}
\includegraphics[clip=true, trim = {0cm 0cm 1cm 0cm},width=4.5cm]{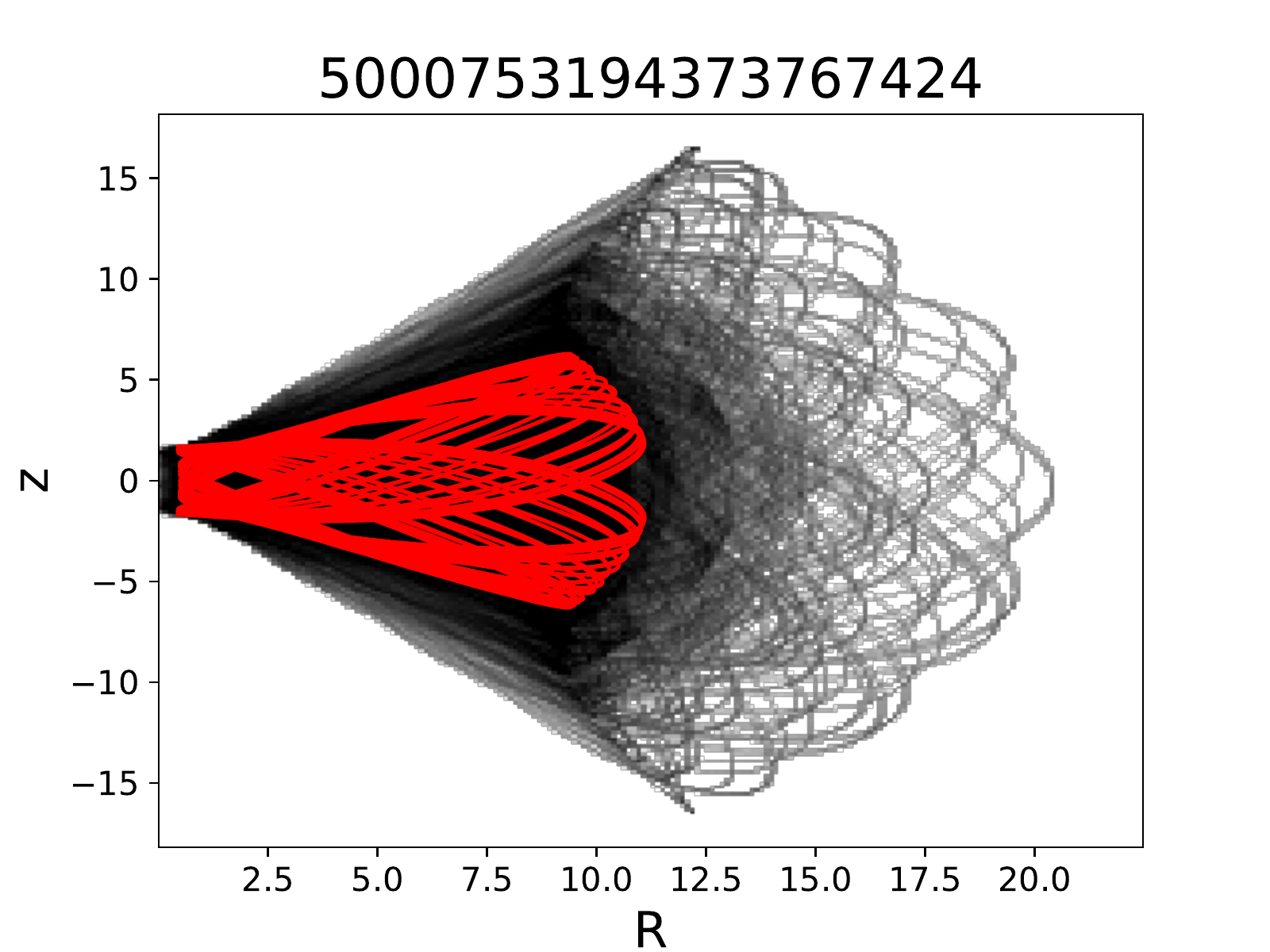}
\includegraphics[clip=true, trim = {0cm 0cm 1cm 0cm},width=4.5cm]{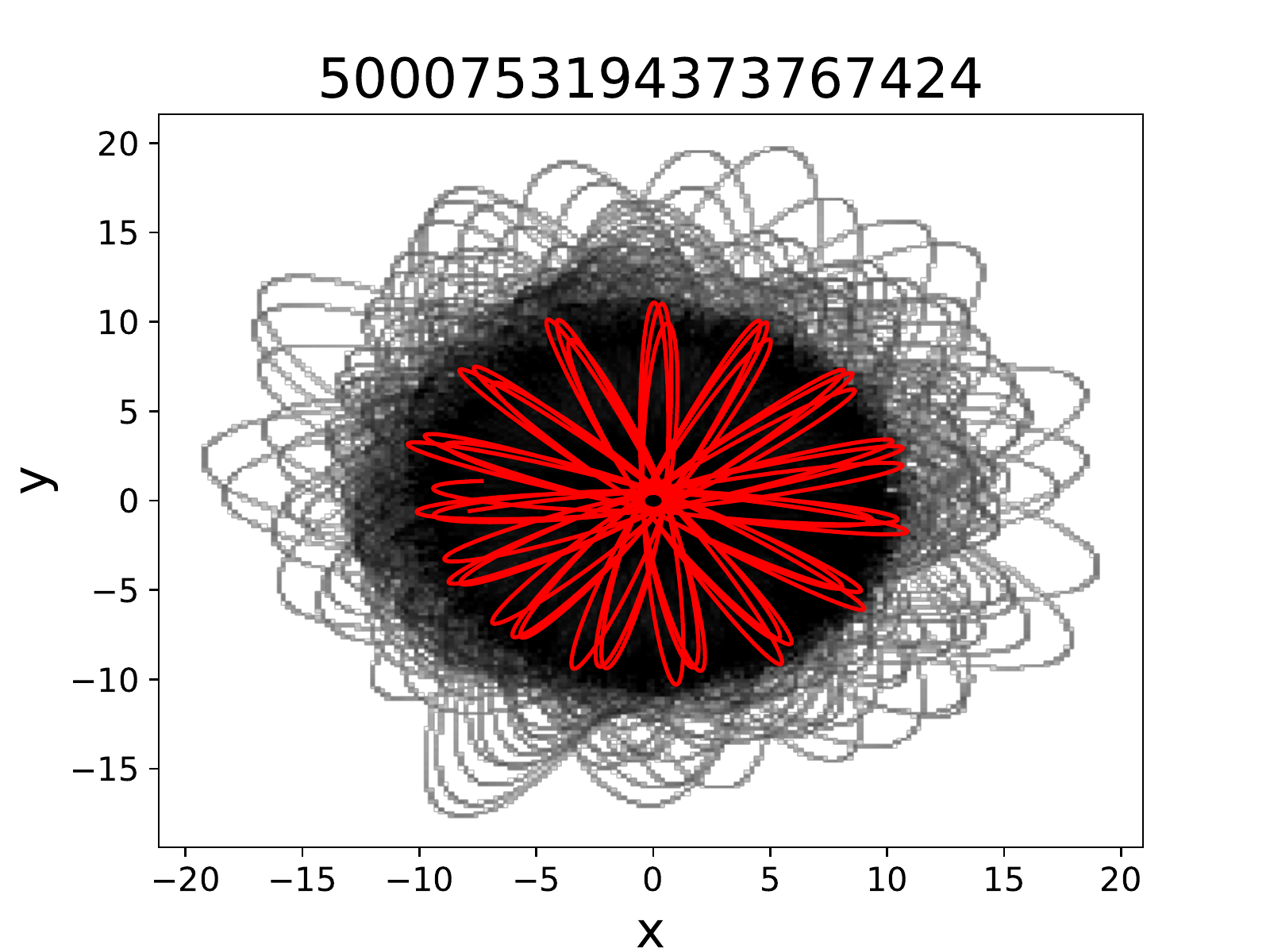}
\includegraphics[clip=true, trim = {0cm 0cm 1cm 0cm},width=4.5cm]{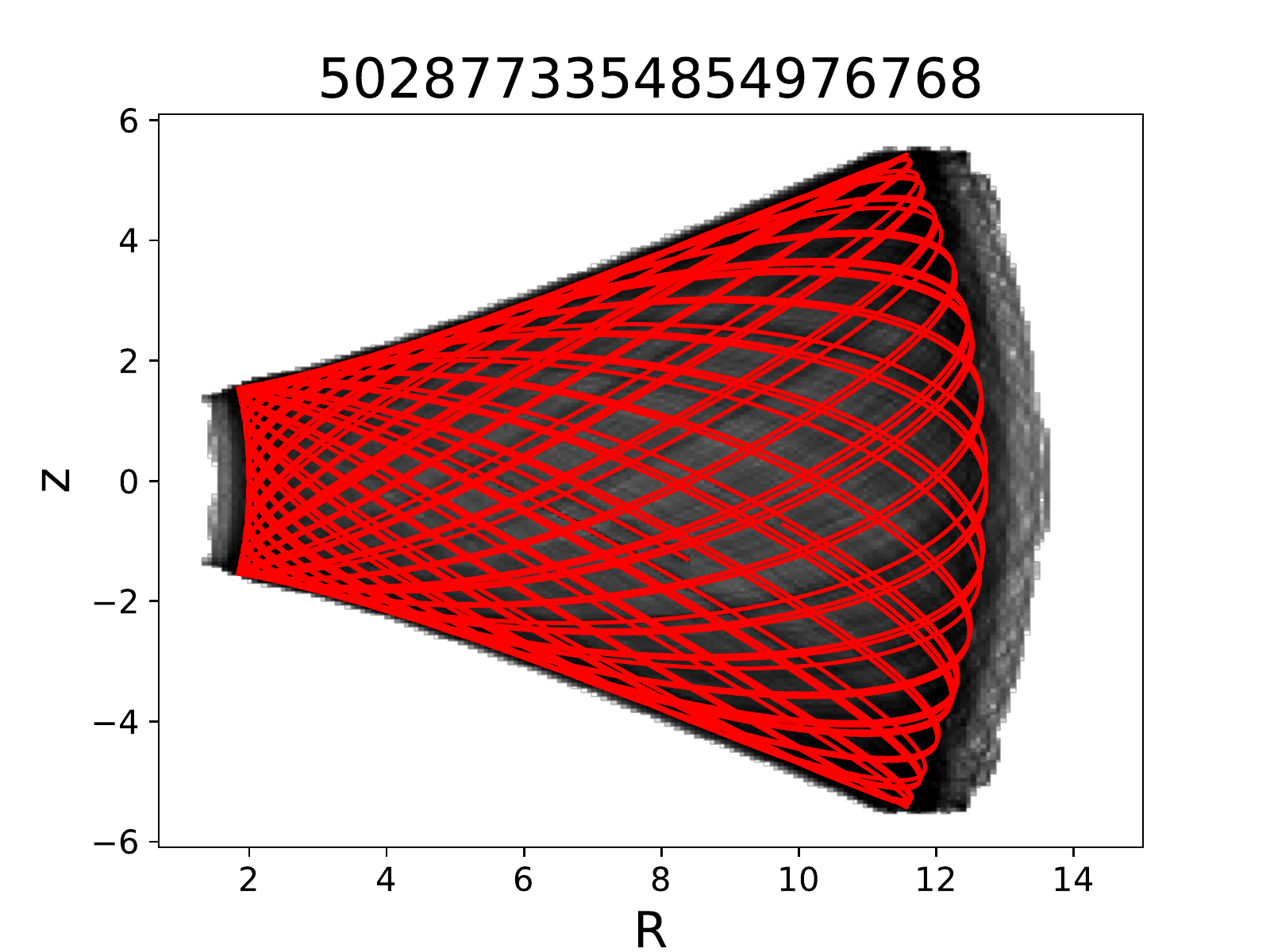}
\includegraphics[clip=true, trim = {0cm 0cm 1cm 0cm},width=4.5cm]{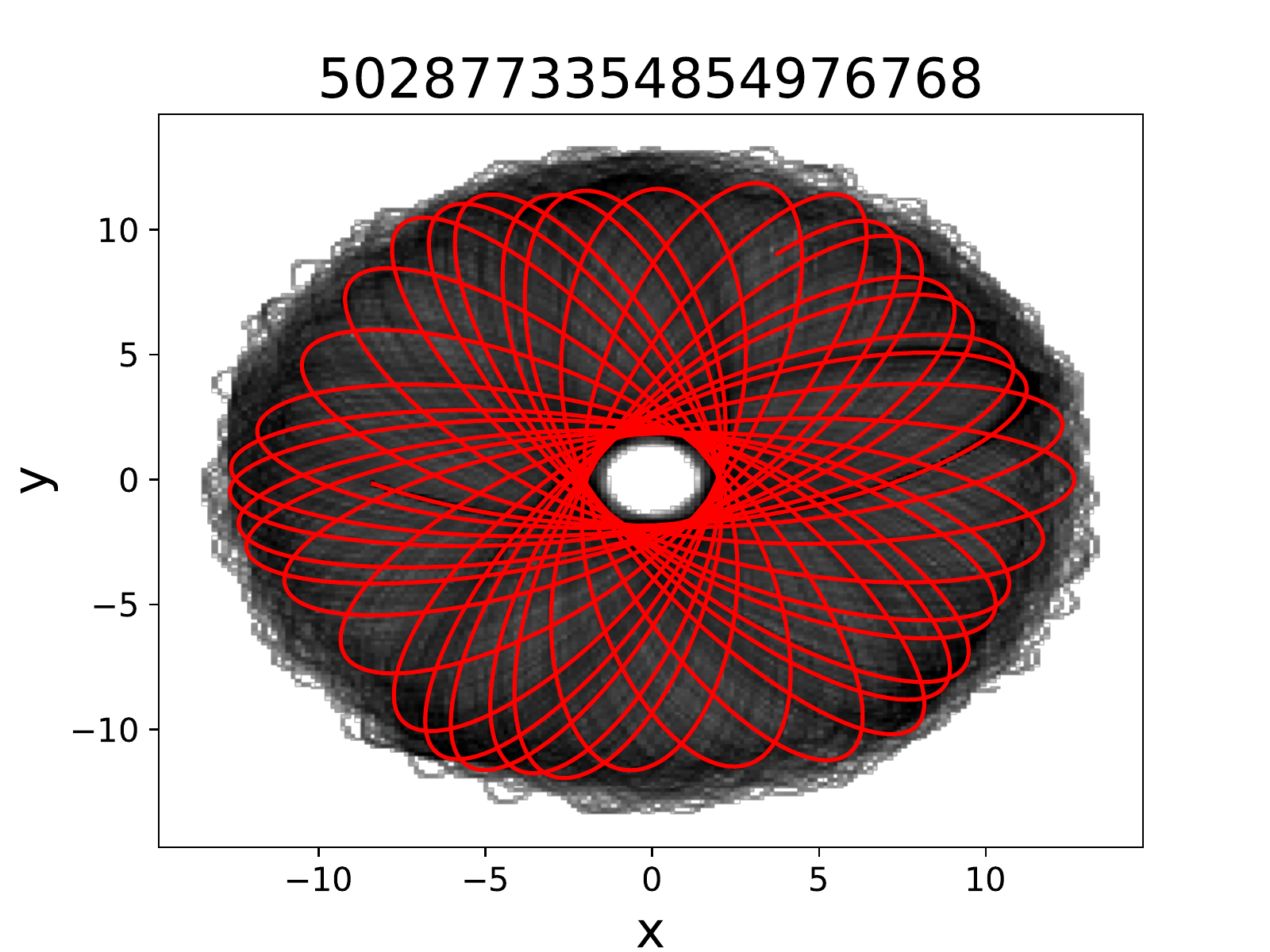}
\includegraphics[clip=true, trim = {0cm 0cm 1cm 0cm},width=4.5cm]{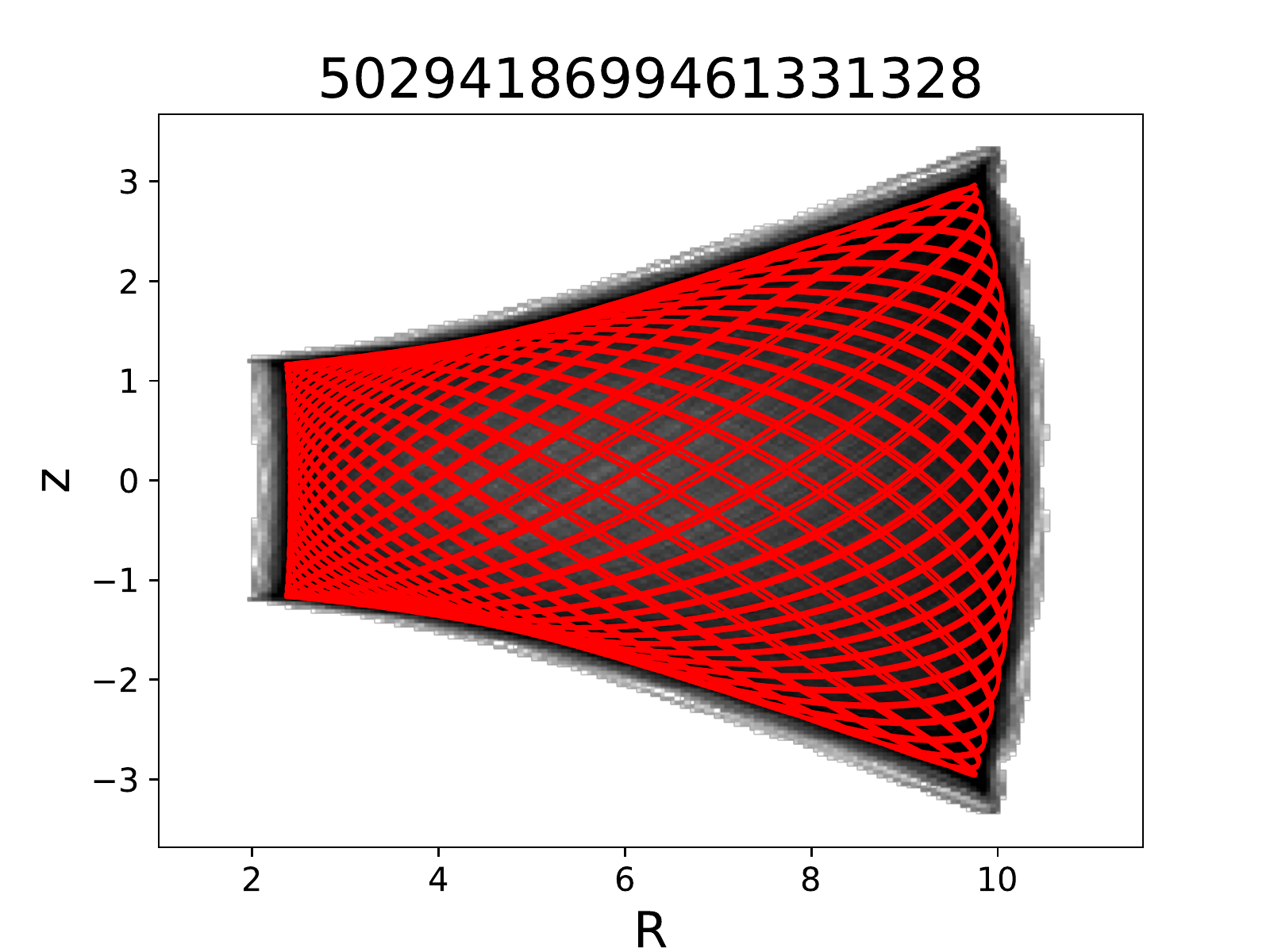}
\includegraphics[clip=true, trim = {0cm 0cm 1cm 0cm},width=4.5cm]{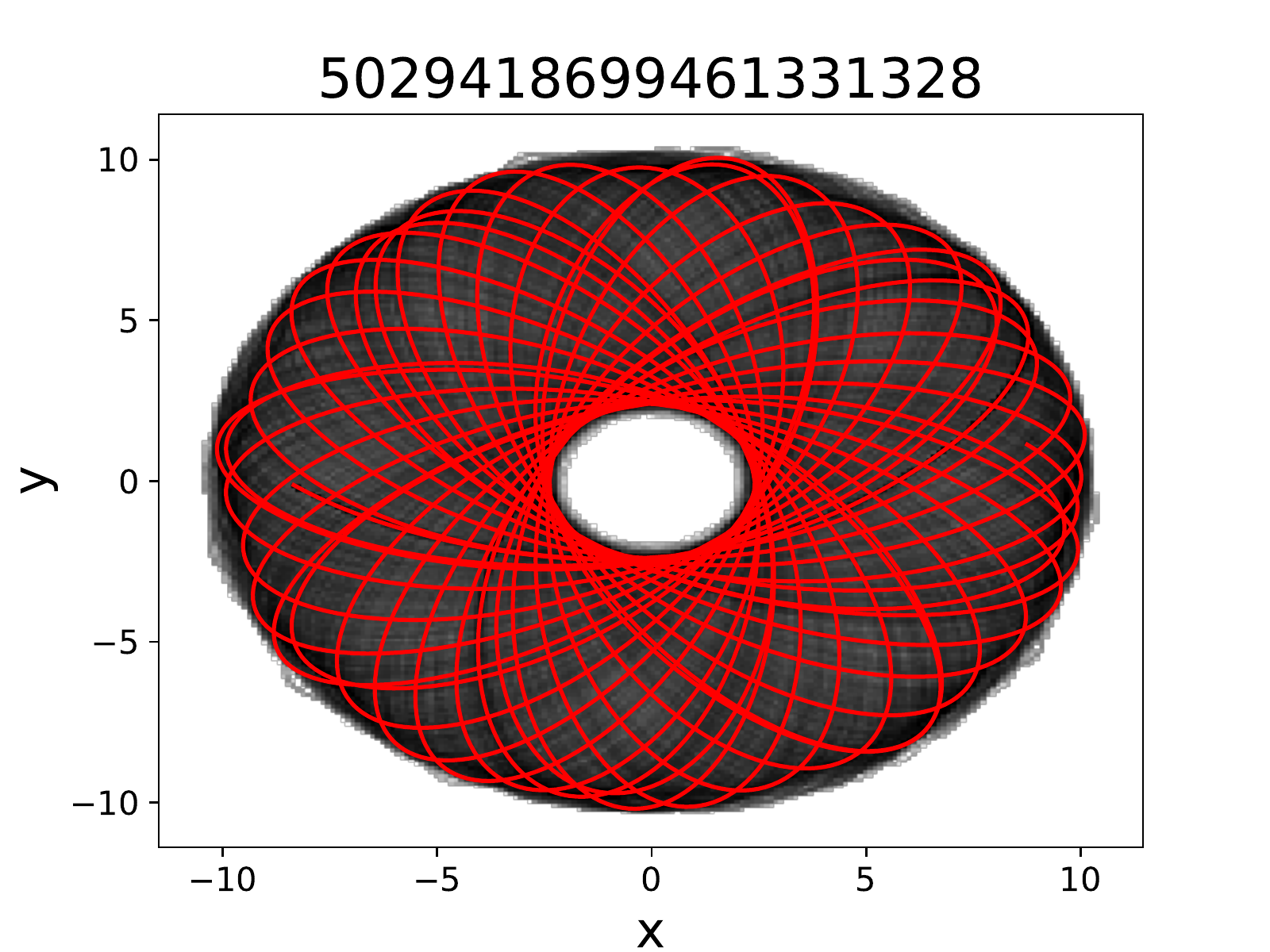}
\includegraphics[clip=true, trim = {0cm 0cm 1cm 0cm},width=4.5cm]{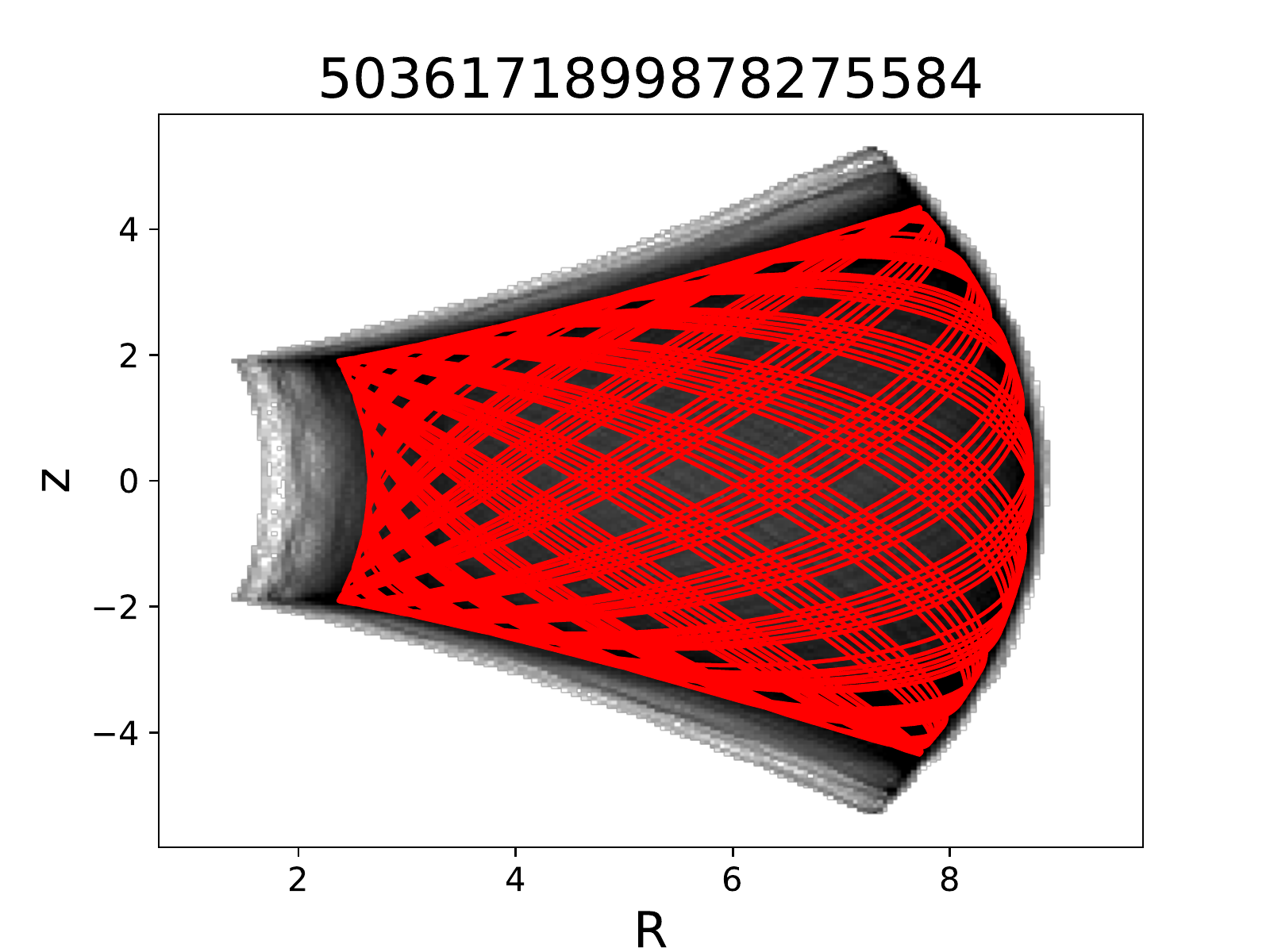}
\includegraphics[clip=true, trim = {0cm 0cm 1cm 0cm},width=4.5cm]{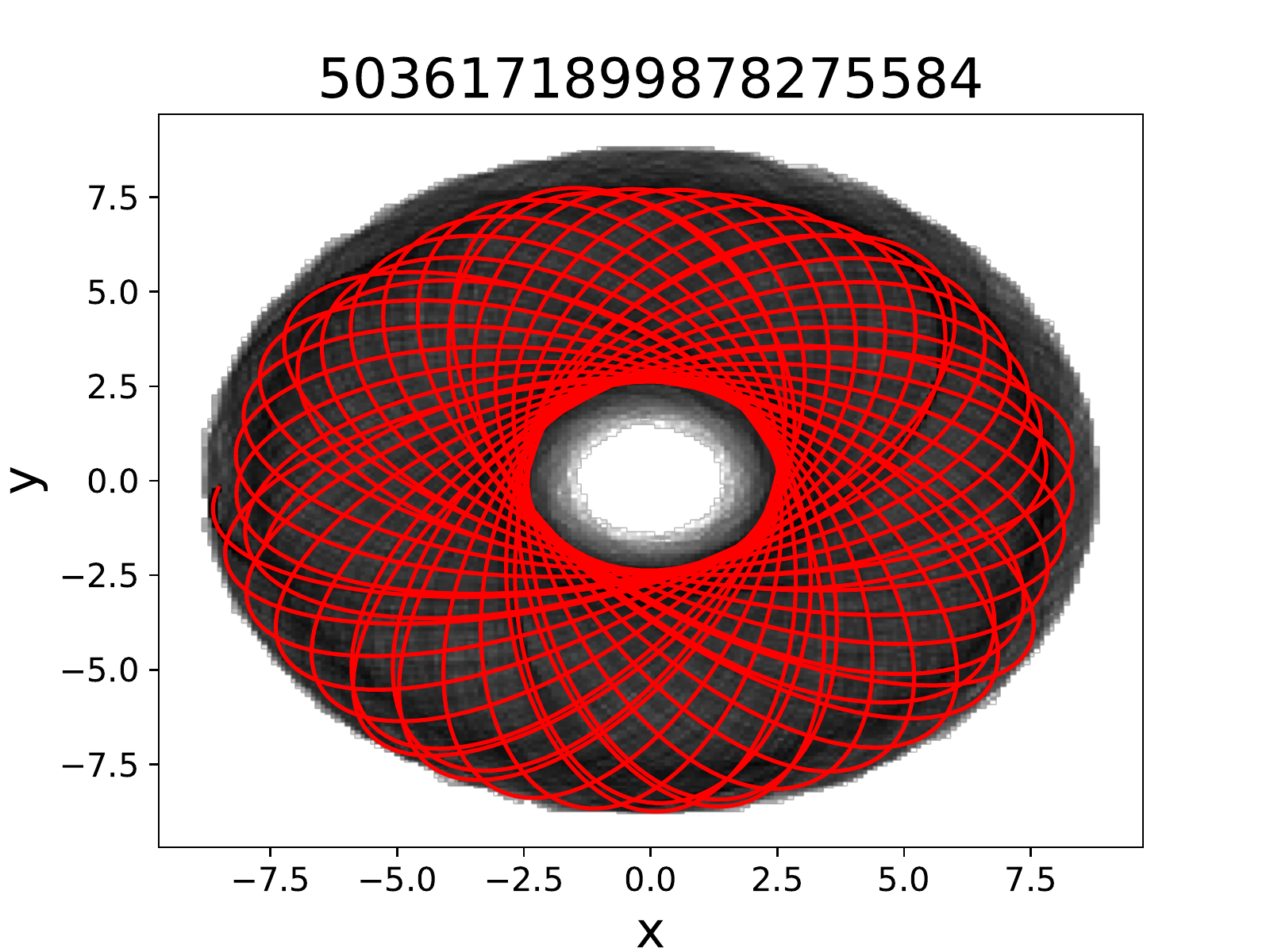}
\includegraphics[clip=true, trim = {0cm 0cm 1cm 0cm},width=4.5cm]{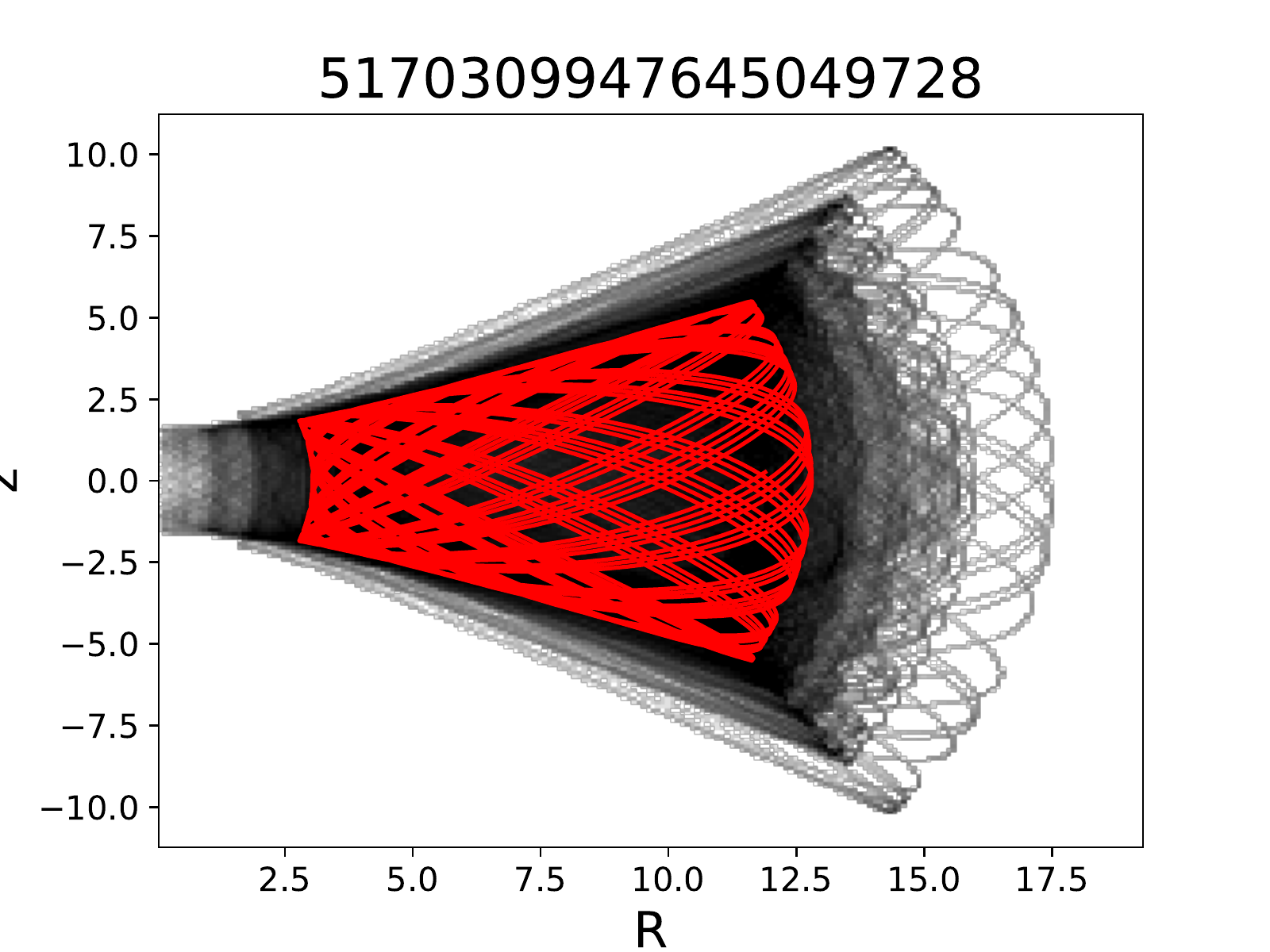}
\includegraphics[clip=true, trim = {0cm 0cm 1cm 0cm},width=4.5cm]{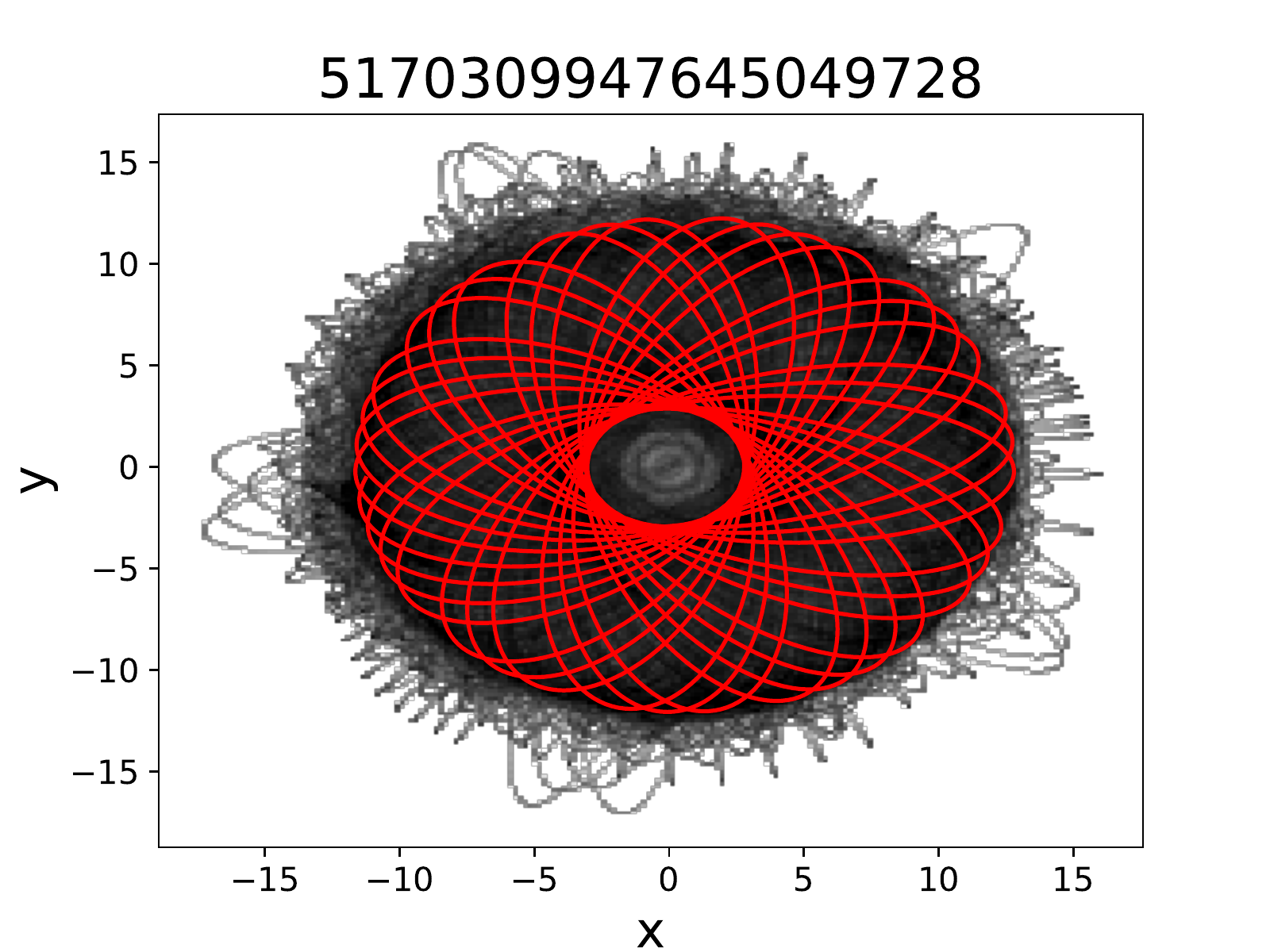}
\includegraphics[clip=true, trim = {0cm 0cm 1cm 0cm},width=4.5cm]{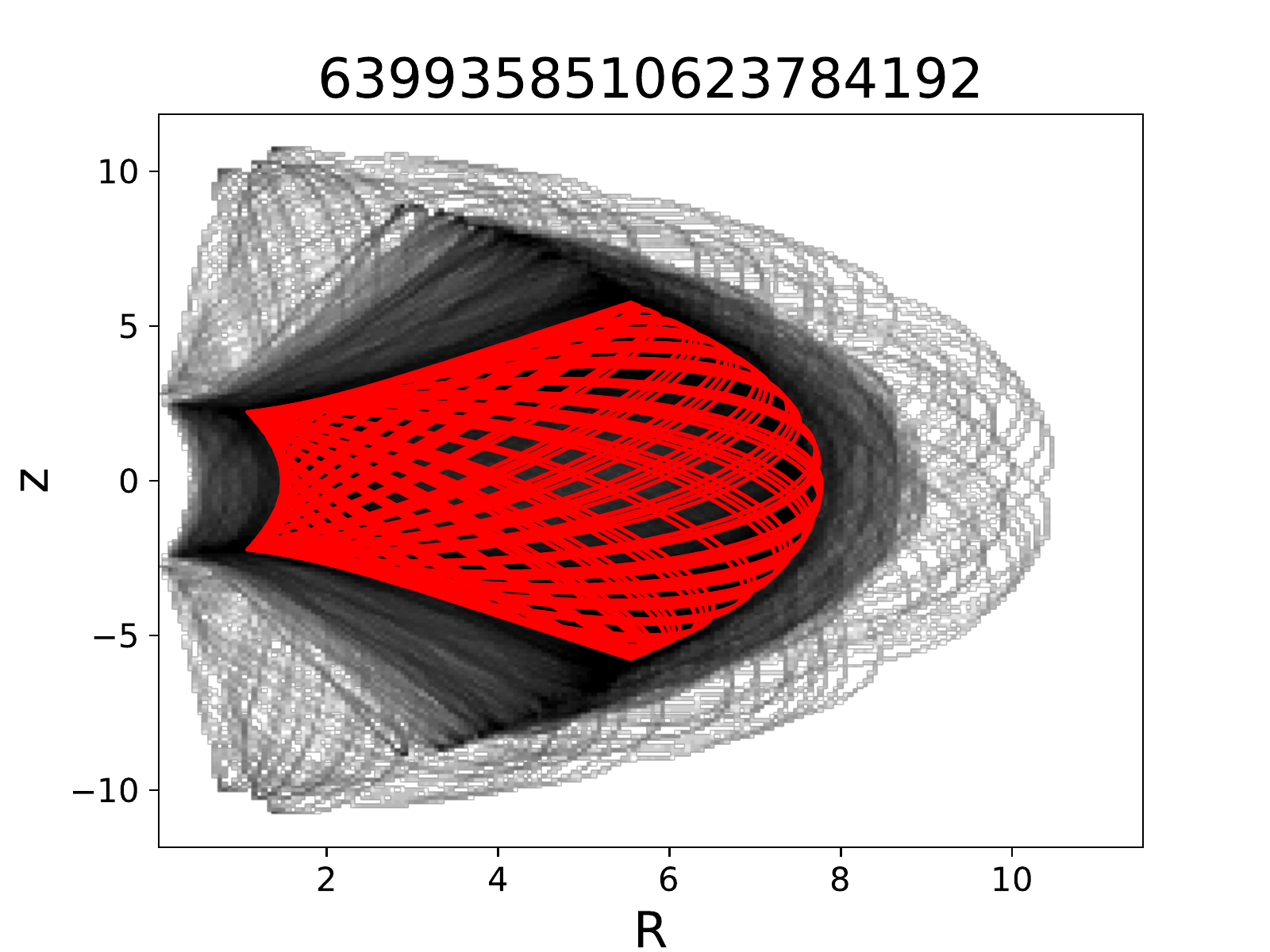}
\includegraphics[clip=true, trim = {0cm 0cm 1cm 0cm},width=4.5cm]{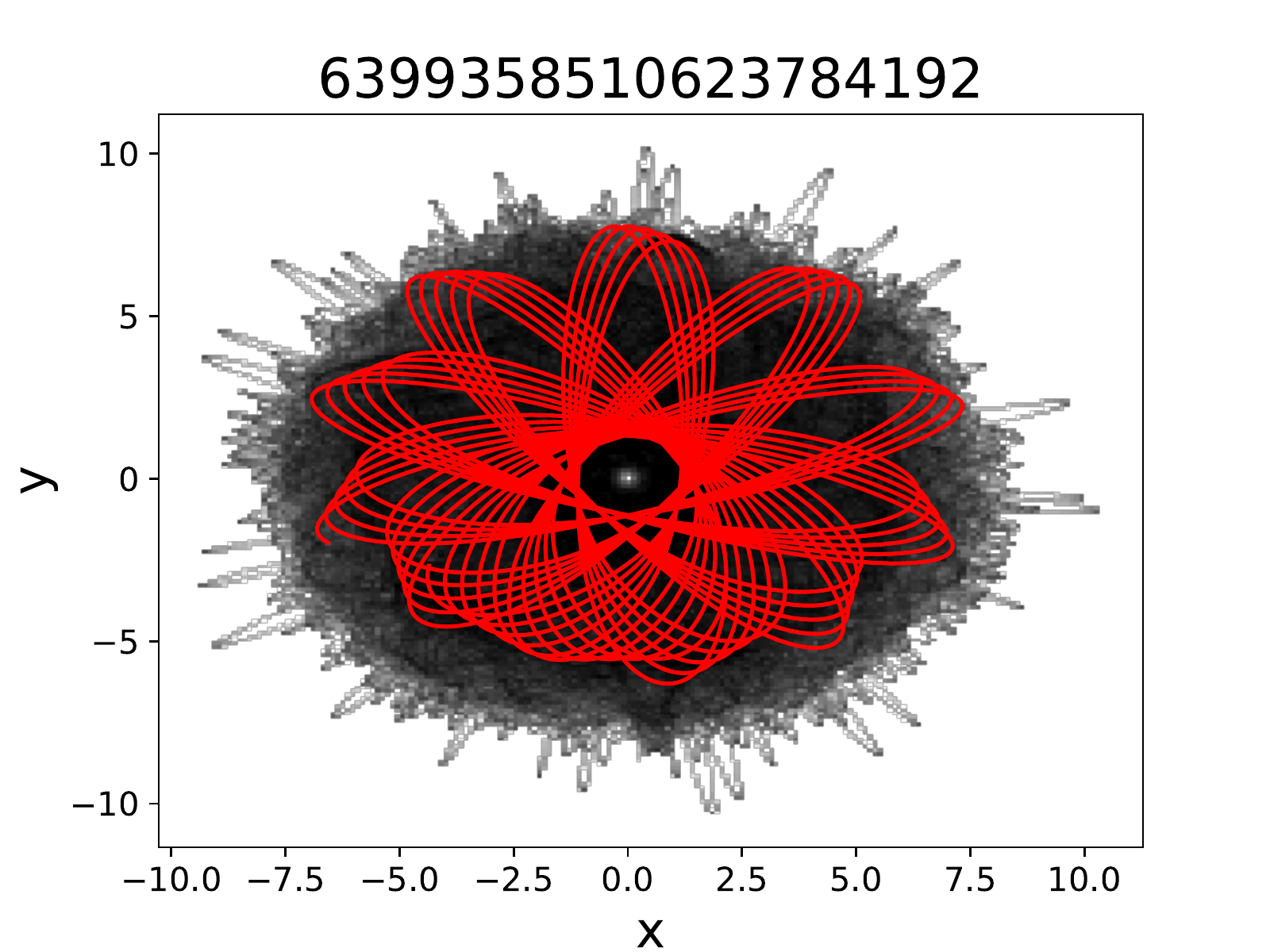}
\includegraphics[clip=true, trim = {0cm 0cm 1cm 0cm},width=4.5cm]{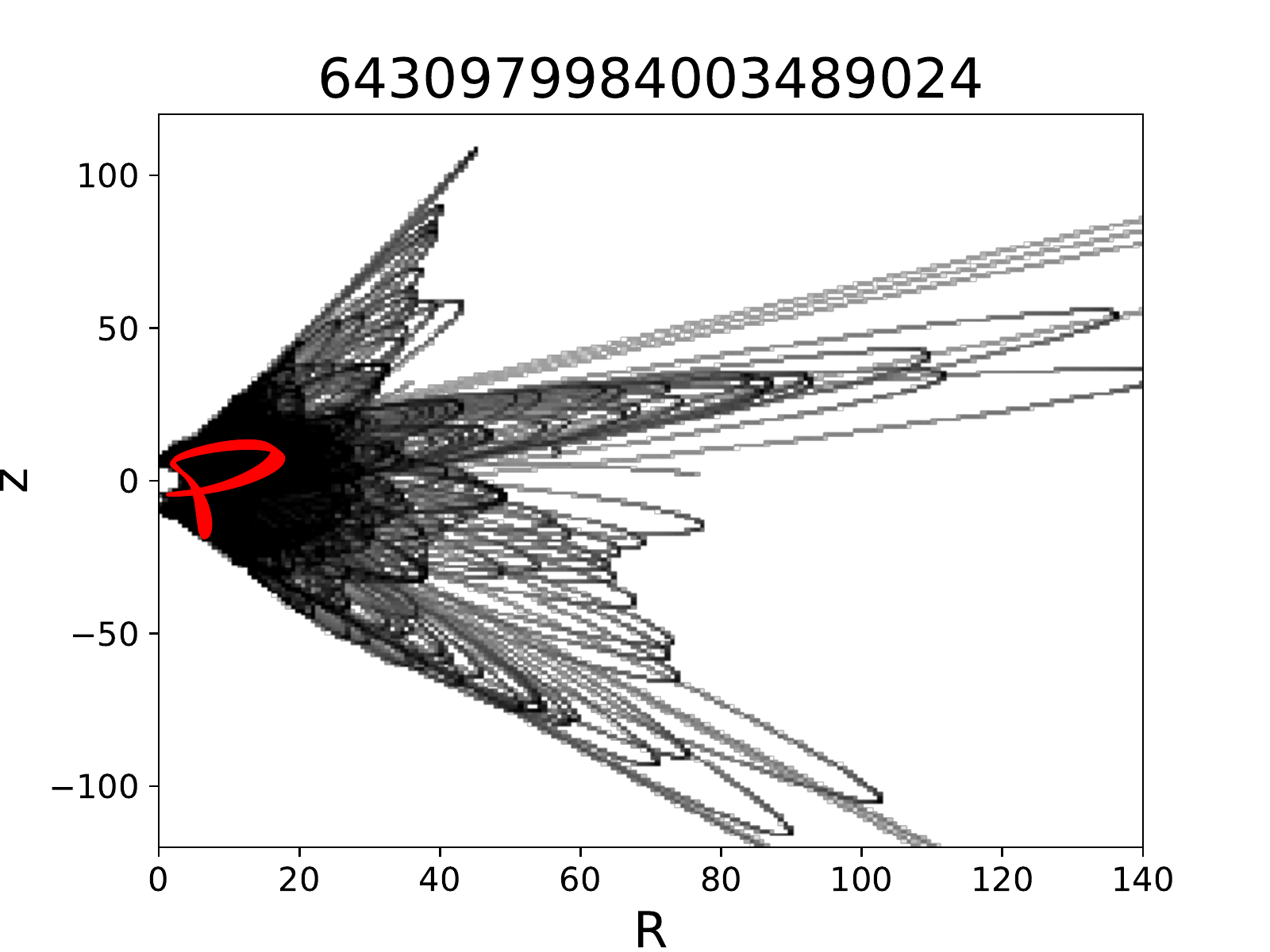}
\includegraphics[clip=true, trim = {0cm 0cm 1cm 0cm},width=4.5cm]{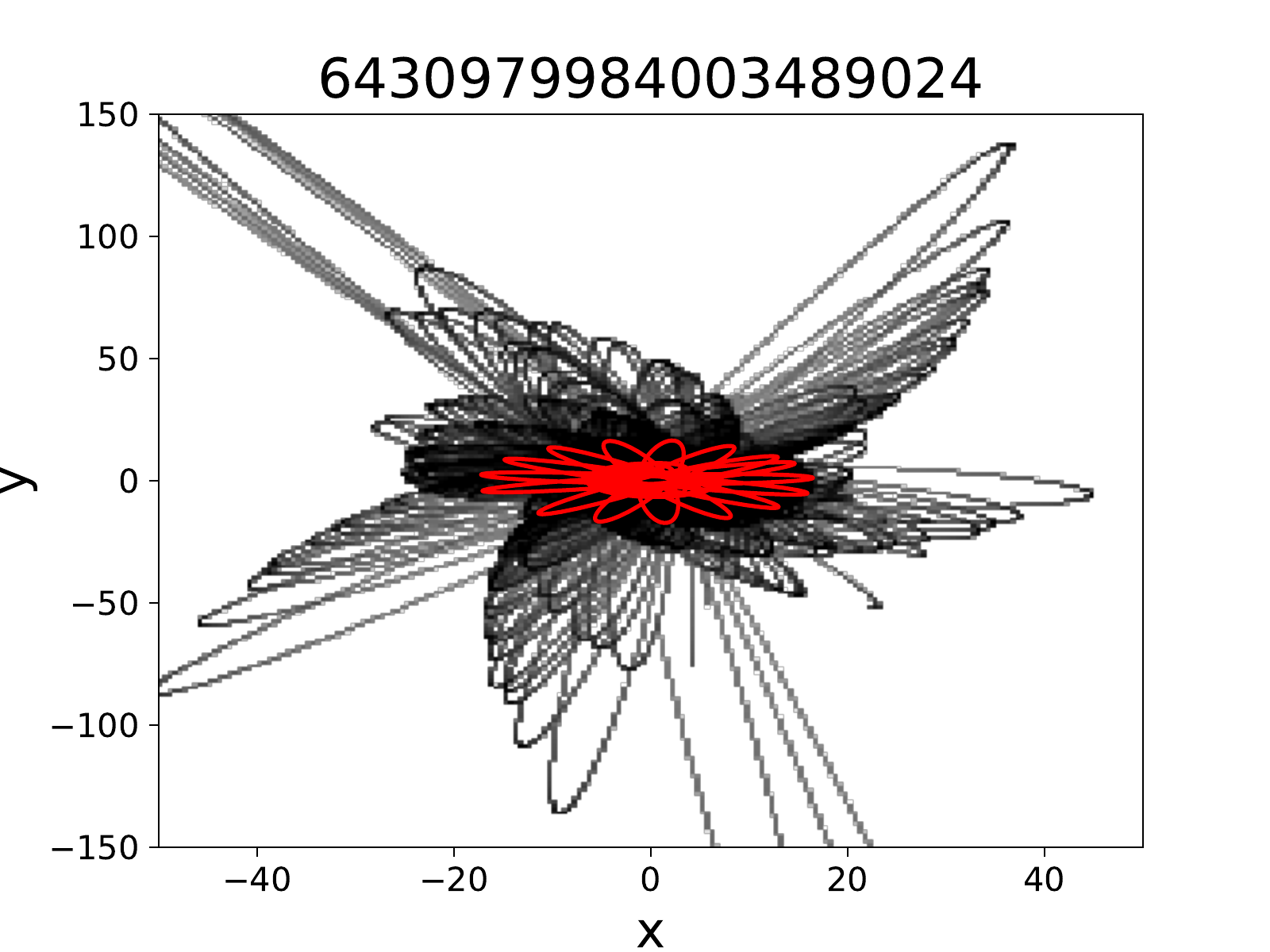}
\includegraphics[clip=true, trim = {0cm 0cm 1cm 0cm},width=4.5cm]{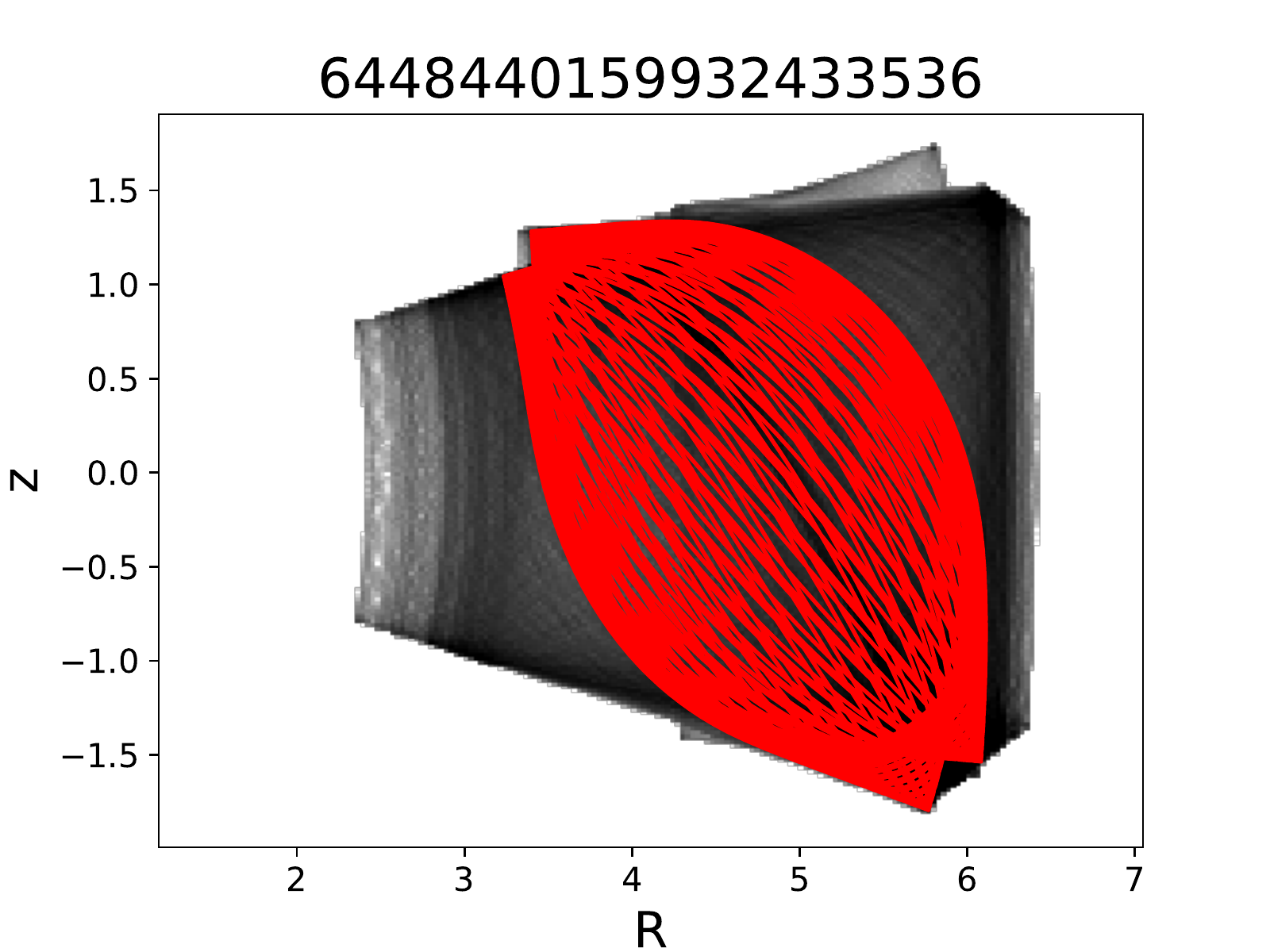}
\includegraphics[clip=true, trim = {0cm 0cm 1cm 0cm},width=4.5cm]{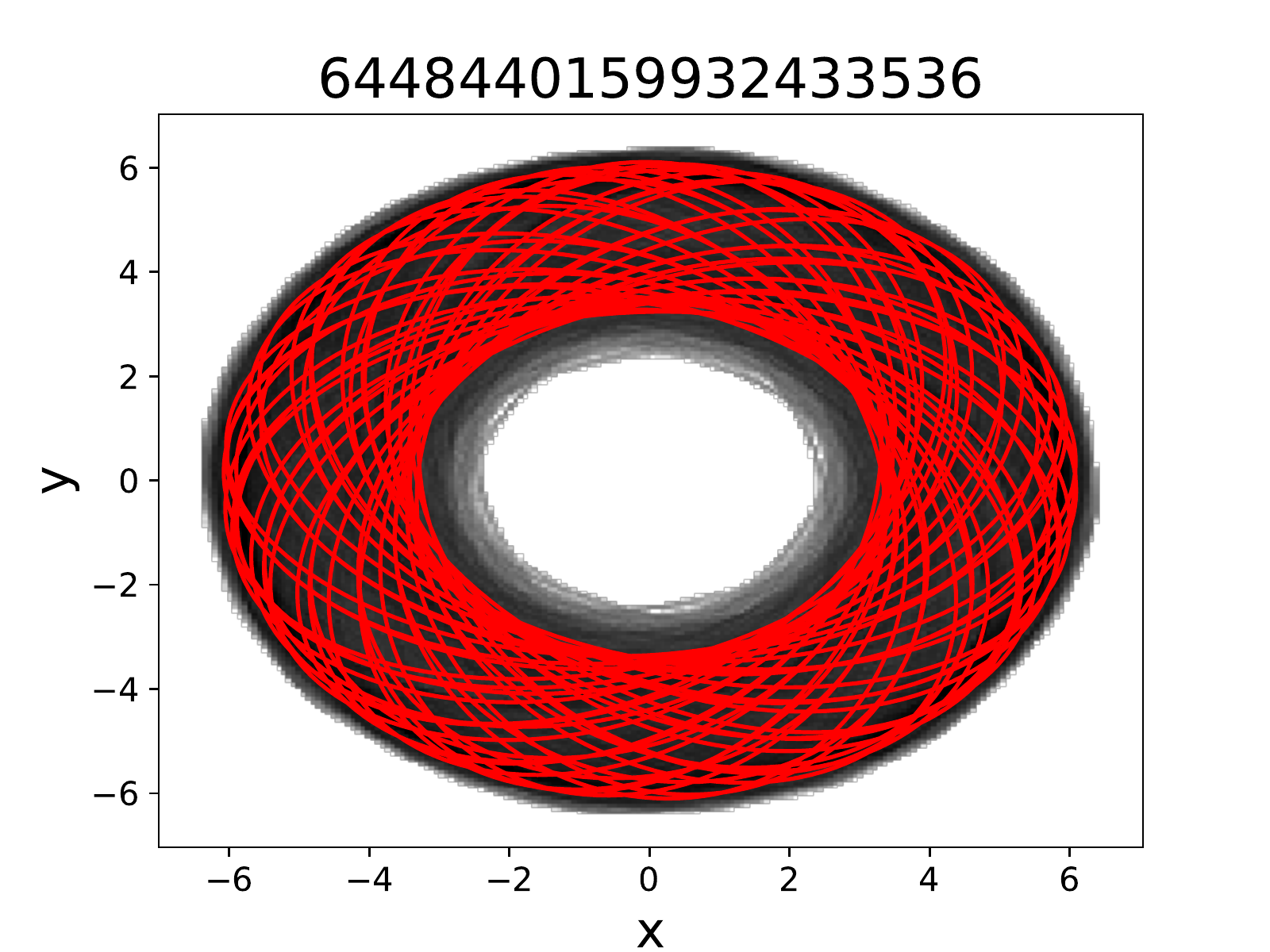}
\includegraphics[clip=true, trim = {0cm 0cm 1cm 0cm},width=4.5cm]{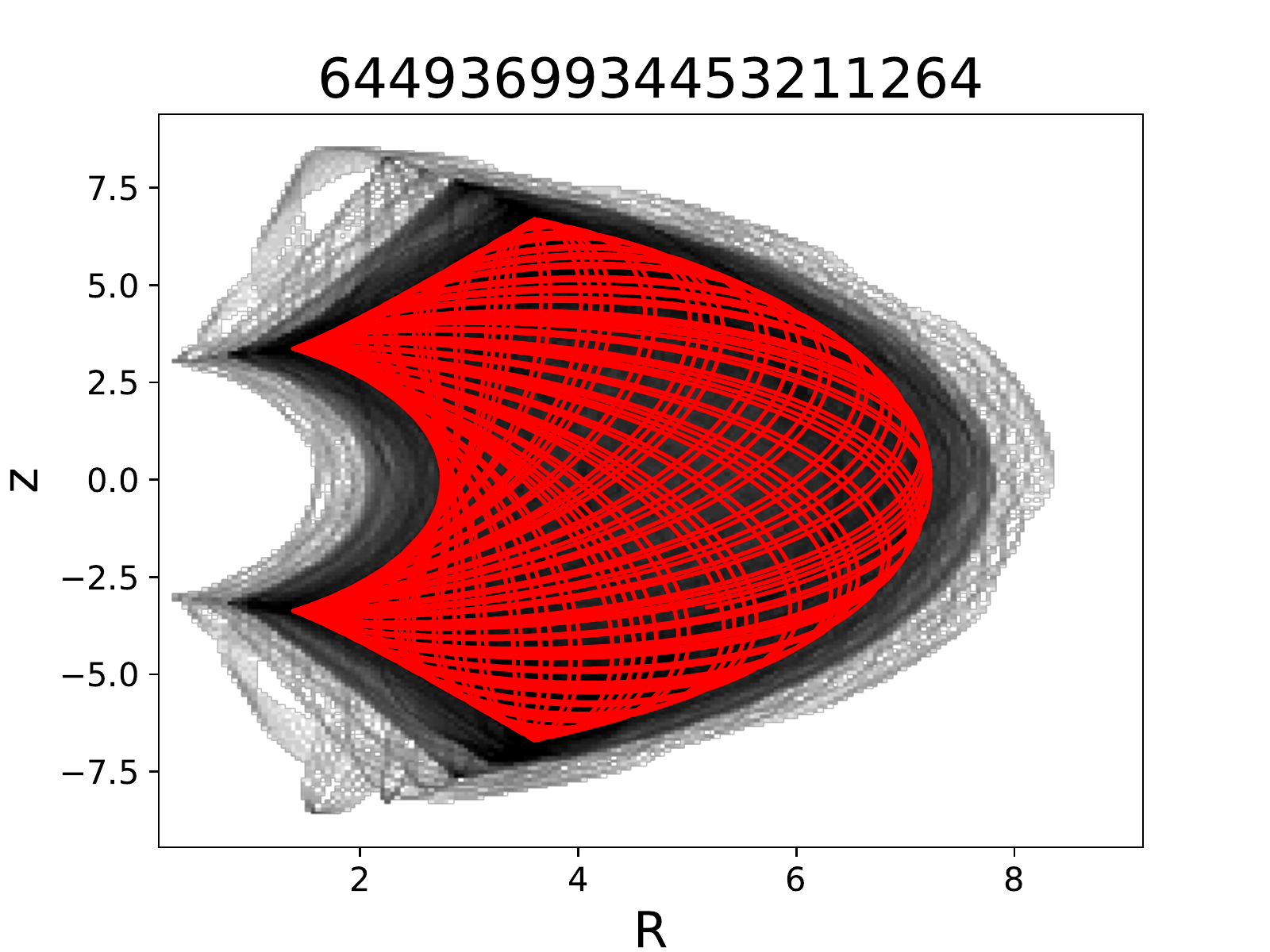}
\includegraphics[clip=true, trim = {0cm 0cm 1cm 0cm},width=4.5cm]{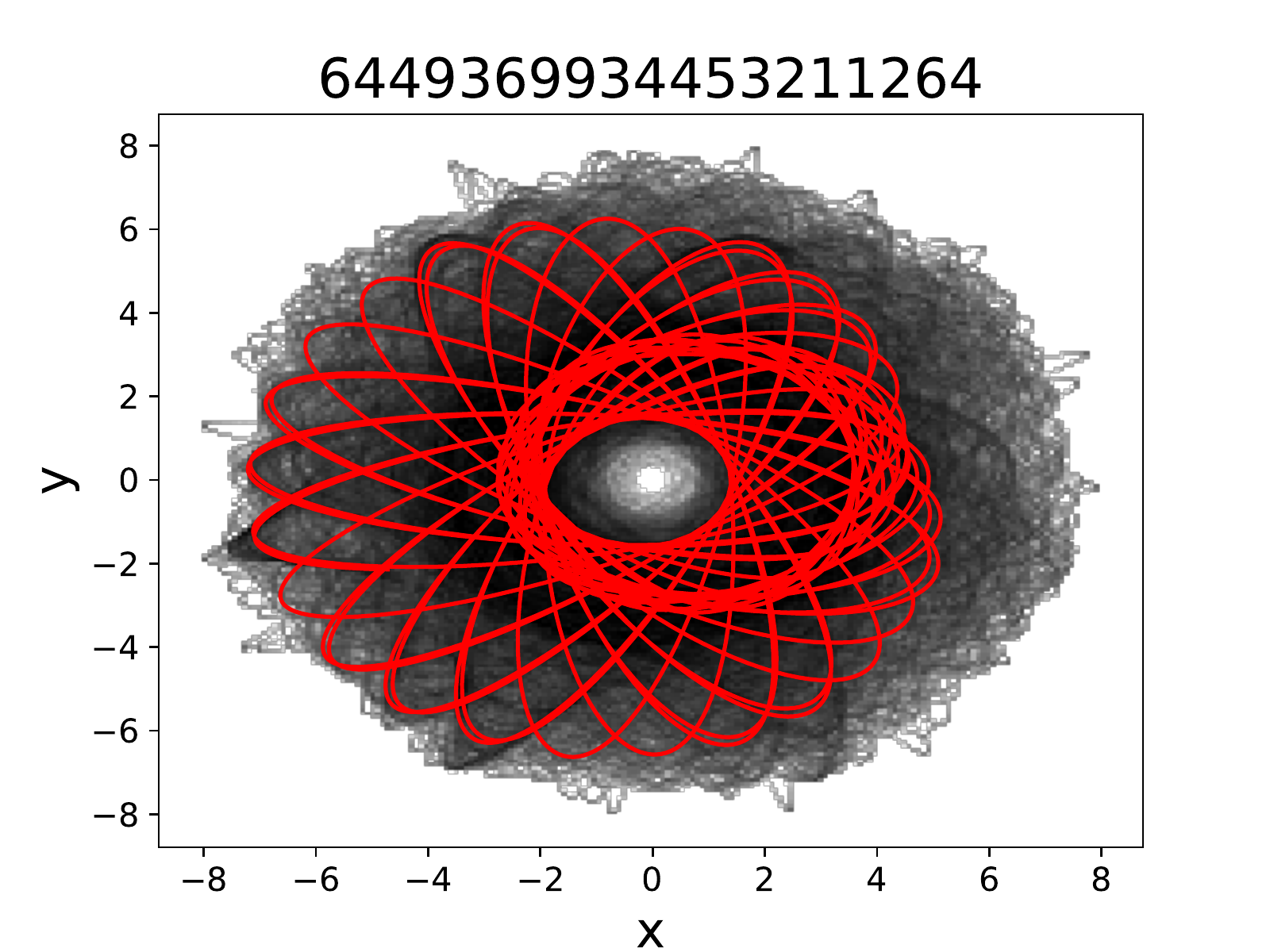}
\includegraphics[clip=true, trim = {0cm 0cm 1cm 0cm},width=4.5cm]{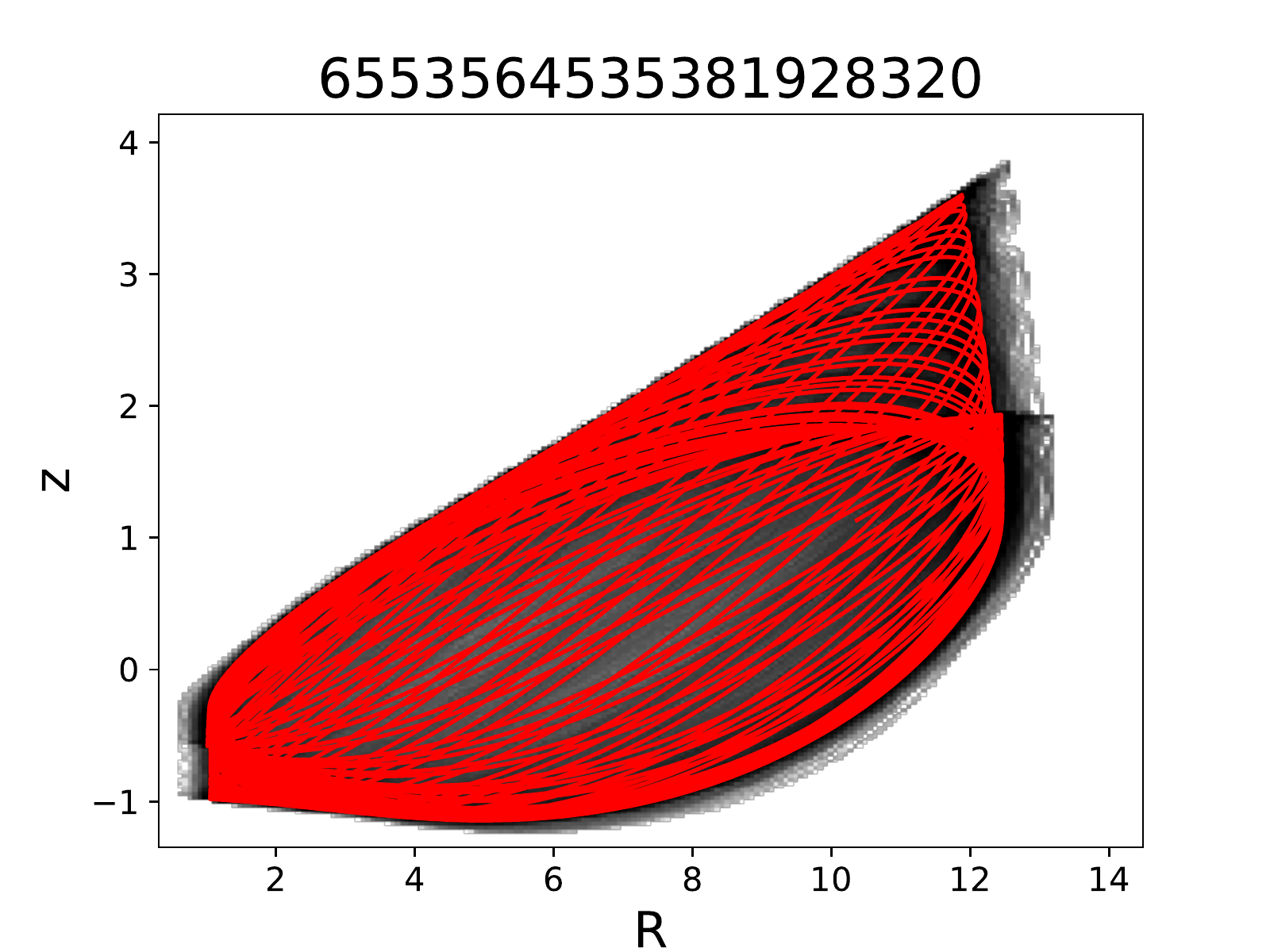}
\includegraphics[clip=true, trim = {0cm 0cm 1cm 0cm},width=4.5cm]{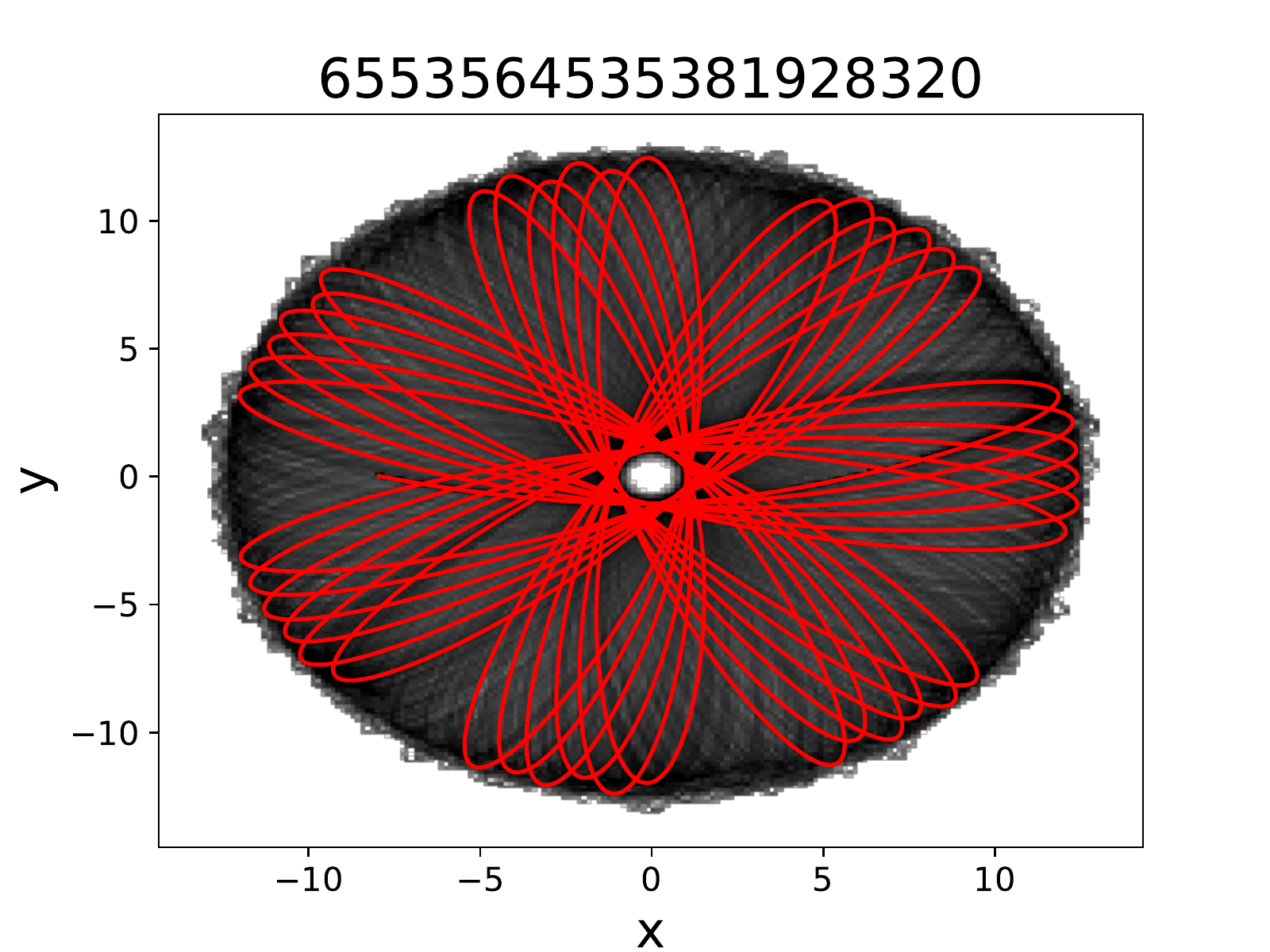}
\includegraphics[clip=true, trim = {0cm 0cm 1cm 0cm},width=4.5cm]{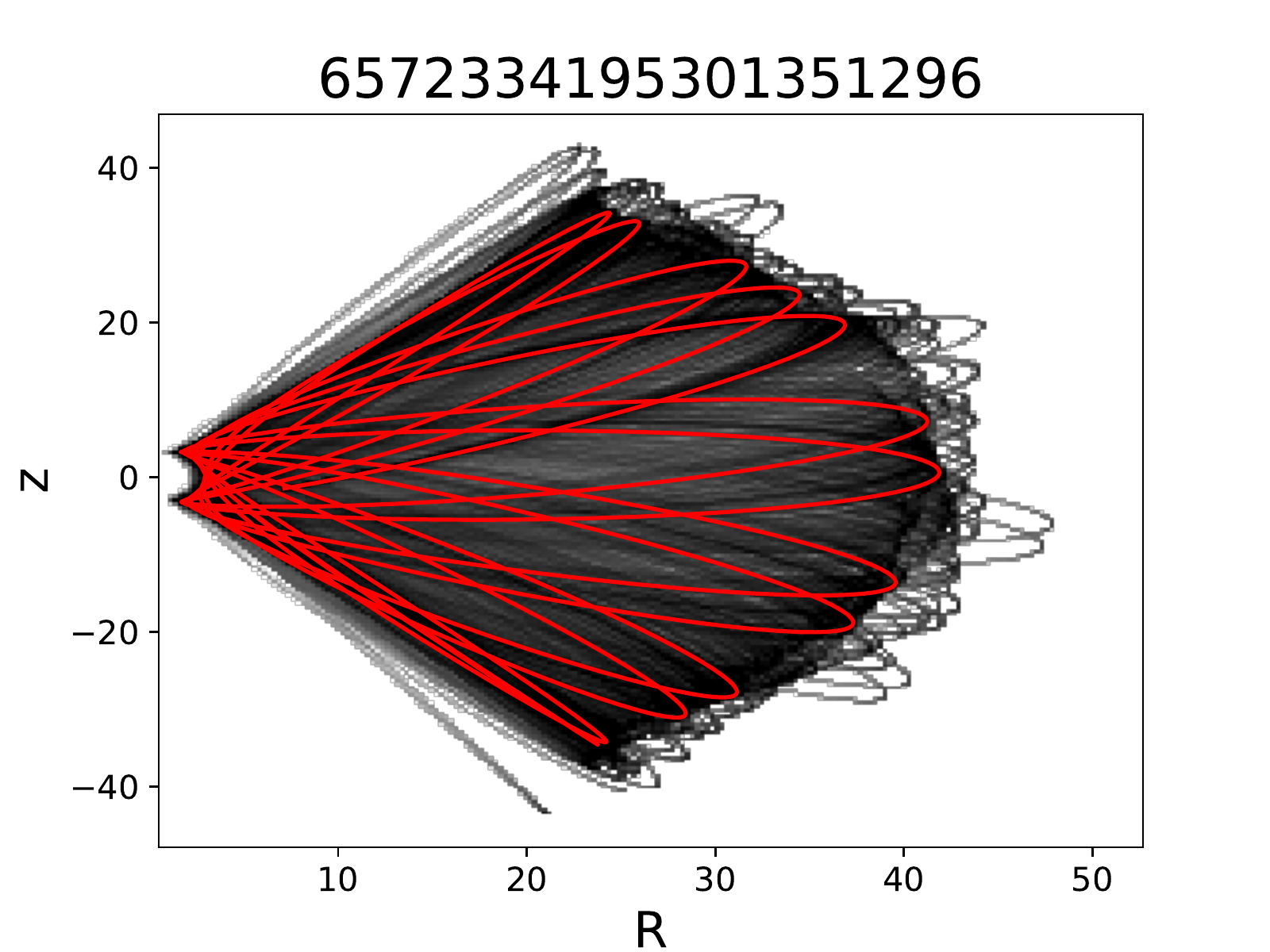}
\includegraphics[clip=true, trim = {0cm 0cm 1cm 0cm},width=4.5cm]{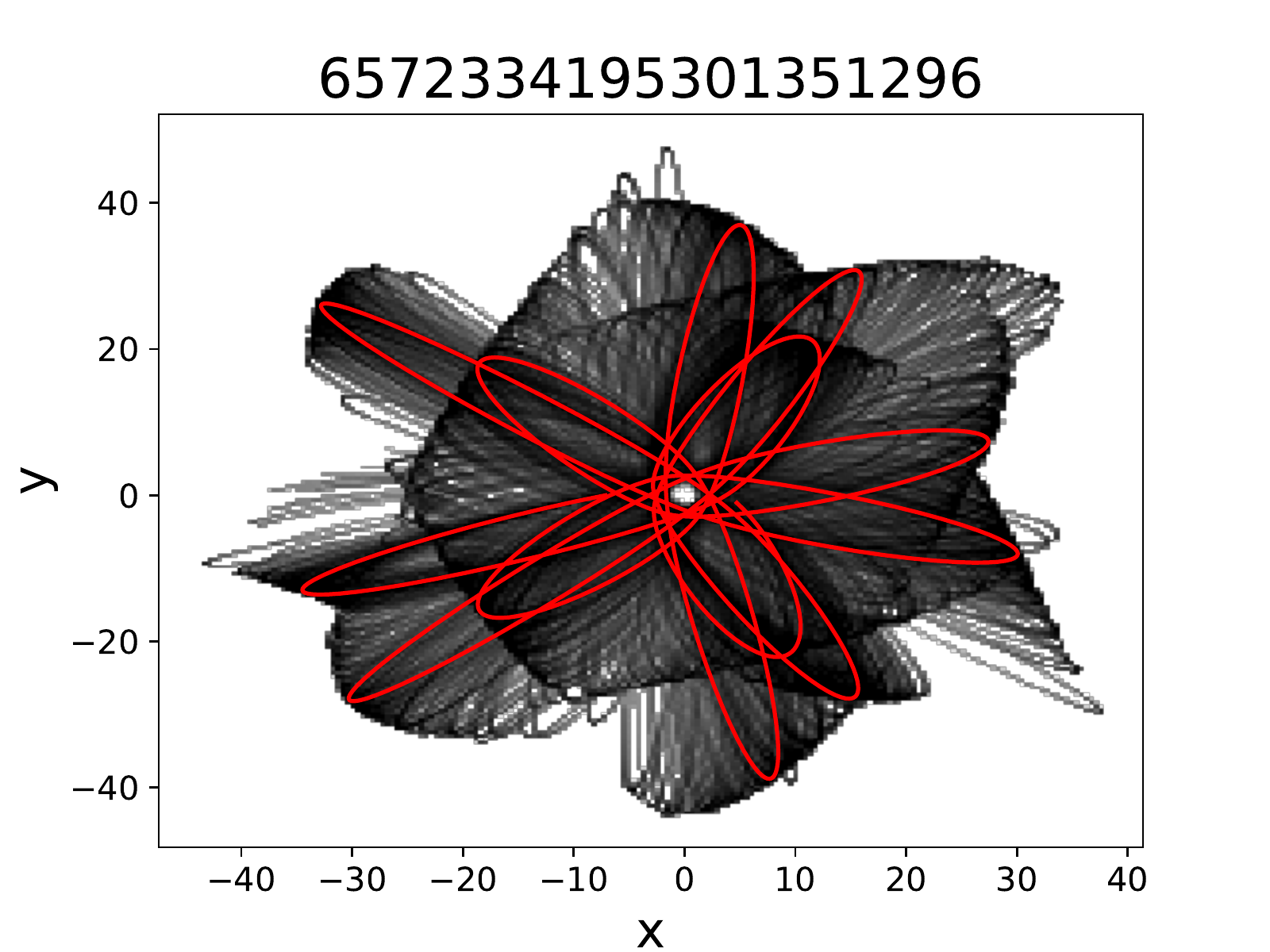}
\includegraphics[clip=true, trim = {0cm 0cm 1cm 0cm},width=4.5cm]{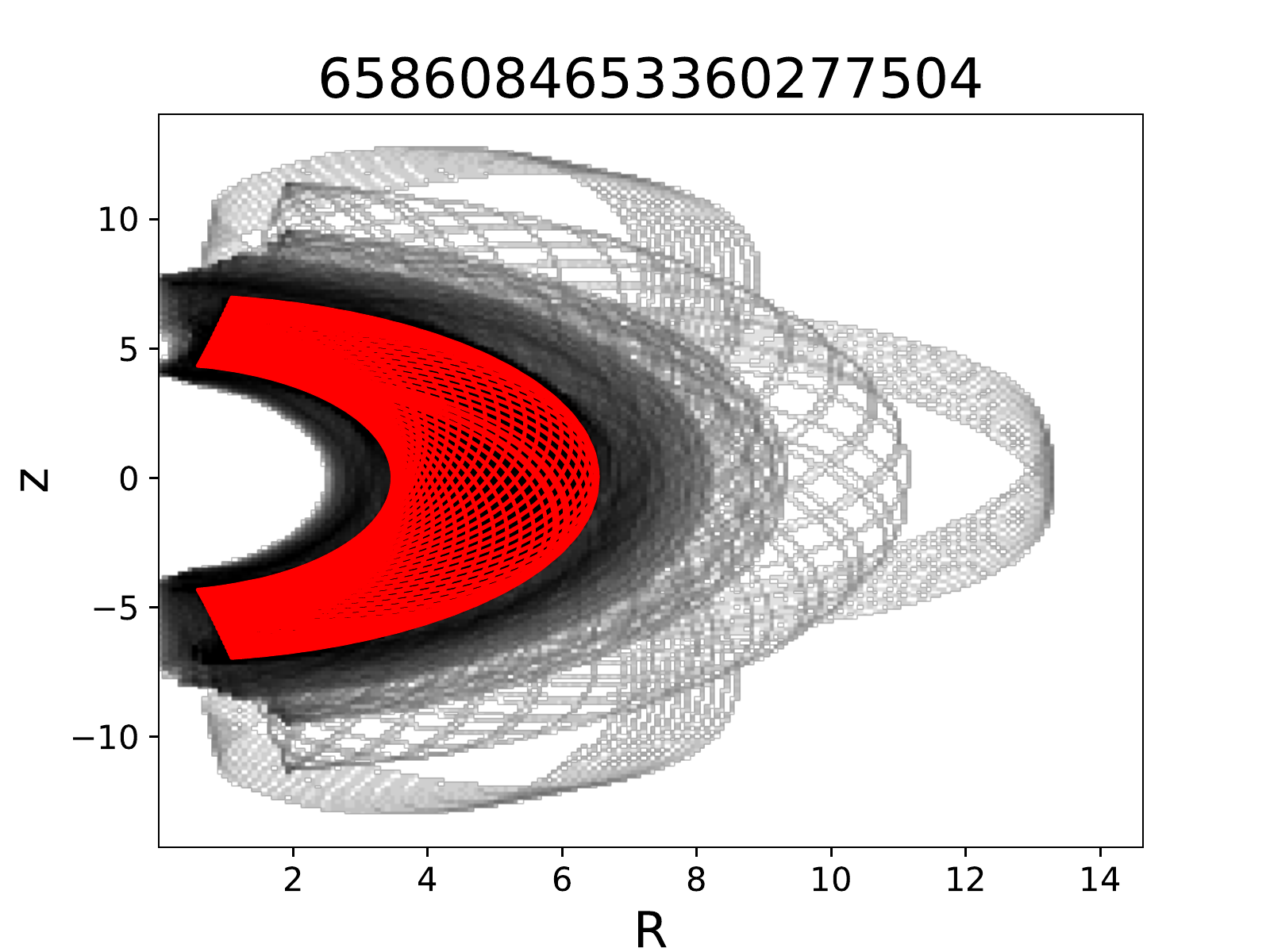}
\includegraphics[clip=true, trim = {0cm 0cm 1cm 0cm},width=4.5cm]{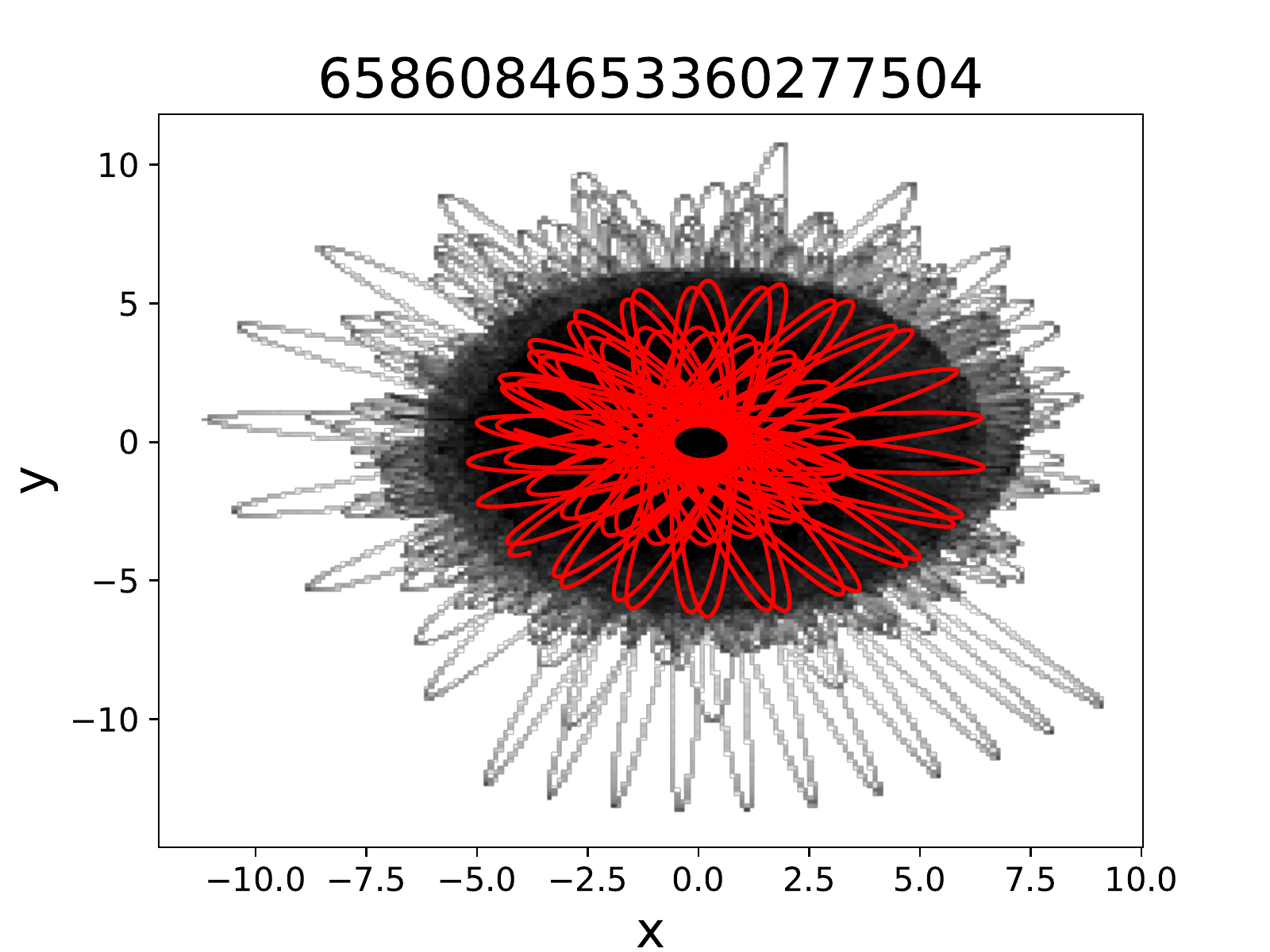}
\caption{Continued from Fig.~\ref{T1}.}
\end{centering}
\end{figure*}

\begin{figure*}
\begin{centering}
\includegraphics[clip=true, trim = {0cm 0cm 1cm 0cm},width=4.5cm]{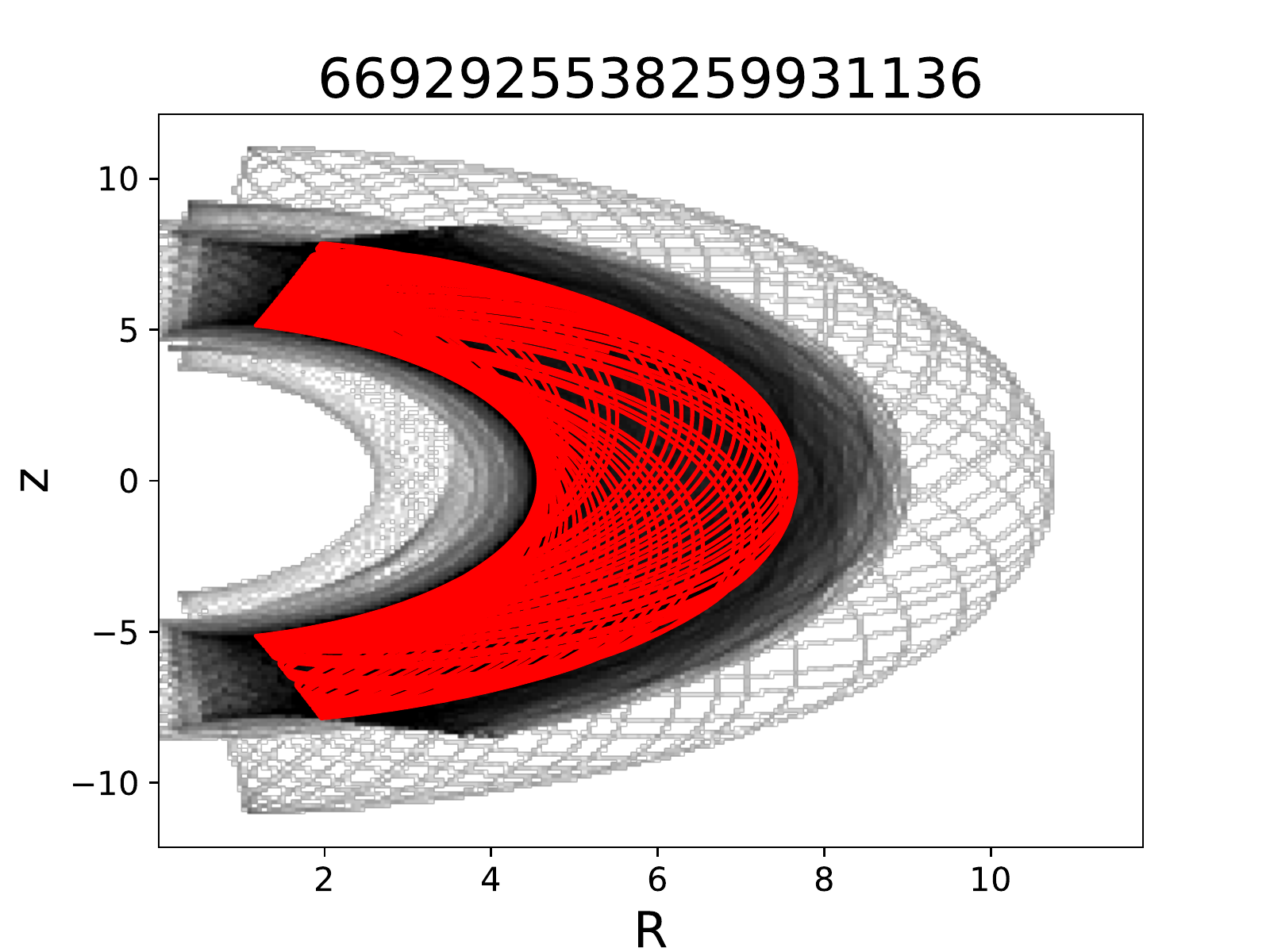}
\includegraphics[clip=true, trim = {0cm 0cm 1cm 0cm},width=4.5cm]{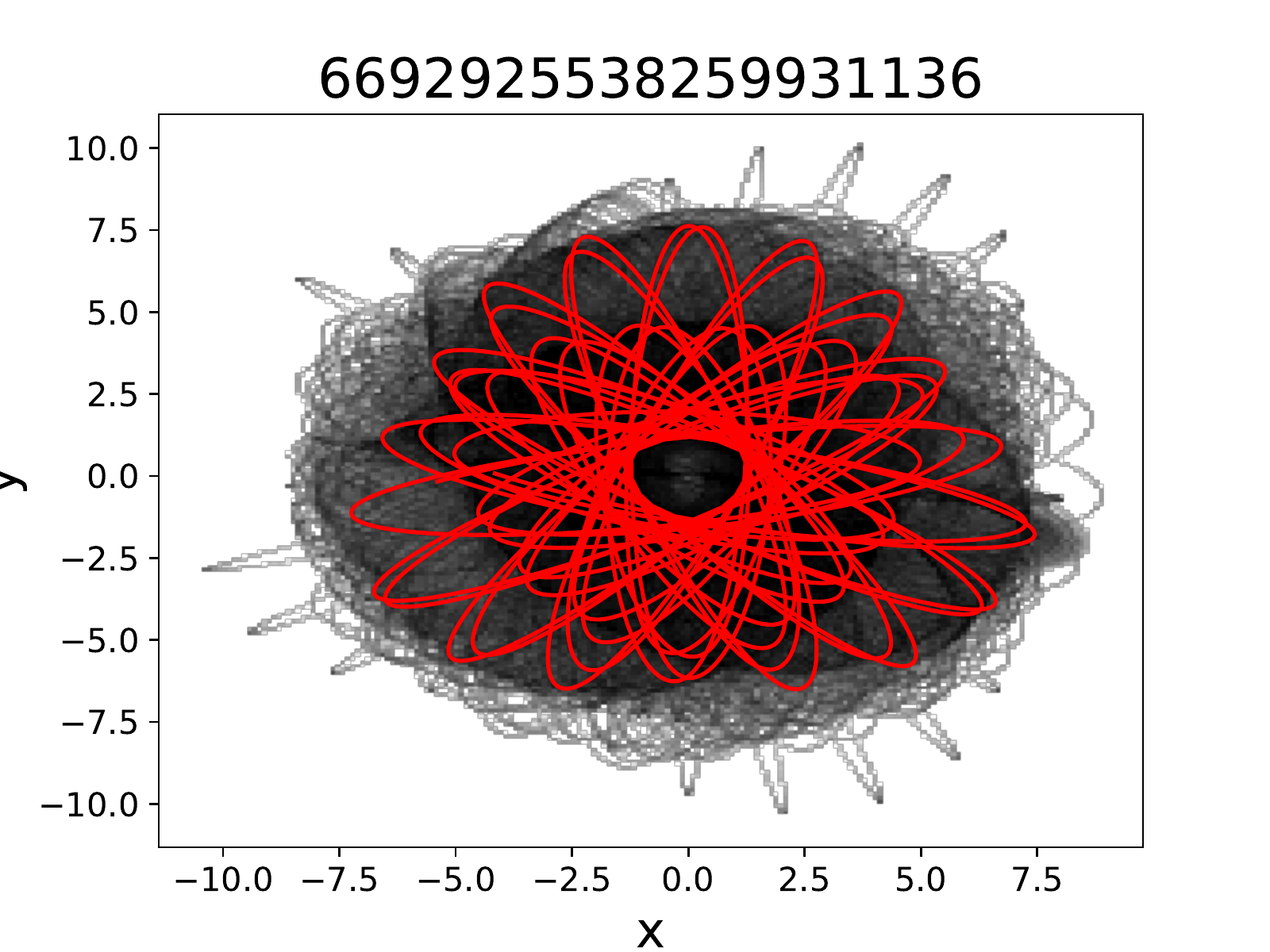}
\includegraphics[clip=true, trim = {0cm 0cm 1cm 0cm},width=4.5cm]{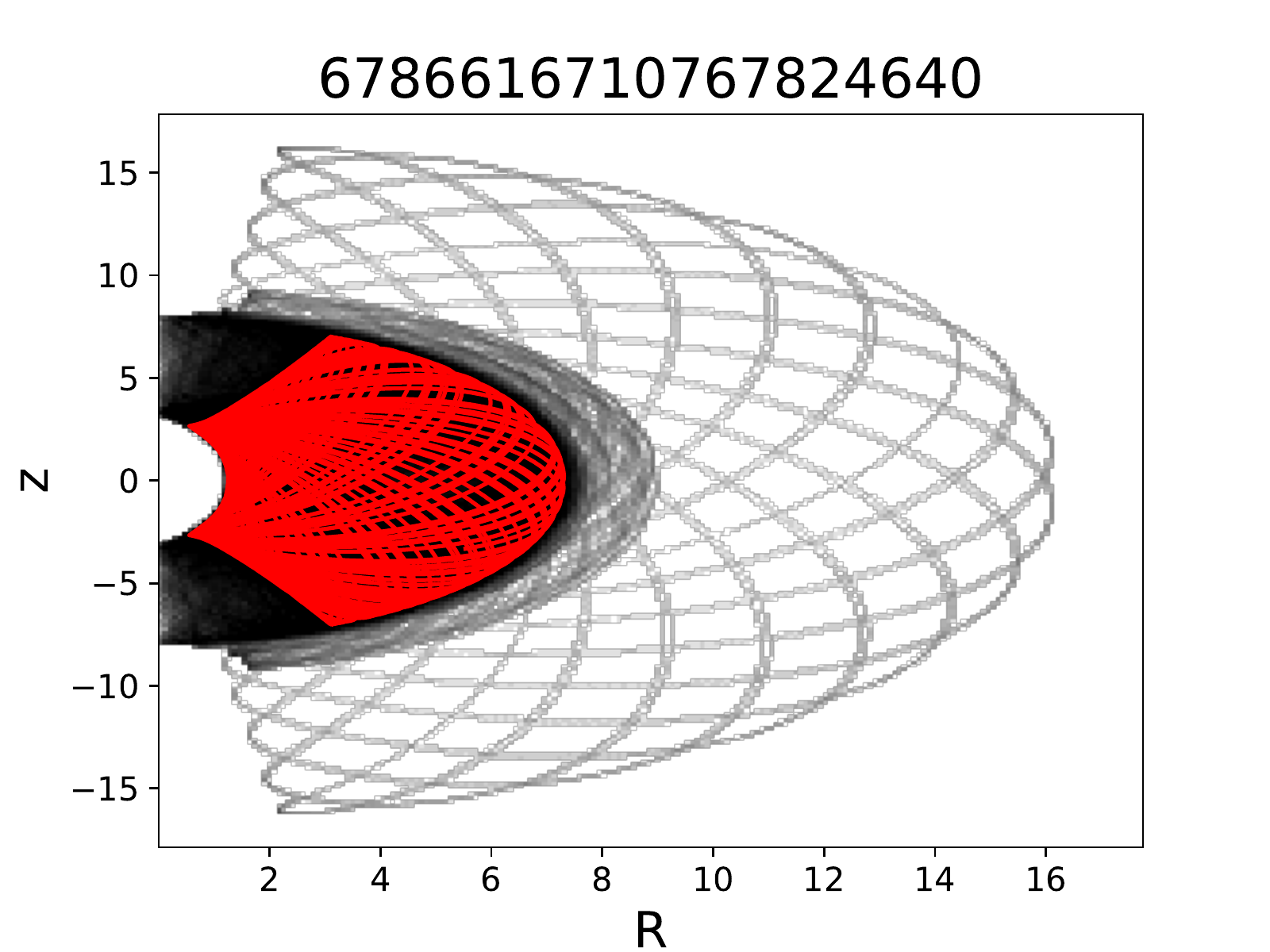}
\includegraphics[clip=true, trim = {0cm 0cm 1cm 0cm},width=4.5cm]{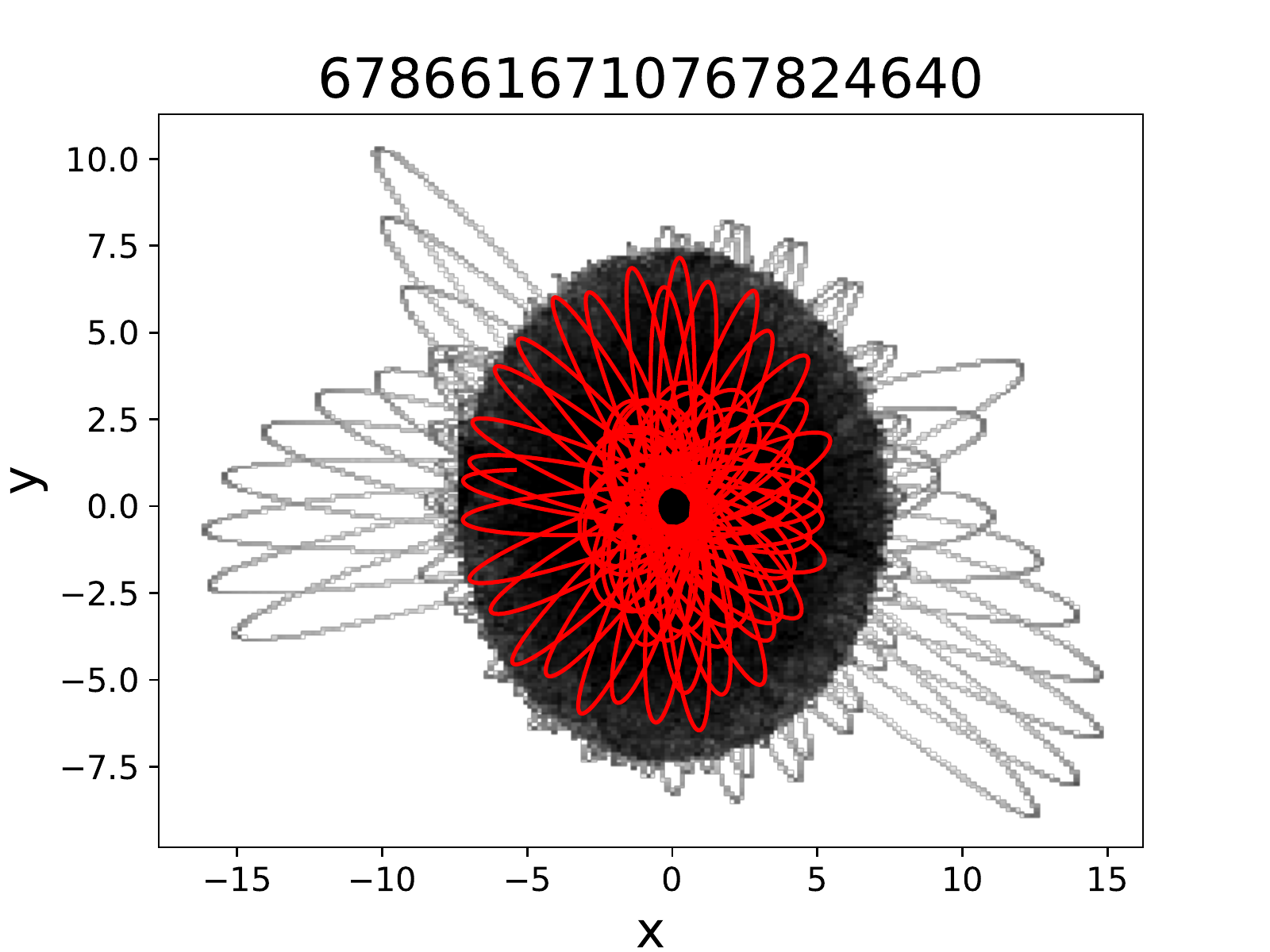}
\includegraphics[clip=true, trim = {0cm 0cm 1cm 0cm},width=4.5cm]{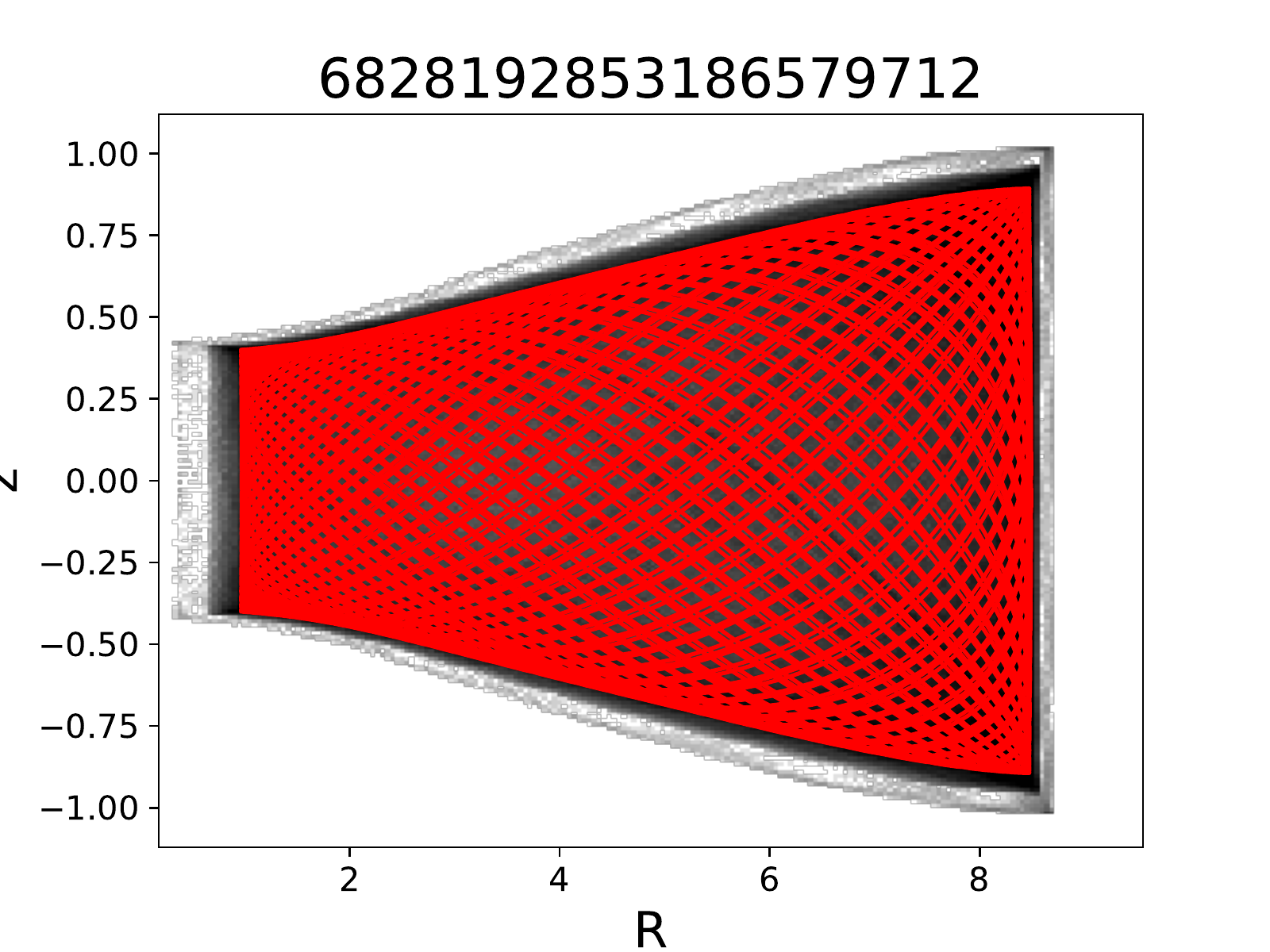}
\includegraphics[clip=true, trim = {0cm 0cm 1cm 0cm},width=4.5cm]{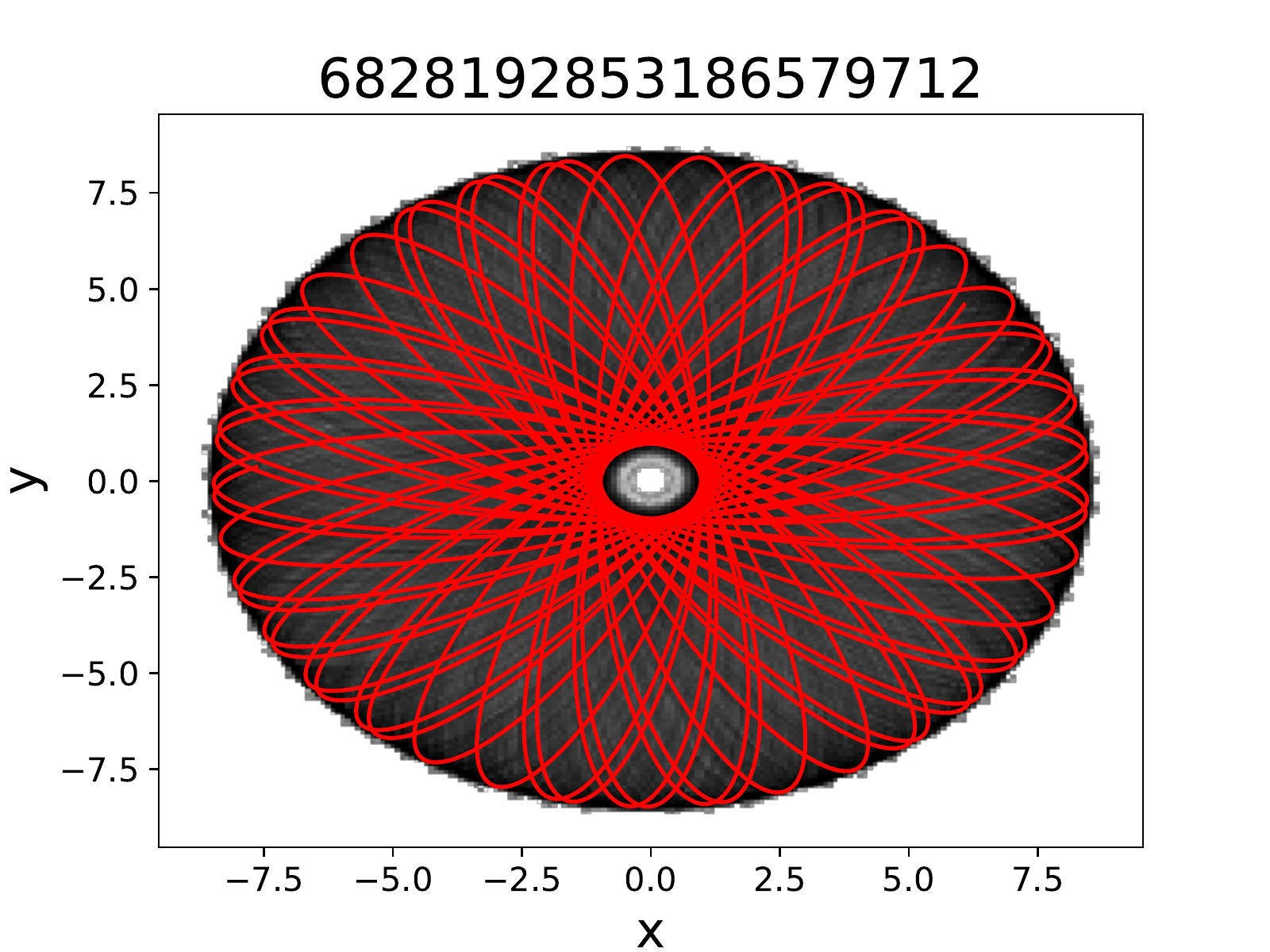}
\includegraphics[clip=true, trim = {0cm 0cm 1cm 0cm},width=4.5cm]{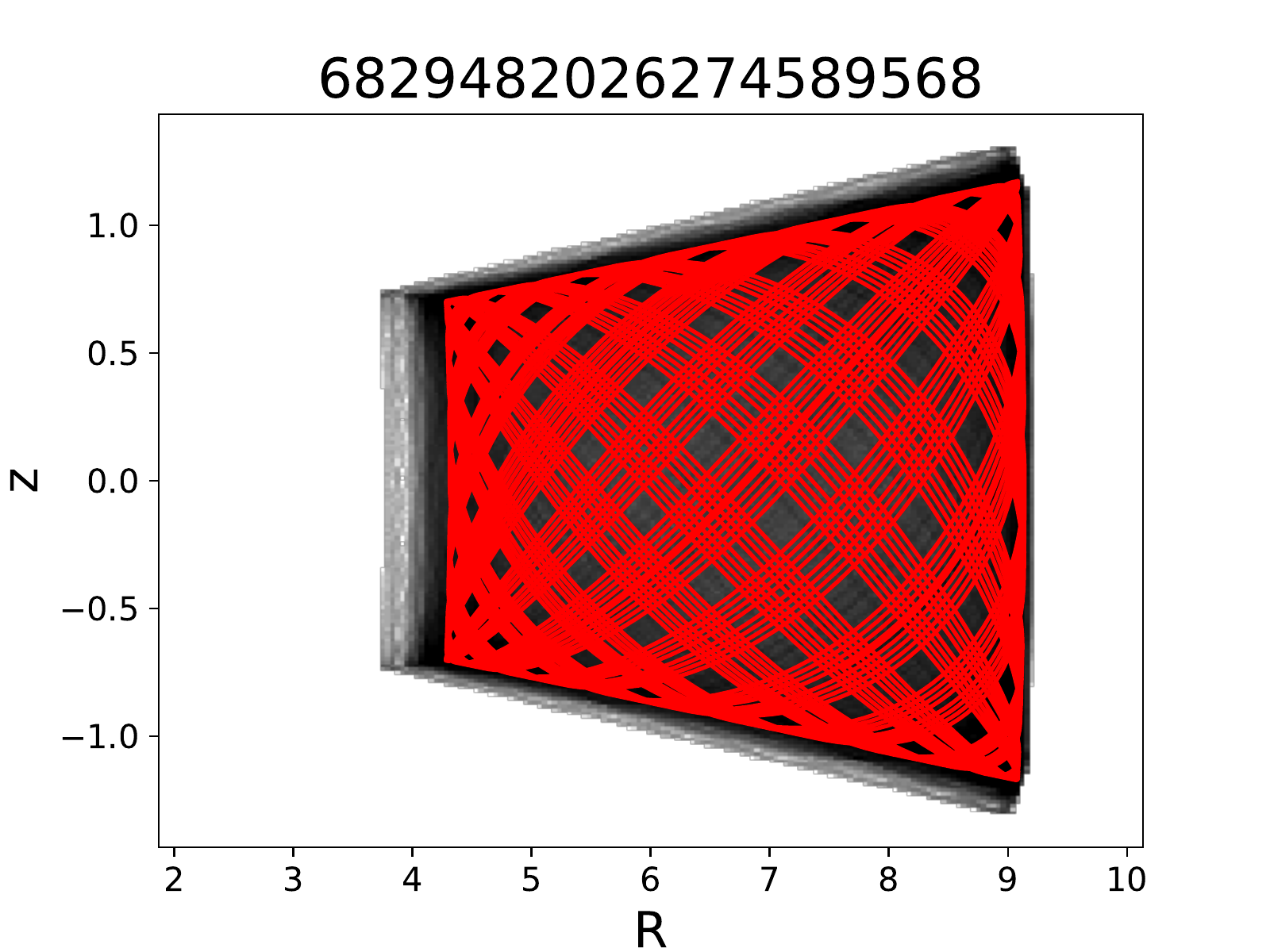}
\includegraphics[clip=true, trim = {0cm 0cm 1cm 0cm},width=4.5cm]{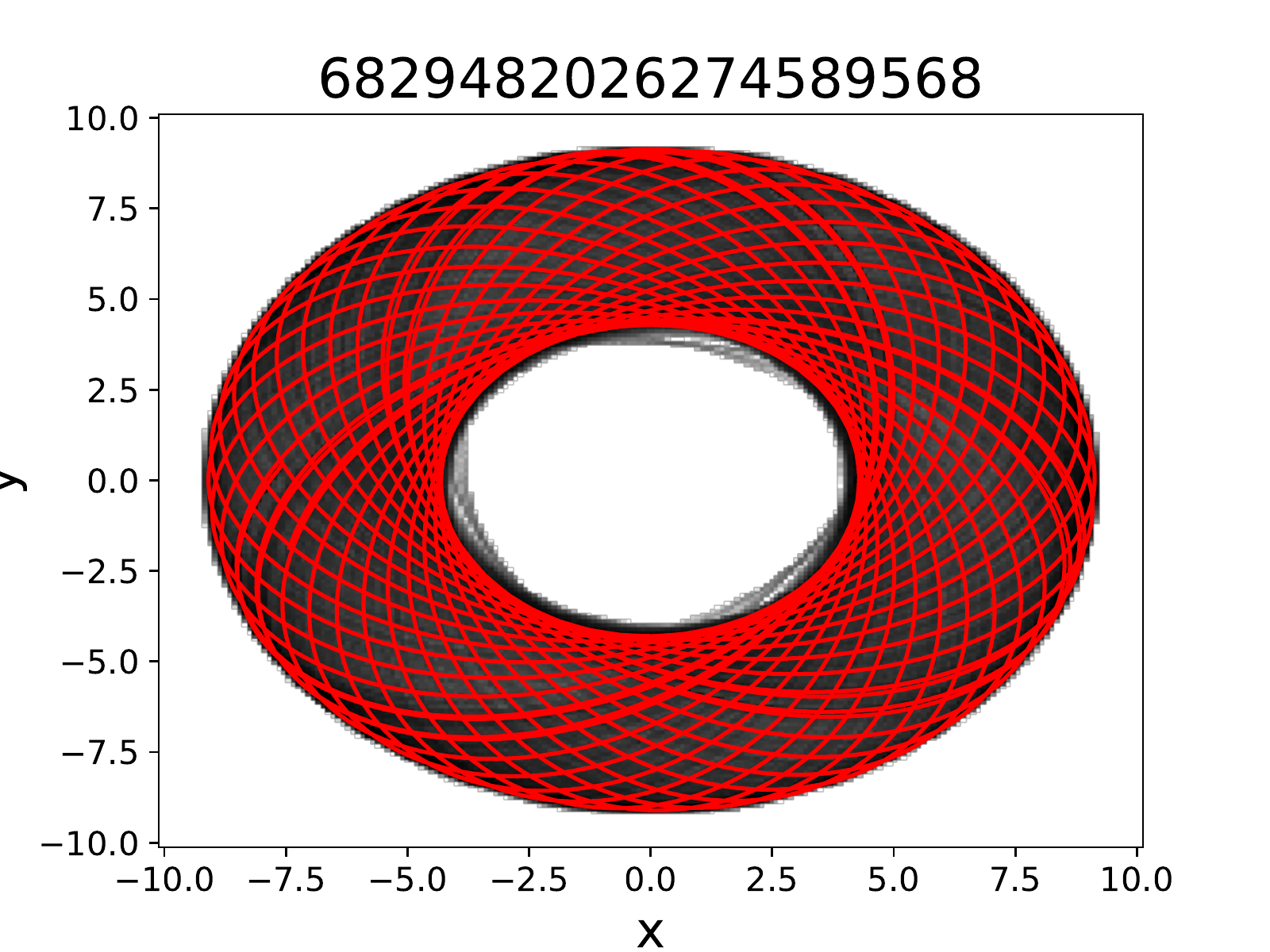}
\includegraphics[clip=true, trim = {0cm 0cm 1cm 0cm},width=4.5cm]{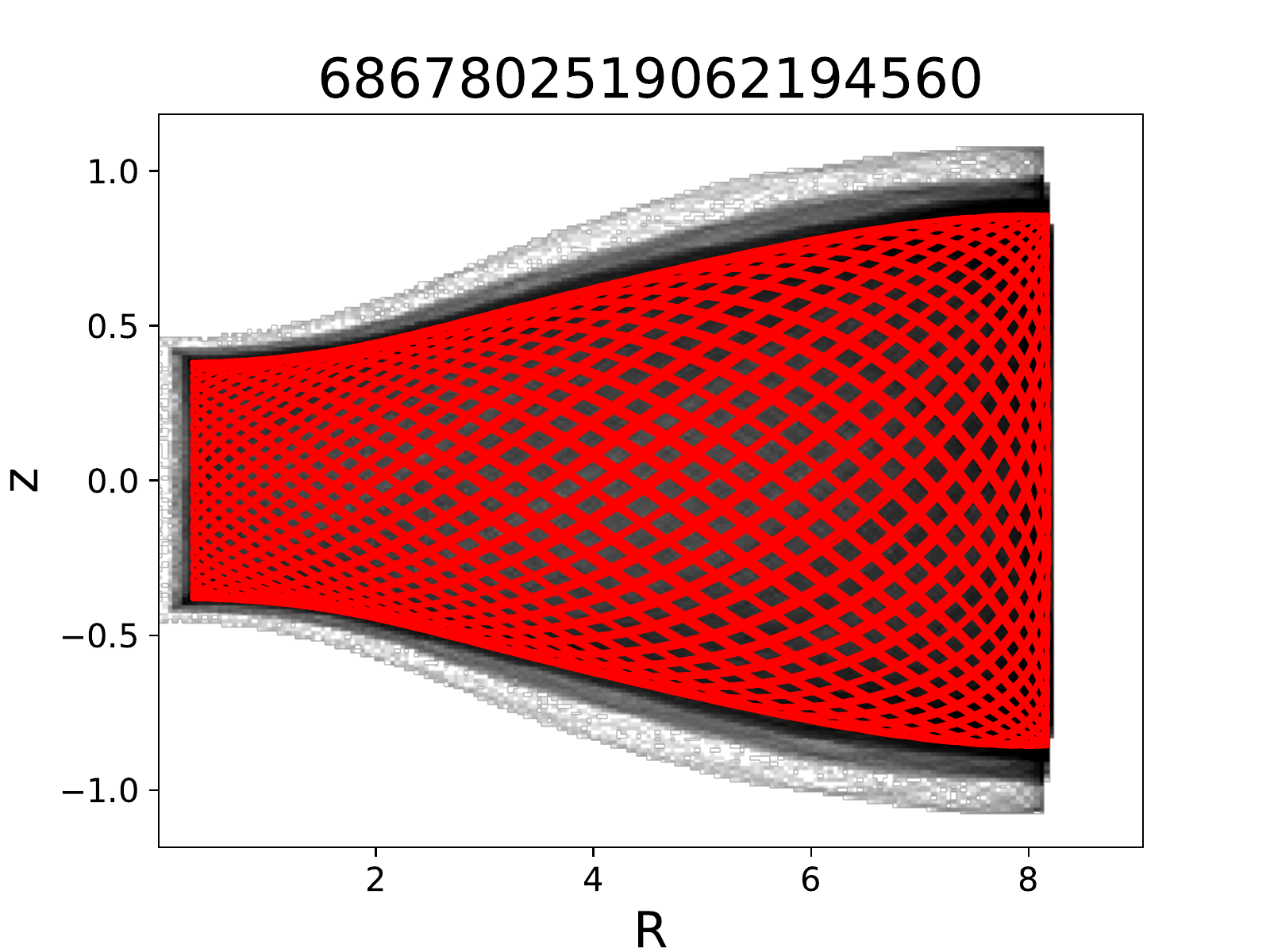}
\includegraphics[clip=true, trim = {0cm 0cm 1cm 0cm},width=4.5cm]{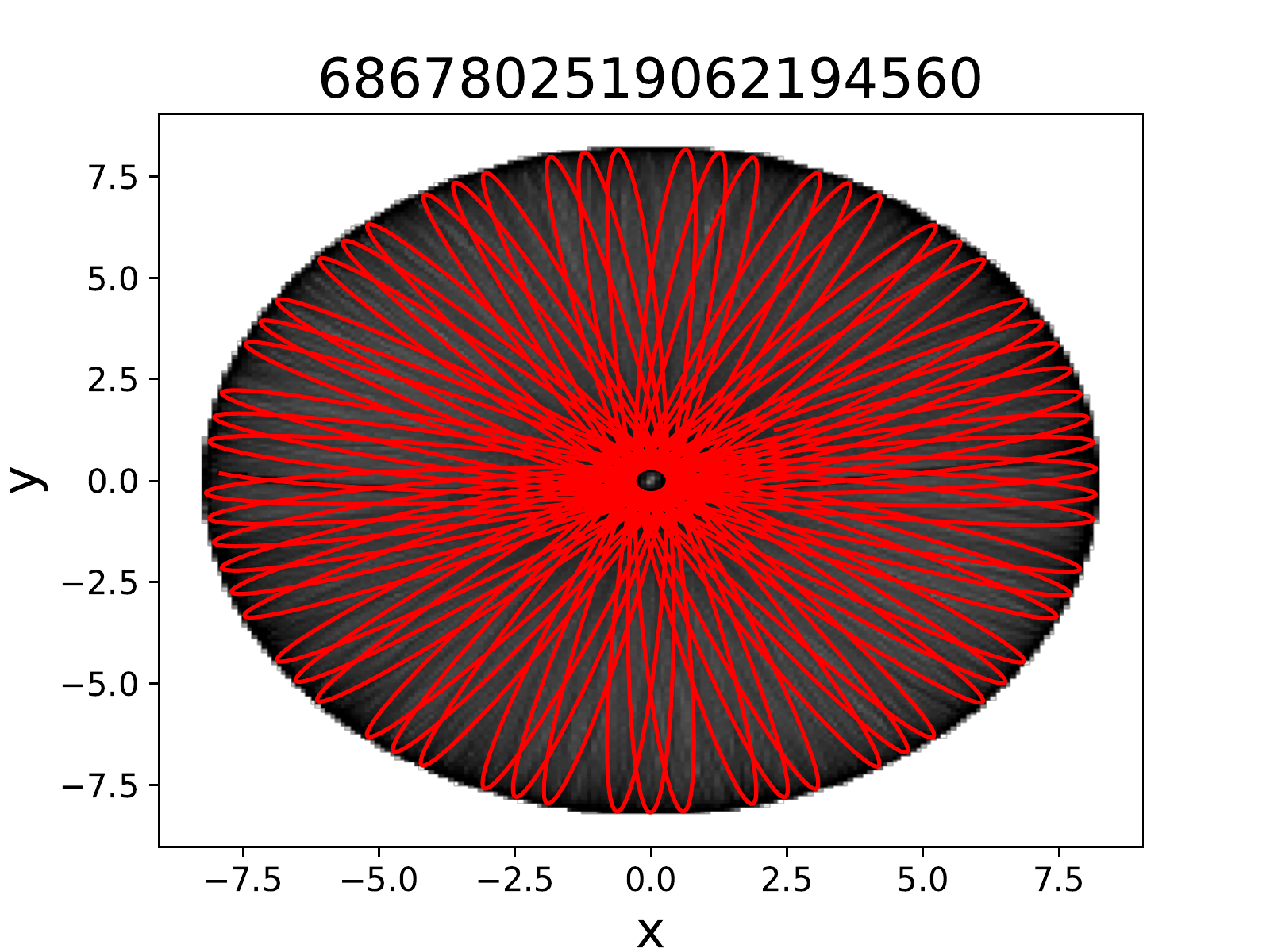}
\includegraphics[clip=true, trim = {0cm 0cm 1cm 0cm},width=4.5cm]{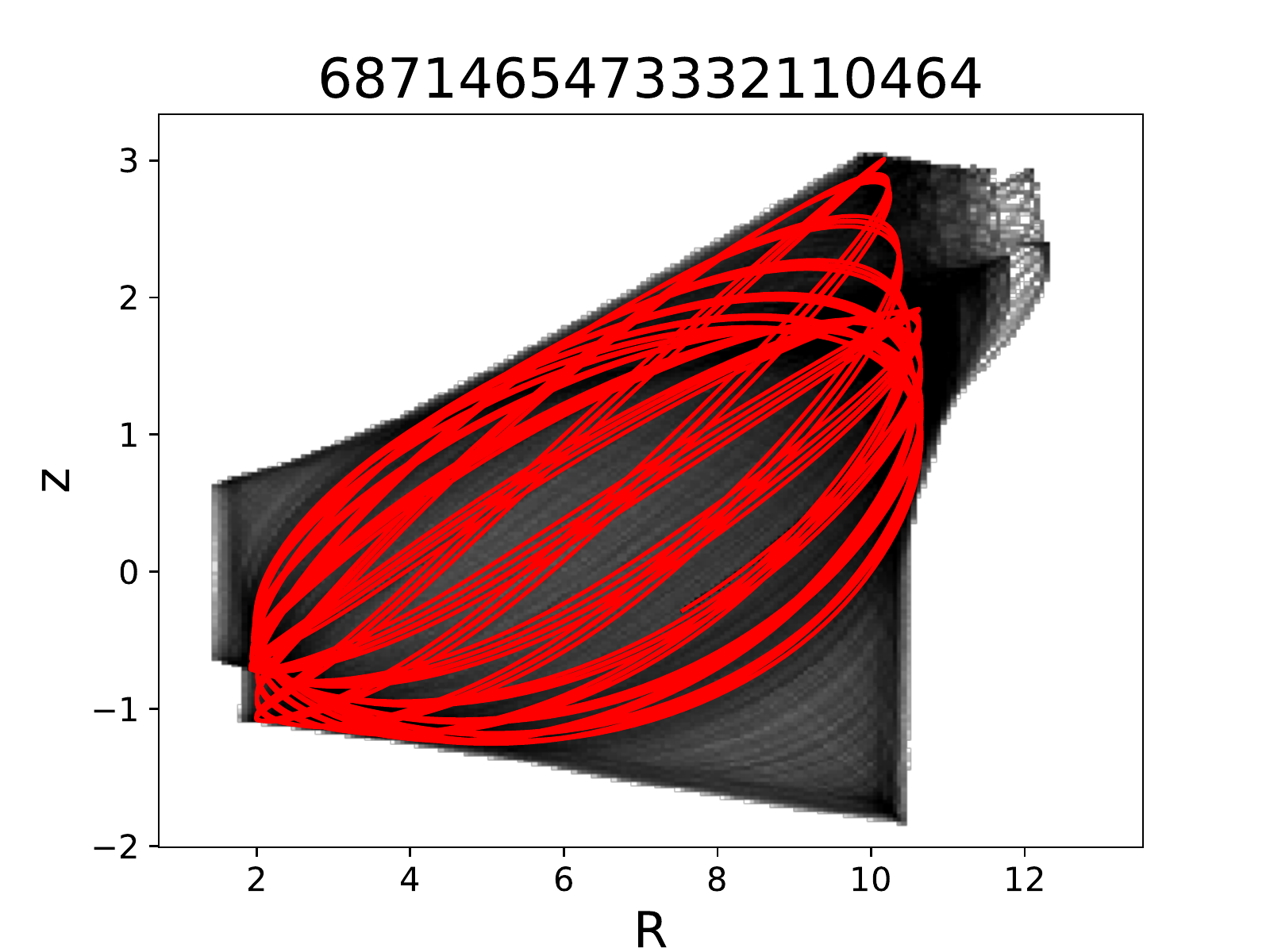}
\includegraphics[clip=true, trim = {0cm 0cm 1cm 0cm},width=4.5cm]{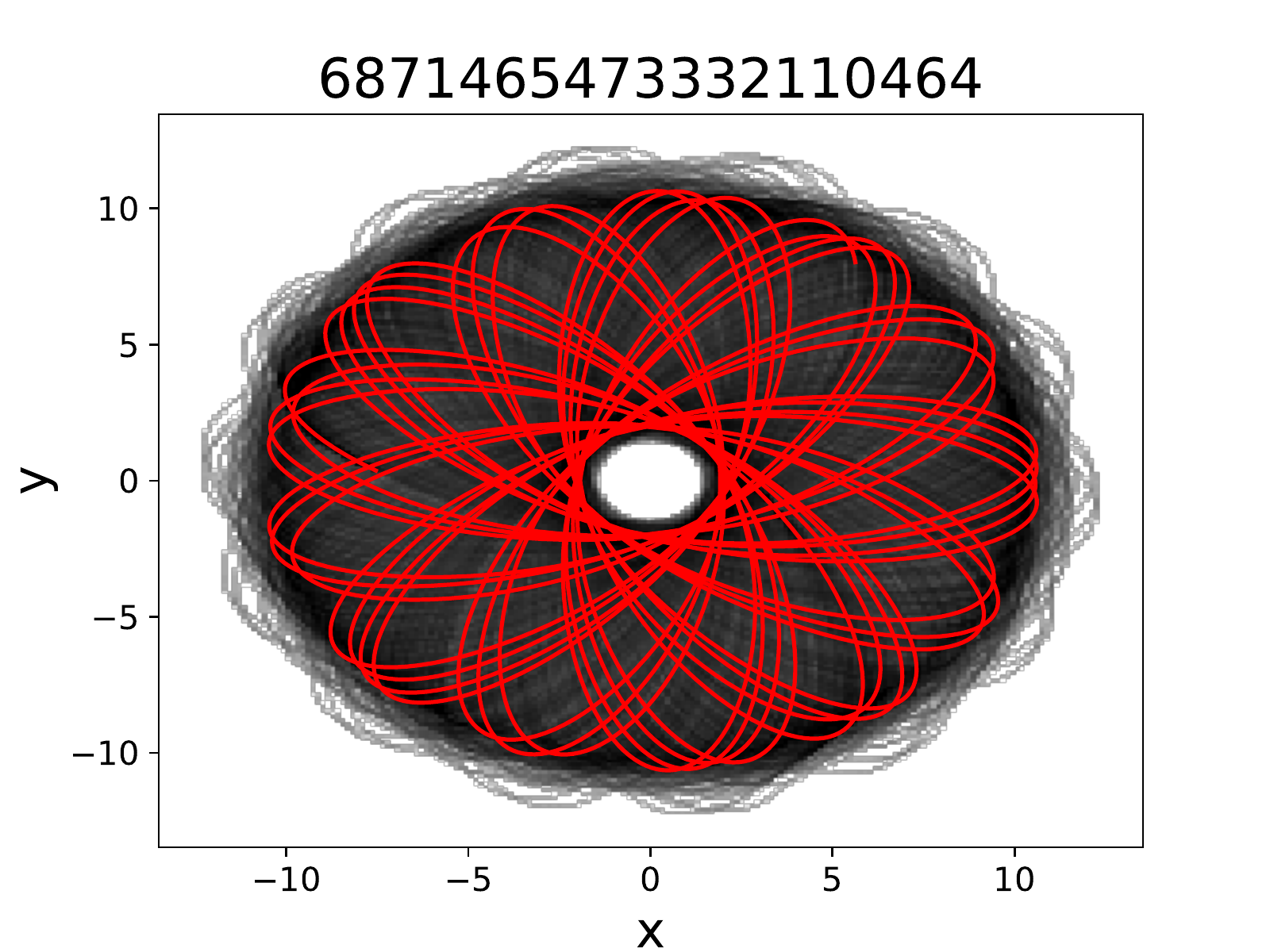}
\caption{Continued from Fig.~\ref{T1}.}
\end{centering}
\end{figure*}

\end{document}